\documentclass[12pt,a4paper,twoside]{book}
\usepackage[pdftex]{graphicx}
\usepackage{url}
\usepackage{color}
\definecolor{astral}{RGB}{46,116,181}
\usepackage[linktocpage=true,colorlinks,linkcolor=astral,citecolor=astral,urlcolor=astral]{hyperref}

\usepackage{amsmath,amsfonts,amsthm,amssymb,mathtools}
\usepackage{dsfont}
\usepackage{bm}
\usepackage{braket}

\usepackage[applemac]{inputenc}
\usepackage{authblk}

\usepackage{tabulary}
\usepackage{setspace}

\usepackage{braket}
\usepackage{lipsum}

\usepackage{sectsty}\chapterfont{\centering}

\usepackage[pagestyles, clearempty]{titlesec}

\usepackage[numbers,square,compress]{natbib}
\usepackage{notoccite}

\usepackage{geometry}

\newgeometry{
    top=2.2cm,
    bottom=2.2cm,
    outer=2.2cm,
    inner=3.5cm,
}

\subsectionfont{\color{black}}
\sectionfont{\color{black}}

\usepackage{fancyhdr}
\titleformat{\chapter}[display]
  {\bfseries\Large}
  {\filleft\Large{\color{black}\chaptertitlename~\thechapter}}
  {2ex}
  {\rule{\textwidth}{0.05cm} \vspace{1ex}\filright}
  [\vspace{0.05ex}\rule{\textwidth}{0.05cm}]

\usepackage[font={small}]{caption}

\usepackage{titletoc}
\titlecontents{chapter}
  [1.0pt]
  {\addvspace{0.5em}\bfseries}
  {\chaptername\ \thecontentslabel:\;}
  {}
  {\mdseries\titlerule*[0.75em]{.}\bfseries\contentspage}
\titlecontents{section}
  [3.5em]
  {\contentslabel{2em}\quad}
  {}
  {}
  {\mdseries\titlerule*[0.75em]{.}\bfseries\contentspage}
\titlecontents{subsection}
  [7.0em]
  {\contentslabel{4em}}
  {}
  {}
  {\mdseries\titlerule*[0.75em]{.}\bfseries\contentspage}

\renewcommand{\thechapter}{\thepart.\arabic{chapter}}

\newcommand*{\defeq}{\mathrel{\vcenter{\baselineskip0.5ex\lineskiplimit0pt\hbox{\scriptsize.}\hbox{\scriptsize.}}}=}

\usepackage{tocbasic}
\DeclareTOCStyleEntry[dynnumwidth]{tocline}{figure}

\newcommand{\ee}{\mathrm{e}}
\newcommand{\ii}{\mathrm{i}}
\newcommand{\dd}{\mathrm{d}}
\newcommand{\Tr}{\mathrm{Tr}}
\renewcommand{\Im}{\,\mathrm{Im}}

\counterwithin{chapter}{part}

\usepackage{lipsum}

\begin{document}

\frontmatter

\newgeometry{top=2.5cm,bottom=2.5cm,right=2.5cm,left=2.5cm}
\begin{titlepage}
  \begin{center}    
    \includegraphics[width=0.6\textwidth]{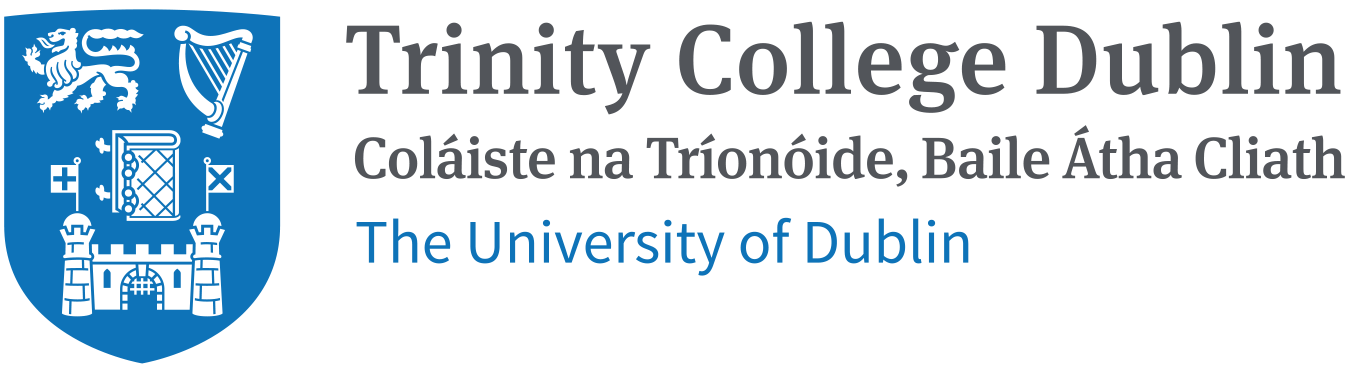} \hfill

    \vspace{4.5cm}
    \begin{spacing}{1.05}
    \huge\bfseries Thermodynamics of interacting many-body quantum systems
    \end{spacing}
  \end{center}
  
  \vspace{1.5cm}

  \vfill

  \centering
  
  {\large Marlon E. \textsc{Brenes Navarro}}\\
  {\large School of Physics \par}
  
  \vspace{2cm}

  A thesis submitted in partial fulfilment of the requirements towards the degree of Doctor of Philosophy in Physics\\
  
  \vspace{2cm}

  {\large
    Hilary Term 2022
  }
\end{titlepage}
 \restoregeometry

\newpage\null\thispagestyle{empty}\newpage

{\setstretch{1.0}
\section*{\Huge{Declaration}}
\setcounter{page}{1}
\vspace{1cm}
I declare that this thesis has not been submitted as an exercise for a degree at this or any other university and it is entirely my own work.

\vspace{1cm}

\noindent
I agree to deposit this thesis in the University's open access institutional repository or allow the library to do so on my behalf, subject to Irish Copyright Legislation and Trinity College Library conditions of use and acknowledgement.

\vspace{3cm}

Signed:~\rule{5cm}{0.3pt}\hfill Date:~\rule{5cm}{0.3pt}
}

\newpage\null\thispagestyle{empty}\newpage

\newpage
\begin{singlespace}
{\centering \textbf{\large Thermodynamics of interacting many-body quantum systems}\\
\vspace{0.2cm}
Marlon Brenes\\
Hilary Term 2022\\
\vspace{0.2cm}
\textbf{Abstract}\\
}

\vspace{0.5cm}

Technological and scientific advances have given rise to an era in which coherent quantum-mechanical phenomena can be probed and experimentally-realised over unprecedented timescales in condensed matter physics. In turn, scientific interest in non-equilibrium dynamics and irreversibility signatures of thermodynamics, such as transport, has taken place in recent decades, particularly in relation to cold-atom platforms and thermoelectric devices. Furthermore, the role of non-linear interactions in quantum thermal machines, whether a hindrance or a resource, has yet to be fully understood particularly in the finite-temperature regime. Diverse numerical and analytical approaches have come to fruition recently, designed to target these problems in certain regimes regulated by microscopic parameters.

This thesis is divided in two parts. Part~\ref{part:one} is devoted to the study of spin/particle transport in strongly correlated systems in the regime of linear response and to the topic of thermalisation. We begin by addressing the role of integrability and its consequences related to transport, which we then use in the context of the {\em single impurity} model, where an integrable model on a one-dimensional lattice is perturbed by an impurity around the centre of the chain. Exhibiting the signatures of quantum chaos, we motivate our work by questioning the nature of transport in this model. We find that despite its chaotic signatures, transport remains ballistic as in the unperturbed model. This result brings us to the question of thermalisation, a topic which is elegantly explained in the context of the eigenstate thermalisation hypothesis (ETH). The ETH postulates that an energy eigenstate encodes the equilibrium ensemble properties in sufficiently complex systems and that local observables in systems initially kept away from equilibrium will eventually thermalise under unitary evolution. Using this framework we find that thermalisation in the single impurity model is anomalous, and the statistical properties of the unperturbed model end up embedded in the perturbed model. We then proceed to investigate the consequences of eigenstate thermalisation in the multipartite entanglement structure of the eigenstates in chaotic Hamiltonians through the quantum Fisher information. We find that the quantum Fisher information can be used to discriminate a pure eigenstate ensemble from a true thermal state. Finally, we address the statistical correlations between matrix elements of local observables in the energy eigenbasis, and provide a connection between these correlations and the timescales of late-time chaos from the out-of-time-order correlators.

In Part~\ref{part:two} we delve into the theory of open quantum systems, particularly in configurations whereby an interacting quantum system is kept out of equilibrium by the action of thermal reservoirs. We begin studying {\em boundary-driven} systems, in which a many-body quantum system is driven out of equilibrium be inducing and removing excitations from the boundaries. This treatment allows us to solidify our results in Part~\ref{part:one}. We criticise boundary-driven configurations from the thermodynamic perspective and argue that such a procedure can only be used to evaluate infinite-temperature properties. Motivated by this fact, we then propose a novel methodology to tractably address finite-temperature transport and thermodynamics in many-body quantum systems, in the context of autonomous thermal machines, overcoming the limitations of boundary-driven configurations.

\end{singlespace}

\newpage\null\thispagestyle{empty}\newpage

{\setstretch{1.0}
{\centering \textbf{\large List of publications}\\
\vspace{0.5cm}
}
\vspace{0.5cm}
\noindent
The results exposed in this thesis are based, primarily, on the following publications:
\begin{enumerate}
\item{\textbf{M. Brenes}, E. Mascarenhas, M. Rigol, and J. Goold,\\ {\em High-temperature coherent transport in the XXZ chain in the presence of an impurity},\\ \href{https://journals.aps.org/prb/abstract/10.1103/PhysRevB.98.235128}{Phys. Rev. B \textbf{98}, 235128 (2018)}}
\item{\textbf{M. Brenes}, S. Pappalardi, J. Goold and A. Silva,\\ {\em Multipartite entanglement structure in the eigenstate thermalization hypothesis},\\ \href{https://journals.aps.org/prl/abstract/10.1103/PhysRevLett.124.040605}{Phys. Rev. Lett. \textbf{124}, 040605 (2020)}}
\item{\textbf{M. Brenes}, T. LeBlond, J. Goold and M. Rigol,\\ {\em Eigenstate thermalization in a locally perturbed integrable system},\\ \href{https://journals.aps.org/prl/abstract/10.1103/PhysRevLett.125.070605}{Phys. Rev. Lett. \textbf{127}, 070605 (2020)}}
\item{\textbf{M. Brenes}, J. J. Mendoza-Arenas, A. Purkayastha, M. T. Mitchison, S. R. Clark and J. Goold,\\ {\em Tensor-network method to simulate strongly interacting quantum thermal machines},\\ \href{https://journals.aps.org/prx/abstract/10.1103/PhysRevX.10.031040}{Phys. Rev. X \textbf{10}, 031040 (2020)}}
\item{\textbf{M. Brenes}, J. Goold and M. Rigol,\\ {\em Low-frequency behavior of off-diagonal matrix elements in the integrable XXZ chain and in a locally perturbed quantum-chaotic XXZ chain},\\ \href{https://journals.aps.org/prb/abstract/10.1103/PhysRevB.102.075127}{Phys. Rev. B \textbf{102}, 075127 (2020)}}
\item{\textbf{M. Brenes}, S. Pappalardi, M. T. Mitchison, J. Goold and A. Silva,\\ {\em Out-of-time-order correlations and the fine structure of eigenstate thermalization},\\ \href{https://journals.aps.org/pre/abstract/10.1103/PhysRevE.104.034120}{Phys. Rev. E \textbf{104}, 034120 (2021)}}
\end{enumerate}
\noindent
The following are other publications of the author which, although relevant to a certain extent to this thesis and completed during the doctoral training, are not covered in detail in the following document:
\begin{enumerate}
	\item{\textbf{M. Brenes}, M. Dalmonte, M. Heyl and A. Scardicchio,\\ {\em Many-body localization dynamics from gauge invariance},\\ \href{https://journals.aps.org/prl/abstract/10.1103/PhysRevLett.120.030601}{Phys. Rev. Lett. \textbf{120}, 030601 (2018)}}
	\item{\textbf{M. Brenes}, V. K. Varma, A. Scardicchio, I. Girotto,\\ {\em Massively parallel implementation and approaches to simulate quantum dynamics using Krylov subspace techniques},\\ \href{https://www.sciencedirect.com/science/article/pii/S0010465518303060}{Comput. Phys. Commun. \textbf{235}, 477-488 (2019)}}
	\item{M. T. Mitchison, A. Purkayastha, \textbf{M. Brenes}, A. Silva and J. Goold,\\ {\em Taking the temperature of a pure quantum state}\\ \href{https://arxiv.org/abs/2103.16601}{(2021), arXiv:2103.16601 [quant-ph]}}
\end{enumerate}
}

{\setstretch{1.0}
\small
\tableofcontents
}

{\setstretch{1.0}
\small
\listoffigures
}

\newpage\null\thispagestyle{empty}\newpage

\mainmatter

\part{Isolated quantum systems}
\label{part:one}

\chapter{Introduction}
\label{chapter:intro_1}

Recent progress has opened up an era in which quantum phenomena can be experimentally observed and controlled on a mesoscopic scale~\cite{Bertini:2021,Eisert:2015,Jepsen:2020,Linke2018}. These developments have stimulated interest in the thermodynamics of quantum systems~\cite{Goold:2016,Benenti2017,Binder2018,Mitchison2019}. Significantly, these studies provide the understanding of the effect of noise on quantum devices, as it could suppress or enact desirable quantum behaviour. Even though the emergence of large-scale thermodynamics has been placed on a firm theoretical and experimental footing, many open questions remain. Further research is needed to determine how the interplay between interactions, external noise and disorder gives rise to observable signatures of irreversibility, e.g., transport. Such questions are not only fundamental for our understanding of many-body physics, but also have applications in the design of tailored quantum matter that could enhance the functionality of future computers and energy-conversion devices~~\cite{Benenti2017}. Among the latter, thermoelectric devices which convert heat into electrical energy highlight the importance to study systems in which this conversion may be favourable~\cite{Benenti2017}. Furthermore, recent advances in noisy intermediate-scale quantum devices~\cite{Preskill2018}, whose backbone constituents are entangled and interacting many-particle systems, call for the need of theoretical understanding of the role of interactions at the fundamental level. 

Isolated quantum systems initially brought away from equilibrium will relax under the underlying microscopic dynamics, in general, through the transport of the conserved quantities dictated by the conservation laws~\cite{Benenti:2009,zotos1997transport}. These conservation laws allow one to make a clear distinction between two different classes of quantum systems in isolated environments. The first class are the ones for which only macroscopic and extensive quantities are conserved, such as energy and number of particles. These quantum systems are usually dubbed {\em generic, chaotic and/or non-integrable}. Some quantum systems, however, present an extensive set of microscopic non-trivial {\em local} conserved quantities which strongly affect how equilibrium is attained. This second class of quantum systems is known as {\em integrable}. Macroscopic irreversibility is manifest by the mechanism under which a quantum system reaches equilibration, through the spread of quantum correlations and transport of conserved quantities. Most interestingly, the degree of control now achievable over devices and experiments at the quantum level allow the exploration of quantum systems that are tuned between the integrable and non-integrable regimes, by controlling microscopic parameters within the system~\cite{Bentsen2019,Jepsen:2020}.

Understanding out-of-equilibration phenomena brings us to the question of transport in quantum systems, which, in spite of theoretical and experimental advances~\cite{Bertini:2021}, still presents challenges particularly in the finite-temperature regime. One-dimensional quantum systems are typically used as prototypes to unravel and to understand these phenomena, which are usually modelled by either spin chains or particles hopping on a lattice. Within these models, spin (particle) currents are the quantities of interest and define regimes of conductivity such as insulators, regular conductors or superconductors~\cite{ShastryKubo2008,RigolShastry2008}. Non-interacting systems typically allow for simple solutions and within the regime of linear response~\cite{Pottier:2010} transport of conserved quantities, such as particle number and energy, is {\em ballistic}, a regime also known as {\em coherent}, which entails currents that do not decay as the size of the system is increased. A well-known exception happens when disorder is modelled in lattice systems, leading to an insulating regime known as Anderson localisation~\cite{Anderson:1958}. 

The introduction of strong interactions into these models gives rise to a rich spectrum in the transport properties. Strong interactions usually lead to complex behaviour and transport is dictated by the pivotal role of integrability~\cite{zotos1997transport}. The archetypical model is the anisotropic Heisenberg (XXZ) model, in which the competition between {\em coherent} and {\em incoherent} effects lead to the aforementioned richness at the level of transport. Even though the model rose as a simple way to describe ferromagnetism in simple materials~\cite{Coleman:2015}, advances in ultracold atoms~\cite{Jepsen:2020} and magnetic materials~\cite{Hess:2019} now allow to directly simulate the microscopic Hamiltonian of the Heisenberg model with an impressive degree of tunability. Transport of conserved quantities and thermodynamics are intricately-related concepts in linear response, through the Onsager relations~\cite{Pottier:2010}. 
   
Integrability in quantum systems is known to be susceptible to perturbations. Given that integrability plays a role in the nature of transport and hence, on the thermodynamics, we are motivated by the following question:
\begin{itemize}
\item{Is the nature of integrability-breaking perturbations relevant to transport?}
\end{itemize}
This simple question is the driving force of Part~\ref{part:one} and from which all of our results and developments follow. 

The role of integrability with respect to linear response transport is clear. While integrable systems display rich transport properties which may result from different microscopic parameters and initial conditions, non-integrable systems contain a high degree of complexity which typically leads to normal diffusion of conserved quantities, described by, for example, Fick's law for particle transport. This universal behaviour is expected to hold as a long as integrability is broken for a given system. 

From this perspective, one may question if the nature of the integrability-breaking perturbation plays a role. For instance, the anisotropic Heisenberg model perturbed by a {\em single magnetic impurity} located around the centre of the spin chain is known to lead to non-integrable signatures~\cite{Santos2020,XotosIncoherentSIXXZ,Pandey:2020}. We are motivated to answer if indeed, the breaking of integrability induced by such a simple perturbation is enough to render a perfect conductor (ballistic) to a normal conductor (diffusive).

We start by describing integrability, chaos and transport in Chapter~\ref{chapter:integrability}. We then proceed to introduce the microscopic models in Chapter~\ref{chapter:models}, with emphasis on global symmetries, continuity equations, expressions for spin currents and a brief survey about experimental realisations. We then proceed to tackle the question of spin transport in the single impurity model in Chapter~\ref{chapter:kubo}, in the regime of linear response. We find that a non-trivial treatment of the conductivity at finite frequencies needs to be carried out to address transport in this system and that, although the model displays the signatures associated to non-integrability, the single magnetic perturbation is insufficient to render a normal diffusive conductor from an unperturbed ballistic model.

This result leads us to the question of thermalisation.

The topic of how macroscopic irreversible behaviour comes into place from the reversible dynamics of the microscopic constituents has been a topic of debate since the inception of statistical mechanics~\cite{Boltzmann1872}. Motivated by advances in experimental realisations, however, a renovated interest in fundamental questions about thermalisation has taken place in recent decades~\cite{Eisert:2015,Polkovnikov2011,Gogolin:2016,Alessio:2016}. An ubiquitous phenomenon with a high degree of universality in sufficiently-complex many-body systems is their tendency to reach thermal equilibrium, in which, symmetries and conservation laws play a pivotal role. Local observables in generic quantum systems which are initially engineered to be away from equilibrium typically equilibrate in the limit of long-times. Moreover, the equilibration value attained at long times coincides with the expectation value evaluated in the ensembles of statistical mechanics, a phenomenon known as {\em thermalisation} which is nowadays understood from the perspective of the {\em eigenstate thermalisation hypothesis} (ETH). The hallmark of this process is that the equilibrium and thermal values do not depend on the initial conditions as long as the energy distribution of the initial state has a well-defined average with a variance that decays as the number of degrees of freedom is increased~\cite{Alessio:2016}. Oblivious to the memory of initial conditions, the dynamics that satisfy the above conditions yield true {\em ergodic} behaviour. Integrable systems on the other hand, do not follow this prescription, their extensive set of non-trivial local conserved quantities preventing them from thermalise in the sense described before. Instead, if equilibration is attained, it is the generalised Gibbs ensemble which describes such equilibration~\cite{vidmar2016}. A known exception to the thermalisation process for generic systems is the one dictated by the dynamics of an interacting system which is perturbed by sufficiently strong disorder. Such systems display a so-called many-body localisation transition, over which ergodicity gets broken~\cite{Basko:2006}. 

Thermalisation in quantum mechanics is typically associated to systems with a certain degree of complexity, in which hydrodynamic behaviour is expected to prevail. In this context, hydrodynamic behaviour refers to transport phenomena that can be described using diffusion equations. Bringing our attention back to the single impurity model, we are interested to investigate if thermalisation is achieved in the sense described above. In Chapter~\ref{chapter:eth} we introduce the fundamental aspects of the eigenstate thermalisation hypothesis, to then evaluate the peculiar occurrence of thermalisation for the single impurity model. We find that the eigenstate thermalisation hypothesis is fully consistent and, moreover, local observables away from the impurity perturbation thermalise to the statistical predictions of the {\em unperturbed XXZ model}. Remarkably, even the total spin current is consistent with this anomalous thermalisation and the model displays both the signatures of ergodicity and {\em coherent} transport.

Establishing these results from the perspective of eigenstate thermalisation brings us to question further details about the fundamental aspects of thermalising systems. In Chapter~\ref{chapter:fine_eth} we introduce advanced topics related to eigenstate thermalisation, a collection of results that we dubbed {\em fine} structure of eigenstate thermalisation. We begin by questioning the entanglement structure of the eigenstates of ergodic systems that satisfy eigenstate thermalisation, to then move to higher order correlation functions and the role of matrix-element correlations in their dynamics.

The ETH poses that local expectation values and two-point correlation functions, the latter of which are most relevant to noise and response functions in linear response, are indistinguishable from their finite-temperature counterparts. The entanglement structure, however, from true thermal ensembles and the corresponding eigenstates to which local measurements thermalise, is in stark contrast. In particular, in Sec.~\ref{sec:entanglement_eth}, we show that that the multipartite entanglement structure in the ETH can be connected to response functions in linear response through the quantum Fisher information~\cite{Pezze:2018}. This observation will allow us to create a hierarchy of the multipartite entanglement structure among different ensembles, including the aforementioned ensembles described by single eigenstates in the context of eigenstate thermalisation.

Finally, our last topic in Part~\ref{part:one} is presented in Sec.~\ref{sec:otoc_eth}. Out-of-time-order correlators (OTOCs) have been introduced to provide perspective about chaotic behaviour from the point of view of {\em information scrambling}~\cite{swingle2018}. OTOCs present a nested time structure, which detects quantum chaos and correlations beyond thermal ones. The ETH imposes a condition on the matrix elements of local observables in the eigenbasis of the Hamiltonian. Crucially, the off-diagonal matrix elements contain a random component. The statistical correlations of the probability distributions of these random variables has been the topic of interest in recent works~\cite{Murthy19,kurchan}. In particular, it has been shown that there exists an energy scale that divides a regime in which statistical correlations are very low, giving rise to random-matrix behaviour from another in which statistical correlations are prevalent~\cite{Richter2020}. In Sec.~\ref{sec:otoc_eth} we provide a thorough study of these statistical correlations, and expose how they are connected to the timescales of late-time chaos from the perspective of the dynamics of high order correlation functions. 
\chapter{Integrability and chaos: Transport}
\label{chapter:integrability}

An interesting and recurring question in the theory of statistical mechanics is: How does hydrodynamic behaviour emerge from the microscopic dynamics and the underlying quantum-mechanical laws? Hydrodynamic behaviour, in this context, refers to transport phenomena that can be described using diffusion equations.

Both in classical physics and in the quantum domain, there are systems for which Hamiltonian dynamics do not lead hydrodynamic behaviour. Such is the case when {\em conservation laws} are at play~\cite{Sethna:2020,zotos1997transport,benenti2013conservation}. In classical physics, hydrodynamic behaviour emerges naturally from complexity. 

Over the past two decades, interest in the dynamics and transport within isolated quantum systems has received a renovated interest. Experimental advances in many-body quantum systems have now led to the observation of quantum-mechanical effects up to unprecedented timescales, long before any decoherent effects from the environment become important. Strides in ultra-cold atom experiments~\cite{Greiner:2002, Kinoshita:2006, Trotzky:2012, Kaufman2016, tang2018thermalization}, in which unitary dynamics dictate quantum effects, have paved the way to new lines of research at the level of thermalisation and transport~\cite{Zelevinsky1996, Polkovnikov:2011, Yukalov:2011, Eisert:2015, Goold:2016, Gogolin:2016, Borgonovi:2016, Alessio:2016}. 

The present chapter introduces the notion of hydrodynamics in classical systems in Sec.~\ref{sec:hydro_classical}, followed by a brief overview of its quantum-mechanical counterpart in Sec.~\ref{sec:hydro_quantum} and the consequences of integrability. Sec.~\ref{sec:thermo_linear_response} then discusses thermodynamics at the level of linear response, attaching the concepts of Sec.~\ref{sec:hydro_classical} and Sec.~\ref{sec:hydro_quantum} into a more complete overview.

\section{Emergence of hydrodynamics in classical systems}
\label{sec:hydro_classical}

The theory of random walks provides an approach to understand the emergence of diffusion in classical systems~\cite{Sethna:2020}.

Consider the probability density $\rho(x, t)$, which we will use to describe the probability of a particle to be located in a given point in space and time\footnote{For this introductory derivation, we consider non-interacting particles. In such case, the probability distribution of one particle can be used to describe an entire ensemble of particles.}.  For simplicity, we consider the one-dimensional problem. 

The starting point is to consider a particle as it moves along a trajectory described by an ensemble of uncorrelated random walks. At each time step $\Delta t$, the particle changes its position $x(t)$ by a single step $l$ in either direction,
\begin{align}
x(t + \Delta t) = x(t) + l(t),
\end{align}  
Since the motion of the particle is described by an ensemble of random walks, we need to introduce a probability distribution $\psi(l)$ that describes the random variable $l(t)$. $\psi(l)$ is a continuous probability distribution, for which we will fix its mean to be zero,
\begin{align}
\int \textrm{d}z \psi(z) \cdot z = 0, 
\end{align}
and its variance to be
\begin{align}
\int \textrm{d}z \psi(z) \cdot z^2 = a^2. 
\end{align}
The question we would like to answer now is: can we obtain a solution for the probability density $\rho(x, t + \Delta t)$, given $\rho(x^{\prime}, t)$? The particle moves from $x^{\prime}$ at time $t$ to $x$ at time $t + \Delta t$, so the step $l(t) = x - x^{\prime}$ occurs with a probability $\psi(x - x^{\prime})$ times the probability density $\rho(x^{\prime}, t)$. If we integrate over all initial positions $x^{\prime}$, we find
\begin{align}
\rho(x, t + \Delta t) = \int_{-\infty}^{+\infty} \textrm{d}x^{\prime} \rho(x^{\prime}, t) \psi(x - x^{\prime})
= \int_{-\infty}^{+\infty} \textrm{d}z \rho(x - z, t) \psi(z).
\end{align}
For small step sizes in the length-scales of $\rho$, we may express $\rho(x - z, t)$ as a Taylor expansion around $x$ for small values of $z$, to obtain 
\begin{align}
\rho(x, t + \Delta t) &\approx \int \textrm{d}z \left[ \rho(x, t) - z\frac{\partial \rho(x,t)}{\partial x} + \frac{z^2}{2}\frac{\partial^2 \rho(x,t)}{\partial x^2} \right] \psi(z) \\
&= \rho(x, t) + \frac{a^2}{2} \frac{\partial^2 \rho(x,t)}{\partial x^2}.
\end{align}
We now assume that $\rho(x,t)$ changes slowly during the time-step, so that we could approximate $\rho(x, t + \Delta t) - \rho(x,t) \approx (\partial \rho(x,t) / \partial t) \Delta t$ to obtain
\begin{align}
\frac{ \partial \rho(x,t)}{\partial t} = \frac{a^2}{2\Delta t}\frac{\partial^2 \rho(x,t)}{\partial x^2},
\end{align}
which corresponds to the diffusion equation with $D = a^2 / (2 \Delta t)$. In such a way, hydrodynamic behaviour naturally emerges from this stochastic process\footnote{For the more general case of interacting particles, one could envisage a {\em correlated} random walk, as opposed to the {\em uncorrelated} version we have employed. It would then be natural to assume that hydrodynamic diffusion, or variations of it, could stem from the nature of these correlations.}.

It is now interesting to think about particle currents. Since the particles are conserved through a section in space, we can use a continuity equation to describe the current flowing through an element $\Delta x$, which can be written down as 
\begin{align}
\frac{\partial \rho(x,t)}{\partial t} &= - \frac{\partial J_1}{\partial x} \\
\label{eq:diffusion}
\implies J_1 &= -D \frac{\partial \rho(x,t)}{\partial x}.
\end{align} 
The last equation corresponds to the {\em linear response regime}, in which the current is directly proportional to a gradient of the density. 

It is remarkable that such a simple stochastic treatment allows one to understand the emergence of hydrodynamic behaviour in classical systems. 

One could question the validity of the model employed here. After all, the dynamics of particles in classical systems are governed by equations of motion with deterministic variables, depending only on initial conditions. However, one can argue that these stochastic processes simulate systems with a certain degree of {\em complexity}. Alternatively, one could think that the stochastic motion of a single particle is the effective result from the elastic collisions with other particles in a closed system with a high degree of complexity. 

\section{Hydrodynamics in quantum systems}
\label{sec:hydro_quantum}

In the quantum regime, understanding how hydrodynamic behaviour emerges is far more complicated. Particularly, the theory of quantum mechanics is best understood in terms of operators and states that live in Hilbert space, and not as much in terms of trajectories (or phase space) due to the undeterministic nature of the wave function. 

The consensus is that, for hydrodynamic behaviour to emerge, one needs non-linear interactions which lead to chaos and, hence, to transport properties that resemble those found in hydrodynamics. Understanding how this complexity emerges brings us to the realm of quantum chaos~\cite{lepri2003thermal, dhar2008heat, chen2014nonintegrability,haake2013quantum,chirikov1997linear}. 

The chaotic behaviour in quantum systems is observed in systems with a certain degree of complexity and in the presence of non-linear interactions. In the discussion of quantum chaotic systems, a categorisation that is typically employed divides quantum systems into two different types: {\em integrable} and {\em non-integrable}.

The concept of integrability is central in the study of transport~\cite{zotos1997transport} and, hence, in the identification of hydrodynamic behaviour. In the most general sense, an integrable system is one for which an extensive set of {\em local} conserved quantities can be identified for the system. For a one-dimensional system embedded on a lattice with a discrete number of sites\footnote{Composed of a finite or countably infinite number of sites.}, a set of conservation laws can be represented by local operators, denoted by $\hat{Q}_n$. The conserved quantities $\hat{Q}_n$ are of the form
\begin{align}
\hat{Q}_n = \sum_{i = 1}^{L} \hat{q}^{n}_{i},
\end{align}
where $q^{n}_{i}$ are local operators involving $n$ sites around site $i$, on a lattice of length $L$. The $\hat{Q}_n$ operators are conserved quantities if
\begin{align}
\begin{cases}
\left[ \hat{Q}_n, \hat{Q}_m \right] &= 0 \quad \forall \, n \neq m, \\
\left[ \hat{Q}_n, \hat{H} \right] &= 0,
\end{cases}
\end{align}
where $[ \cdot, \cdot ]$ denotes the quantum-mechanical commutator and $\hat{H}$ is the total Hamiltonian of the system under investigation.

The concept of hydrodynamics in quantum systems can be understood from these objects. In particular, {\em Mazur's inequality}~\cite{Mazur:1969}, was proposed to make a connection between the existence of these conserved quantities and transport in quantum systems. The inequality states that
\begin{align}
\label{eq:mazur}
\lim_{\tau \to \infty} \frac{1}{\tau}\int_{0}^{\tau} \textrm{d}t \langle \hat{J}(t) \hat{J}(0) \rangle \geq \sum_{n} \frac{\langle \hat{J}(0) \hat{Q}_n \rangle^2}{\langle \hat{Q}_n^2 \rangle},
\end{align}
where $\langle \cdot \rangle$ denotes a thermodynamic average and the sum is taken over a subset of conserved quantities $\hat{Q}_n$ which are orthogonal to each other, i.e., $\langle \hat{Q}_n \hat{Q}_m \rangle = \delta_{n,m} \langle \hat{Q}_n^2 \rangle$. $\hat{J}(t)$ is the current operator, which may refer to thermal or particle transport and it is written in the Heisenberg picture: $\hat{J}(t) = \hat{U}^{\dagger} \hat{J} \hat{U} = e^{\textrm{i}\hat{H}t} \hat{J} e^{-\textrm{i}\hat{H}t}$, satisfying $\langle \hat{J}(t) \rangle = 0$ and $\hat{J}^{\dagger} = \hat{J}$.

If one is interested in the transport regime of a given system, one could investigate the quantum-mechanical operator of the current $\hat{J}$ through Mazur's inequality. In this scenario, the behaviour one is interested in relates to the long-time decay-value of the correlation function $\langle \hat{J}(t) \hat{J}(0) \rangle$.

Evaluating the right-hand side of Eq.~\eqref{eq:mazur} is, in general, a formidable task. Not only does it involve the construction of the operators representing the conserved quantities $\hat{Q}_n$, but the evaluation of the overlap with a given operator of interest. If achieved, however, the statements one can guarantee about a system with a given microscopic Hamiltonian description are indeed very powerful. It suffices to find that the overlap between $\hat{J}(0)$ and {\em one} of the $\hat{Q}_n$ is non-vanishing, to claim that in the long-time limit the correlation function $\langle \hat{J}(t) \hat{J}(0) \rangle$ does not decay to zero.

Therein lies the importance of integrability at the level of transport. The hallmark of integrable systems, which possess an extensive number of conserved quantities $\hat{Q}_n$, is the observation that correlation functions in time of the form $\langle \hat{J}(t) \hat{J}(0) \rangle$ do not decay to zero in the limit of infinite time. The quantum-mechanical conservation laws prevent the dynamics from ever decaying. This translates, according to Mazur's inequality, to transport properties associated to the {\em ballistic regime}, in which the expectation value of a given current operator $\hat{J}$ does not decay in the thermodynamic limit; i.e., $L \to \infty$. 

The consensus is that non-integrable systems behave in the opposite way. Since non-integrable systems possess no extensive set of conserved quantities, the dynamics of two-point correlation functions in time will, in general, decay to zero in the limit of infinite time. This implies that the transport properties associated to such systems can either behave according to hydrodynamics, i.e., following the diffusion equation [Eq.~\eqref{eq:diffusion} with respect to a given current $\hat{J}$]; or according to anomalous diffusion. Anomalous transport could be of different types, such as sub- or super-diffusion. Anomalous or regular diffusion can be understood from the mean-square displacement of an initially-localised perturbation as a function of time. As the perturbation propagates within the system, its mean-square displacement can be expressed as $\langle \Delta x^2 \rangle = 2Dt^{2\alpha}$ where $0 < \alpha \leq 1$~\cite{LiScaling2003}. Crucially, one can provide a connection between the mean-square displacement and the decay of the expectation value of the current operator as a function of the system size, $\langle \hat{J} \rangle \propto 1 / L^{\nu}$, where $\alpha = 1 / (1 + \nu)$~\cite{LiScaling2003}. From this perspective: Normal diffusion corresponds to $\nu = 1$, sub-diffusion to $\nu > 1$ and super-diffusion to $\nu < 1$. Perfect and non-decaying (ballistic) currents are characterised by $\nu = 0$. 

A quantum system that behaves hydrodynamically is one for which the expectation value of a given current operator $\hat{J}$ can be expressed to be directly proportional to the gradient of a driving field $\epsilon$~\cite{Prosen:2009},
\begin{align}
\label{eq:diffusion_q}
\langle \hat{J} \rangle = -\eta \nabla \epsilon.
\end{align}
On physical grounds, the decay of the correlation function $\langle \hat{J}(t) \hat{J}(0) \rangle$ in the limit of infinite time for non-integrable systems is expected for systems that display normal conduction, in which a perturbation propagates through the system and decays in time due to scattering or non-elastic interactions~\cite{Prosen:2009}. Such behaviour is described by the diffusion equation Eq.~\eqref{eq:diffusion_q}

\subsection{Indicators of integrability}

In many-body quantum systems, identifying the presence of conserved quantities (or lack thereof) that may be responsible for distinct transport regimes, following Marzur's inequality, is usually a very complicated task. It is then common to use diagnostic tools to try to identify if a system is integrable. The following are some of the most common diagnostics used for this purpose. 

\subsubsection{Level spacing statistics}
\label{sec:level_spacing_statistics}

\begin{figure}[t]
\fontsize{13}{10}\selectfont 
\centering
\includegraphics[width=0.6\columnwidth]{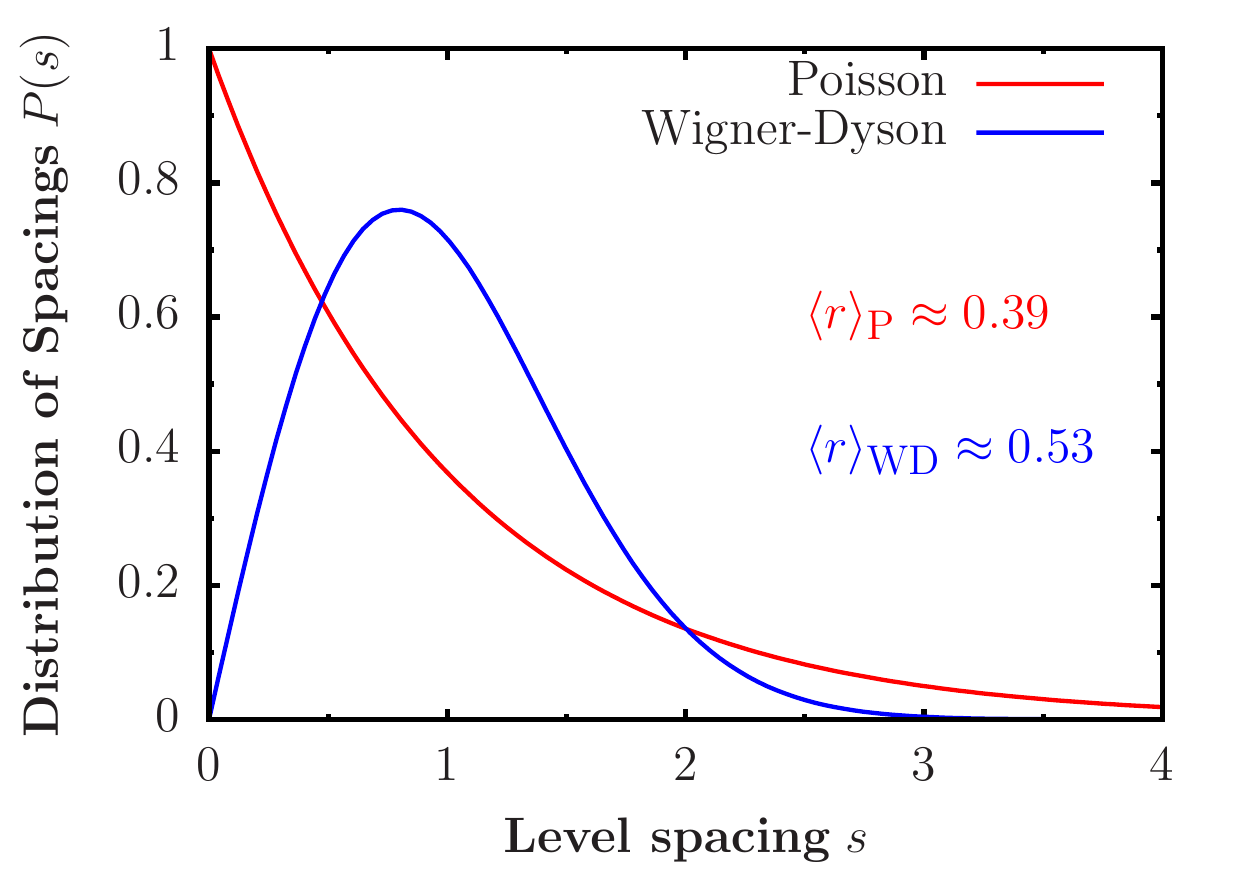}
\caption{Poisson and Wigner-Dyson level spacing distributions}
\label{fig:1.2.1}
\end{figure}

Spectral properties can be used as a diagnostic for integrability breaking. To investigate these, the objective is to target the solution of the eigenvalue problem corresponding to the time-independent Schr\"odinger equation ($\hbar \defeq 1$)
\begin{align}
\hat{H}\ket{\Psi} = E \ket{\Psi}.
\end{align}
Such a solution is equivalent to finding a rotation matrix $U$ that renders the Hamiltonian diagonal, $\tilde{H} = \hat{U}^{\dagger} \hat{H} \hat{U}$. $\tilde{H}$ is a diagonal matrix whose elements correspond to the $\mathcal{D}$ eigenvalues $\lambda_{\alpha}$, where $\mathcal{D}$ is the dimension of the Hilbert space.

The probability distribution $P(s_{\alpha})$ of spacings $s_{\alpha}$ of neighbouring energy levels shows different behaviour depending on whether a quantum system is chaotic or integrable~\cite{Alessio:2016}, where $s_{\alpha} = \lambda_{\alpha + 1} - \lambda_{\alpha}$, assuming the $\lambda_{\alpha}$ are sorted in ascending order.

For an integrable system, energy levels are expected to be independent from each other and crossings are not prohibited from occurring. This follows from the fact that local conserved quantities would typically translate into degenerate energy levels. Therefore, the statistics of the levels in this case is Poissonian,
\begin{align}
\label{eqn:poisson}
P(s) = e^{-s}.
\end{align}

On the other hand, a hallmark of quantum chaos is that energy levels repel each other and become correlated. As obtained from random matrix theory~\cite{Mehta:2004}, the level spacings of quantum chaotic systems with time-reversal invariance exhibit a Wigner-Dyson distribution given by 
\begin{align}
\label{eq:wddistro}
P(s) = \frac{\pi s}{2}e^{-\frac{\pi s^2}{4}}.
\end{align} 

These distributions are exposed in Fig.~\ref{fig:1.2.1}. In the Poissonian distribution, there exist a high probability to find neighbouring energy levels with a vanishing spacing between them, while the opposite is true for the Wigner-Dyson distribution. Asides from the distributions themselves, it is common to study a quantifier of the distribution. The distributions can be probed 
by studying the mean ratio of adjacent level spacings, defined as
\begin{align}
\braket{r} \defeq \frac{1}{M}\sum_{\alpha} \frac{\textrm{min}\{ s_{\alpha}, s_{\alpha + 1} \}}{\textrm{max}\{ s_{\alpha}, s_{\alpha + 1} \}},
\end{align}
where $M$ is the dimension of a subspace of the Hilbert space. It is common to restrict the Hilbert space to $M$, in order to avoid possible finite-size effects found close to the edges of the spectrum. Typically, however, $M \approx \mathcal{D}$, where $\mathcal{D}$ is the dimension of the Hilbert space. Poissonian distributions posses $\braket{r}_{\small\textrm{P}} \approx 0.39$, while for Wigner-Dyson distributions $\braket{r}_{\small\textrm{WD}} \approx 0.53$. The mean ratio of adjacent level spacings is useful to identify cross-over points, for which there might be a transition between the two distributions as a function of a free parameter of the Hamiltonian.

It is crucial to remark that these quantifiers or the distributions themselves can only be used as means of a diagnostic, since they only probe the local chaotic properties~\cite{Prosen:2013lss,Santos:2004, santos2011domain, torres2014local, XotosIncoherentSIXXZ,JuanThesis:2014,Huang2013}. 

\subsubsection{Spectral form factor}

Level spacing statistics, although widely used as a diagnostic, suffers from several shortcomings to fully characterise quantum chaos. To be useful in general applications, requires a procedure known as {\em spectral unfolding} and a clear distinction between symmetry sub-sectors of the model, given that the spectra of different symmetry sectors are uncorrelated~\cite{Santos:2010b, Santos:2010}. Furthermore, it naturally probes only the {\em local} characteristics of chaotic eigenstates.

To address a more complete picture of chaotic eigenstates and chaotic dynamics, the spectral form factors have been studied as yet a more reliable diagnostic~\cite{Mehta:2004}, first introduced in the context of high-energy, black-hole physics and Sachdev-Ye-Kitaev models~\cite{Cotler:2017,Papadodimas:2015}. In the limit of infinite temperature, the spectral form factor is defined as
\begin{align}
\left| Z(t) \right|^2 = \left| \textrm{Tr}[ e^{-\textrm{i}\hat{H}t} ] \right|^2 = \sum_{\alpha \beta} e^{\textrm{i} (\lambda_{\alpha} - \lambda_{\beta}) t}.
\end{align}
Most commonly, however, $\left| Z(t) \right|^2$ is normalised and averaged over different samples and energy regimes, in the form of $\langle \left| Z(t) \right|^2 \rangle$. In this context, $\langle \cdot \rangle$ denotes an average over different samples. For instance, evaluating $\left| Z(t) \right|^2$ for random matrices entails averaging over different samples of random matrices~\cite{Prakash:2021}. The sample-averaged spectral form factor as a function of time will display different signatures depending on the random matrix ensemble considered. In particular, the dynamics of the spectral form factor will depend on the probability distributions from which the elements of the random matrices are drawn~\cite{Prakash:2021}. 

The dynamics of $\langle \left| Z(t) \right|^2 \rangle$ computed within known ensembles could be used to compare against the ones obtained for a given physical system, to then conclude whether a system behaves according to chaotic predictions. Furthermore, the dynamics of the spectral from factor may serve to provide connection to transport by extracting the timescales relevant to hydrodynamics, such as the Thouless timescale~\cite{Alessio:2016,Suntajs:2020}.

\subsubsection{Adiabatic gauge potential}

Another more recent approach to diagnose quantum chaos was suggested by Pandey {\em et al.}~\cite{Pandey:2020}, based on the adiabatic eigenstate deformations. In this context, one considers a parameter-dependent Hamiltonian, $\hat{H}(\kappa)$, to study the adiabatic evolution of its eigenstates generated by the adiabatic gauge potential
\begin{align}
\mathcal{A}_{\kappa} \ket{\lambda(\kappa)} = \textrm{i} \partial_{\kappa} \ket{\lambda(\kappa)},
\end{align}
where the $\ket{\lambda(\kappa)}$ are the $\kappa$-dependent eigenstates of $\hat{H}(\kappa)$, i.e., $\hat{H}(\kappa) \ket{\lambda(\kappa)} = \lambda(\kappa) \ket{\lambda(\kappa)}$. 

The adiabatic gauge potential can be used as a diagnostic of integrability, by studying the scaling as a function of the system size of the $L_2$-norm of $\mathcal{A}_{\kappa}$, given by
\begin{align}
|| \mathcal{A}_{\kappa} ||^2 = \frac{1}{\mathcal{D}} \sum_{\alpha} \sum_{\alpha \neq \beta} | \braket{\lambda_{\alpha} | \mathcal{A}_{\kappa} | \lambda_{\beta}} |^2,
\end{align}
where $\mathcal{D}$ is the dimension of the Hilbert space and the $\braket{\lambda_{\alpha} | \mathcal{A}_{\kappa} | \lambda_{\beta}}$ matrix elements can be shown to be
\begin{align}
\braket{\lambda_{\alpha} | \mathcal{A}_{\kappa} | \lambda_{\beta}} = -\frac{\textrm{i}}{\omega_{\alpha \beta}} \braket{\lambda_{\alpha} | \partial_{\kappa} \hat{H}(\kappa) | \lambda_{\beta}},
\end{align}
with $\omega_{\alpha \beta} \defeq \lambda_{\alpha}(\kappa) - \lambda_{\beta}(\kappa)$.

The $L_2$-norm of $\mathcal{A}_{\kappa}$ scales exponentially with the system size in non-integrable systems, $|| \mathcal{A}_{\kappa} ||^2 \sim \textrm{exp}[S]$, where $S$ is the thermodynamic entropy of the system. In practice, the adiabatic gauge potential is utilised as a diagnostic in its regularised form (see Refs.~\cite{Pandey:2020,Kolodrubetz:2017,Claeys:2019} for further details). Leaving the latter technicality aside, the power of using the adiabatic gauge potential as diagnostic of integrability is that it is extremely sensible to infinitesimal integrability-breaking perturbations, making it a very promising approach to detect integrability in a wide range of physical systems. 

\section{Thermodynamics in linear response}
\label{sec:thermo_linear_response}

The discussion pertaining to integrability above, as we shall see here, has significant consequences for the thermodynamics of quantum systems in general.

In linear response, thermodynamics in isolated quantum systems can be understood from the interplay between particle and energy currents, both of which are proportional to external perturbations. We start by considering a quantum system coupled to two thermal reservoirs, say, on the left and on the right sides of the system. The left (right) reservoir is characterised by a temperature $T_L$ ($T_R$) and a chemical potential $\mu_L$ ($\mu_R$). In the steady state, i.e., the state approached as time reaches the infinite limit, a constant flow of particles and energy is induced. Linear response concerns the physics of such a system-environment configuration, when the temperature difference $\Delta T \defeq T_L - T_R$ and the chemical potential difference $\Delta \mu \defeq \mu_L - \mu_R$ are small compared to, say, their average values.    

The expectation values of induced particle current $\hat{J}_1$ and thermal current $\hat{J}_2$ depend on the temperature gradient $\nabla T$ and the chemical potential gradient $\nabla \mu$ by the following relation (we set the electric chage $e \defeq 1$)
\begin{align}
\label{eq:onsager}
\begin{pmatrix} \langle \hat{J}_1 \rangle \\ \langle \hat{J}_2 \rangle \end{pmatrix} = \begin{pmatrix} \mathcal{L}_{11} & \mathcal{L}_{12} \\  \mathcal{L}_{21} & \mathcal{L}_{22} \end{pmatrix} \begin{pmatrix}  \nabla \mu \\ -\nabla T \end{pmatrix},
\end{align} 
where the $\mathcal{L}_{ij}$ $(i, j = 1,2)$ are the transport coefficients. While the diagonal elements represent transport of either $\hat{J}_1$ or $\hat{J}_2$ from direct perturbations, the off-diagonal elements refer to induction of either a particle or a thermal current indirectly. Namely, $\mathcal{L}_{12}$ refers to a contribution to the particle current from a temperature gradient; while $\mathcal{L}_{21}$ refers to a contribution to the thermal current from a chemical potential gradient~\cite{MeisnerThesis:2005,JuanThesis:2014}. 

The second law of thermodynamics may be written in terms of the entropy production from the current operators and their corresponding gradients $X_i$,
\begin{align}
\label{eq:entropy_production}
\frac{\partial S}{\partial t} = \sum_{i} \langle \hat{J}_i \rangle X_i.
\end{align}
One then requires that the elements of the Onsager matrix need to satisfy
\begin{align}
\mathcal{L}_{11} \geq 0 \quad \textrm{and} \quad \mathcal{L}_{22} \geq \frac{(\mathcal{L}_{12} + \mathcal{L}_{21})^2}{\mathcal{L}_{11}} \geq 0,
\end{align}
which can be shown by enforcing positivity of the entropy production rate [Eq.~\eqref{eq:entropy_production}] subject to physical constraints~\cite{Benenti:2017}. The Onsager relation 
\begin{align}
\mathcal{L}_{21} = T \mathcal{L}_{12}
\end{align}
is satisfied in linear response. As can be expected from physical grounds, particle and thermal currents are not independent, but intertwined physical quantities. 

More familiar transport properties, such as {\em conductivity}, can be expressed in terms of the transport coefficients. For instance, the electric conductivity, measured in the condition $\nabla T = 0$ is
\begin{align}
\sigma = \left( \frac{\langle \hat{J}_1 \rangle}{\nabla \mu} \right)_{\nabla T = 0} = \mathcal{L}_{11},
\end{align}
while the thermal conductivity, defined with the condition of zero particle flow $\langle \hat{J}_1 \rangle = 0$ and typically written down as $\langle \hat{J}_2 \rangle = -\kappa \nabla T$, is given by
\begin{align}
\kappa = \left( \frac{\langle \hat{J}_2 \rangle}{\nabla T} \right)_{\langle \hat{J}_1 \rangle = 0} = \mathcal{L}_{22} - \frac{1}{T}\frac{\mathcal{L}_{21} \mathcal{L}_{12}}{\mathcal{L}_{11}}.
\end{align}
Another important quantity is the thermopower $R = \mathcal{L}_{12} / \mathcal{L}_{11}$, defined under the condition $\langle \hat{J}_1 \rangle = 0$. The phenomenon is known as the Seebeck effect, in which a chemical potential gradient arises from a temperature gradient. In the reversed problem, under the condition $\nabla T = 0$, a particle current drives a thermal current where $\langle \hat{J}_2 \rangle = \Pi \langle \hat{J}_1 \rangle$, with $\Pi = \mathcal{L}_{21} / \mathcal{L}_{11}$.

\begin{figure}[t]
\fontsize{13}{10}\selectfont 
\centering
\includegraphics[width=0.8\columnwidth]{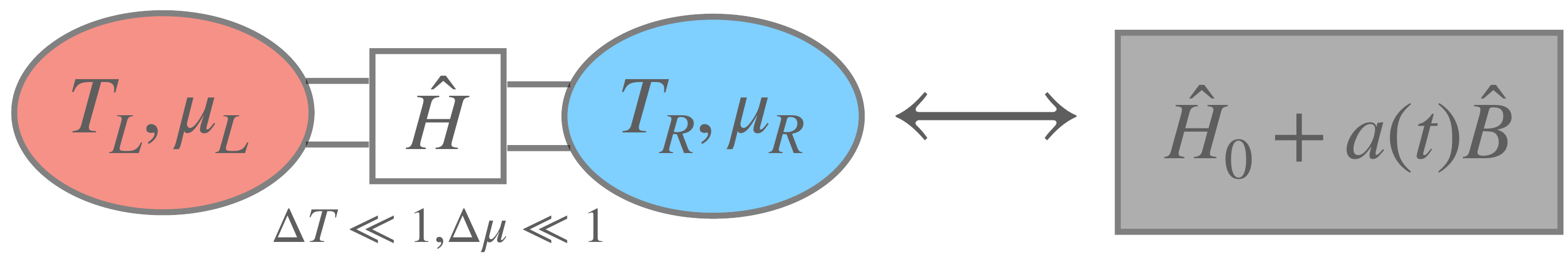}
\caption[Induced currents $\hat{J}_1$ and $\hat{J}_2$ studied from the steady-properties of a sample Hamiltonian $\hat{H}$]{In linear response, the induced currents $\hat{J}_1$ and $\hat{J}_2$ can be studied from the steady-properties of a sample Hamiltonian $\hat{H}$ connected to two thermal reservoirs, or from the dynamics generated by a perturbation into the isolated sample.}
\label{fig:1.2.2}
\end{figure}

Several thermodynamic relations can be obtained from the expressions above~\cite{Benenti:2017}. Regardless, a sobering fact can be deduced: all the thermodynamic properties of the configuration (as described in Fig.~\ref{fig:1.2.2}) can be computed from the Onsager matrix in Eq.~\eqref{eq:onsager}. By extension, it suffices to understand the electric and thermal conductance of a given system to understand its thermodynamic properties. 

So far, our discussion has referred to a system coupled to thermal reservoirs. Such type of configurations are best understood in the framework of open systems theory and are discussed in Part~\ref{part:two} of this thesis. However, the electric and thermal conductance can also be understood from the perspective of isolated systems. Moreover, both approaches are equivalent to each other under minimal assumptions as shown in Ref.~\cite{Purkayastha:2019}.

In the isolated-system language, one considers an equilibrium Hamiltonian $\hat{H}_0$ perturbed in the following way:
\begin{align}
\hat{H}(t) = \hat{H}_0 + a(t)\hat{B}.
\end{align}
$\hat{B}$ is some Hermitian operator and $a(t)$ is a weakly-perturbing field. Placing our attention now to a physical operator $\hat{A}$, we want to determine the response of $\hat{A}$ to the weakly-perturbing field from its equilibrium value, i.e.,
\begin{align}
\delta \langle \hat{A}(t) \rangle = \langle \hat{A}(t) \rangle - \langle \hat{A} \rangle_{\textrm{eq}}, 
\end{align}
where the $\langle \hat{A} \rangle_{\textrm{eq}}$ represents the equilibrium average in a statistical mechanics ensemble. It is common to consider the canonical ensemble, in which $\langle \cdot \rangle_{\textrm{eq}} = \textrm{Tr}[\hat{\rho} \cdot]$. In this particular case, $\hat{\rho}$ is the density operator of the canonical ensemble, expressed as $\hat{\rho} = e^{-\beta \hat{H}} / Z$, where $k_B \defeq 1$, $\beta \defeq 1 / T$ is the inverse temperature and $Z = \textrm{Tr}[e^{-\beta \hat{H}}]$ is the partition function. In the linear range, the response to the weakly-perturbing field is of the form~\cite{Pottier:2010,Russomanno:2013}
\begin{align}
\delta \langle \hat{A}(t) \rangle = \int_{-\infty}^{+\infty} \textrm{d}t^{\prime} \chi_{AB}(t, t^{\prime})a(t^{\prime}).
\end{align}

The linear response function, $\chi_{AB}(t,t^{\prime})$ depends {\em only} on the properties of the unperturbed system and can be shown to be directly connected to the matrix elements $\mathcal{L}_{ij}$ of the Onsager matrix in Eq.~\eqref{eq:onsager}. Due to time-translational invariance~\cite{Pottier:2010}, $\chi_{AB}(t,t^{\prime})$ depends not on the two arguments $t$ and $t^{\prime}$, but on the difference $t - t^{\prime}$. For simplicity in the notation, we henceforth write it as $\chi_{AB}(t)$. The response function can be written as ($\hbar \defeq 1$)~\cite{Kubo:1991} 
\begin{align}
\chi_{AB}(t) \defeq -\textrm{i}\theta(t) \langle [\hat{A}(t), \hat{B}] \rangle_{\textrm{eq}},
\end{align}
where $[\cdot, \cdot]$ represents the quantum-mechanical commutator and $\theta(t)$ is the Heaviside step function. It is simple to see that $\chi_{AB}(t)$ is real if $\hat{A}$ and $\hat{B}$ are Hermitian\footnote{$\langle \hat{A}(t) \hat{B} \rangle = \textrm{Tr}[\hat{\rho} \hat{A}(t)\hat{B}] =  \textrm{Tr}[\hat{\rho} \hat{U}^{\dagger} \hat{A} \hat{U} \hat{B}] $. Since $\hat{A}$ and $\hat{B}$ are Hermitian, $\langle \hat{A}(t) \hat{B} \rangle = \textrm{Tr}[\hat{B} \hat{U}^{\dagger} \hat{A} \hat{U} \hat{\rho}]^{*} = \langle \hat{B} \hat{A}(t) \rangle^{*}$. Therefore, if $\langle \hat{A}(t) \hat{B} \rangle \defeq a + \textrm{i}b$, then $\langle [\hat{A}(t), \hat{B}] \rangle = 2\textrm{i}b$.}. More generally, $\chi_{AB}(t)$ relates to the imaginary part of the correlation function
\begin{align}
C_{AB}(t) \defeq \langle A(t) B \rangle,
\end{align}
while the real part of $C_{AB}(t)$ is related to the so-called symmetrised noise $S(t) \defeq \langle \{\hat{A}(t), \hat{B}\} \rangle$, where we use the notation $\{ \cdot, \cdot \}$ for the quantum anti-commutator. These correlation functions are studied in greater detail within the context of eigenstate thermalisation in Chapter~\ref{chapter:eth}.

At the level of transport, $C_{AB}(t)$ connects to the Kubo correlation function~\cite{Bertini:2021,Kubo:1991}
\begin{align}
K_{AB} = \frac{1}{\beta} \int_{0}^{\beta} d\lambda \langle \hat{B} \hat{A}(t + \textrm{i}\lambda) \rangle.
\end{align}
Furthermore, within linear response theory, the coefficients of the Onsager matrix $\mathcal{L}_{ij}$ in frequency space are directly related to the Kubo correlation function~\cite{MeisnerThesis:2005}
\begin{align}
\mathcal{L}_{ij}(\omega) = \beta^{r} \lim_{t \to \infty} \lim_{L \to \infty} \frac{1}{L} \int_{0}^{t} \textrm{d}\tau e^{\textrm{i}\omega\tau} K_{\hat{J}_i \hat{J}_j}(\tau),
\end{align} 
where $r = 1$ for $j = 1$ and $r = 2$ for $j = 2$. The order of the limits is crucial, since taking the limit $t \to \infty$ first would reflect finite-size effects as opposed to the physical properties of the bulk of the sample\footnote{Most interestingly, in the open systems language discussed at the beginning of this section, the order of the limits needs to be reversed to obtain agreement. See Ref.~\cite{Purkayastha:2019} for a discussion.}.

\subsection{Relevance of integrability in connection to transport}
\label{sec:relevance_of_integrability}

In linear response, as gathered from the discussion above, the transport properties of a given system depend on the time-dependent correlation functions of the relevant current operators. In the frequency domain, it is common to split the real part of $\mathcal{L}_{ij}(\omega)$ into the zero-frequency and finite frequency contributions:
\begin{align}
\label{eq:drude_decomp}
\textrm{Re}[\mathcal{L}_{ij}(\omega)] = 2\pi D_{ij} \delta(\omega) + \mathcal{L}_{ij}^{\textrm{reg}}(\omega),
\end{align} 
where $D_{ij}$ is the so-called Drude weight and refers to the zero-frequency contribution to $\mathcal{L}_{ij}(\omega)$ via the delta function $\delta(\omega)$, while $\mathcal{L}_{ij}^{\textrm{reg}}(\omega)$ refers to the finite-frequency contribution and it is known as the regular part of $\mathcal{L}_{ij}(\omega)$. 

A closed-form expression can be obtained for $D_{ij}$ from Eq.~\eqref{eq:drude_decomp} in terms of the matrix elements of $\hat{J}_i$, by expressing symmetrised noise $\langle \{ \hat{J}_i (t), \hat{J}_j \} \rangle$ in the eigenbasis of $\hat{H}$ and taking the Fourier transform. Such expressions will be most relevant and studied in Chapter \ref{chapter:kubo}. For the sake of the present discussion, it suffices to realise that since the Drude weight refers to the zero-frequency contribution of $\mathcal{L}_{ij}(\omega)$, then, such quantity is relevant to the long-time value of the correlation function $\langle \hat{J}_i (t) \hat{J}_j \rangle$. In particular~\cite{Bertini:2021,zotos1997transport,MeisnerThesis:2005},
\begin{align}
\label{eq:drude_corr}
D_{ij} = \frac{\beta^r}{2L} \lim_{t \to \infty} \langle \hat{J}_i (t) \hat{J}_j \rangle.
\end{align}
Let us now focus on the two cases $i = j$, i.e., direct particle and thermal transport. We can first decompose the correlation function
\begin{align}
\label{eq:j_decomp}
\langle \hat{J}_i (t) \hat{J}_i \rangle = C_{\hat{J}_i \hat{J}_i} + C(t),
\end{align}
i.e., into a sum of time-independent factor with its corresponding time dependence\footnote{This, again, follows from the expressions of the symmetrised noise in the eigenbasis of the Hamiltonian, which will be further analysed in Chapter~\ref{chapter:kubo} and Chapter~\ref{chapter:eth}.}. If, and only if, $\langle \hat{J}_i (t) \hat{J}_i \rangle$ possesses non-singular low-frequency behaviour, then
\begin{align}
\lim_{\tau \to \infty} \frac{1}{\tau} \int_{0}^{\tau} \textrm{d}t C(t) = 0 
\end{align}
and 
\begin{align}
C_{\hat{J}_i \hat{J}_i} = \lim_{t \to \infty} \langle \hat{J}_i (t) \hat{J}_i \rangle.
\end{align}
This object is precisely the physical quantity bounded by Mazur's inequality in Eq.~\eqref{eq:mazur}, and so, from Eq.~\eqref{eq:drude_corr}
\begin{align}
D_{ii} \geq \frac{\beta^r}{2L} \sum_{n} \frac{\langle \hat{J}_i \hat{Q}_n \rangle^2}{\langle \hat{Q}_n^2 \rangle},
\end{align} 
where the $\hat{Q}_n$ are the conserved quantities discussed in Sec.~\ref{sec:hydro_quantum}. 

It is then, from this analysis and our previous discussion in Sec.~\ref{sec:hydro_quantum}, that we conclude the pivotal role of integrability in the thermodynamics of quantum systems. The decomposition written down in Eq.~\eqref{eq:j_decomp} will be absolutely crucial, as we explore the notion of integrability and transport for physical systems in Chapter~\ref{chapter:kubo}. In particular, the concept of translational invariance plays a central role in the analysis of transport from the perspective of the Drude weight. In general, however, the overlaps between a current operator and the conserved quantities of a given system provide a lower bound for the Drude weight.

Physically, a finite Drude weight implies that the current correlation function does not decay in the limit of infinite time and transport is ballistic, as discussed in Sec.~\ref{sec:hydro_quantum}, while a vanishing Drude weight is associated to hydrodynamical or anomalous behaviour~\cite{Bertini:2021}. These regimes, in linear response, regulate the entire thermodynamic behaviour of a given physical system.

Even though several authors (see, for instance, Refs.~\cite{zotos1997transport,benenti2013conservation,XotosIncoherentSIXXZ,Ilievski:2013}) have argued about transport properties from the perspective of integrability, we would like to note that doing so is particularly challenging for systems that are not solvable by the Bethe ansatz~\cite{Bertini:2021}. For many particular systems of interest, this leaves one with the choice of estimating the correlation functions themselves in the time/frequency domain. The latter approach, does not come without its own challenges, for which one must face problems whose dimension increases exponentially with the system size, probing long-time dynamics at the same time. A detailed discussion of these issues and how to overcome them is one of the topics of interest in this thesis. 
 
\chapter{The 1D anisotropic Heisenberg model}
\label{chapter:models}

The Heisenberg model is one of the simplest models devised to understand ferromagnetic behaviour. In its most simple form, the model describes the interplay between {\em coherent} effects, physically described by quantum-mechanical transitions between quantum states, and {\em incoherent} effects, which are the understood as the result of scattering processes or non-elastic interactions. The model is visualised as a collection of interacting spins embedded on a lattice. The spin-1/2 version of the Heisenberg model was the first physical system treated with the Bethe ansatz~\cite{Bethe:1931,Takahashi:1999, ShastryBethe1990, Cazalilla:2011}. It has been the subject of fervent research throughout the past century and, even now, some questions remain despite impressive progress, particularly relating to finite-temperature transport~\cite{Bertini:2021}. In many ways, the model is the perfect testbed for studies in statistical mechanics and the effect of interactions. We introduce the model in Sec.~\ref{sec:model_xxz} and discuss the global symmetries pertaining to it, which relates to the rich transport properties of the model. Continuity equations and expressions for current operators are introduced in Sec.~\ref{sec:continuity_transport}. A mathematically-equivalent form using spinless fermions via the Jordan-Wigner transformation is discussed in Sec.~\ref{sec:jordan_wigner}. Integrability breaking in terms of local and global perturbations is discussed in Sec~\ref{sec:integrability_breaking}. We finalise our discussion by providing a short overview of experimental realisations in Sec~\ref{sec:experiments_xxz}.

\section{The model}
\label{sec:model_xxz}

For the specific case of spin-1/2 systems in one dimension, the Hamiltonian of the anisotropic Heisenberg model is expressed as
\begin{align}
\label{eq:h_xxz}
\hat{H}_{\textrm{XXZ}} = \sum_{i}\left[\alpha\left(\hat{\sigma}^x_{i}\hat{\sigma}^x_{i+1} + \hat{\sigma}^y_{i}\hat{\sigma}^y_{i+1}\right) + \Delta\,\hat{\sigma}^z_{i}\hat{\sigma}^z_{i+1}\right],
\end{align} 
where $\hat{\sigma}^\nu_{i}$, $\nu = x,y,z$, correspond to Pauli matrices in the $\nu$ direction at site $i$ in a one-dimensional lattice with $L$ sites. The Pauli matrices satisfy
\begin{align}
\label{eq:pauli}
[\hat{\sigma}^{\mu}_i, \hat{\sigma}^{\nu}_j] = 2\textrm{i}\epsilon_{\mu \nu \gamma} \hat{\sigma}^{\gamma}_i \delta_{ij},
\end{align}
where $\epsilon_{\mu \nu \gamma}$ represents the Levi-Civita tensor, $\delta_{ij} = 1 \; \forall i = j$ and $\delta_{ij} = 0 \; \forall i \neq j$. The model can be studied under different boundary conditions. Boundary conditions are specified as {\em open} if the sum in Eq.~\eqref{eq:h_xxz} includes all the sites but the last one ($L-1$) and {\em periodic} if it includes all the sites ($L$), with $L + 1 \to 1$. $\Delta$ is known as the anisotropy parameter. For the particular case $\Delta=\alpha$, the Hamiltonian~\eqref{eq:h_xxz} is the Hamiltonian of the spin-1/2 Heisenberg chain. The model is also known as the spin-1/2 XXZ chain, it is integrable and exactly solvable via Bethe ansatz~\cite{ShastryBethe1990, Cazalilla:2011}. 

The Hamiltonian is only composed of {\em local} interactions and couplings, in the sense that it is the sum of terms of the form
\begin{align}
\hat{h}^{\textrm{XXZ}}_{i,i+1} = \alpha\left(\hat{\sigma}^x_{i}\hat{\sigma}^x_{i+1} + \hat{\sigma}^y_{i}\hat{\sigma}^y_{i+1}\right) + \Delta\,\hat{\sigma}^z_{i}\hat{\sigma}^z_{i+1},
\end{align} 
where $\hat{H}_{\textrm{XXZ}} = \sum_{i} \hat{h}^{\textrm{XXZ}}_i$. Locality has significant consequences in the properties of the model. 

\subsection{Global symmetries}
\label{sec:global_symmetries}

Being an integrable system, the Hamiltonian $\hat{H}_{\textrm{XXZ}}$ possesses an extensive set of non-trivial {\em local} and {\em quasi-local} conserved quantities that affect transport in the model~\cite{Bertini:2021}. These conserved quantities can be exploited to solve the model, to obtain solutions to the eigenvalue problem and even correlation functions. In this section, however, we refer to the known {\em global} symmetries of the model which are most relevant to numerical approaches. These symmetries have been pedagogically described in the works of L. Santos {\em et al.} The following is a brief survey of the relevant global symmetries. We refer the reader to Refs.~\cite{Gubin:2012,Joel:2013} for further details. 

A didactic way to understand the global symmetries of the anisotropic Heisenberg model is to consider the basis states corresponding to the eigenstates of $\bigotimes_{i=1}^{L} \hat{\sigma}^{z}_i$, i.e., the $\mathcal{D} = 2^{L}$ set of up/down spins which constitute a complete basis in the Hilbert space. It is common to represent each of these states with up/down arrows $\ket{\uparrow_1 \uparrow_2 \cdots \uparrow_L}$, where there are $\mathcal{D}$ independent states corresponding to all the possible $2^L$ combinations. Naturally, the Hamiltonian~\eqref{eq:h_xxz} is not diagonal in this basis, but allows for a simple description to express $\hat{H}_{\textrm{XXZ}}$ as a matrix operator in Hilbert space in that basis. It is common to refer to this basis as the {\em computational basis}. 

To understand the effect of $\hat{H}_{\textrm{XXZ}}$ onto the computational basis states, it is useful to introduce Pauli spin raising and lowering operators, defined as
\begin{align}
\hat{\sigma}^{+}_i \defeq \frac{1}{2} \left( \hat{\sigma}^x_i + \textrm{i}\hat{\sigma}^y_i \right), \quad \hat{\sigma}^{-}_i \defeq \frac{1}{2} \left( \hat{\sigma}^x_i - \textrm{i}\hat{\sigma}^y_i \right).
\end{align} 
Using these operators, $\hat{H}_{\textrm{XXZ}}$ can be re-written as
\begin{align}
\label{eq:h_xxz_pm}
\hat{H}_{\textrm{XXZ}} = \sum_{i}\left[2\alpha\left(\hat{\sigma}^+_{i}\hat{\sigma}^-_{i+1} + \hat{\sigma}^-_{i}\hat{\sigma}^+_{i+1}\right) + \Delta\,\hat{\sigma}^z_{i}\hat{\sigma}^z_{i+1}\right].
\end{align}
From this expression one can observe that the effect of the first term on up/down states is to move neighbouring excitations through the chain, in the form
\begin{align}
2\alpha\left(\hat{\sigma}^+_{i}\hat{\sigma}^-_{i+1} + \hat{\sigma}^-_{i}\hat{\sigma}^+_{i+1}\right) \ket{\cdots \uparrow_i \downarrow_{i+1} \cdots} = 2\alpha \ket{\cdots \downarrow_i \uparrow_{i+1} \cdots},
\end{align} 
while the second term introduces an attraction term if the $z$-alignment of neighbouring states are anti-aligned, as
\begin{align}
\Delta\,\hat{\sigma}^z_{i}\hat{\sigma}^z_{i+1} \ket{\cdots \uparrow_i \downarrow_{i+1} \cdots} = -\Delta \ket{\cdots \uparrow_i \downarrow_{i+1} \cdots},
\end{align}
similarly for $\ket{\cdots \downarrow_i \uparrow_{i+1} \cdots}$. A repulsive term is introduced if neighbouring states are aligned
\begin{align}
\Delta\,\hat{\sigma}^z_{i}\hat{\sigma}^z_{i+1} \ket{\cdots \uparrow_i \uparrow_{i+1} \cdots} = +\Delta \ket{\cdots \uparrow_i \uparrow_{i+1} \cdots},
\end{align}
equivalently for $\ket{\cdots \downarrow_i \downarrow_{i+1} \cdots}$. In this language, $\hat{H}_{\textrm{XXZ}}$ represents the interplay between neighbouring excitation hopping, modulated by $\alpha$, and neighbouring interactions modulated by $\Delta$. Most notably, $\hat{H}_{\textrm{XXZ}}$ only moves excitations through the chain and does not create nor destroy them. This observation has significant consequences that manifest as global symmetries in the model.

\subsubsection{Conservation of total $z$-magnetisation} 

\begin{figure}[t]
\fontsize{13}{10}\selectfont 
\centering
\includegraphics[width=0.53\columnwidth]{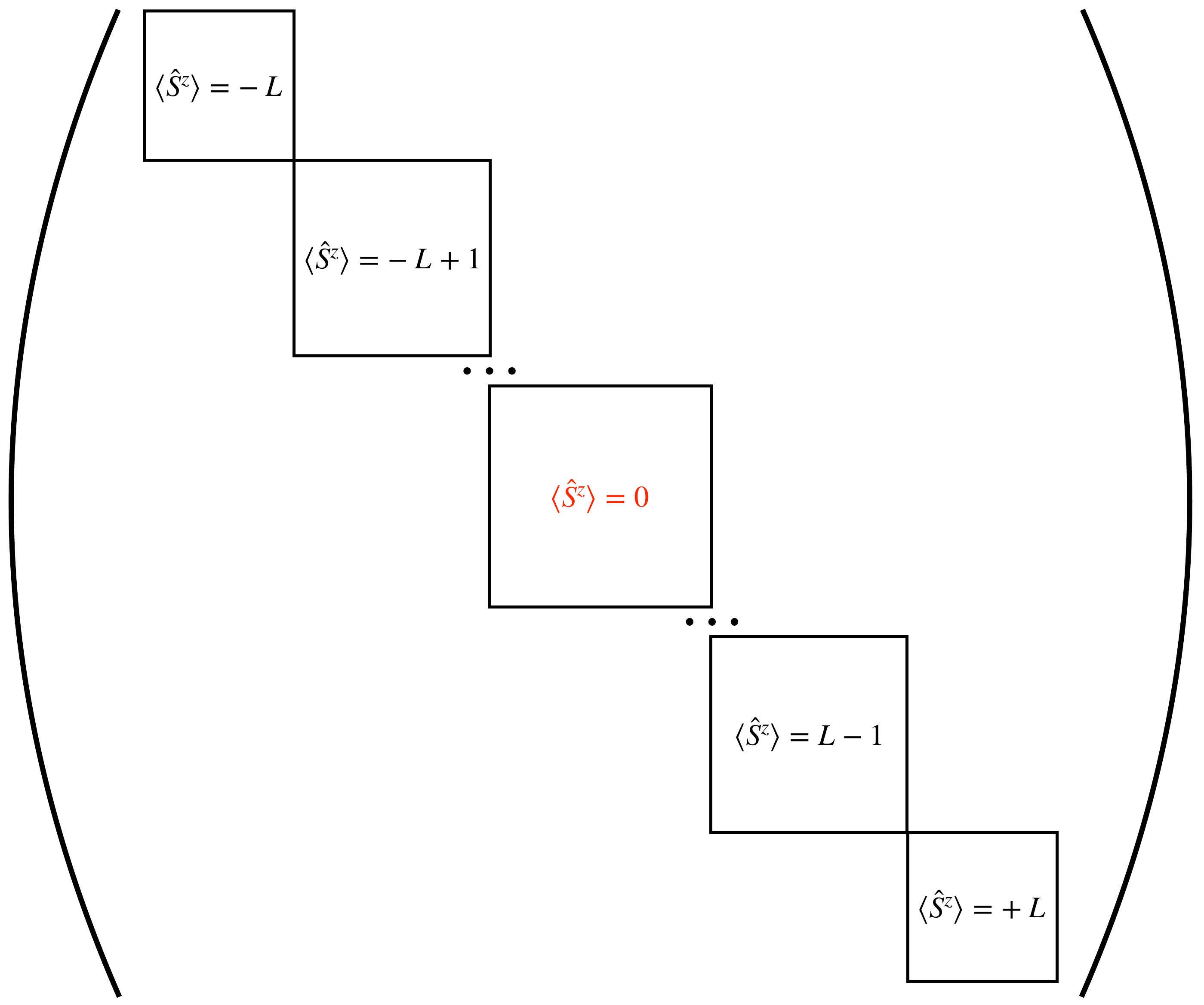}
\caption[Block diagonal structure of the Hamiltonian matrix operator $\hat{H}_{\textrm{XXZ}}$]{Block diagonal structure of the Hamiltonian matrix operator $\hat{H}_{\textrm{XXZ}}$. Note that only even $L$ contains a $\langle \hat{S}^z \rangle = 0$ sub-sector.}
\label{fig:1.3.1}
\end{figure}

One of the most relevant symmetries concerning transport in the anisotropic Heisenberg model is conservation of total magnetisation in the $z$ direction. As noted above, the Hamiltonian generates the {\em swapping} of excitations through the chain, without creating or destroying them. This is an indication of a global symmetry. In fact, if one defines the total magnetisation in the $z$ direction as
\begin{align}
\hat{S}^z \defeq \sum_{i} \hat{\sigma}^z_i,
\end{align}
it is straightforward to show that
\begin{align}
[ \hat{H}_{\textrm{XXZ}}, \hat{S}^z ] = 0,
\end{align}
which implies that $\hat{S}^z$ is conserved\footnote{The isotropic model, $\Delta = \alpha$, conserves total spin $\hat{S}^2 = \left( \sum_{i=1}^L \vec{\sigma}_i \right)^2$, i.e, $[\hat{H}_{\textrm{XXZ}}, \hat{S}^2] = 0$. This symmetry is known as SU(2) symmetry.}. It is quite important to remark since the number of excitations are conserved, the Hamiltonian and other operators in Hilbert space such as the generator of the dynamics $\hat{U}(t) = \exp({-\textrm{i}\hat{H}t})$ do not admix different excitation sectors. In common linear algebra terms, conservation of $\hat{S}^z$ implies that $\hat{H}_{\textrm{XXZ}}$ can be represented as a block diagonal matrix operator, each block pertaining to a magnetisation sector (Fig.~\ref{fig:1.3.1}). This symmetry is known as U(1) symmetry.

The dimension of each magnetisation sector depends on the number of excitations and it is given by the all possible spin up/down combinations that preserve $\langle \hat{S}^z \rangle = N$. Each block has dimension
\begin{align}
\mathcal{D}_{\langle \hat{S}^z \rangle = N} = \begin{pmatrix} L \\ N \end{pmatrix} = \frac{L!}{N! (L - N)!}.
\end{align}
The largest sub-sector is the one for which the number of excitations $N = L / 2$ for even $L$. In this configuration, the number of spins up and spins down are the same and so, the configurational space is the largest possible. For odd $L$, the sub-sectors $\langle \hat{S}^z \rangle = +1$ and $\langle \hat{S}^z \rangle = -1$ are of the same size and correspond to the largest magnetisation sub-sectors in that case. Conservation of total magnetisation, as we shall describe, has significant consequences for transport. Furthermore, this symmetry may co-exist with others depending on the configuration of the Heisenberg chain and the parameters of the Hamiltonian.

\subsubsection{Translation invariance}

The specific case of periodic boundary conditions, where the sum in Eq.~\eqref{eq:h_xxz} runs up to $L$, results in another global symmetry known as translational invariance. The corresponding conservation law associated to this symmetry is conservation of momentum.

Translational symmetry can be understood by first considering the $\langle \hat{S}^z \rangle = -L$ magnetisation sub-sector, which contains only one state and may be written in the computational basis as
\begin{align}
\ket{0} \defeq \ket{\downarrow \cdots \downarrow}.
\end{align} 
The states for the subsequent sub-sector $\langle \hat{S}^z \rangle = -L + 1$ may be generated by the spin-excitation operator
\begin{align}
\ket{l} \defeq \hat{\sigma}^+_l \ket{0}, \quad l = 1, \cdots, L.
\end{align}
These states are not eigenstates of the Hamiltonian, but translational-symmetric eigenstates can be constructed from a linear combination as
\begin{align}
\ket{\psi_k} = \frac{1}{\sqrt{L}} \sum_{l = 1}^{L} e^{\textrm{i}kl} \ket{l},
\end{align}
where $k = 2\pi m/L \;(m = 0,\cdots,L-1)$ is the wave number. These states are eigenstates of the translation operator $\hat{T}$, which moves forward an excitation in space, such that\footnote{This can be shown by noticing that $\hat{T} \ket{\psi_k} = \sum_{l=1}^L e^{\textrm{i}kl} \hat{T} \ket{l} = \sum_{l=1}^{L-1} e^{\textrm{i}kl} \ket{l + 1} + e^{\textrm{i}kL} \ket{1}$, then $\sum_{l=1}^L e^{\textrm{i}kl} \hat{T} \ket{l} = \sum_{l=1}^L e^{\textrm{i}k(l-1)}\ket{l}$.}
\begin{align}
\hat{T} \ket{\psi_k} = e^{-\textrm{i}k} \ket{\psi_k}.
\end{align}
For a translation-invariant Hamiltonian, $\ket{\psi_k}$ are both eigenstates of the translation operator and of the Hamiltonian. We observe, then, that translational invariance divides the Hamiltonian into $k$-momentum sub-sectors. The generalisation to sub-sectors with a higher number of excitations follows from the above approach. In such case, translational invariance divides the Hamiltonian magnetisation sub-sectors into additional $L$ sub-sectors which share the same $k$ parameter, dubbed $k$-{\em quasi-momentum} sectors.

\subsubsection{Conservation of parity}

An additional global symmetry in $\hat{H}_{\textrm{XXZ}}$ is conservation of parity, which can be understood by considering a mirror located at one of the edges of the chain for a system with open boundary conditions. The parity operator is defined as~\cite{Joel:2013}
\begin{align}
\hat{\Pi} = \begin{cases} \hat{\mathcal{P}}_{1,L}\hat{\mathcal{P}}_{2,L-1}\cdots\hat{\mathcal{P}}_{\frac{L}{2},\frac{L+2}{2}} \; \forall \; \textrm{even} \; L \\  \hat{\mathcal{P}}_{1,L}\hat{\mathcal{P}}_{2,L-1}\cdots\hat{\mathcal{P}}_{\frac{L-1}{2},\frac{L+3}{2}} \; \forall \; \textrm{odd} \; L,\end{cases}
\end{align}
where $\hat{\mathcal{P}}_{i,j} = (\hat{\sigma}^x_i\hat{\sigma}^x_j + \hat{\sigma}^y_i\hat{\sigma}^y_j + \hat{\sigma}^z_i\hat{\sigma}^z_j + \mathds{1}) / 2$ permutes the $i$-th and $j$-th spins. Conservation of parity implies $[\hat{H}_{\textrm{XXZ}}, \hat{\Pi}] = 0$. Since $\hat{H}_{\textrm{XXZ}}$ conserves parity, its eigenstates may have even or odd parity
\begin{align}
\hat{\Pi} \ket{\psi_j} &= +\ket{\psi_j} \; \textrm{even parity},\\
\hat{\Pi} \ket{\psi_j} &= -\ket{\psi_j} \; \textrm{odd parity},
\end{align}  
it follows that the amplitudes $c_i$ in the expansion of a given eigenstate in the computational basis $\ket{\psi_j} = \sum_i c_i \ket{i}$ will be equal for the $\ket{i}$ that are equivalent under a parity permutation (up to a factor of $-1$ for odd parity). Just as before, this symmetry can be used to reduce the effective Hilbert space dimension by constructing superposition states from computational basis states that are equivalent under a parity permutation. Moreover, some numerical treatments require this symmetry to be resolved, given that different symmetry sub-sectors are independent and therefore uncorrelated. Such is the case, for instance, when computing spectral level distributions as discussed in Chapter~\ref{chapter:integrability}.

\subsubsection{Reflection symmetry}

The last global symmetry of the model relates to the symmetry under a global $\pi$ rotation in the $x$ direction. For $\hat{H}_{\textrm{XXZ}}$, this symmetry is only present in the zero magnetisation sub-sector, i.e., $\langle \hat{S}^z \rangle = 0$. This operation can be written as
\begin{align}
\hat{R}_{\pi}^{x} = \prod_{i=1}^L \hat{\sigma}^x_i,
\end{align}
naturally, the symmetry is implied from the fact that $[\hat{H}_{\textrm{XXZ}}, \hat{R}_{\pi}^{x}] = 0$. This symmetry is also knows as spin inversion, or $\mathds{Z}_2$ symmetry.

\section{Continuity equations and transport}
\label{sec:continuity_transport}

In a similar fashion to how a particle current was defined in classical hydrodynamics in Sec.~\ref{sec:hydro_classical}, conservation laws yield continuity equations and definitions of current operators in the quantum realm. Akin to the local conservation of particles through a section in space described in Sec.~\ref{sec:hydro_classical}, if a quantum operator is a sum of local operators, the local density of this quantity moves through a section in space from one side to another and continuity is satisfied. On the other hand, even if there exists a conservation law for a given operator which is not composed of a sum of local operators, a continuity equation for the transport of such quantity is somewhat meaningless. Following our discussion of global symmetries before, it is meaningful to discuss transport of spin excitations, while meaningless to try to describe a continuity equation for the conservation of parity~\cite{KimThesis:2014}. In this section we define continuity equations for spin excitations in the $z$ direction and energy, which lead to forms of current operators in terms of Pauli matrices.

The total magnetisation operator $\hat{S}^z$ is a conserved quantity. We can then write down a continuity equation for the local site magnetisation in the $z$ direction as follows
\begin{align}
\frac{\textrm{d} \langle \hat{\sigma}^z_i \rangle}{\textrm{d} t} = \textrm{i} \langle [\hat{H}, \hat{\sigma}^z_i] \rangle.
\end{align} 
Using Eq.~\eqref{eq:pauli}, it is straightforward to show
\begin{align}
\textrm{i} \langle [\hat{H}, \hat{\sigma}^z_i] \rangle = \langle \hat{j}^{\textrm{P}}_{i-1} \rangle - \langle \hat{j}^{\textrm{P}}_i \rangle,  
\end{align}
where the expectation value is assumed to be taken over one of the ensembles of statistical mechanics and
\begin{align}
\label{eq:j1_xxz}
\hat{j}^{\textrm{P}}_i \defeq 2\alpha \left( \hat{\sigma}^x_i \hat{\sigma}^y_{i+1} - \hat{\sigma}^y_i \hat{\sigma}^x_{i+1} \right).
\end{align}
In the language introduced in Chapter~\ref{chapter:integrability},
\begin{align}
\hat{J}_1 = \sum_i \hat{j}^{\textrm{P}}_i.
\end{align}
Note that these definitions apply only to the bulk of the sample, currents on the boundaries for an open Heisenberg chain are ill-defined, although in the thermodynamic limit $L \to \infty$, boundary effects are negligible. Eq.~\eqref{eq:j1_xxz} is an explicit form of the current operator in the $z$ direction. It should be noted that in the non-interacting Hamiltonian, for which $\Delta = 0$, $\hat{J}_1$ is a conserved quantity itself. This implies that the magnetisation gradient in the $z$ direction for the non-interacting system is zero, and the current never decays. Following our discussion from Chapter~\ref{chapter:integrability}, transport of spin excitations in the $z$ direction is ballistic for the non-interacting case $\Delta = 0$.

For the interacting Hamiltonian, $\Delta \neq 0$, transport is far more complicated and a great amount of effort has been devoted to characterise it. Away from the zero-magnetisation sector, a finite lower bound for the spin Drude weight exists for any value of $\Delta$ at infinite- and finite-temperature~\cite{zotos1997transport}, which entails that spin transport is ballistic. Furthermore, for the zero-magnetisation section in the so-called weakly-interacting regime $0 < \Delta < 1$ (assuming $\alpha = 1$) the presence of quasi-local conserved quantities has been identified and established~\cite{Prosen:2011,Prosen:2013,Pereira:2014}. A finite lower bound on the spin Drude weight exists in that particular case at infinite temperature, indicating ballistic transport as well. Although its full temperature dependence has yet to be fully characterised, there exists evidence that suggests the Drude weight vanishes at finite temperature~\cite{Bertini:2021}. 

In the strongly-interacting regime $ \Delta > 1$ (again, assuming $\alpha = 1$), though a formal proof is lacking~\cite{Bertini:2021}, overwhelming numerical evidence suggests a vanishing spin Drude weight and diffusive transport from the perspective of dynamical typicality approaches~\cite{Steinigeweg:2014}, open quantum systems~\cite{Znidaric:2011} and generalised hydrodynamics~\cite{Ilievski:2017}. There exists strong numerical evidence to suggest that spin transport in the isotropic model, $\Delta = \alpha$, is super-diffusive with a known transport exponent~\cite{Znidaric:2011,Bertini:2021} at infinite-temperature, although a formal proof is lacking and its temperature dependence remains an open question. An in-depth analysis of spin transport in the anisotropic Heisenberg model is provided in Ref.~\cite{Bertini:2021}.

As spin transport, thermal transport follows from a conservation law and a continuity equation. Since the system is isolated, total energy is conserved and the local nature of the Hamiltonian allows one to write the following continuity equation:
\begin{align}
\frac{\textrm{d} \langle \hat{h}^{\textrm{XXZ}}_{i,i+1} \rangle}{\textrm{d} t} &= \textrm{i} \langle [\hat{H}, \hat{h}^{\textrm{XXZ}}_{i,i+1}] \rangle \nonumber \\
&= \langle [\hat{h}^{\textrm{XXZ}}_{i-1,i}, \hat{h}^{\textrm{XXZ}}_{i,i+1}] \rangle + \langle [\hat{h}^{\textrm{XXZ}}_{i+1,i+2}, \hat{h}^{\textrm{XXZ}}_{i,i+1}] \rangle.
\end{align}
It is straightforward, albeit a bit cumbersome, to show
\begin{align}
\frac{\textrm{d} \langle \hat{h}^{\textrm{XXZ}}_{i,i+1} \rangle}{\textrm{d} t} = \langle \hat{j}^{\textrm{E}}_i \rangle - \langle \hat{j}^{\textrm{E}}_{i+1} \rangle,
\end{align}
where
\begin{align}
\hat{j}^{\textrm{E}}_i &= 2\alpha^2 \left( \hat{\sigma}^y_{i-1} \hat{\sigma}^z_{i} \hat{\sigma}^x_{i+1} - \hat{\sigma}^x_{i-1} \hat{\sigma}^z_{i} \hat{\sigma}^y_{i+1} \right) \nonumber \\
&+ 2\alpha \Delta \left( \hat{\sigma}^x_{i-1} \hat{\sigma}^y_{i} \hat{\sigma}^z_{i+1} -  \hat{\sigma}^z_{i-1} \hat{\sigma}^y_{i} \hat{\sigma}^x_{i+1}\right) \nonumber \\
&+ 2\alpha \Delta \left( \hat{\sigma}^z_{i-1} \hat{\sigma}^x_{i} \hat{\sigma}^y_{i+1} -  \hat{\sigma}^y_{i-1} \hat{\sigma}^x_{i} \hat{\sigma}^z_{i+1}\right).
\end{align}
The total thermal current can be defined from these local objects as 
\begin{align}
\hat{J}_2 = \sum_i \hat{j}^{\textrm{E}}_i.
\end{align}
Most interestingly, $\hat{J}_2$ is a non-trivial conserved quantity of the Hamiltonian $\hat{H}_{\textrm{XXZ}}$, commonly referred to as $\hat{Q}_3$, i.e., $[\hat{H}_{\textrm{XXZ}}, \hat{J}_2] = 0$~\cite{zotos1997transport}. It is then clear that, as opposed to spin transport, thermal transport is rather simple. The vanishing commutator implies that correlation functions of the form $\langle \hat{J}_2(t) \hat{J}_2(0) \rangle$ are independent of time, which leads to a diverging thermal conductivity, i.e., ballistic thermal transport. 

\section{Representation in spinless fermions}
\label{sec:jordan_wigner}

Although not true in general, in one dimension spin-1/2 systems behave like fermions. The Pauli algebra carries over to the domain of the singly-occupied fermionic level. The many-body problem, however, involving more than a single spin degree of freedom is more complicated. This is due to the fact that spin degrees of freedom commute on different lattices sites according to Eq.~\eqref{eq:pauli}, while fermionic degrees of freedom require anti-commutation for appropriate statistics to be obtained. It was the realisation by Jordan and Wigner~\cite{Jordan:1928,Coleman:2015}, that a mapping can be introduced in the many-body problem by associating a spin degree of freedom with a fermionic degree of freedom coupled to a phase factor called a {\em string}, allowing for the appropriate fermionic statistics.

Such mapping can be achieved by associating
\begin{align}
\label{eq:jordan_wigner}
\hat{\sigma}^+_i &\rightarrow \hat{c}^{\dagger}_i e^{\textrm{i}\hat{\phi}_i}, \\
\hat{\sigma}^-_i &\rightarrow \hat{c}_i e^{-\textrm{i}\hat{\phi}_i}, \\
\hat{\sigma}^z_i &\rightarrow 2\hat{n}_i - 1 = 2\hat{c}^{\dagger}_i \hat{c}_i - 1 = 2\hat{\sigma}^+_i \hat{\sigma}^-_i - 1, 
\end{align} 
where we have identified $\hat{c}^{\dagger}_i$ as the fermionic creation operator, $\hat{c}_i$ as the fermionic annihilation operator, $\hat{n}_i \defeq \hat{c}^{\dagger}_i \hat{c}_i$ as the counting operator and 
\begin{align}
\hat{\phi}_i = \pi \sum_{l < i} \hat{n}_l = \pi \sum_{l < i} \frac{1}{2}(1 + \hat{\sigma}^z_l)
\end{align}
as the phase operator. The string operator we referred to before is $e^{\textrm{i}\hat{\phi}_i}$. Note that the sum for the string only counts the fields on the left side of a given lattice site $i$. With the string association to each fermionic field, the creation and annihilation operators follow the appropriate anti-commutation relations
\begin{align}
\{ \hat{c}^{\dagger}_i, \hat{c}_j \} = \delta_{i,j}.
\end{align} 
Eq.~\eqref{eq:jordan_wigner} is a transformation that allows us to re-write $\hat{H}_{\textrm{XXZ}}$ in terms of fermionic fields. After invoking the transformation and reorganising terms, it follows from Eq.~\eqref{eq:h_xxz_pm} that
\begin{align}
\hat{H}^f_{\textrm{XXZ}} = 4 \sum_{i} \left[\frac{\alpha}{2}\left(\hat{c}^{\dagger}_{i}\hat{c}_{i+1} + \hat{c}^{\dagger}_{i+1}\hat{c}_{i}\right) + \Delta\,\left( \hat{n}_i - \frac{1}{2} \right)\left( \hat{n}_{i+1} - \frac{1}{2} \right) \right].
\end{align}
The mapping allows us to identify the first term as a hopping factor in the Hamiltonian or a {\em kinetic} term, that allows fermions to move through the chain. The second factor corresponds to a nearest-neighbour interaction, which introduces an energetic penalty if two fermions are next to each other in the lattice. All the global symmetries introduced in Sec.~\ref{sec:global_symmetries} follow naturally, although they usually show up under different names in the literature, such as particle-hole symmetry for reflection symmetry, momentum conservation for translational invariance and total particle number conservation for the conservation of total $z$-magnetisation. Equivalence between the spin-1/2 chain and the fermionic model is one of the reasons behind the theoretical interest in the anisotropic Heisenberg model, as it allows for different experimental realisations. 

Out of the global symmetries described in Sec.~\ref{sec:global_symmetries}, it is important to remark conservation of total $z$-magnetisation, which translates to conservation of total particle number in the fermionic language $[\hat{H}^f_\textrm{XXZ}, \sum_i \hat{n}_i] = 0$. This expression leads to a continuity equation and a local particle transport operator which follows from Eq.~\eqref{eq:j1_xxz}. In the fermionic language, the particle transport operator can be shown to be\footnote{Either from a continuity equation in the fermionic language, or by a Jordan-Wigner transformation of Eq.~\eqref{eq:j1_xxz}.}
\begin{align}
\hat{j}^f_1 = 4\textrm{i}\alpha \left( \hat{c}^{\dagger}_{i}\hat{c}_{i+1} - \hat{c}_{i+1}^{\dagger}\hat{c}_{i} \right),
\end{align}
which remarks that transport can be studied from the perspective of either physical system, spin-1/2 chains or spinless fermions hopping in a one-dimensional lattice. 

\section{Integrability breaking}
\label{sec:integrability_breaking}

The anisotropic Heisenberg model is one of the quintessential integrable models solved by the Bethe ansatz~\cite{ShastryBethe1990, Cazalilla:2011}. This feature, however, can be broken in different ways by introducing terms into $\hat{H}_{\textrm{XXZ}}$ which destroy the non-trivial conserved quantities associated with integrability. Following our discussion from Chapter~\ref{chapter:integrability}, it is expected that integrable systems in the presence of a significant integrability-breaking perturbation will display normal (diffusive) conduction. A known exception is given by an integrability breaking perturbation induced by weak disorder, which leads to anomalous diffusive behaviour according to numerical evidence~\cite{vznidarivc2016diffusive, vznidarivc2017dephasing, mendoza2018asymmetry, schulz2018energy}, although normal conduction appears to be recovered once the perturbation becomes sufficiently strong.

\begin{figure}[t]
\fontsize{13}{10}\selectfont 
\centering
\includegraphics[width=0.5\columnwidth]{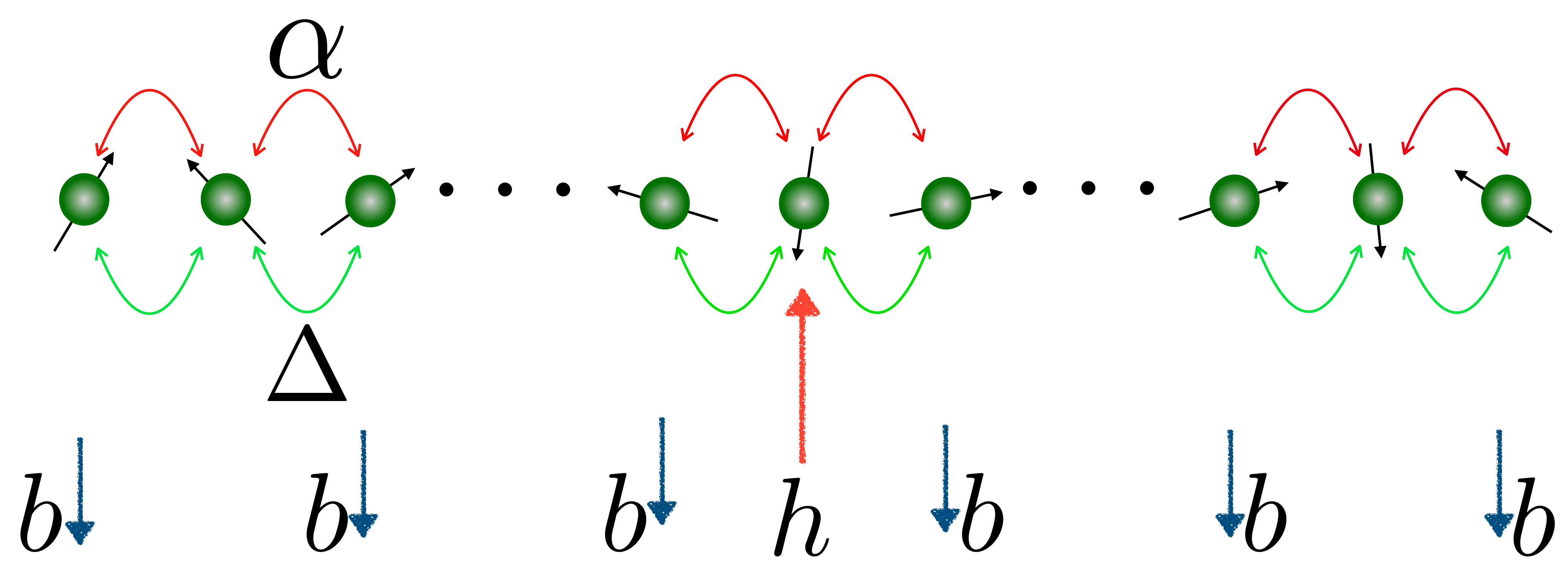}
\caption[Diagrammatic depiction of the single impurity model $\hat{H}_{\textrm{SI}}$ and the staggered field model $\hat{H}_{\textrm{SI}}$]{Diagrammatic depiction of the single impurity model $\hat{H}_{\textrm{SI}}$ and the staggered field model $\hat{H}_{\textrm{SI}}$, both of which break the integrability of the anisotropic Heisenberg model $\hat{H}_{\textrm{XXZ}}$.}
\label{fig:1.3.2}
\end{figure}

In this section we introduce two different models that break the integrability of anisotropic Heisenberg spin-1/2 chain from the perspective of the level spacing statistics, by either adding a staggered magnetic field or a single magnetic perturbation near the centre of a chain with open boundary conditions. The addition of staggered magnetic field constitutes a {\em global} form of integrability breaking perturbation, one for which the perturbation extends over the entire support of the spin chain. On the other hand, as we shall see, a single magnetic impurity located in the vicinity of the centre of the chain breaks the integrability of the Heisenberg chain. This form of integrability breaking is {\em local}, in the sense that it does not scale with the size of the system. A diagrammatic depiction is shown in Fig.~\ref{fig:1.3.2}. The form of these integrability breaking perturbations, as we shall see in Chapter~\ref{chapter:kubo}, has significant consequences at the level of spin transport.

\subsubsection{The staggered field model}

A {\em global} form of integrability breaking is obtained by adding a staggered magnetic field in the $z$ direction to the anisotropic Heisenberg model. The {\em staggered field} model is described by the Hamiltonian
\begin{align}
\label{eq:h_sf_3}
\hat{H}_{\textrm{SF}} = \hat{H}_{\textrm{XXZ}} + b\,\sum_{i\,\textrm{odd}} \hat{\sigma}^z_{i}\,.
\end{align} 
For the purposes of the analysis to follow, we consider spin chains with even number of lattice sites $L$. It is important to remark that a constant magnetic field in the $z$ direction added over the entire support of the chain, does not break integrability and merely shifts the eigenenergies following the direction of the field. 

With respect to the global symmetries introduced in Sec.~\ref{sec:global_symmetries}, the addition of the staggered magnetic field does not break translational invariance or conservation of total $z$-magnetisation, the latter manifest in the form of $[\hat{H}_{\textrm{SF}}, \sum_i \hat{\sigma}^z_i] = 0$. Translational invariance, however, is broken for the open chain. A relevant symmetry related to parity and reflection remains present in the zero magnetisation sub-sector. 

\subsubsection{The single-impurity model}

On the other hand, a {\em local} form of integrability breaking is obtained by the addition of a single magnetic defect around the centre of the chain
\begin{align}
\label{eq:h_si_3}
\hat{H}_{\textrm{SI}} = \hat{H}_{\textrm{XXZ}} + h\, \hat{\sigma}^z_{L/2}\,,
\end{align}
where we have assumed an even number of lattice sites $L$ and the defect is located in the left-most centre of the chain. This model is known in the literature to lead to quantum chaos by integrability breaking~\cite{Santos:2004, santos2011domain, torres2014local, XotosIncoherentSIXXZ}. It is very interesting that a single defect located near the edges of the chain does not break integrability~\cite{Santos:2004}. We refer to $\hat{H}_{\textrm{SI}}$ as the {\em single impurity} model.

Some of the underlying global symmetries present in the anisotropic Heisenberg model are broken by the effect of the impurity. Both parity and reflection symmetries are broken in any total magnetisation $\langle \hat{S}^z \rangle$ sub-sector. Conservation of total $z$-magnetisation remains even with the addition of the single defect, i.e., $[\hat{H}_{\textrm{SI}}, \sum_i \hat{\sigma}^z_i] = 0$, which allows the Hamiltonian to be sub-divided in different total magnetisation blocks. Most interestingly, translational invariance is broken irrespective if the model is defined with periodic or open boundary conditions. This illuminating fact will become very important in Chapter~\ref{chapter:kubo}, when we analyse transport in the model. 

\subsubsection{Level spacing statistics}

To understand if the models described above break the underlying integrability of the spin-1/2 Heisenberg model, we now turn to a level spacing statistics analysis introduced in Sec.~\ref{sec:level_spacing_statistics}.

A cross-over between an integrable system and a non-integrable system, is signalled by the statistical distribution of the spacings $s$ between neighbouring energy levels. Recalling, distributions of level spacings in integrable systems are characterised by Poisson distributions
\begin{align}
P(s) = e^{-s},
\end{align}
while the distributions in non-integrable systems follow a Wigner-Dyson distribution
\begin{align}
P(s) = \frac{\pi s}{2}e^{-\frac{\pi s^2}{4}}.
\end{align} 

\begin{figure}[t]
\fontsize{13}{10}\selectfont 
\centering
\includegraphics[width=0.6\columnwidth]{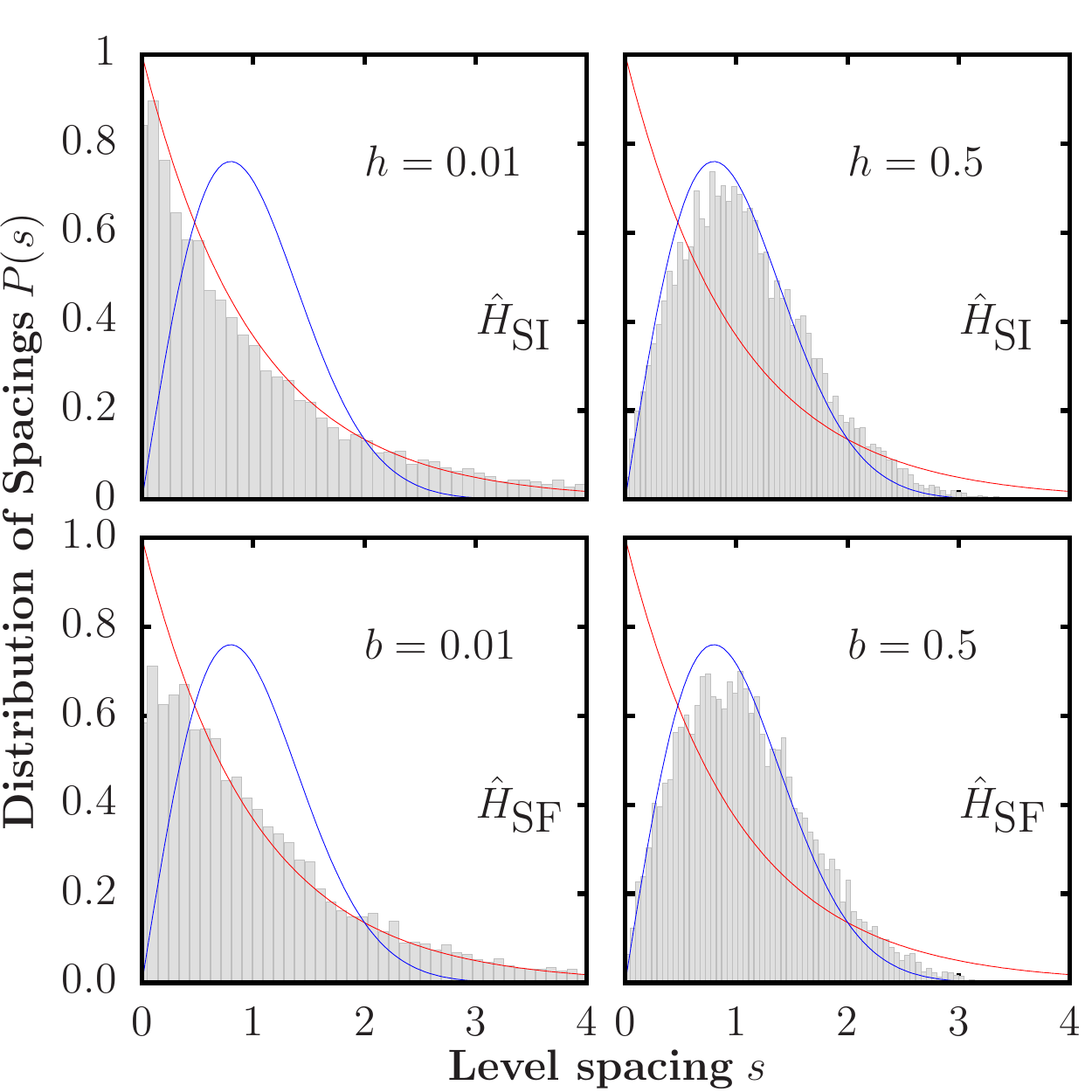}
\caption[Level spacing distribution $P(s)$ for the anisotropic Heisenberg model in the presence of a single magnetic impurity and in the presence of a staggered magnetic field]{Level spacing distribution $P(s)$ for the anisotropic Heisenberg model (top) in the presence of a single magnetic impurity [see Eq.~\eqref{eq:h_si_3}], and (bottom) in the presence of a staggered magnetic field [see Eq.~\eqref{eq:h_sf_3}]. The red line corresponds to a Poisson distribution [Eq.~\eqref{eqn:poisson}], while the blue line depicts a Wigner-Dyson distribution [Eq.~\eqref{eq:wddistro}]. The results shown are for chains with open boundary conditions, $L = 16$, $\Delta = 0.5$, $\sum_{j=1}^{N} \langle \hat{\sigma}^z_j \rangle = 0$, and two values of $h$ and $b$.}
\label{fig:1.3.3}
\end{figure}

In Fig.~\ref{fig:1.3.3}, we show the behaviour of the distribution $P(s_n)$ for both the $\hat{H}_{\textrm{SI}}$ model in Eq.~\eqref{eq:h_si_3}, for different strengths of the impurity, and for the $\hat{H}_{\textrm{SF}}$ model in Eq.~\eqref{eq:h_sf_3}, for different strengths of the staggered field. The calculations were done in the zero magnetisation sector, $\sum_{j=1}^{N} \langle \hat{\sigma}^z_j \rangle = 0$, whose Hilbert space dimension is given by
\begin{align}
\mathcal{D} = \frac{L!}{(L/2)! (L/2)!},
\end{align}
in chains with $L=16$ sites and open boundary conditions. These results confirm that, as previously observed for $\hat{H}_{\textrm{SI}}$~\cite{Santos:2004, santos2011domain, torres2014local, XotosIncoherentSIXXZ} and for $\hat{H}_{\textrm{SF}}$~\cite{JuanThesis:2014,Huang2013}, the level spacing distribution becomes Wigner-Dyson as one increases the magnitude of $h$ and $b$, respectively, without changing $\Delta$ or $L$. 

For the single impurity model, at fixed $\Delta$ and $L$, the probability distribution of energy spacings was shown in Ref.~\cite{XotosIncoherentSIXXZ} to be of the Wigner-Dyson type for a wide range of values of $h$. It was also shown therein that, increasing $L$ at fixed $\Delta$ increases the range of values of $h$ for which quantum chaotic behaviour occurs. As for systems in which integrability is broken by means of global perturbations~\cite{Santos:2010b, Santos:2010}, for $\Delta\neq0$ in the thermodynamic limit one expects quantum chaotic behaviour to occur whenever $h\neq0$ and $h\neq\infty$.

In order to obtain the correct level spacing distribution, an {\em unfolding procedure} of the spectrum needs to be used in which one locally rescales the energies $\lambda_{\alpha}$, so that the local density of states (LDOS) is normalised to 1. The symmetries of the model have to be taken into account as well, given that energy levels from different symmetry sub-sectors (subspaces of the Hilbert space) are independent from each other and therefore uncorrelated \cite{Santos:2010b, Santos:2010}. 

For $\hat{H}_{\textrm{SI}}$, the reflection symmetry of the XXZ model is broken by the impurity, while for $\hat{H}_{\textrm{SF}}$ there is a related remaining symmetry that needs to be resolved. The key point to be emphasised is that both integrability breaking perturbations, a {\em local} one in $\hat{H}_{\textrm{SI}}$ and a {\em global} one in $\hat{H}_{\textrm{SF}}$, lead to the same quantum chaotic behaviour of the level spacing distributions.

Following the discussion from Chapter~\ref{chapter:integrability}, it would be interesting to investigate whether a single magnetic impurity suffices to render the system diffusive, given that it renders the anisotropic Heisenberg model non-integrable. We will explore this question in Chapter~\ref{chapter:kubo} from the perspective of linear response. 

\section{Experimental realisations}
\label{sec:experiments_xxz}

The discussion of this thesis involves theoretical descriptions of thermodynamics and transport in integrable and non-integrable one dimensional latices, however, extraordinary experimental advances have driven the field as well. This section attempts to exemplify these advances by introducing some experimental results in the platforms of ultra-cold atoms and magnetic materials. For extensive recent reviews, we refer the reader to Refs.~\cite{Chien:2015, Krinner:2017} for ultra-cold atom experiments and Ref.~\cite{Hess:2019} for magnetic material experiments.

\subsection{Ultra-cold atoms in optical lattices}

Ultra-cold atoms trapped in optical lattices correspond to one of the most prominent platforms to realise spin Hamiltonians of the form $\hat{H}_{\textrm{XXZ}}$ [Eq.~\eqref{eq:h_xxz}]. Consequently, experiments in these platforms not only allow the degree of tunability and control required to realise short-range interactions, but provide a promising route to realise quantum simulators by means of local control gates that can be implemented~\cite{Bernien:2017}. Technological advances allow to study the unitary dynamics of a given isolated quantum system over timescales long enough such that decoherence effects of the environment are unimportant. Such advances have led to the renewed interest in theoretical investigation of unitary dynamics and transport~\cite{Zelevinsky1996, Polkovnikov:2011, Yukalov:2011, Eisert:2015, Goold:2016, Gogolin:2016, Borgonovi:2016, Alessio:2016}. 

A particularly relevant example is given by the work of Jepsen {\em et al.}, in which $^7$Li atoms are trapped in optical lattices to realise the anisotropic Heisenberg Hamiltonian, with a tunable anisotropy parameter $\Delta$ made possible by applied magnetic fields in a given direction. The lattice depth can be used to dictate the spin-exchange term in the Hamiltonian [the first term in Eq.~\eqref{eq:h_xxz}], which is the term responsible for the transport of spin excitations in the $z$ direction. Remarkably, the authors in Ref.~\cite{Jepsen:2020} were able to tune the anisotropy parameter $\Delta$ over a wider range of values than the works that pre-dated theirs. The initial state is relevant to the dynamics of the system. In the experiment, a spin helix was engineered. The first step is to realise such state, each local spin in the lattice pointing in a helicoidal direction in the $x$ and $y$ directions, after which, unitary dynamics are observed from the propagator defined from $\exp({-\textrm{i}\hat{H}_{\textrm{XXZ}}t})$. This corresponds to a quench experiment, in which an eigenstate of a different Hamiltonian is evolved under the dynamics dictated by $\hat{H}_{\textrm{XXZ}}$. To study transport, the authors then evaluate the decay timescale $\tau$ as a function of a modulation lengthscale $\lambda$, which is a free parameter of the initial state. In practice, this procedure amounts to studying transport from the perspective of the decay of an initial state under the unitary dynamics dictated by the Hamiltonian, a procedure which is often used numerically as well (see Secs.~IV and IX in Ref.~\cite{Bertini:2021}).

\begin{figure}[t]
\fontsize{13}{10}\selectfont 
\centering
\includegraphics[width=0.7\columnwidth]{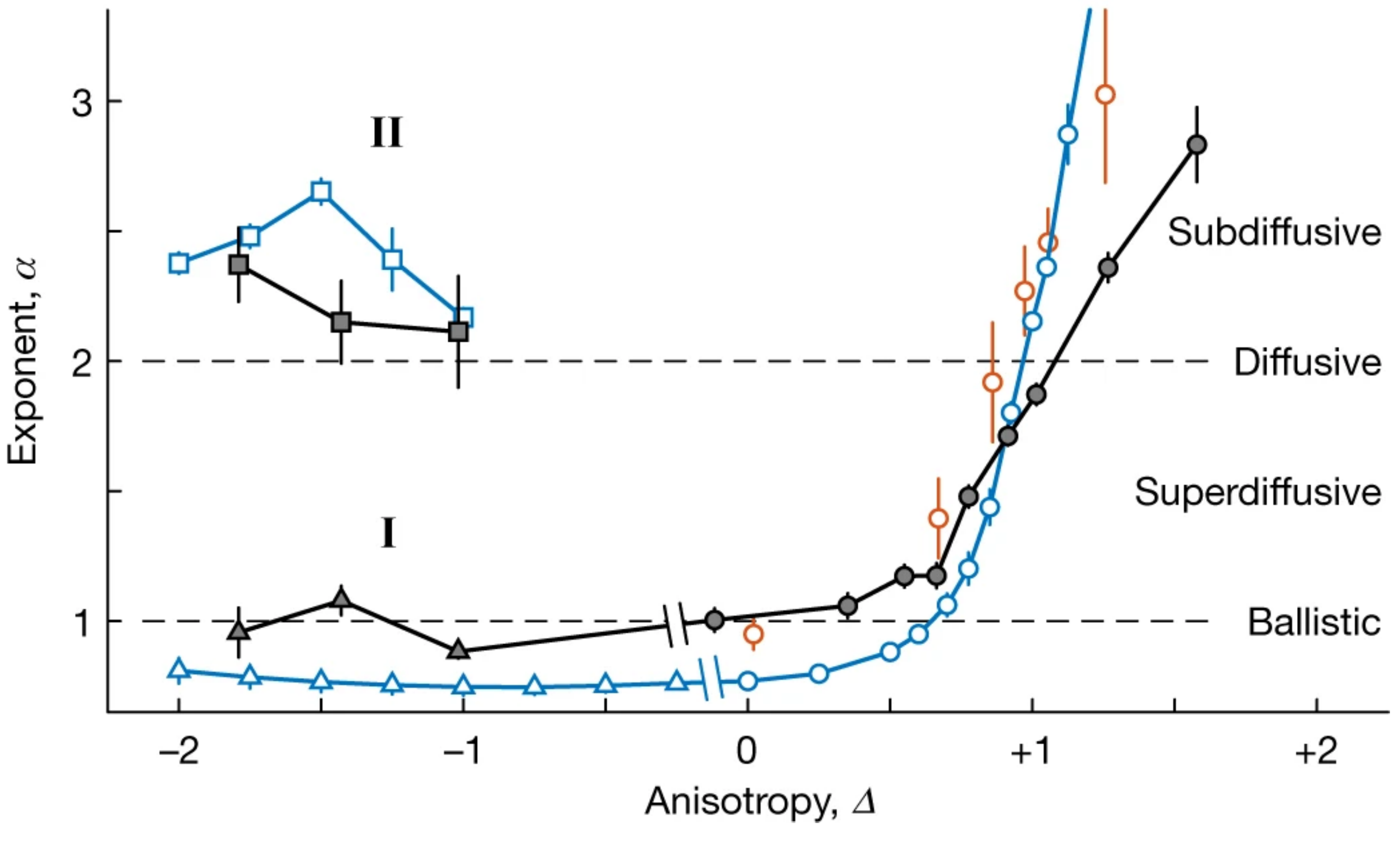}
\caption[Power-law transport exponent $\alpha$ as a function of the anisotropy parameter $\Delta$ in the anisotropic Heisenberg model realised with ultra-cold atoms.]{Power-law transport exponent $\alpha$ as a function of the anisotropy parameter $\Delta$ in the anisotropic Heisenberg model realised with ultra-cold atoms. Filled symbols correspond to transport exponents fitted from experimental results, while open symbols correspond to theoretical predictions. Reprinted by permission from Springer Nature Customer Service Centre GmbH: Springer Nature Ref.~\cite{Jepsen:2020}, Copyright Springer Nature (2020).}
\label{fig:1.3.4}
\end{figure}

The central results of the experiment can be summarised in Fig.~\ref{fig:1.3.4}, in which the quench dynamics illustrated before gives rise to an effective infinite-temperature transport regime. The decay timescale $\tau$ of the engineered helicoidal state is related to the modulation lengthscale $\lambda$. Ballistic transport is manifest in a linear dependence $\tau \propto \lambda$, while diffusion scales quadratically $\tau \propto \lambda^2$. Anomalous behaviour is signalled by a different exponent $\tau \propto \lambda^{\alpha}$. In Fig.~\ref{fig:1.3.4}, the transport exponent is shown as a function of the anisotropy parameter $\Delta$. Specifically, the curves denoted in the I section of Fig.~\ref{fig:1.3.4} correspond to: theoretical predictions from quench procedures in the $\hat{H}_{\textrm{XXZ}}$ model evaluated numerically denoted with blue open symbols, theoretical predictions on a relevant fermionic model with high particle/lattice occupation (see Sec.~\ref{sec:jordan_wigner}) and experimental results with filled symbols. Section II of Fig.~\ref{fig:1.3.4} denotes the long-time behaviour of the decay of the initial state, pointing towards diffusion for $\Delta < 0$ (the coupling strength, first term in Eq.\eqref{eq:h_xxz} is fixed throughout the experiment). The authors expose that spin transport points towards ballistic regimes for $\Delta < 0$ at short times, and diffusion for long times. Furthermore, sub-diffusion is observed for $\Delta > 1$. This transport behaviour has not yet been identified numerically or theoretically, except in the case of quench dynamics under disordered Hamiltonians~\cite{vznidarivc2017dephasing,mendoza2018asymmetry}. The authors then argue that this behaviour could be related to the far-from-equilibrium configuration achieved with the helicoidal initial states. 

These exciting results validate some previous theoretical studies and give rise to open questions at the theoretical level, particularly when far-from-equilibrium dynamics and finite-temperatures are at play. The discussion of Part~\ref{part:two} of this thesis is most relevant to addressing these open questions. 

\subsection{Magnetic materials}

Several magnetic materials can be modelled according to the Hamiltonian $\hat{H}_{\textrm{XXZ}}$~\cite{Hess:2019}, although, as opposed to experiments in ultra-cold atomic platforms, the anisotropy $\Delta$ is difficult to tune. Spin chains composed of Sr$_2$CuO$_3$ have been realised experimentally and modelled according to the isotropic Heisenberg chain, i.e., $\Delta = \alpha$ in Eq.~\eqref{eq:h_xxz}. The crystalline structure associated with antiferromagnetic Sr$_2$CuO$_3$ materials can be observed to be arranged in interconnected longitudinal chains~\cite{Ami:1995}, in which interactions in the longitudinal direction are favoured with respect to transversal interactions~\cite{Hess:2019}. This structure is associated with quasi-one-dimensional behaviour, since interactions along a given direction are suppressed due to electronic structure. The same behaviour can be observed experimentally in other magnetic materials modelled by spin-1/2 quasi-one-dimensional chains, such as CaCu$_2$O$_3$~\cite{Kiryukhin:2001}. 
 
Experiments in cuprate materials have reported large thermal conductivities~\cite{Sologubenko:2001}, which support the theoretical integrability picture described in Sec.~\ref{sec:continuity_transport} that energy transport behaves ballistically and thermal conductivities diverge. Such ballistic heat transport has been observed in spin chain compounds Sr$_2$CuO$_3$ and SrCuO$_2$ and are considered to be excellent experimental realisations of the spin-1/2 Heisenberg model~\cite{Sologubenko:2000}. However, thermal conductivity has also been shown to be extremely sensible to chemical impurities~\cite{Hess:2019}. Furthermore, in most applications, the effect of disorder and phononic contributions are typically too strong to be neglected~\cite{Rozhkov:2005}, leading to finite thermal conductivities and, moreover, the actual measurement procedure of thermal transport makes the introduction of spin-phonon coupling effects necessary in the models. Therefore, establishing a connection between integrability and the observed thermal conductivities is more complicated~\cite{Bertini:2021}. Understanding this phenomenology then, requires more complicated theoretical models that account for phonon interactions. 

Spin transport measurement and analysis is more complicated in magnetic material platforms than their counterpart in ultra-cold atoms. Indirect measurements have been carried out from nuclear magnetic resonance (NMR) which gives information about the decay of spin-spin correlation functions~\cite{Thurber:2001,Kuhne:2009}, leading to diffusive relaxation in cuprate materials. Recent experiments using muon spin-resonance techniques have reported both ballistic and diffusive spin dynamics in aqueous pyrimidine materials~\cite{Huddart:2021,Bertini:2021}, which support the theoretical picture that spin dynamics strongly depends on the anisotropy parameter $\Delta$ as described in Sec.~\ref{sec:continuity_transport}. 
\chapter{Spin transport in the single impurity model}
\label{chapter:kubo}

In Chapter~\ref{chapter:models}, we introduced the anisotropic Heisenberg model and discussed two different forms of perturbations that break integrability.  A global form of integrability breaking given by the staggered field model
\begin{align}
\label{eq:h_sf}
\hat{H}_{\textrm{SF}} = \hat{H}_{\textrm{XXZ}} + b\,\sum_{i\,\textrm{odd}} \hat{\sigma}^z_{i}\,,
\end{align} 
and a local perturbation in the single impurity model
\begin{align}
\label{eq:h_si}
\hat{H}_{\textrm{SI}} = \hat{H}_{\textrm{XXZ}} + h\, \hat{\sigma}^z_{L/2}\,.
\end{align}
The statistics of level spacings in both models follow the Wigner-Dyson distributions found in non-integrable systems, as shown in Sec.~\ref{sec:integrability_breaking}. Following our discussion from Chapter~\ref{chapter:integrability}, one would be tempted to make the assumption that spin transport has to be incoherent (normal conduction, i.e., diffusion) in both models. Since both of them are non-integrable, one can then invoke Mazur's inequality in Eq.~\eqref{eq:mazur}, which would yield a vanishing lower bound for the spin Drude weight for a system without non-trivial conserved quantities, pointing towards incoherent transport.

Compelling numerical evidence indicates that the staggered field model displays normal spin conduction~\cite{Prosen:2009,JuanThesis:2014} for a sufficiently strong magnetic field $b$, even in the regime where $\alpha = 1$ and $|\Delta| < 1$, in which the unperturbed Hamiltonian displays ballistic transport as discussed in Chapter~\ref{chapter:models}. We will revisit these results from the perspective of open quantum systems in Part~\ref{part:two} of this thesis. 

Given that both the single impurity model and the staggered field model appear to be non-integrable from the perspective of level spacing statistics, this chapter is driven by the following question:
\begin{itemize}
	\item{Is a {\em local} perturbation enough to destroy ballistic spin conduction and render transport incoherent?}
\end{itemize}
This is a natural question to ask, particularly in light of recent results that highlight the non-integrability of the single impurity model from the perspective of adiabatic deformations~\cite{Pandey:2020}, asides from the Wigner-Dyson distributions of level spacings exposed in Sec.~\ref{sec:integrability_breaking}. One could think about tuning between ballistic and an incoherent spin transport from local operations routinely realised in, for instance, ultra-cold atoms experiments if a single perturbation were enough to destroy coherence.

In this chapter, we focus on the regime $\alpha = 1$ and $\Delta = 0.5$ [Eq.~\eqref{eq:h_xxz}] in the zero-magnetisation sub-sector, for which the XXZ model is known to display ballistic transport. We then introduce a single magnetic impurity $h = 0.5$ and study spin transport from the perspective of linear response as exposed in Chapter~\ref{chapter:integrability}, in the high-temperature regime. In Sec.~\ref{sec:lin_response_trans} we introduce the importance of considering the breaking of translational invariance symmetry in the evaluation of the spin conductivity. We then analyse spin conductivity for non-interacting systems in Sec.~\ref{sec:non_interacting_regime_impurity} and for interacting systems in Sec.~\ref{sec:interacting_regime_impurity}. We finalise with a discussion in Sec.~\ref{sec:discussion_4}. Our results for the single impurity model will be solidified in Part~\ref{part:two} of this thesis, when we revisit the model from the perspective of open quantum systems.

\section{Linear response theory and translational invariance}
\label{sec:lin_response_trans}

Within linear response theory, the real part of the conductivity can be written as ($\hbar = 1$ and $k_B = 1$) \cite{kubo1957statistical, kubo1957statistical2, ShastryKubo2008, RigolShastry2008}
\begin{align}
\label{eq:kuboformula}
\textrm{Re}[\sigma_L(\omega)] = \pi D_L\delta(\omega) + \frac{\pi}{L}\left(\frac{1-e^{-\beta \omega}}{\omega}\right)\sum_{\epsilon_n \neq \epsilon_m} p_n|J_{nm}|^2\delta(\epsilon_m - \epsilon_n - \omega),
\end{align}
where $D_L$ is the Drude weight or spin stiffness at finite size $L$, $\beta$ is the inverse temperature, $p_n= e^{-\beta\epsilon_n} / Z$ is the Boltzmann weight of eigenstate $\ket{n}$ with energy $\epsilon_n$, and $Z$ is the partition function. $J_{nm} \defeq \braket{n | \hat{J} | m}$ are the matrix elements of the {\em total} spin current operator in the energy eigenbasis. In the language of Chapter~\ref{chapter:models}
\begin{align}
\label{eq:totalcurrent_4}
\hat{J} = \hat{J}_1 = \sum_i \hat{j}^{\textrm{P}}_i,
\end{align}
with the sum adjusted properly depending on whether the system has periodic or open boundary conditions. Here, $\hat{j}_i^{\textrm{P}}$ is the local spin current operator from Eq.~\eqref{eq:j1_xxz},
\begin{align}
\label{eq:spincurrent}
\hat{j}^{\textrm{P}}_i \defeq 2\alpha \left( \hat{\sigma}^x_i \hat{\sigma}^y_{i+1} - \hat{\sigma}^y_i \hat{\sigma}^x_{i+1} \right).
\end{align} 
The Kubo formula in Eq.~\eqref{eq:kuboformula} follows from the zero- and finite-frequency contributions of the correlation function $\langle \hat{J}_1(t) \hat{J}_1(0) \rangle$, related to direct spin transport from linear response in the frequency domain. The expectation value, in this case, is taken on the canonical ensemble.

The Drude weight can be calculated using the expression
\begin{align}
\label{eq:drude1}
D_L = \frac{1}{L}\left[\langle -\hat{\Gamma} \rangle - \sum_{\epsilon_n \neq \epsilon_m} \frac{p_n - p_m}{\epsilon_m - \epsilon_n}|J_{nm}|^2\right],
\end{align}
where $\hat{\Gamma}$ is the so-called stress tensor operator~\cite{ShastrySum2006}, which is identical to the kinetic energy operator 
\begin{align}
\label{eq:kinetic_operator}
\hat{T} = \sum_i \alpha\,\left( \hat{\sigma}^x_{i}\hat{\sigma}^x_{i+1} + \hat{\sigma}^y_{i}\hat{\sigma}^y_{i+1} \right),
\end{align}
in both the $\hat{H}_{\textrm{XXZ}}$ and $\hat{H}_{\textrm{SI}}$ models from Eqs.~\eqref{eq:h_xxz} and \eqref{eq:h_xxz}, respectively. In one dimension and for sufficiently high temperatures (in the absence of superconductivity \cite{ShastryKubo2008, RigolShastry2008}), the Drude weight can also be obtained using the expression~\cite{Heidrich-Meisner03}
\begin{align}
\label{eq:drude2}
\bar{D}_L = \frac{\beta}{L}\sum_{\epsilon_n = \epsilon_m} p_n|J_{nm}|^2.
\end{align}

In the thermodynamic limit, Eq.~\eqref{eq:kuboformula} leads to the decomposition
\begin{align}
\textrm{Re}[\sigma_{\infty}(\omega)] = \pi D_{\infty}\delta(\omega) + \sigma_{\textrm{reg}}(\omega),
\end{align}
where $D_{\infty} = \lim_{L \rightarrow \infty}D_L = \lim_{L \rightarrow \infty}\bar{D}_L$ and $\sigma_{\textrm{reg}}(\omega)$ is the regular part of the conductivity. A non-zero $D_{\infty}$ signals that transport is ballistic, the current-current correlation function does not vanish in the limit of infinite time. This is a property of integrable systems as highlighted in Chapter~\ref{chapter:integrability}. In systems that display diffusive transport, expected for non-integrable systems, $D_{\infty}=0$. 

Equations~\eqref{eq:kuboformula} through~\eqref{eq:drude2} are usually evaluated in systems with translation invariance. In systems with open boundary conditions, where translational invariance is not present, obtaining $D_{\infty}$ is subtle. In such systems, the position operator
\begin{align}
\label{eq:position}
\hat{X} \defeq \sum_k k\,\hat{\sigma}^{+}_k\hat{\sigma}^{-}_k
\end{align}
is well-defined \cite{ShastrySum2006}.  The position operator can be used to define the total current operator as 
\begin{align}
\hat{J} = \textrm{i}[\hat{X}, \hat{H}],
\end{align}
and the stress tensor operator as~\cite{RigolShastry2008}
\begin{align}
\hat{\Gamma} = -\textrm{i}[\hat{X}, \hat{J}].
\end{align}
If one uses these relations to evaluate the matrix elements of the total current operator, one finds that
\begin{align}
J_{nm} = \textrm{i}\braket{n | \hat{X}\hat{H} | m} - \textrm{i}\braket{n | \hat{H}\hat{X} | m} = \textrm{i}(\epsilon_m - \epsilon_n)\braket{n | \hat{X} | m},
\end{align}
which implies that $\bar{D}_{L}$ from Eq.~\eqref{eq:drude2} is exactly zero, since the sum only counts terms for which $\epsilon_n = \epsilon_m$. Furthermore, this also implies that $D_L$ in Eq.~\eqref{eq:drude1} is exactly zero, since both quantities are equivalent in the thermodynamic limit~\cite{RigolShastry2008}.

This implies that in systems with open boundary conditions, lack of translational invariance implies
\begin{align}
\lim_{L \rightarrow \infty}D_L = \lim_{L \rightarrow \infty}\bar{D}_L=0
\end{align} 
irrespective of whether the system is integrable or not, in disagreement with what is known for translationally-invariant systems. 

Such a disagreement may lead one to question whether the Drude weight obtained from this picture [Eqs.~\eqref{eq:kuboformula}--\eqref{eq:drude2}] is a meaningful thermodynamic quantity. The fact that it is was argued for in Ref.~\cite{RigolShastry2008}.

A central finding of Ref.~\cite{RigolShastry2008} is that, in order to obtain $D_{\infty}\neq0$ in integrable systems with open boundary conditions in which translational invariance is broken, and conciliate the result with the one obtained in translationally-invariant systems with periodic boundary conditions, one needs to study the behaviour of the finite frequency part of the Kubo formula [the second term in Eq.~\eqref{eq:kuboformula}]. In the thermodynamic limit, a peak develops at zero frequency from the collapse of peaks located at finite (size-dependent) frequencies in finite-size systems.

The main point of our previous discussion is that the single impurity model, $\hat{H}_{\textrm{SI}}$, breaks the translational invariance of the XXZ model $\hat{H}_{\textrm{XXZ}}$ irrespective if one considers open or periodic boundary conditions. This is due to the fact that the introduction of the perturbation allows one to physically allocate a position for the perturbation and appropriately define a position operator. Furthermore, an impurity with a very strong field ($h\rightarrow\infty$) is equivalent to considering open boundary conditions. Moreover, in the non-interacting limit ($\Delta=0$) for which transport must be ballistic, the presence of the impurity breaks the $k,-k$ degeneracy in the single-particle spectrum resulting in $D_L = \bar{D}_L=0$. The latter remains true for $\Delta\neq0$.

\subsubsection{Numerical procedure}

Since $D_L = \bar{D}_L=0$ in systems without translation invariance, one needs to directly evaluate the behaviour of the finite-frequency contribution of the real part of the spin conductivity to infer transport regimes.

From the second term in Eq.~\eqref{eq:kuboformula}, it can be observed that this contribution depends on the matrix elements of the total spin current operator in the energy eigenbasis $J_{nm}$ and on the Boltzmann weights of the eigenstates at a given inverse temperature $\beta$. A simple, yet costly, numerical procedure known as exact diagonalisation can be used to evaluate the finite-frequency term in the conductivity. It suffices to use a computational basis representation introduced in Chapter~\ref{chapter:models} to represent the Hamiltonian as matrix operator. One then needs to find the transformation $\hat{U}$ that renders the Hamiltonian diagonal
\begin{align}
\tilde{H} = \hat{U}^{\dagger} \hat{H} \hat{U},
\end{align}
with eigenvalues $\epsilon_n$ as diagonal matrix elements. This allows one to express $p_n = e^{-\beta \epsilon_n} / \sum_n (\epsilon_n e^{-\beta \epsilon_n})$ and $J_{nm} = \braket{n | \hat{U}^{\dagger} \hat{J} \hat{U} | m}$, where $\hat{J}$ is the total current operator from Eq.~\eqref{eq:totalcurrent_4} written in the computational basis. To evaluate the finite-frequency contribution of the conductivity, one then proceeds to create a histogram using frequency bins of a given width to yield a finite-frequency signal of the conductivity, following Eq.~\eqref{eq:kuboformula}. This procedure is very costly since the dimension of the Hilbert space $\mathcal{D}$ increases exponentially with the system size, allowing one to study only moderate system sizes even in one dimension.  

\section{The non-interacting regime: $\Delta = 0$}
\label{sec:non_interacting_regime_impurity}

We begin our analysis of the spin conductivity for the non-interacting case, i.e., the case for  which $\Delta = 0$ in the XXZ model in Eq~\eqref{eq:h_xxz} and the single impurity model in Eq.~\eqref{eq:h_si}. We refer to the non-interacting case of the XXZ model as the XX model, with Hamiltonian $
\hat{H}_{\textrm{XX}} = [\hat{H}_{\textrm{XXZ}}]_{\Delta = 0}$. 

The real part of the conductivity in Eq.~\eqref{eq:kuboformula} satisfies the following sum rule~\cite{Pottier:2010,RigolShastry2008}
\begin{align}
\label{eq:sumrule1}
\int^{\infty}_0\textrm{Re}[\sigma(\omega)]d\omega = \frac{\pi \langle -\hat{T} \rangle}{2L},
\end{align}
where $\hat{T}$ is the kinetic energy operator in Eq.~\eqref{eq:kinetic_operator}. It is quite useful to consider the above sum rule for the analysis of the conductivity. Following our previous discussion, lack of translational invariance implies that, in the thermodynamic limit, $\lim_{L \rightarrow \infty}D_L = \lim_{L \rightarrow \infty}\bar{D}_L=0$. It follows that the sum rule is fully accounted for in the finite-frequency regime of the conductivity, i.e., the second term in Eq.~\eqref{eq:kuboformula}. In translationally-invariant systems, however, the sum rule comes from both the zero- and finite-frequency contributions in general.  

For translational-invariant models in the non-interacting regime, such as the $\hat{H}_{\textrm{XX}}$ model, the properties of the total current operator $\hat{J}$ [Eq.~\eqref{eq:totalcurrent_4}] can be calculated analytically. In the free-fermion representation~\cite{Cazalilla:2011}, the eigenstates of the single-particle Hamiltonian are plane waves
\begin{align}
\label{eq:planewaves}
\ket{m} = \frac{1}{\sqrt{L}}\sum_{j}e^{\textrm{i}k_mj}c^{\dagger}_j\ket{0},
\end{align}
where $\ket{m}$ is the $m$-th eigenstate, with energy $\epsilon_m = -4\alpha\cos{(k_m)}$, $c^{\dagger}_j$ is the fermionic creation operator on site $j$, $\ket{0}$ is the vacuum state, and $k_m = 2\pi m / L$ with $m = -L/2 + 1, \cdots, L/2$. From this, the matrix elements of the total current operator are given by 
\begin{align}
|J_{nm}|^{2} = [4\alpha\sin{(k_m)}]^2\delta_{nm},
\end{align}
i.e., the total current operator is diagonal in the energy eigenbasis. This implies that the second term in Eq.~\eqref{eq:drude1} is zero, and we obtain $D_L / (\langle -\hat{T} \rangle / L) = 1$ for any value of $L$.

On the other hand, as discussed in Sec.~\ref{sec:lin_response_trans}, chains with open boundary conditions have $D_L=\bar D_L=0$ irrespective of the presence or absence of interactions~\cite{RigolShastry2008}. Remarkably, $D_L=\bar D_L=0$ for the single impurity model in the non-interacting limit even in systems with periodic boundary conditions. This is the case because the impurity breaks the degeneracies between the single-particle $k$ and $-k$ eigenkets present in the translationally invariant case. Since the non-interacting limits of the XXZ and single impurity models are trivially integrable and must exhibit coherent transport, it is already apparent in this limit that the finite frequency part of Eq.~\eqref{eq:kuboformula} needs to be studied to compute the Drude weight~\cite{RigolShastry2008}.

\begin{figure}[t]
\fontsize{13}{10}\selectfont 
\centering
\includegraphics[width=0.9\columnwidth]{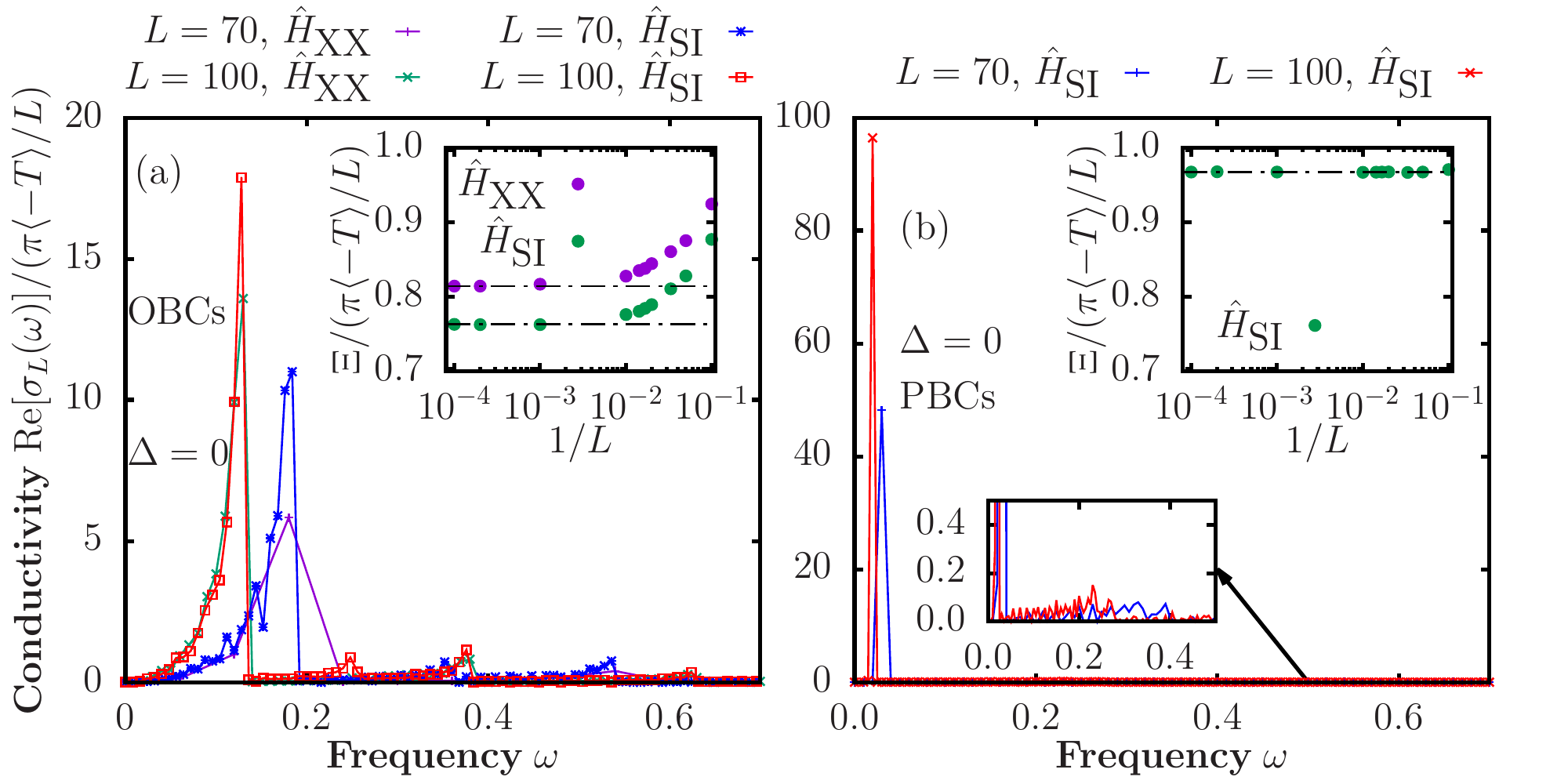}
\caption[Finite frequency part of the conductivity in non-interacting systems ($\Delta=0$)]{Finite frequency part of the conductivity $\textrm{Re}[\sigma_L(\omega)]$ in non-interacting systems ($\Delta=0$). (a) Non-interacting limits of the $\hat{H}_{\textrm{XXZ}}$ and $\hat{H}_{\textrm{SI}}$ models with open boundary conditions. The inset in (a) shows the weight of the lowest frequency peak as a function of the system size. (b) Non-interacting limit of the $\hat{H}_\text{SI}$ model with periodic boundary conditions. The top inset in (b) shows the weight of the lowest frequency peak as a function of the system size. The results were obtained at very high temperature $\beta = 0.001$.}
\label{fig:1.4.1}
\end{figure}

In Fig.~\ref{fig:1.4.1}(a), we show the finite-frequency part of the conductivity in the non-interacting limit of the $\hat{H}_{\textrm{XXZ}}$ and $\hat{H}_{\textrm{SI}}$ models with open boundary conditions. In the non-interacting regime, the numerical procedure described in Sec.~\ref{sec:lin_response_trans} can be used with linear scaling with the system size to evaluate the conductivity for these models, by using a free-fermion representation following the Jordan-Wigner transformation from Sec.~\ref{sec:jordan_wigner} in the single-particle sub-sector.

In all cases, the conductivity is normalised by the sum rule. Since $D_{L} = 0$ in both cases, the sum rule in Eq.~\eqref{eq:sumrule1} is fully accounted for by the finite-frequency part of the conductivity. Figure~\ref{fig:1.4.1}(a) shows that, with increasing system size in both models, the peaks present at finite frequency move toward $\omega=0$ (their frequency is $\omega \propto 1/L$~\cite{RigolShastry2008}) and become sharper. The weight of the peaks converge to a non-vanishing and size-independent value with increasing system size. The inset in Fig.~\ref{fig:1.4.1}(a) shows the weight $\Xi$ as a function of system size. $\Xi$ is defined as the area limited by the lowest-frequency peak, located at $\omega \approx 4\pi/L$. The weight $\Xi$ is two times the area under the lowest-frequency peak, since 
\begin{align}
\int^{\infty}_0\textrm{Re}[\sigma(\omega)]d\omega &= \frac{\pi \langle -\hat{T} \rangle}{2L} \nonumber \\
\implies &\frac{L}{\pi \langle -\hat{T} \rangle} \int^{\infty}_0\textrm{Re}[\sigma(\omega)]d\omega = \frac{1}{2},
\end{align}
These results show that, in the thermodynamic limit, the systems develop a peak at $\omega = 0$ stemming from the collapse of peaks present at finite frequencies in finite systems. The weight of such a zero-frequency peak in systems with open boundary conditions is exactly the Drude weight predicted in systems with periodic boundary conditions~\cite{RigolShastry2008}.

Figure~\ref{fig:1.4.1}(b), and its bottom inset, show the finite-frequency part of the conductivity in the non-interacting limit of the $\hat{H}_{\textrm{SI}}$ model with periodic boundary conditions. The top inset in Fig.~\ref{fig:1.4.1}(b) shows the scaling of the weight of the lowest frequency peak as a function of system size. The same conclusions drawn for chains with open boundary conditions apply for chains with periodic boundary conditions. The lowest frequency peak, however, is much closer to $\omega=0$ and is much sharper in chains with periodic boundary conditions. Furthermore, the weight of the lowest frequency peak is higher for periodic boundary conditions [see the top inset in Fig.~\ref{fig:1.4.1}(b) vs the inset in Fig.~\ref{fig:1.4.1}(a)]. In the thermodynamic limit, the lowest frequency peak almost accounts for the Drude weight in chains with periodic boundary conditions. 

The results for non-interacting systems discussed here, given the trivial nature of their coherent transport, highlight the subtleties discussed in Sec.~\ref{sec:lin_response_trans} when dealing with Kubo's linear response theory in systems without translational invariance. One needs to study the finite-frequency response in such systems in order to be able to determine whether transport is coherent or incoherent.

\section{The interacting regime $\Delta \neq 0$}
\label{sec:interacting_regime_impurity}

We now turn to the evaluation of the spin conductivity in the interacting regime. For these calculations we focus on the parameters $\alpha = 1$, $\Delta = 0.5$ and $h = 0.5$. In this parameter range, as exposed in Chapter~\ref{chapter:models}, the XXZ model displays ballistic transport. We then address the linear response conductivity in the presence of a single impurity, to understand if the local perturbation is enough to render transport incoherent in the high-temperature regime.

We compute the finite-frequency part of Eq.~\eqref{eq:kuboformula} within the grand-canonical ensemble (at zero chemical potential), for which finite-size effects are expected to be the smallest in the presence of translational invariance~\cite{iyer_srednicki_15}. We only study chains with an even number of lattice sites given the known presence of strong even-odd effects at high temperature~\cite{BrenigTypicality2015}.  Since we are interested in the high temperature regime (we take $\beta = 0.001$ in all our calculations), the calculation requires the evaluation of all the eigen-energies and eigen-vectors of the Hamiltonian. This is achieved using full exact diagonalisation, for which the accessible system sizes with our computational resources are $L\lesssim18$.

\begin{figure}[t]
\fontsize{13}{10}\selectfont 
\centering
\includegraphics[width=0.9\columnwidth]{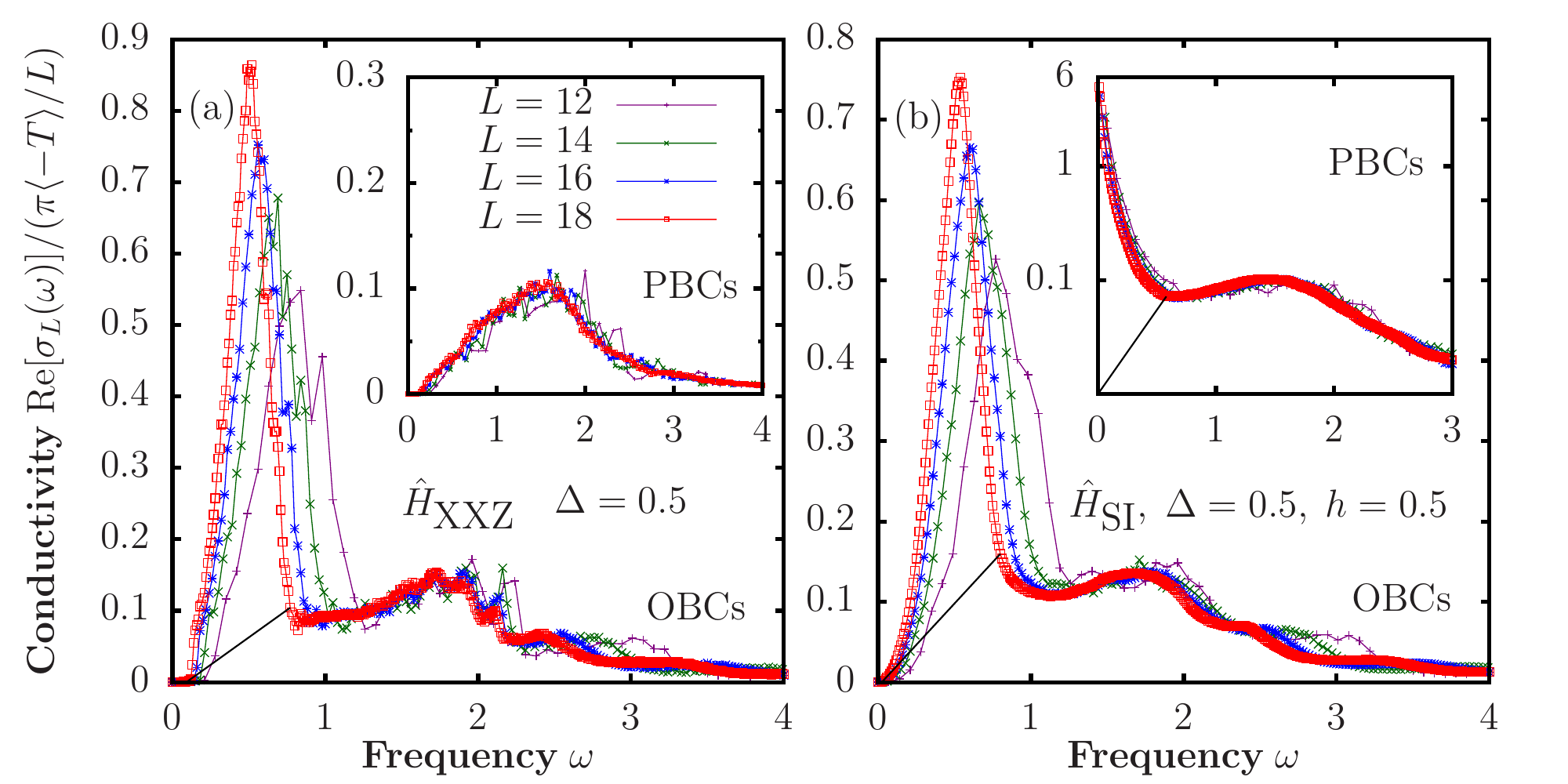}
\caption[Finite frequency part of the conductivity in interacting systems ($\Delta=0.5$)]{Finite-frequency part of the spin conductivity [the second term in Eq.~\eqref{eq:kuboformula}]. (a) Integrable $\hat{H}_{\textrm{XXZ}}$ model at $\Delta = 0.5$, in chains with (main panel) open boundary conditions and (inset) periodic boundary conditions. (b) Single impurity model $\hat{H}_{\textrm{SI}}$, for $\Delta = 0.5$ and $h=0.5$, in chains with (main panel) open boundary conditions and (inset) periodic boundary conditions (linear-log scale). The results were obtained at very high temperature $\beta = 0.001$. The straight lines in the main panels and in the inset in (b), shown only for $L=18$, are approximate delimiters for the bottom of the large low-frequency peak as suggested by the smooth curves in the inset in (a).}
\label{fig:1.4.2}
\end{figure}

In Fig.~\ref{fig:1.4.2}(a) and its inset, we show the finite-frequency part of the conductivity for XXZ chains with open and periodic boundary conditions, respectively. A binning procedure was used in order to obtain smooth curves. The size of the frequency bins is selected to be large enough so that the bins contain a large enough number of the discrete frequencies of the system, but small enough so that the results are robust against changes of the bin size. In our simulations, we used bin sizes of 0.001-0.1 depending on the dimension of the Hilbert space for each magnetisation sub-sector. The curves are normalised to satisfy the sum rule Eq.~\eqref{eq:sumrule1}, so that the area under the curves is $1 / 2$.

\begin{figure}[t]
\fontsize{13}{10}\selectfont 
\centering
\includegraphics[width=0.9\columnwidth]{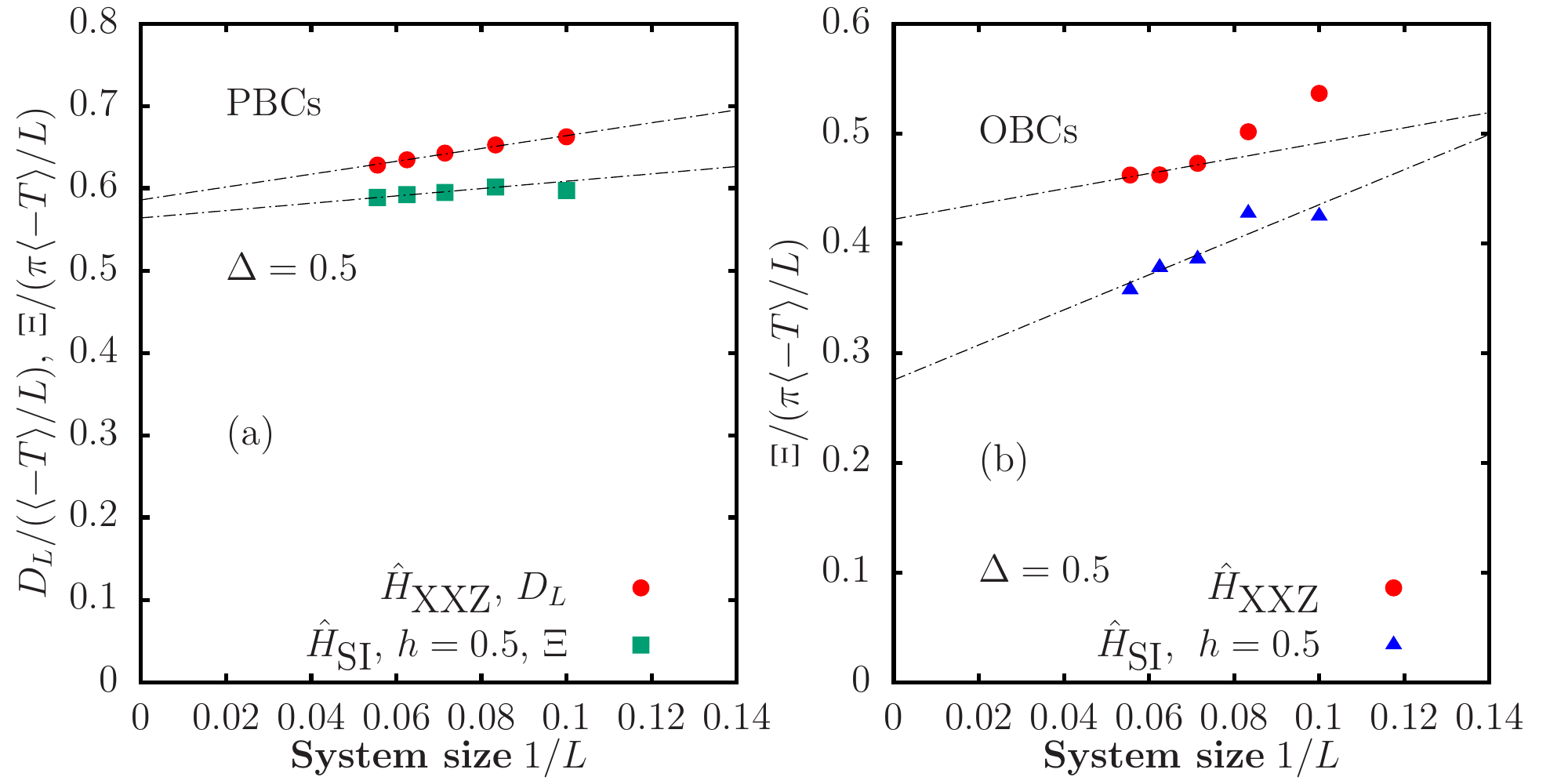}
\caption[Finite-size scaling analysis of the Drude weight and of the weight of the lowest frequency peak $\Xi$ in interacting systems]{Finite-size scaling analysis (up to $L = 18$) of (a) the Drude weight for the $\hat{H}_{\textrm{XXZ}}$ model and the weight of the lowest frequency peak $\Xi$ for the $\hat{H}_{\textrm{SI}}$ model in chains with periodic boundary conditions, and (b) the weight of the lowest frequency peak $\Xi$ for the $\hat{H}_{\textrm{XXZ}}$ and $\hat{H}_{\textrm{SI}}$ models in chains with open boundary conditions. All results were obtained at very high temperature $\beta = 0.001$.}
\label{fig:1.4.3}
\end{figure}

The main panel and the inset in Fig.~\ref{fig:1.4.2}(a) show that there is a stark contrast between the finite-frequency part of $\textrm{Re}[\sigma_L(\omega)]$ in the integrable XXZ model depending of whether the chains have open or periodic boundary conditions (see also Fig.~1 in Ref.~\cite{RigolShastry2008}). For periodic boundary conditions, the finite-frequency part exhibits a smooth behaviour that is nearly size-independent. The Drude weight in that case, shown in Fig~\ref{fig:1.4.3}(a), extrapolates to a non-zero value in the thermodynamic limit. 

For open boundary conditions, a large sharp peak can be seen at low frequencies (smaller sharp peaks occur at higher frequencies) on top of an otherwise smooth part that resembles that of the system with periodic boundary conditions. This sharp peak moves toward smaller frequencies with increasing system size ($\omega_\text{peak}\propto 1/L$, so one expects it to be at zero frequency in the thermodynamic limit. The area under this peak, and above the smooth curve seen in the system with periodic boundary conditions, extrapolates to a finite value in the thermodynamic limit. The latter is shown in Fig~\ref{fig:1.4.3}(b), where $\Xi$ is two times the area under the peak and above of the straight line in Fig.~\ref{fig:1.4.2}(a). The extrapolated value obtained for $\Xi$ in the thermodynamic limit is smaller than the one obtained for $D_\infty$ in systems with periodic boundary conditions in Fig~\ref{fig:1.4.3}(a). The expectation for systems with open boundary conditions is that other peaks at higher frequencies, which are also proportional to $1/L$, will collapse to $\omega=0$ in the thermodynamic limit, and their added weight will be identical to the Drude weight obtained in systems with periodic boundary conditions~(see Ref.~\cite{RigolShastry2008}). This is how a non-vanishing Drude weight appears in systems with open boundary conditions, for which $D_L=\bar D_L=0$ for any $L$.

In the main panel of Fig~\ref{fig:1.4.2}(b), we show the finite-frequency part of $\textrm{Re}[\sigma_L(\omega)]$ in the single-impurity model for chains with open boundary conditions. The curves are very similar to those obtained for the integrable XXZ model in Fig~\ref{fig:1.4.2}(a). Also, the extrapolation shown in Fig~\ref{fig:1.4.3}(b) suggests that the area under the large low-frequency peak is finite in the thermodynamic limit as for the integrable XXZ model. The inset in Fig~\ref{fig:1.4.2}(b) shows the results for the finite-frequency part of $\textrm{Re}[\sigma_L(\omega)]$ in the $\hat{H}_{\textrm{SI}}$ model for chains with periodic boundary conditions. They are in stark contrast to those for the XXZ chain in systems with periodic boundary conditions, and have features present in the results for chains with open boundary conditions. A smooth, nearly system-size independent, part is seen at frequencies $\omega>0.5$, and a sharp peak is seen about $\omega=0$. The width of the sharp peak decreases with increasing system size, while its area extrapolates to a finite value in the thermodynamic limit. In Fig.~\ref{fig:1.4.3}(a) we show the extrapolation of $\Xi$, which gives a result in the thermodynamic limit that is very close to the Drude weight obtained in systems with periodic boundary conditions in the absence of the impurity. This suggests that, in the thermodynamic limit, the low-frequency peak collapses to $\omega=0$ resulting in a non-zero Drude weight. Our results for the $\hat{H}_{\textrm{SI}}$ model, both in systems with open and periodic boundary conditions, indicate that transport in the $\hat{H}_{\textrm{SI}}$ model is coherent, as in the unperturbed model $\hat{H}_{\textrm{XXZ}}$.

It should be remarked that there exists an earlier study of the finite-frequency part of $\textrm{Re}[\sigma_L(\omega)]$ in the $\hat{H}_{\textrm{SI}}$ model for chains with periodic boundary conditions~\cite{XotosIncoherentSIXXZ}. The results reported in that work are similar to those reported in the inset in Fig~\ref{fig:1.4.2}(b). However, the low-frequency peak whose width vanishes with increasing system size was interpreted as indicating incoherent transport with a relaxation time $\tau\propto L$.  Similar results and conclusions to those in Ref.~\cite{XotosIncoherentSIXXZ} were reported in Refs.~\cite{Metavitsiadis2010, Metavitsiadis2011} for energy transport in the presence of an impurity.

\section{Discussion and outlook}
\label{sec:discussion_4}

Integrability is known to be fragile against perturbations. It is still remarkable that a single impurity can break integrability in an $L\rightarrow\infty$ chain~\cite{Santos:2004, santos2011domain, torres2014local, XotosIncoherentSIXXZ,Pandey:2020}. This can be understood in view of the fact that an $O(1)$ local integrability-breaking perturbation can mix exponentially many extended eigenstates of an integrable model and produce a Wigner-Dyson level spacing distribution typical of quantum chaotic models. Since the quantum chaotic models studied to date exhibit incoherent transport, a Wigner-Dyson level spacing distribution is usually assumed to mean incoherent transport. 

In this chapter we have studied a model, the first one known to us, for which this intuition does not apply. We showed that, while a single impurity in the XXZ model changes the level spacing distribution from Poisson to Wigner-Dyson in Chapter~\ref{chapter:models}, it does not change the nature of spin transport in the chain from coherent (for $0<\Delta<1$) to incoherent. We argued that this has to be the case from the perspective of linear response theory in the high-temperature regime and our argument applies to chains with open and periodic boundary conditions.

It is not surprising that a single magnetic perturbation is not enough to render transport incoherent. On physical grounds, transmission of spin excitations (or, equivalently, spinless fermions) from one end of the chain to the other should be unaffected as the system size increases in a Hamiltonian which is composed of local terms. More surprising, however, is the fact that the conventional picture of random matrices predicts the emergence of hydrodynamics at certain lengthscales, as discussed in Chapter~\ref{chapter:integrability}. This seemingly-contradicting fact can be conciliated if one understands that repulsion in level spacings is exponentially small in the system size. The same observation holds for the chaotic-behaviour prediction from adiabatic deformations of the single impurity model (see Ref.~\cite{Pandey:2020}), since the chaotic properties probed by such analysis refers to timescales which increase exponentially with the system size. The relevant timescales for transport, however, are the ones dictated by the dynamics at timescales which increase polynomially in system size. 

In Sec.~\ref{sec:relevance_of_integrability}, we provided a connection between transport linear response theory and Mazur's inequality. For the non-integrable single impurity model, Mazur's inequality would predict a vanishing Drude weight, pointing towards incoherent transport. The main point of this chapter is to stress the fact that the vanishing Drude weight could not necessarily stem from true hydrodynamic behaviour, but from the structure of the global symmetries presented here. In other words, Mazur's inequality prediction of a vanishing Drude weight for the single impurity model is not inconsistent, being that the model does exhibit a vanishing Drude weight. Our point is that a vanishing Drude weight does not necessarily imply incoherent transport and it may, for instance, stem from lack of translational invariance even in a system that displays coherent transport. Furthermore, this indicates that the order of limits in the evaluation of the conductivity, the thermodynamic limit ($L \to \infty$) first and the long-time limit ($t \to \infty$) second, is crucial and not interchangeable to obtain correct predictions.    

The results presented in this chapter open new questions about hydrodynamics and its emergence from quantum dynamics. 

It is interesting to consider the onset of diffusion for systems in which integrability is broken not by a single impurity but by an increasing number of impurities that, e.g., interpolate between the single impurity model and the staggered field model. The latter can be shown to exhibit the expected incoherent transport for a quantum chaotic model (see Part~\ref{part:two} of this thesis). This question has already been addressed in Ref.~\cite{Znidaric:2020} following some of our non-trivial observations. In such a scenario, the density of the number of staggering impurities gives rise to anomalous scaling of the diffusion constant at small densities and, most interestingly, there exists a regime for which adding impurities could give rise to enhanced transport.

Another interesting question is related to equilibration and thermalisation, which will drive the rest of Part~\ref{part:one} of this thesis. Our results hint that the equilibration properties of the single impurity model should be anomalous. The fact that models with single impurities can display anomalies in their approach to equilibrium is a topic that has started to be explored recently~\cite{fagottiImpurity17, Bastianello:2019}. 
\chapter{Eigenstate thermalisation}
\label{chapter:eth}

Thermalisation is a phenomenon in many-body physics that occurs with a high degree of universality~\cite{Gallavotti:2013}. The question of how and why thermalisation emerges from unitary quantum time evolution was posed even in the inception of quantum theory by some of its founding fathers~\cite{Schrodinger:1927,Vonneumann:1929,Goldstein:2010}. Nature shows us that the evolution of a pure, thermally isolated system typically results in an asymptotic state that is indistinguishable from a finite temperature Gibbs ensemble by either local or linear response measurements. However, a complete understanding of this indistinguishably is still very much an active research question~\cite{Eisert:2015}. 

One predictive framework for understanding thermalisation from quantum dynamics is the Eigenstate Thermalisation Hypothesis (ETH). It synthesises the conditions to be satisfied by the matrix elements of an operator $\hat{O}$ in the energy eigenbasis, to have expectation values and correlation functions indistinguishable from their corresponding finite-temperature counterparts. 

Inspired by early works by Berry~\cite{Berry:1977,Berry:1977b}, later formulated by Deutsch~\cite{Deutsch:1991}, ETH was fully established by Srednicki as a condition on matrix elements of generic operators $\hat O$ in the energy eigenbasis~\cite{Srednicki:1994,Srednicki:1996,Srednicki:1999}. Subsequently, ETH has motivated a considerable body of numerical work over the past decade~\cite{Rigol:2008,Polkovnikov:2011,Alessio:2016}.
Far from being an academic issue, thermalisation in closed quantum systems is now regularly scrutinised in laboratories worldwide where advances in the field of ultra-cold atom physics have allowed for probing quantum dynamics on unprecedented timescales in condensed matter physics~\cite{Kinoshita:2006,Polkovnikov:2011,Lewenstein:2007,langen2015ultracold,Bloch:2012}. In particular, seminal experiments have demonstrated that integrability inhibits thermalisation~\cite{Kinoshita:2006}, and that integrability-breaking perturbations can be used to controllably bring a system to thermal equilibrium~\cite{Tang:2018}. The latter experimental results are consistent with the expectation that generic isolated quantum systems thermalise to a microcanonical distribution consistent with their energy density. The accepted mechanism for this is given by the ETH. 

For a local observable $\hat{O}$, the ETH for the matrix elements $O_{nm} \defeq \langle n|\hat{O}|m\rangle$ in the energy eigenbasis ($\hat H|m\rangle=E_m|m\rangle$) is formulated by the relation
\begin{equation}
\label{eq:eth}
O_{n m} = O(\bar{E}) \delta_{n m} + e^{-S(\bar{E}) / 2}f_{\hat{O}}(\bar{E}, \omega)R_{n m},
\end{equation}
where $\bar{E} \defeq(E_{n} + E_{m}) / 2$ and $\omega \defeq E_{m} - E_{n}$. $S(\bar{E})$ is the thermodynamic entropy at energy $\bar{E}$, $R_{n m}$ is a random variable with zero mean and unit variance ($\overline{R_{nm}^2} = 1$), and $O(\bar{E})$ and $f_{\hat{O}}(\bar{E}, \omega)$ are smooth functions. The value $O(\bar{E})$ corresponds to the expectation value of $\hat{O}$ in the microcanonical ensemble at energy $\bar{E}$. $R_{n m}$ could be a complex random variable, in which case the only assumption about it remains to be its zero mean and unit variance ($\overline{|R_{nm}|^2} = 1$). The first term in Eq.~\eqref{eq:eth} advances that the diagonal matrix elements of observables are smooth functions of the energy $E_n$, which can be inferred from the fact that the eigenstate to eigenstate fluctuations are exponentially small in the size of the system~\cite{Steinigeweg:2013, Kim_Huse:2014, beugeling2014finite, Mondaini:2016, yoshizawa2018numerical, Vidmar_Fabian_19, Leblond:2019}. From the second term, it can be observed that the off-diagonal matrix elements are exponentially small in the system size (because of $e^{-S(\bar{E}) / 2}$) and that, up to random fluctuations, they are characterised by smooth functions $f_{\hat{O}}(\bar{E}, \omega)$~\cite{Alessio:2016, Vidmar_Fabian_19, Leblond:2019, Khatami:2013, Moessner:2015, Mondaini:2017}. Those functions carry important information on fluctuation dissipation relations~\cite{Srednicki:1999, Khatami:2013, Alessio:2016} and, hence, on thermodynamics.

Integrable systems, which possess extensive sets of non-trivial local conserved quantities, do not follow the ETH. The diagonal matrix elements of observables exhibit eigenstate to eigenstate fluctuations that do not vanish in the thermodynamic limit~\cite{Rigol:2008, rigol2009breakdown, rigol_offd_int1, Santos:2010, Steinigeweg:2013, beugeling2014finite, vidmar2016, Leblond:2019}, while their variance vanishes as a power law in the system size~\cite{biroli2010effect, ikeda2013finite, alba2015, Leblond:2019}. Because of this, in general, integrable systems do not thermalise~\cite{rigol_16}. They do equilibrate and, after equilibration, they are described by generalised Gibbs ensembles (GGEs)~\cite{vidmar2016, rigol_dunjko_07, essler_fagotti_review_16, caux_review_review_16}. The physical consequences with respect to equilibration in systems that are described by the GGE, is that the equilibrium state will carry some information pertaining to the initial state of the system, unlike ergodic systems, for which the equilibrium state depends only on macroscopic quantities. For the off-diagonal matrix elements of observables in interacting integrable systems, it was recently shown that their variance is a well-defined (exponentially small in the system size) function of the average energy and the energy difference of the eigenstates involved~\cite{mallayya, Leblond:2019}, like in systems that satisfy the ETH.

Integrability leads to the breaking of ergodicity and thermalisation. As in interacting integrable systems, the presence of strong disordered perturbations is known to lead to the breaking of ergodicity in systems that display a many-body localisation transition~\cite{Basko:2006, Abanin:2019}. Disordered-interacting systems display a transition between an ergodic phase at low disorder and a non-ergodic phase for sufficiently strong disorder in interacting systems. This physical scenario is considered to be the generalisation for interacting systems of the metallic-insulating transition known as Anderson localisation~\cite{Anderson:1958}. In the strongly-disordered regime, the emergence of an extensive set of quasi-local conserved quantities is considered to give rise to the breaking of ergodicity, akin to the mechanism for ergodicity breaking in integrable systems~\cite{Huse:2014}. Even though the presence of said transition has been the topic of recent debate~\cite{Bertini:2018}, many-body localisation effects have been observed experimentally~\cite{Schreiber842, Choi1547, Rubio:2019}. The presence of quantum many-body scars gives rise to non-thermal behaviour as well~\cite{Turner2018}.

From Chapter~\ref{chapter:kubo}, we concluded that integrability is unstable to local perturbations. Remarkably, from linear response theory (and open-system dynamics in Part~\ref{part:two}) we learnt that even though a single magnetic impurity suffices to induce level repulsion and random matrix statistics in the spectrum~\cite{Santos:2004, santos2011domain, torres2014local, Torres_Herrera_2015, XotosIncoherentSIXXZ, Metavitsiadis2010}, it does not induce the incoherent spin transport associated to non-integrable models. Our motivation for this chapter is to understand the properties related to thermalisation in non-integrable systems in general, to then study thermalisation in the single impurity model. This simple problem will then lead us to introduce entanglement structure and high-order correlation functions in Chapter~\ref{chapter:fine_eth}, when we dig deeper into the consequences of the ETH.

In Sec.~\ref{sec:thermalisation} we review the main concepts related to thermalisation within the ETH, while Sec.~\ref{sec:eth_correlation_functions} describes two-point correlation functions at thermal equilibrium and statistical measures of off-diagonal elements for systems that satisfy the ETH, as well as the fluctuation-dissipation relation. In Sec.~\ref{sec:microwave_chicken} we introduce a numerical experiment that illustrates thermalisation from the ETH. Finally, in Sec.~\ref{sec:impurity_eth} we visit the problem of thermalisation in the single impurity model.

\section{Equilibration and thermalisation}
\label{sec:thermalisation}

Thermalisation in interacting, isolated quantum systems can be very easily mistaken for equilibration. Equilibration is a property of certain quantum systems isolated from their environment and refers to the long-time behaviour of the expectation value of observables. Most naturally, within a certain quantum system, the expectation values of a certain {\em local} observable $\hat{O}$ may relax and converge to a given long-time value. Whether or not a quantum system reaches equilibration depends on the fine details of the setting, such as the initial state and the properties of the microscopic Hamiltonian describing the system. 

We are interested in isolated systems with $L \gg 1$ degrees of freedom, whose microscopic Hamiltonian $\hat{H}$ description does not allow for the presence of non-trivial local conserved quantities, i.e., the system is non-integrable. There may be some global symmetries associated to the Hamiltonian description, however, as discussed in Chapter~\ref{chapter:models}. The effect of these symmetries is to reduce the Hilbert space into independent sectors which cannot be connected coherently thought unitary dynamics. If there exists some global symmetries associated to the specific system, our analysis will carry independently for each sub-sector defined by the symmetry. The system is initially prepared in the initial state $\ket{\psi_0}$ at time $t = 0$ and we will assume such initial state to be pure and not an eigenstate of the Hamiltonian $\hat{H}$.

The non-stationary initial state will then undergo unitary dynamical evolution. We then consider the time evolution the operator $\hat{O}$ in the Schr\"odinger picture, which can be done since the system is isolated from its environment, through the time-evolving wave function
\begin{align}
\ket{\psi(t)} = \sum_m c_m e^{-\textrm{i}E_m t} \ket{m},
\end{align} 
where $c_m = \braket{m | \psi_0}$. We then may write the time evolution of $\hat{O}$ in the eigenbasis of the Hamiltonian $\hat{H}$ as
\begin{align}
\label{eq:o_energy_basis_time}
\langle \hat{O}(t) \rangle = \braket{\psi(t) | \hat{O} | \psi(t)} = \sum_n |c_n|^2 O_{nn} + \sum_{n, m \neq n} c_n^{*}c_m e^{\textrm{i}(E_n - E_m)t} O_{nm},
\end{align}
where $O_{nm} \defeq \braket{n | \hat{O} | m}$. If the system equilibrates, we may study the long-time average of the expectation value of $\hat{O}$ in the form 
\begin{align}
\overline{O} \defeq \lim_{\tau \to \infty} \frac{1}{\tau} \int_0^{\tau} \textrm{d}t \langle \hat{O}(t) \rangle.
\end{align}
An illuminating fact can be observed in the long-time average from Eq.~\eqref{eq:o_energy_basis_time}. If we assume that there are no degeneracies in the spectrum or, alternatively, if that there is no extensive amount of degeneracies present, the second term in Eq.~\eqref{eq:o_energy_basis_time} averages to zero in the equilibration value $\overline{O}$, i.e., 
\begin{align}
\label{eq:o_de_def}
\overline{O} = \sum_n |c_n|^2 O_{nn} = \textrm{Tr}[\hat{O} \hat{\rho}_{\textrm{DE}}] \defeq \langle \hat{O} \rangle_{\textrm{DE}}
\end{align}
where
\begin{align}
\hat{\rho}_{\textrm{DE}} = \sum_n |c_n|^2 \ket{n}\bra{n}
\end{align}
is the density matrix of the diagonal ensemble that contains the information about the initial state in the eigenbasis of the Hamiltonian. A non-degenerate spectrum is not a strong assumption about a generic chaotic Hamiltonian, once any trivial global symmetries have been categorised in different symmetry sub-sectors. 

The above statements have no implications about thermalisation, only about equilibration which translates into a well-defined asymptotic value of $\langle \hat{O}(t) \rangle$. Thermalisation in statistical mechanics implies that the asymptotic value can be obtained from an ensemble average of the quantity of interest, which means
\begin{align}
\overline{O} \rightarrow O_{\textrm{MC}}(E) =  \textrm{Tr}[\hat{O} \hat{\rho}_{\textrm{MC}}],
\end{align}
where $O_{\textrm{MC}}$ is the equilibrium expectation value of the operator $\hat{O}$ in the microcanonical ensemble, a paramater-dependent quantity given by
\begin{align}
O_{\textrm{MC}}(E) = \frac{1}{\Omega(E)}\sum_n O_{nn} \delta(E-E_n),
\end{align}
where $\Omega(E) = \sum_n\delta(E-E_n)$ is the density of states. Note that this ensemble is characterised by a single parameter which can be associated to the microcanonical temperature $\beta(E) = \textrm{d}S/ \textrm{d}E$, through Boltzmann's relation $S(E) = \ln W(E)$, where $W(E) = \Omega(E)\textrm{d} E$ corresponds to the number of microstates in a small energy interval $\textrm{d} E$. Generic systems, however, typically satisfy ensemble equivalence, in which case any ensemble in statistical mechanics can be used to describe the thermal expectation value. We note, however, that a known instance in which the equivalence of ensembles does not hold is found whenever long-range interactions are at play~\cite{Barre:2001,Caseti:2007}.

Crucially, the ETH states that the first term in Eq.~\eqref{eq:eth} is directly related to $\overline{O}$ from Eq.~\eqref{eq:o_de_def}, i.e.,
\begin{align}
\label{eq:thermalisation_1}
\overline{O} = \sum_n |c_n|^2 O_{nn} &= \sum_n |c_n|^2 \left[ O(E_n) + e^{-S(\bar{E}) / 2}f_{\hat{O}}(\bar{E}, 0)R_{n n} \right] \nonumber \\
&\approx \sum_n |c_n|^2 O(E_n),
\end{align}
where the second term is suppressed exponentially due to $e^{-S(\bar{E}) / 2}$. We now see that, if the ETH in Eq.~\eqref{eq:eth} is satisfied by the matrix elements of the operator $\hat{O}$ in the eigenbasis of the Hamiltonian, we have that
\begin{align}
\label{eq:thermalisation_2}
\overline{O} = O_{\textrm{MC}}(\langle E \rangle) = O(\langle E \rangle).
\end{align}
If a system fulfils the above condition, we state that observable $\hat{O}$ {\em thermalises}. We further require $|c_n|^2$ to be narrowly-peaked around $\langle E \rangle = \braket{\psi_0 | \hat{H} | \psi_0}$ for the above condition to be satisfied\footnote{Interestingly, even though energy distributions after quenches are expected to be smooth, this is not a necessary requirement for thermalisation according to the ETH~\cite{Alessio:2016}.}. This last requirement, is a necessary, but not sufficient, condition for thermalisation~\cite{Alessio:2016}. For the physical configuration to thermalise in the sense above, we have to assume that the microcanonical expectation value is well defined. Since such quantity is an averaged value over a statistical ensemble, the averaged quantity is only physically relevant if the fluctuations around the average are small and decay as the number of degrees of freedom is increased, as it is usually argued in classical statistical physics~\cite{Sethna:2020}. As expected from statistical mechanics, then, thermalisation is satisfied for a system with $L$ degrees of freedom if Eq.~\eqref{eq:thermalisation_2} holds, while 
\begin{align}
\label{eq:fluctuations_eth_energy}
\langle E \rangle = \braket{\psi_0 | \hat{H} | \psi_0} \propto L, \quad \frac{(\delta E)^2}{\langle E \rangle^2} \propto \frac{1}{L},
\end{align}
where $(\delta E)^2 = \braket{\psi_0 | \hat{H}^2 | \psi_0} - \braket{\psi_0 | \hat{H} | \psi_0}^2$ are the ensemble energy fluctuations. Such conditions are expected to hold in generic systems with short-range interactions. It can be shown that Eq.~\eqref{eq:fluctuations_eth_energy} ensures that fluctuations around the microcanonical expectation value $O_{\textrm{MC}}(\langle E \rangle)$ are sub-extensive as the size of the system grows~\cite{Alessio:2016}. This follows from from the fact that $\delta E$ is finite. 

In fact, one can quantify the difference between the microcanonical expectation value and the long-time averaged value of equilibration. If we expand the smooth function $O(\bar{E})$ from Eq.~\eqref{eq:eth} around the mean energy $\langle E \rangle$ in a Taylor series, we find
\begin{align}
O_{nn} \approx O(\langle E \rangle) + (E_n - \langle E \rangle))\left[ \frac{\textrm{d}O(\bar{E})}{\textrm{d}\bar{E}} \right]_{\langle E \rangle} + \frac{1}{2} (E_n - \langle E \rangle))^2 \left[ \frac{\textrm{d}^2O(\bar{E})}{\textrm{d}\bar{E}^2} \right]_{\langle E \rangle},
\end{align}
which, upon substitution in Eq.~\eqref{eq:o_de_def}, yields~\cite{Alessio:2016}
\begin{align}
\overline{O} &\approx O(\langle E \rangle) + \frac{1}{2} (\delta E)^2 \left[ \frac{\textrm{d}^2O(\bar{E})}{\textrm{d}\bar{E}^2} \right]_{\langle E \rangle} \nonumber \\
&\approx O_{\textrm{MC}}(\langle E \rangle) + \frac{1}{2} [(\delta E)^2 - (\delta E_{\textrm{MC}})^2] \left[ \frac{\textrm{d}^2O(\bar{E})}{\textrm{d}\bar{E}^2} \right]_{\langle E \rangle},
\end{align}
where $\delta E_{\textrm{MC}}$ are the sub-extensive energy fluctuations in the microcanonical ensemble. We then see that thermalisation in the sense of ETH, yields thermal expectation values which agree with long-time averaged values up to a sub-extensive correction.

Moreover, the ETH allows one to describe the temporal fluctuations of the long-time average throughout the dynamics of $\langle \hat{O}(t) \rangle$. We have~\cite{Alessio:2016}
\begin{align}
\sigma_{\hat{O}}^2 &\defeq \lim_{\tau \to \infty} \int_{0}^{\tau} \textrm{d}t [\langle \hat{O}(t) \rangle]^2 - \overline{O}^2 \nonumber \\
&= \sum_{n, m \neq n} |c_n|^2 |c_m|^2 |O_{nm}|^2 \nonumber \\
&\leq \textrm{max}[|O_{nm}|^2] \propto e^{-S(E)},
\end{align} 
which then implies that the time fluctuations of the expectation value $\langle \hat{O}(t) \rangle$ are exponentially small in the size of the system $L$, given that the entropy is extensive.

In practical applications, it is usual to consider the eigenstate-to-eigenstate fluctuations of $O_{nn}$, i.e., the diagonal part of the ETH, as a necessary condition for thermalisation. Given that a requirement to be met for thermalisation is that the microcanonical expectation value is a well-defined smooth function of the energy, the eigenstate-to-eigenstate fluctuations
\begin{align}
\label{eq:ete_fluctuations}
\overline{|\delta O_{nn} |} \defeq \overline{| O_{nn} - O_{n+1 n+1}|},
\end{align}
must decay as the number of degrees of freedom is increased and, moreover, they should decay exponentially with the system size as first suggested by Kim {\em et al.} in Ref.~\cite{Kim_Huse:2014}. This follows from the fact that adjacent eigenstates in non-integrable systems  should display an energy difference that decays exponentially with the size of the system. In Eq.~\eqref{eq:ete_fluctuations} we have assumed that the eigenstates $\ket{n}$ have been sorted in ascending order according to their eigenvalue $E_n$. Further studies have confirmed that this is indeed the case for non-integrable systems that thermalise according to the ETH through compelling numerical evidence using different physical models~\cite{Mondaini:2016,Vidmar_Fabian_19,Leblond:2019}.

\subsubsection{Connection between temperature and average energy}

An illuminating fact can be inferred from the fact that the expectation value of $\hat{O}$ from statistical mechanics in non-integrable systems becomes a smooth function of the energy in the thermodynamic limit, related to $O(\bar{E})$ in Eq.~\eqref{eq:eth}.

Consider the canonical thermal average of $\hat{O}$
\begin{align}
\langle \hat{O} \rangle_{\beta} = \frac{\textrm{Tr}[\hat{O} e^{-\beta \hat{H}}]}{\textrm{Tr}[e^{-\beta \hat{H}}]},
\end{align}
where $\beta$ is the canonical inverse temperature. We can estimate this expectation value from eigenstate thermalisation by expanding the calculation in the energy eigenbasis in the continuum limit~\cite{Srednicki:1999} 
\begin{align}
\sum_n \rightarrow \int \textrm{d}\bar{E}e^{S(\bar{E})},
\end{align}
where $\sum_n$ denotes the summation over energy eigenstates and $S(\bar{E})$ is the thermodynamic entropy. Invoking Eq.~\eqref{eq:eth}, we note
\begin{align}
\langle \hat{O} \rangle_{\beta} = \frac{\int \textrm{d}\bar{E}e^{S(\bar{E}) - \beta \bar{E}}O(\bar{E})}{\int \textrm{d}\bar{E}e^{S(\bar{E}) - \beta \bar{E}}} + \mathcal{O}(e^{-S / 2}).
\end{align}
This expression can be approximated by noting that the thermodynamic entropy in statistical mechanics is extensive for when $L \gg 1$. One can then apply a steepest-descent procedure to obtain~\cite{Srednicki:1999} 
\begin{align}
\langle \hat{O} \rangle_{\beta} = O(\bar{E}) + \mathcal{O}(L^{-1}) + \mathcal{O}(e^{-S/2}),
\end{align}
where the last two terms are sub-extensive and, therefore, sub-leading as $L \to \infty$. In particular, the last term comes from the second part of the expression in Eq.~\eqref{eq:eth}, and it is exponentially suppressed in the system size through the thermodynamic entropy. In finite-size systems, such contributions are typically non-negligible, though they decay fast as the number of degrees of freedom is increased. Crucially, $\bar{E}$ is now fixed in terms of $\beta$ by the steepest-descent condition $\partial S/ \partial \bar{E} = \beta$.   

The above also implies that the total energy in the isolated quantum system 
\begin{align}
\langle E \rangle = \textrm{Tr}[\hat{H}\hat{\rho}],
\end{align}
where $\hat{\rho}$ is a canonical state $\hat{\rho} \defeq e^{-\beta \hat{H}} / Z$, $\beta$ the canonical inverse temperature and $Z = \textrm{Tr}[e^{-\beta \hat{H}}]$, can be estimated on a single eigenstate in the thermodynamic limit, since the calculation only involves the diagonal part $O_{nn}$ in Eq.~\eqref{eq:eth}. Explicitly so,
\begin{align}
\langle E \rangle &= \textrm{Tr}[\hat{H}\hat{\rho}] \nonumber \\
&= \bar{E} = E_n.
\end{align}
This implies that a single energy eigenstate in non-integrable chaotic systems determines the thermal properties of an entire statistical ensemble in the thermodynamic limit at temperature $\beta$, by associating the canonical temperature $\beta$ to an energy eigenstate $\ket{n}$.

\subsection{Thermalisation and diagonal elements of local observables in chaotic models}
\label{sec:thermalisation_chaotic_models}

The occurrence of eigenstate thermalisation has been established in different non-integrable models numerically.  The usual analyses found in the literature~\cite{Alessio:2016} involve the study of the diagonal matrix elements of a given local operator $\hat{O}$ in the energy eigenbasis, i.e., $O_{nn}$ in Eq.~\eqref{eq:eth}. Evidence for eigenstate thermalisation has been observed for interacting spin chains in one dimension~\cite{Santos:2010,Steinigeweg:2013,Kim_Huse:2014,Moessner:2015,rigol2009breakdown,Rigol:2010,Lux:2014,Khodja:2015}, two-dimensional Ising model with transversed fields~\cite{Mondaini:2016} and interacting hard-core bosons in two-dimensional lattices~\cite{Rigol:2008}. In all cases, eigenstate thermalisation appears to be ubiquitous as long as the physical systems are located in non-integrable regimes regulated, usually, by the spatial configuration or the Hamiltonian parameters. 

Let us reconsider the non-integrable model introduced in Chapter~\ref{chapter:models}, the anisotropic Heisenberg model in the presence of a staggered magnetic field on a one-dimensional lattice of length $L$ with open boundary conditions, with Hamiltonian
\begin{align}
\label{eq:h_sf_5}
\hat{H}_{\textrm{SF}} = \hat{H}_{\textrm{XXZ}} + b\,\sum_{i\,\textrm{odd}} \hat{\sigma}^z_{i}\,,
\end{align} 
where $\hat{H}_{\textrm{XXZ}}$ is the Hamiltonian of the XXZ model in Eq.~\eqref{eq:h_xxz} and $b$ is the strength of the staggered magnetic field. Placing our attention in the parameter regime given by $\Delta = 0.55\alpha$ and $b = \alpha$, we proceed to numerically find the unitary transformation $\hat{U}$ that renders the Hamiltonian $\hat{H}_{\textrm{SF}}$ diagonal. We focus on the zero-magnetisation sector of the XXZ Hamiltonian, and introduce a small magnetic perturbation on the first site of the chain $\delta \hat{\sigma}^z_1$ with $\delta = 0.1$ to avoid parity or reflection symmetries present in the model. Observables that exhibit the occurrence of eigenstate thermalisation can either be local such as
\begin{align}
\label{eq:a_sf_5}
\hat{A}_{\textrm{SF}} = \hat{\sigma}^{z}_{\frac{L}{2}} \hat{\sigma}^{z}_{\frac{L}{2} + 1},
\end{align}
or sums of local operators such as the total staggered magnetisation
\begin{align}
\label{eq:b_sf_5}
\hat{B}_{\textrm{SF}} = \frac{1}{L} \sum_i (-1)^i \hat{\sigma}^z_{i}.
\end{align}
The diagonal matrix elements of these observables in the eigenbasis of the Hamiltonian are given by $O_{nn} = [\hat{U}^{\dagger} \hat{O} \hat{U}]_{nn}$.

\begin{figure}[t]
\fontsize{13}{10}\selectfont 
\centering
\includegraphics[width=0.55\columnwidth]{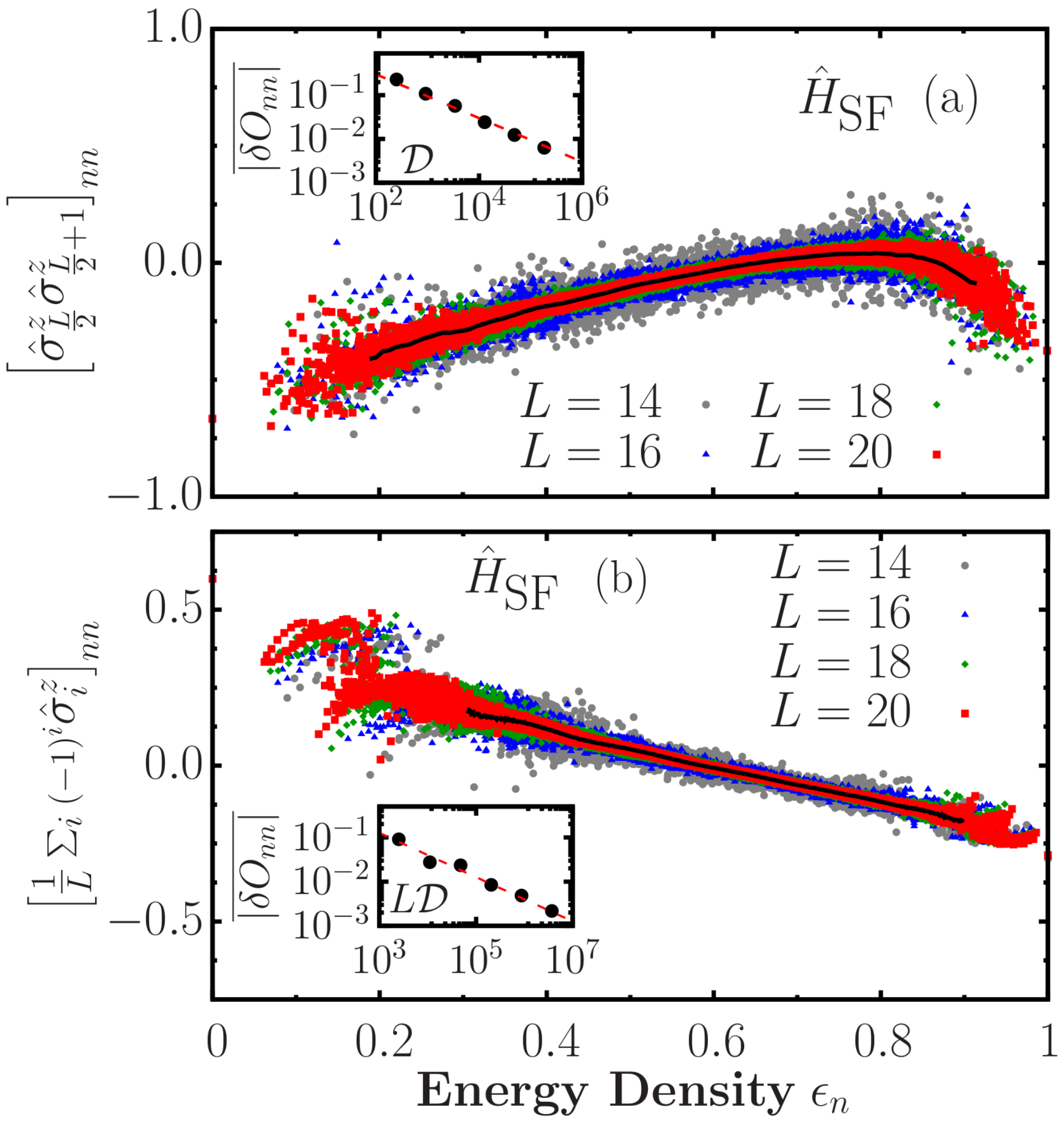}
\caption[Diagonal matrix elements of $\hat{A}_{\textrm{SF}}$ (a) and $\hat{B}_{\textrm{SF}}$ (b) as a function of the energy density $\epsilon_n \defeq (E_n - E_{\textrm{min}}) / (E_{\textrm{max}} - E_{\textrm{min}})$ and of the system size $L$]{Diagonal matrix elements of $\hat{A}_{\textrm{SF}}$ (a) and $\hat{B}_{\textrm{SF}}$ (b) as a function of the energy density $\epsilon_n \defeq (E_n - E_{\textrm{min}}) / (E_{\textrm{max}} - E_{\textrm{min}})$ and of the system size $L$. The black lines depict an approximation of the smooth function $O(\bar{E})$ obtained from a coarse-grained average of the data for the largest system size. The insets show the eigenstate-to-eigenstate fluctuations for different systems sizes, obtained from the eigenvalues in the central region. The dashed lines on the insets show the (a) $\mathcal{D}^{-1/2}$ and (b) $(L \mathcal{D})^{-1/2}$ scalings. The parameters of the Hamiltonian $\hat{H}_{\textrm{SF}}$ were selected as $\Delta = 0.55\alpha$ and $b = \alpha$.}
\label{fig:1.5.1}
\end{figure}

A strong indication of eigenstate thermalisation is the behaviour of diagonal matrix elements of observables in the eigenbasis of the Hamiltonian~\cite{Alessio:2016,Leblond:2019}. In Fig.~\ref{fig:1.5.1} we show the diagonal matrix elements of $\hat{A}_{\textrm{SF}}$ [panel (a)] and of $\hat{B}_{\textrm{SF}}$ [panel (b)] for the non-integrable staggered field model. We defined the energy density $\epsilon_n \defeq (E_n - E_{\textrm{min}}) / (E_{\textrm{max}} - E_{\textrm{min}})$ and computed all the matrix elements in the eigenbasis of the Hamiltonian by full diagonalisation. It can be observed that, as the system size $L$ is increased, the support over which the matrix elements exist shrink. This observation strongly suggests that in the thermodynamic limit, the diagonal matrix elements can be described by a smooth function $O(\bar{E})$ corresponding to the microcanonical prediction (note that the second term in Eq.~\eqref{eq:eth} is exponentially suppressed in Hilbert space dimension $\mathcal{D}$). The black lines in Fig.~\ref{fig:1.5.1} depict an approximation of the smooth function $O(\bar{E})$, obtained from a coarse-grained average of the data for the largest system size $L = 20$. 

The insets in Fig.~\ref{fig:1.5.1} highlight the trend of the absolute value of the eigenstate-to-eigenstate fluctuations, computed for 20\% of the total eigenvalues in the centre of the spectrum. The dashed lines on the insets in Fig.~\ref{fig:1.5.1} correspond to the scaling $\mathcal{D}^{-1/2}$ [panel (a)] and $L\mathcal{D}^{-1/2}$ [panel (b)], expected in the high-temperature regime (corresponding to the states around the centre of the spectrum) of non-integrable models. We remark that the eigenstate-to-eigenstate fluctuations for sums of local observables scales like $L\mathcal{D}^{-1/2}$, as opposed to the more common $\mathcal{D}^{-1/2}$ scaling observed for local observables. This behaviour can be attributed to the $1 / \sqrt{L}$ scaling of the Schmidt norm for this class of observables~\cite{Vidmar:2019,Vidmar2:2020,Leblond:2019}.

The results shown in Fig.~\ref{fig:1.5.1} indicate that the ETH is obeyed at the level of thermalisation, i.e., diagonal matrix elements for the staggered field model, the observables considered and the Hamiltonian parameters selected. Non-generic features can be appreciated at the extrema of the spectrum, which is often the case in physical systems near low temperatures~\cite{Alessio:2016}. 

The physical implications of this observation have been described in Sec.~\ref{sec:thermalisation}: the long-time behaviour of the expectation value $\langle \hat{O}(t) \rangle = \braket{\psi(t) | \hat{O} | \psi(t)}$, will relax to an asymptotic value $\overline{O}$, which will coincide with $O(\bar{E}) = O(E_n)$ from the ETH, where $E_n$ is the eigenvalue of the energy state $\ket{n}$ whose total energy coincides with the initial energy of the system from its statistical expectation value $\langle E \rangle_{\beta}$. Indeed, the correspondence between the thermal expectation values of local observables and those computed at the level of single eigenstates can be observed in physical systems with a Hamiltonian description.

Consider now the expectation value of the local magnetisation at the centre of the chain, $\hat{O} = \hat{\sigma}^z_{L/2}$ in the staggered field model. Our starting point is to evaluate the expectation value of $\hat{O}$ in the canonical ensemble
\begin{align}
\langle \hat{\sigma}^z_{L/2} \rangle = \textrm{Tr}[\hat{\rho} \hat{\sigma}^z_{L/2}],
\end{align}
where $\hat{\rho} \defeq e^{-\beta \hat{H}_{\textrm{SF}}} / Z$ is the canonical state, $Z = \textrm{Tr}[e^{-\beta \hat{H}_{\textrm{SF}}}]$. This expectation value can be computed at temperature $T = 1 / \beta$ from the full diagonalisation of $\hat{H}_{\textrm{SF}}$, by representing both $\hat{\rho}$ and $\hat{O} = \hat{\sigma}^z_{L/2}$ in the energy eigenbasis. We can then compare this expectation value with the ETH prediction. In the thermodynamic limit, a single energy eigenstate comprises the entire statistical ensemble, in which the total energy $\langle E \rangle = \textrm{Tr}[\hat{\rho} \hat{H}_{\textrm{SF}}]$ gives rise to the average energy $\bar{E} = E_n$ corresponding to a single eigenstate $\ket{n}$ in Eq.~\eqref{eq:eth}. Our analysis, however, has to be carried out in systems with a finite size $L$, for which eigenstate-to-eigenstate fluctuations are present and non-negligible. Instead, one can then consider the expectation value from the average between the matrix elements $O_{nn}$ within a certain energy window centred at $E_n$, to average these fluctuations. 

\begin{figure}[t]
\fontsize{13}{10}\selectfont 
\centering
\includegraphics[width=0.85\columnwidth]{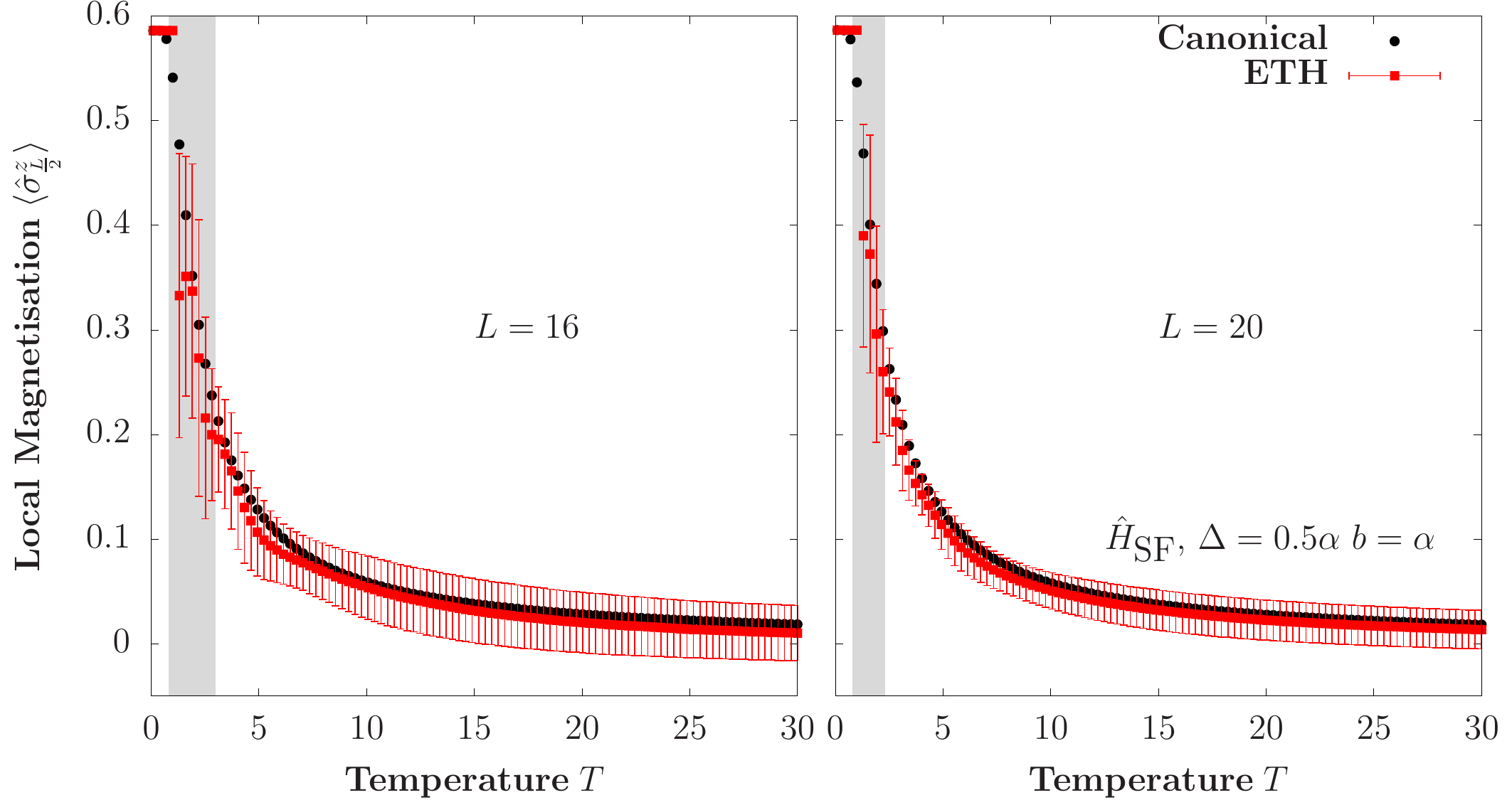}
\caption[Expectation value of the local magnetisation in the centre of the chain as a function of temperature in both the canonical ensemble and the corresponding ETH prediction in the staggered field model]{Expectation value of the local magnetisation in the centre of the chain as a function of temperature in both the canonical ensemble and the corresponding ETH prediction for and $L = 16$ (left panel) $L = 20$ (right panel) in the staggered field model.}
\label{fig:1.5.2}
\end{figure}

The results are shown in Fig.~\ref{fig:1.5.2}. The expectation value in the ETH framework is computed from the average of a small window of energies centred around $\langle E \rangle_{\beta} \approx \bar{E} = E_n$ of width $0.1\epsilon$, where $\epsilon = E_{\textrm{max}} - E_{\textrm{min}}$ is the bandwidth of the Hamiltonian for a given system size $L$. Note that the total energy, which depends on $\beta$, has been associated to a given energy eigenstate through $\bar{E} = E_n$. $\langle E \rangle_{\beta}$ is an average energy resulting from the expectation value of energy in the canonical ensemble with inverse temperature $\beta$, i.e, $E = \text{Tr}(\hat{H}\, e^{-\beta \hat H})/Z$, where $\hat{H}$ is the Hamiltonian of the staggered field model.

Fig.~\ref{fig:1.5.2} shows $\langle \hat{O} \rangle_{\beta}$ as a function of temperature for two different system sizes, including $L=20$, the largest system we have access to from exact diagonalisation (Hilbert space dimension $\mathcal{D} = L!/[(L/2)!(L/2)!] = 184\;756$). The results exhibit the expected behaviour predicted from ETH for finite-size systems: the thermal expectation value is well approximated away from the edges of the spectrum (low temperature, section highlighted in grey on Fig.~\ref{fig:1.5.2}), and, moreover, the canonical expectation value is better approximated as the system size increases, which can be observed from the smaller fluctuations of the predicted expectation value at a given temperature.

\section{Temporal correlation functions and linear response}
\label{sec:eth_correlation_functions}

The ETH in Eq.~\eqref{eq:eth} provides the form of the off-diagonal matrix elements required to predict the dynamics of two-point correlation functions at thermal equilibrium. We are interested in correlation functions of the form
\begin{align}
\label{eq:f2}
F_2(t) \defeq \langle \hat{O}(t) \hat{O}(0) \rangle_c \defeq \langle \hat{O}(t) \hat{O}(0) \rangle - \langle \hat{O}(t) \rangle \langle \hat{O}(0) \rangle,
\end{align}
where the expectation values are evaluated in one of the ensembles of statistical mechanics. One could, for instance, consider the canonical ensemble. For an operator $\hat{O}$ in such case, we have that $\langle \hat{O} \rangle = \textrm{Tr}[\hat{\rho} \hat{O}]$, where $\hat{\rho} = e^{-\beta \hat{H}} / Z$ is the density matrix operator for a system with Hamiltonian $\hat{H}$, partition function $Z = \textrm{Tr}[e^{-\beta \hat{H}}]$ and inverse temperature $\beta = 1 / T$. In Eq.~\eqref{eq:f2}, the operators are written in the Heisenberg picture $\hat{O}(t) = e^{\textrm{i} \hat{H} t} \hat{O}(0) e^{-\textrm{i} \hat{H} t}$.

On the other hand, eigenstate thermalisation suggests that such expectation values could be evaluated for a {\em single eigenstate} $\ket{n}$ where, as described before, the eigenstate with energy $E_n$ is chosen to be compatible with the total energy $\langle E \rangle_{\beta}$. In this case
\begin{align}
F_2(E_n, t) \defeq \braket{ n | \hat{O}(t) \hat{O}(0) | n} - \braket{n | \hat{O}(t) | n} \braket{n | \hat{O}(0) | n}.
\end{align}
Note that we differentiate between $F_2(t)$, the two-point temporal correlation function computed in the statistical mechanics ensembles, and $F_2(E_n, t)$, as the one computed on a single eigenstate. We want to evaluate this expression using the ETH in Eq.~\eqref{eq:eth}.

We start by decomposing the $\hat{O}(t) \hat{O}(0)$ in the eigenbasis of the Hamiltonian, which yields
\begin{align}
[ \hat{O}(t) \hat{O}(0) ]_{jl} \defeq \braket{j | \hat{O}(t) \hat{O}(0) | l} = \sum_{k} e^{\textrm{i}(E_j - E_k)t}O_{jk}O_{kl},
\end{align}
which implies
\begin{align}
\braket{n | \hat{O}(t) \hat{O}(0) | n} = \sum_k e^{i(E_n - E_k)t} |O_{nk}|^2.
\end{align}
ETH provides the general form of the $O_{nk}$ elements. Using Eq.~\eqref{eq:eth} and noting $\braket{n | \hat{O}(t) | n}\braket{n | \hat{O}(0) | n} = |O_{nn}|^2$, we find that
\begin{align}
\label{eq:rnm_averaging}
F_2(E_n, t) \stackrel{\textrm{ETH}}{=} \sum_{k \neq n} e^{-\textrm{i}\omega t}e^{-S(E_n + \frac{\omega}{2})}|f_{O}(E_n + \frac{\omega}{2}, \omega)|^2 |R_{nk}|^2,
\end{align}
where $\bar{E} = E_n + \frac{\omega}{2}$ and $\omega = E_{k} - E_n$. Since ETH requires $f_{\hat{O}}(E_n, \omega)$ to be a smooth function of its arguments, the fluctuations $|R_{\lambda k}|^2$ average to unity in the sum, and the summation over states $k$ can be replaced by integration over $\omega$ as 
\begin{align}
\sum_k \rightarrow \int d \omega \Omega(E_n + \omega) = \int d \omega e^{S(E_n + \omega)},
\end{align}
where $\Omega(E_n + \omega)$ is the density of states. We may then write
\begin{align}
F_2(E_n, t) = \int_{-\infty}^{\infty} d \omega e^{-i \omega t} e^{S(E_n + \omega) - S(E_n + \frac{\omega}{2})} |f_{\hat{O}}(E_n + \frac{\omega}{2}, \omega)|^2.
\end{align}
We can Taylor expand the entropy term around $\omega = 0$\footnote{This expansion is typically justifiable on the grounds of numerical observations of the behaviour of the two-point correlation function in physical systems with short-range interactions. The approximation is typically not severe, since $|f_{\hat{O}}(E_n, \omega)|^2$ normally decays exponentially away from the features observed near $\omega \to 0$. See, for instance, Ref.~\cite{Mondaini:2017}.} as
\begin{align}
S \left( E_n + \omega \right) - S \left( E_n + \frac{\omega}{2} \right) = \frac{\beta \omega}{2} + \frac{3 \omega ^2}{8} \frac{\partial \beta}{\partial E_n} + \cdots,
\end{align} 
and the smooth function $f_{\hat{O}}$ as well
\begin{align}
|f_{\hat{O}}(E_n + \frac{\omega}{2}, \omega)|^2 = |f_{\hat{O}}(E_n, \omega)|^2 + \frac{\omega}{2} \left[ \frac{\partial |f_{\hat{O}}(\bar{E}, \omega)|^2}{\partial \bar{E}} \right]_{E_n} + \cdots.
\end{align}
Keeping terms up to linear order in $\omega$ yields
\begin{align}
F_2(E_n, t) \approx  \int_{-\infty}^{\infty} d \omega e^{\frac{\beta \omega}{2} - i\omega t} \left\{ |f_{\hat{O}}(E_n, \omega)|^2 + \frac{\omega}{2} \left[ \frac{\partial |f_O(\bar{E}, \omega)|^2}{\partial \bar{E}} \right]_{E_n}  \right\}.
\end{align}
We can then take the Fourier transform of $F_2(E_n, t)$, defined as
\begin{align}
F_2(E_n, \omega) = \int_{-\infty}^{\infty}dt e^{\textrm{i}\omega t}F_2(E_n, t)
\end{align}
to obtain
\begin{align}
\label{eq:c_lambda}
F_2(E_n, \omega) = 2\pi e^{\frac{\beta \omega}{2}} \left\{ |f_{\hat{O}}(E_n, \omega)|^2 + \frac{\omega}{2} \left[ \frac{\partial |f_O(\bar{E}, \omega)|^2}{\partial \bar{E}} \right]_{E_n}  \right\}.
\end{align}
Eq.~\eqref{eq:c_lambda} is the two-point correlation function in frequency domain of the operator $\hat{O}$ obtained from ETH at the level of a single eigenstate $\ket{n}$ of the Hamiltonian. We can then see that the dynamics of two-point correlation functions are dictated by the function $f_{\hat{O}}(E_n, \omega)$.

\subsubsection{Finite-size effects}

The temporal correlation function evaluated on a thermal state $F_2(t)$ [Eq.~\eqref{eq:f2}] differs from the one computed on a single eigenstate $F_2(E_n, t)$ by a sub-leading term that vanishes in the thermodynamic limit~\cite{Alessio:2016}.

The expectation value computed in the canonical ensemble is defined, in the eigenbasis of the Hamiltonian, as 
\begin{align}
\langle \cdot \rangle \defeq \textrm{Tr}[\hat \rho \, \cdot ] = \sum_n p_n \bra{n} \cdot \ket{n}, 
\end{align}
where in the case of a canonical density matrix one has  $p_n = e^{-\beta E_n}/Z$. In general, other ensembles can be considered, provided that the distribution of the $p_n$ is sufficiently peaked around some average energy $\langle E \rangle = \textrm{Tr}[{\hat{\rho}\hat{H}}]$ with a well-defined variance $(\delta E)^2= \braket{\hat H^2}-\braket{\hat{H}}^2$, such that $(\delta E)^2 / \langle E \rangle ^2 \sim 1/L$.
Defining $O_{nm} \defeq \bra{n} \hat{O} \ket{m}$, the two-point function in the eigenbasis of the Hamiltonian can be expressed as
\begin{align}
\label{eq:F2Exp}
        F_2(t) 
        & = \sum_{nm} p_n e^{-\textrm{i}(E_m-E_n)t}\, O_{nm} O_{mn} - \left (\sum_n p_n O_{nn}\right )^2 
        \nonumber \\
        & = \sum_{n\neq m} p_n e^{-\textrm{i}(E_m-E_n)t}\, O_{nm} O_{mn} + \braket{\hat{O}^2} - \braket{\hat{O}}^2 
        \nonumber \\
        & = \sum_n p_n \, \bra{n}\hat O(t) \hat O \ket{n}_c\, + (\delta \hat{O})^2 \ ,
\end{align}
where in the second line we have identified $\langle \hat{O}^2\rangle=\sum_n p_n [O_{nn}]^2$,  $\langle \hat{O} \rangle=\sum_n p_n O_{nn}$ and defined $(\delta \hat{O})^2 \defeq \braket{\hat{O}^2} - \braket{\hat{O}}^2$. 
The first term in Eq.~\eqref{eq:F2Exp} coincides with the two-point function evaluated on a single eigenstate $F_2(E_n, t)$, while the second one is a time-independent quantity that can be shown to be sub-leading, i.e.,
\[
F_2(t) \sim F_2(E_n, t) + \mathcal O(1/L) \quad \text{for}  \quad L\gg 1 \ .
\]
Using the ETH in Eq.~\eqref{eq:eth} and the fact that $p_n$ is peaked around energy $\langle E \rangle$, we may proceed to write down a Taylor expansion of the microcanonical function $O(\bar{E})$ around $\langle E \rangle$. For the diagonal elements $O_{nn} = O(E_n)$, we have that
\begin{equation}
O(E_n) = O(\langle E \rangle) + (E_n-\langle E \rangle) \left. \frac{\partial O}{\partial E_n} \right|_{\langle E \rangle} + \frac {(E_n-\langle E \rangle)^2} 2 \left. \frac{\partial^2 O}{\partial E_n^2} \right|_{\langle E \rangle} + \dots
\end{equation}
This expansion can be used to express $(\delta \hat{O})^2$ with respect to its leading order. We have
\begin{equation}
(\delta O)^2 = \left( \frac{\partial O}{\partial E_n}\right)^2 (\delta E)^2.
\end{equation}
The fact that $O(E_n)$ is required to be a smooth function of the energy, implies that fluctuations must decay for the microcanonical average to be well-defined. As we detailed before, this implies $(\delta E)^2 / \langle E \rangle^2 \sim 1/L$, then $(\delta O)^2$ is sub-leading. For finite-size systems, however, this term is expected to be present and amounts to a difference between the ensemble and the ETH predictions. Such difference, as noted above, should decrease as the system size and the number of degrees of freedom in the sample is increased.

\subsubsection{Estimation of $f_{\hat{O}}(E_n, \omega)$ in chaotic systems}

The function $f_{\hat{O}}(E_n, \omega)$ in the ETH plays a pivotal role in the dynamics of the system. It encodes the information required to predict the behaviour of two-point correlation functions as noted above. 

In this section, we study the off-diagonal elements of local observables in the energy eigenbasis as a function of $\omega = E_m - E_n$, where $E_k$ labels the $k$-th energy eigenvalue of the Hamiltonian. The appropriate analysis of these elements leads to the smooth function $e^{-S(\bar{E})/2}f_{\hat{O}}(\bar{E},\omega)$, from which the non-equal correlation functions in time can be estimated in the ETH predicition.

According to the ETH, the off-diagonal matrix elements of observables in the energy eigenbasis are described according to
\begin{align}
O_{n m} = e^{-S(\bar{E}) / 2}f_{\hat{O}}(\bar{E}, \omega)R_{n m} \quad \forall n \neq m.
\end{align}
We shall study the matrix elements of local observables obtained numerically in the staggered field model to understand their behaviour.

Our starting point is to select a target energy $\langle E \rangle$, such that $\langle E \rangle = \braket{n | \hat{H} | n} = \text{Tr}(\hat {H}\, e^{-\beta \hat H})/Z$, where $\hat{H}$ is the Hamiltonian of the staggered field model. In the thermodynamic limit, a single eigenstate $\ket{n}$ and its corresponding off-diagonal overlaps with $\hat{O}$ suffice to compute $f_{\hat{O}}(E_n, \omega)$ and, hence, the correlation functions according to the ETH prediction. For finite-size systems, however, we focus on a small window of energies centred around the target energy $\langle E \rangle$ of width $0.1\epsilon$, where $\epsilon = E_{\textrm{max}} - E_{\textrm{min}}$ corresponds to the bandwidth of the Hamiltonian at a given system size. Presumably, all the eigenstates in this energy window contain approximately the same average energy. This procedure is carried out to average eigenstate-to-eigenstate fluctuations~\cite{Mondaini:2017}. 

\begin{figure}[t]
\fontsize{13}{10}\selectfont 
\centering
\includegraphics[width=0.9\columnwidth]{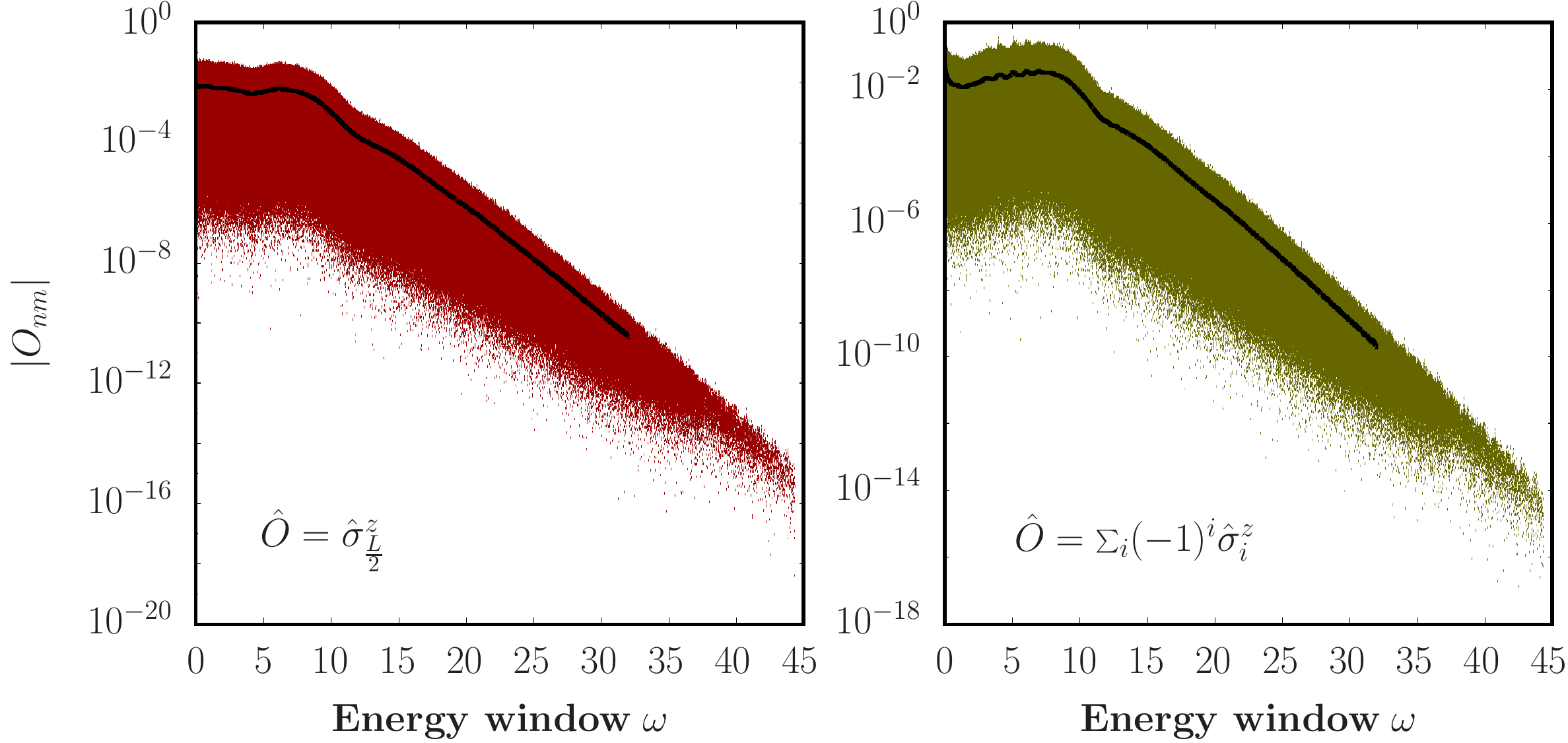}
\caption[Absolute value of the off-diagonal elements in the energy eigenbasis of the local magnetisation in the middle of the chain and the total staggered magnetisation for the staggered field model]{Absolute value of the off-diagonal elements in the energy eigenbasis of the local magnetisation in the middle of the chain (left) and the total staggered magnetisation (right) as a function of $\omega$ for $T = 5\alpha$ and $L = 18$ for the staggered field model with $\Delta = 0.5\alpha$ and $b = \alpha$ in Eq.~\eqref{eq:h_sf_5}. The black lines correspond to binned averages.}
\label{fig:1.5.3}
\end{figure}

To extract $e^{-S(\bar{E})/2}f_{\hat{O}}(\bar{E},\omega)$, we compute the binned average of the off-diagonal matrix elements $O_{nm}$ within the energy window mentioned above. The binned average is computed using, in turn, small frequency windows $\delta \omega$. The size of these windows is selected such that a smooth curve is obtained from the average and the resulting function is not sensitive to the particular choice of $\delta \omega$. This window of frequencies typically changes depending on the dimension of the magnetisation sub-sector studied in our spin model~\cite{Mondaini:2017}.

In Fig.~\ref{fig:1.5.3} we present the absolute value of the off-diagonal elements of both the local magnetisation operator in the middle of the chain and the total staggered magnetisation. These matrix elements were computed for $T = 5\alpha$, $L = 18$ and an energy window of width $0.1\epsilon$. The temperature sets the average energy $\langle E \rangle$ that locates the eigenstate $\ket{n}$, around which the off-diagonal matrix elements are computed. The smooth black lines shown are binned averages for each corresponding observable. This average corresponds to $e^{-S(\bar{E})/2}f_{\hat{O}}(\bar{E},\omega)$. Note that this procedure yields $f_{\hat{O}}(\bar{E},\omega)$ up to a factor that does not depend on $\omega$. This factor can be estimated from the density of states or from additional properties of the correlation functions, as we shall see in Sec.~\ref{sec:fluctuation_dissipation}. Irrespective of how this term is estimated, the described treatment allows us to focus on the $\omega$ dependence of $f_{\hat{O}}(\bar{E},\omega)$.

The binned average of the local observables from Fig.~\ref{fig:1.5.3} exhibits an interesting exponential decay behaviour at high frequencies, which has been observed in other chaotic models~\cite{Mondaini:2017, Alessio:2016,Moessner:2015} and has been argued to be related to the universal exponential decay of two-point correlation functions in time for chaotic systems with a bounded spectrum~\cite{Huse2006,murthy2019}. On the opposite side of the spectrum, at low frequencies, $f_{\hat{O}}(\bar{E},\omega)$ contains important features relevant to the long-time behaviour of correlation functions. These frequencies are the most relevant for the response functions associated with linear response.

\subsubsection{Statistical distribution of off-diagonal matrix elements in chaotic systems}

We have understood how the structure of the off-diagonal matrix elements in the energy eigenbasis in chaotic gives rise to a smooth function $f_{\hat{O}}(\bar{E},\omega)$ in the ETH. In the procedure described above, averaging was employed to obtain the smooth function from a set of off-diagonal matrix elements within a given energy window. As predicted by the ETH, however, the off-diagonal matrix elements are described by the smooth function $f_{\hat{O}}(\bar{E},\omega)$ in conjunction with a statistical random matrix $R_{nm}$ with a well-defined mean and variance, which render the off-diagonal matrix elements effectively random around the smooth function $f_{\hat{O}}(\bar{E},\omega)$. This can already be observed for local observables in the staggered field model from Fig.~\ref{fig:1.5.3}.

We are now interested in investigating the probability distribution associated to the random matrix $R_{nm}$. The first step towards a statistical characterisation of $\hat{O}$ is to understand the distribution of its individual matrix elements.

Since the number of matrix elements is very large even at small system sizes, we begin by studying the distribution of off-diagonal matrix elements $O_{nm} = \braket{n | \hat{O} | m}$ in a small frequency-resolved window $\omega \lesssim 0.05$ and a finite temperature $T = \beta^{-1} = 5\alpha$ ($k_B\defeq 1$). As before, the temperature is calculated by associating the average energy $\bar{E}$ with a canonical density matrix $\hat{\rho} = e^{-\beta\hat{H}}/Z$ as $\bar{E} = \textrm{Tr}[\hat{\rho} \hat{H}]$, with $Z = \textrm{Tr}[e^{-\beta\hat{H}}]$. The probability distribution can then be inferred by creating a histogram of all the matrix elements that satisfy $\bar{E} = \textrm{Tr}[\hat{\rho} \hat{H}]$ and $\omega < 0.05$.

\begin{figure}[t]
\fontsize{13}{10}\selectfont 
\centering
\includegraphics[width=0.7\columnwidth]{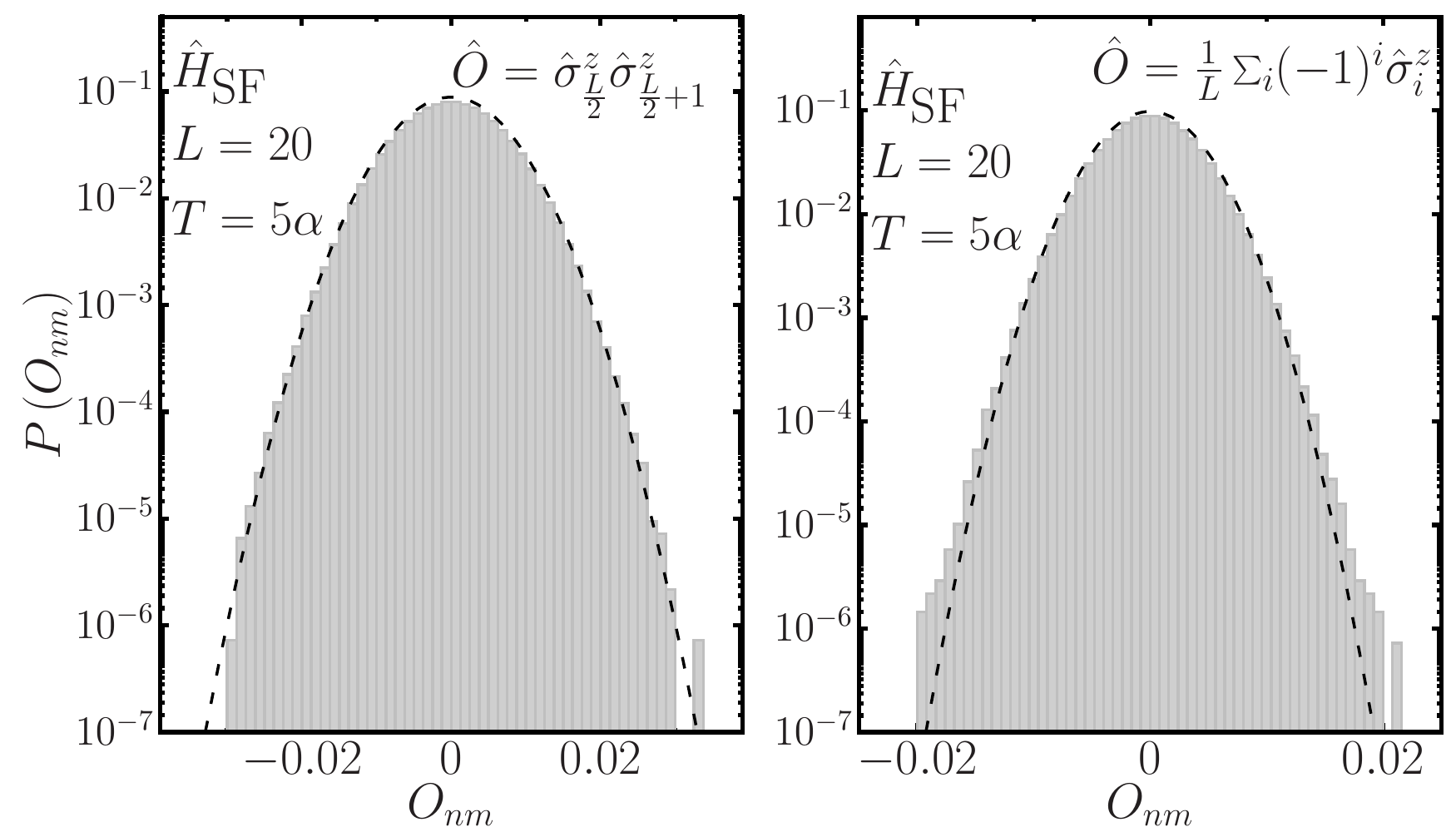}
\caption[Probability distributions of off-diagonal matrix elements in a small frequency range $\omega \lesssim 0.05$ for the staggered field model]{Probability distributions of off-diagonal matrix elements in a small frequency range $\omega \lesssim 0.05$ for the staggered field model with $\Delta = 0.5\alpha$ and $b = \alpha$ in Eq.~\eqref{eq:h_sf_5}. The average energy $\bar{E}$ selected is consistent with a finite canonical temperature $T = 5\alpha$. The distributions are shown in (a) for $\hat{A}_{\textrm{SF}}$ and in (b) for $\hat{B}_{\textrm{SF}}$. Results obtained for finite-sized systems of $L = 20$. Dashed lines depict a Gaussian distribution with the same mean and variance.}
\label{fig:1.5.4}
\end{figure}

In Fig.~\ref{fig:1.5.4} we show the probability distributions obtained by this procedure. The matrix elements are Gaussian-distributed for both the local and sums of local operators in the staggered field model, as has previously been found for other models and observables in the infinite temperature regime~\cite{Beugeling2015, Leblond:2019, Khaymovich2019, Leblond:2020, Santos2020}. Most interestingly, Gaussianity of the off-diagonal matrix elements has now started to be considered an identifier of quantum chaos~\cite{Richter2020,Leblond:2019,Leblond:2020,Santos2020}. 

Our previous analysis reveals the probability distribution of off-diagonal matrix elements of local observables in the $\omega \sim 0$ regime. In order to understand if this property pertains to the entire spectrum away from zero frequency and if the same distributions are observed at all temperatures where the ETH is expected to hold, we proceed to evaluate the frequency-dependent ratio~\cite{Leblond:2019}
\begin{align}
\label{eq:gamma_1}
\Gamma_{\hat{O}}(\omega) \defeq \overline{|O_{nm}|^2} / \overline{|O_{nm}|}^2,
\end{align}
where the averages are performed over a small frequency window $\delta\omega = 0.05$. Should the individual matrix elements be Gaussian-distributed with zero mean at a given value of $\omega$, then $\Gamma_{\hat{O}}(\omega) = \pi / 2$. 
For this particular analysis, we consider $\omega = E_m - E_n$ over the entire spectrum, while the average energy $\bar{E} = (E_n + E_m) / 2$ is chosen to be compatible with a corresponding canonical temperature. The quantity is computed over small bins in $\omega$ of a given size and within a small energy window of width $0.05\epsilon$, where $\epsilon \defeq E_{\textrm{max}} - E_{\textrm{min}}$ is the bandwidth of the Hamiltonian. The average over the small energy window is carried out to account for finite-size eigenstate-to-eigenstate fluctuations.  

\begin{figure}
\fontsize{13}{10}\selectfont 
\centering
\includegraphics[width=0.605\columnwidth]{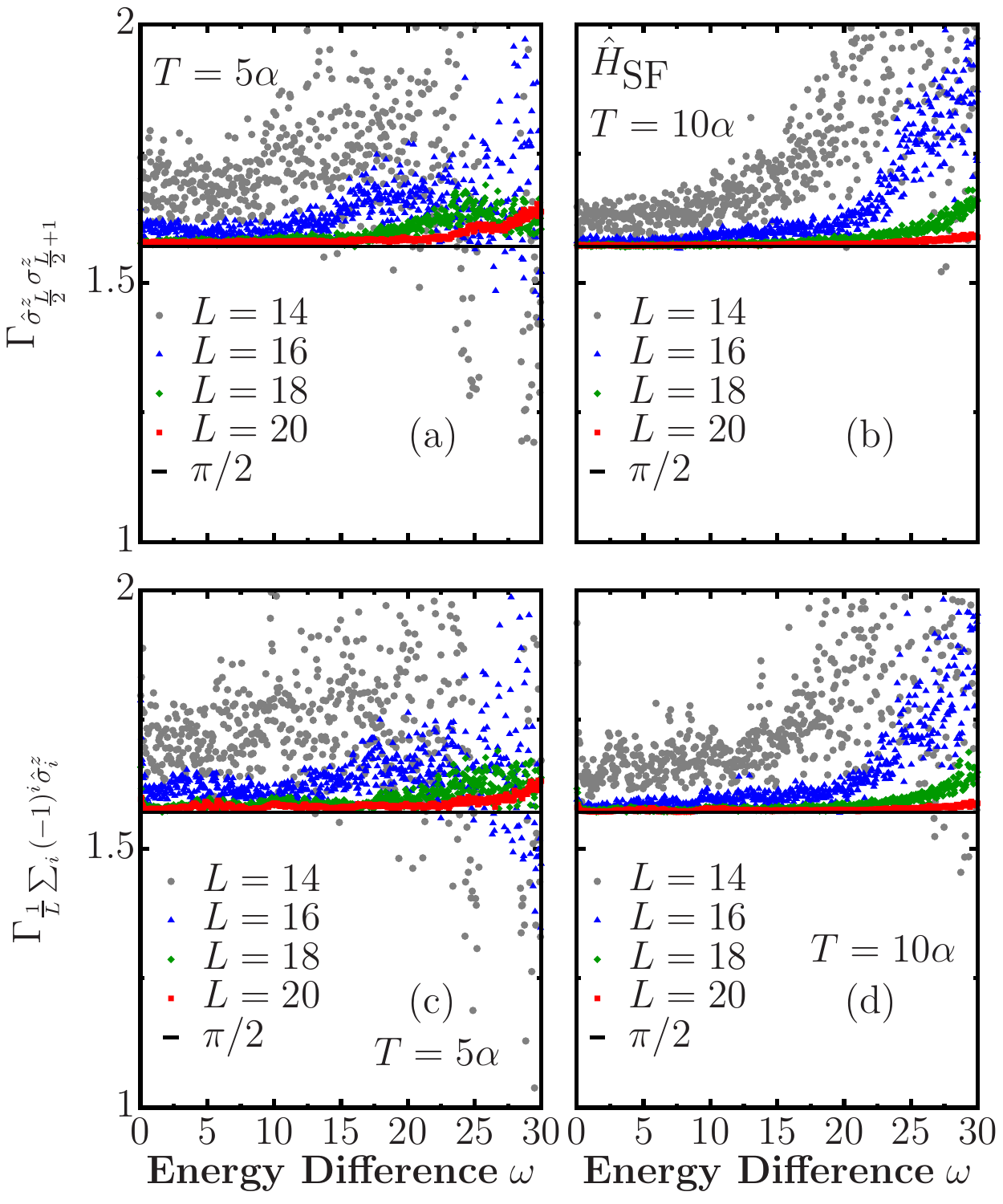}
\caption[$\Gamma_{\hat{O}}(\omega)$, from Eq.~\eqref{eq:gamma_1}, for operators $\hat{A}_{\textrm{SF}}$ and $\hat{B}_{\textrm{SF}}$ in the eigenbasis of $\hat{H}_{\textrm{SF}}$]{$\Gamma_{\hat{O}}(\omega)$, from Eq.~\eqref{eq:gamma_1}, for operators $\hat{A}_{\textrm{SF}}$ [(a) and (b)] and $\hat{B}_{\textrm{SF}}$ [(c) and (d)] in the eigenbasis of $\hat{H}_{\textrm{SF}}$ with $\Delta = 0.5\alpha$ and $b = \alpha$ in Eq.~\eqref{eq:h_sf_5}. Two different finite temperatures were chosen, $T = 5\alpha$ [(a) and (c)] and $T = 10\alpha$ [(b) and (d)]. The black horizontal line shows the value $\Gamma_{\hat{O}}(\omega) = \pi / 2$. The matrix elements were computed in a small energy window $0.05\epsilon$ where $\epsilon \defeq E_{\textrm{max}} - E_{\textrm{min}}$, and a frequency window $\delta \omega = 0.05$.}
\label{fig:1.5.5}
\end{figure}

In Fig.~\ref{fig:1.5.5} we show the $\Gamma_{\hat{O}}(\omega)$ ratio as a function of $\omega$ and of the system size $L$ for $\hat{H}_{\textrm{SF}}$ evaluated for the local observable $\hat{A}_{\textrm{SF}}$ [panels (a) and (b)] and the staggered magnetisation $\hat{B}_{\textrm{SF}}$ [panels (c) and (d)] and for two different temperatures $T = 5\alpha$ and $T = 10\alpha$. We have chosen to display our results for values of temperature away from the infinite-temperature regime.
Gaussian statistics emerge at all frequencies, i.e.~ $\Gamma_{\hat{O}} \approx \pi / 2$ for increasing values of $\omega$ as the system size increases. These findings, together with recent results that have highlighted normality in the distributions of off-diagonal matrix elements in the high-temperature limit~\cite{Leblond:2019,Leblond:2020,Richter2020}, strongly suggest that Gaussianity is ubiquitous in non-integrable models for which the ETH is expected to hold, even at finite temperature.

It is important to remark that even though the off-diagonal matrix elements of local observables in the energy eigenbasis are Gaussian-distributed, this does not imply that they are uncorrelated. In other words, distributions observed are normal, but these does not guarantee that $R_{nm}$ is a statistical matrix composed of identical- and independently-distributed random variables. Correlations and its consequences for dynamical quantities will be studied in Chapter~\ref{chapter:fine_eth}.

\subsection{The fluctuation-dissipation relation}
\label{sec:fluctuation_dissipation}

The quantum analog of the celebrated fluctuation-dissipation theorem~\cite{Sethna:2020} can be derived from eigenstate thermalisation.

The evaluation involves the symmetric and anti-symmetric response functions which yield, respectively, the real and imaginary parts of $F_2(t)$. Written in such fashion we have
\begin{align}
S^{+}_{\hat{O}}(E_n, t) &\defeq \braket{n | \{ \hat{O}(t), \hat{O}(0) \} | n}_c = 2\,\textrm{Re}[F_2(E_n, t)]\nonumber \\
S^{-}_{\hat{O}}(E_n, t) &\defeq \braket{n | \,[ \hat{O}(t), \hat{O}(0) \,] | n}_c = 2\textrm{i}\, \textrm{Im}[F_2(E_n, t)],
\end{align}
where $\{\cdot, \cdot \}$ and $[ \cdot, \cdot ]$ stand for the anti-commutator and commutator, respectively. In this notation, $F_2(t)$ is the one considered in the canonical ensemble, while $F_2(E_n, t)$ is the one evaluated for a single energy eigenstate. 

The symmetric and anti-symmetric response functions follow from the temporal correlation function in the frequency domain $F_2(E_n, \omega)$, 
\begin{align}
S^{+}_{\hat{O}}(E_n, \omega) &=  F_2(E_n, \omega) + F_2(E_n, -\omega), \nonumber \\
S^{-}_{\hat{O}}(E_n, \omega) &=  F_2(E_n, \omega) - F_2(E_n, -\omega).
\end{align}
From Eq.~\eqref{eq:c_lambda}, where we derived $F_2(E_n, \omega)$ from the ETH up to linear order in $\omega$, we can express
\begin{align}
S^{+}_{\hat{O}}(E_n, \omega) &=  4\pi \left\{ \cosh \left(\frac{\beta \omega}{2} \right) |f_{\hat{O}}(E_n, \omega)|^2 + \frac{\omega}{2} \sinh \left(\frac{\beta \omega}{2} \right) \left[ \frac{\partial |f_{\hat{O}}(\bar{E}, \omega)|^2}{\partial \bar{E}} \right]_{E_n}  \right\}, \nonumber \\
S^{-}_{\hat{O}}(E_n, \omega) &=  4\pi \left\{ \sinh \left(\frac{\beta \omega}{2} \right) |f_{\hat{O}}(E_n, \omega)|^2 + \frac{\omega}{2} \cosh \left(\frac{\beta \omega}{2} \right) \left[ \frac{\partial |f_{\hat{O}}(\bar{E}, \omega)|^2}{\partial \bar{E}} \right]_{E_n}  \right\}.
\end{align}
The linear term in $\omega$ is sub-leading for both local and sums of local operators. This follows from the fact that $|f_{\hat{O}}(E_n, \omega)|^2$ is sub-extensive for local operators, i.e., $\mathcal{O}(1)$, while the energy $\bar{E}$ is extensive and of order $\mathcal{O}(L)$. It then follows that $\partial |f_{\hat{O}}(\bar{E}, \omega)|^2 / \partial \bar{E} \sim 1 / L$. For sums of local operators, $|f_{\hat{O}}(E_n, \omega)|^2 \sim L$ and, still, $\bar{E} \sim L$, which then implies that even in this case $\partial |f_{\hat{O}}(\bar{E}, \omega)|^2 / \partial \bar{E}$ is a sub-leading term in the expression. 

Given that both the symmetric and anti-symmetric response functions are connected via $|f_{\hat{O}}(E_n, \omega)|^2$, we have that 
\begin{align}
S^{-}_{\hat{O}}(E_n, \omega) = \tanh \left(\frac{\beta \omega}{2} \right) S^{+}_{\hat{O}}(E_n, \omega),
\end{align}
which is valid for correlation functions evaluated in the ensembles of statistical mechanics and for individual eigenstates in systems that satisfy $L \gg 1$. This fluctuation-dissipation relation provides a connection between noise and dissipative response at the level of a single eigenstate.
Most importantly, the anti-symmetric correlation function is directly connected to Kubo's linear response susceptibility, defined in the frequency domain as
\begin{align}
\chi_{\hat{O}}(\omega) = \textrm{i}\int_{0}^{\infty} \textrm{d}t e^{\textrm{i} \omega t} \langle [\hat{O}(t), \hat{O}(0)] \rangle.
\end{align}
Using a similar procedure to the one we used before, one can show~\cite{Alessio:2016} that the imaginary part of $\chi_{\hat{O}}(\omega)$ is connected to the anti-symmetric correlation function through $f_{\hat{O}}(\bar{E}, \omega)$ for a single eigenstate,
\begin{align}
\textrm{Im}[\chi_{\hat{O}}(E_n, \omega)] = \frac{1}{2} S^{-}_{\hat{O}}(E_n, \omega).
\end{align}
It then follows, that Kubo's linear response theory, in turn, is also directly encoded in $f_{\hat{O}}(\bar{E}, \omega)$. 

\subsubsection{Response functions in chaotic systems}

Dynamical quantities such as response functions, temporal auto-correlation functions and Kubo's linear susceptibility are all encoded within the function $f_{\hat{O}}(\bar{E}, \omega)$. Having established a framework to extract this function from the off-diagonal matrix elements of observables in the energy eigenbasis in Sec.~\ref{sec:eth_correlation_functions}, we can proceed to numerically evaluate these dynamical quantities in the framework of the ETH. For such an evaluation, we return to our non-integrable staggered field model.

Following our derivation above~[See Refs.~\cite{Srednicki:1999,Alessio:2016,Vidmar:2020} for further details] from the ETH, one obtains the correlation functions in frequency domain in the thermodynamic limit to leading order
\begin{align}
\label{eq:f2_w_re_im}
S^{+}_{\hat{O}}(E_n, \omega) &\approx 4\pi \cosh(\beta \omega /2) |f_{\hat{O}}(E_n, \omega)|^2, \nonumber \\
S^{-}_{\hat{O}}(E_n, \omega) &\approx 4\pi \sinh(\beta \omega /2) |f_{\hat{O}}(E_n, \omega)|^2.
\end{align}
Given that these relations are symmetric and anti-symmetric, respectively, their Fourier transforms to yield the correlation functions in the time domain are simplified to
\begin{align}
\label{eq:f2_t_re_im}
\textrm{Re}[F_2(E_n, t)] = \int_0^{\infty} d\omega \cos(\omega t) S^{+}_{\hat{O}}(E_n, \omega), \nonumber \\
\textrm{Im}[F_2(E_n, t)] = \int_0^{\infty} d\omega \sin(\omega t) S^{-}_{\hat{O}}(E_n, \omega).
\end{align}
At this point is important to make two observations with respect to Eq.~\eqref{eq:f2_t_re_im}. First, in the thermodynamic limit we expect $F_2(E_n, t) = F_2(t)$. This immediately follows from the association of the energy $E_n$ to a corresponding canonical inverse temperature $\beta$ by assigning $E_n = \langle E \rangle = \textrm{Tr}[\hat{H} e^{-\beta \hat{H}}] / Z$. Second, in Eq.~\eqref{eq:f2_w_re_im}, there is no dependency of the random variable $R_{nm}$. This follows from the fact that this term enters the dynamical correlations in the form of the average of $|R_{nm}|^2$, which is unity by assumption~\cite{Srednicki:1999,Alessio:2016}. Indeed, it suffices that this random variable has a well-defined variance for the $|R_{nm}|^2$ term to vanish from the final expressions [see Eq.~\eqref{eq:rnm_averaging}].

The equivalency between the dynamics of two-point correlation functions in statistical mechanics and the corresponding dynamics of the same object predicted by the ETH can be observed in generic systems. Following Eqs.~\eqref{eq:f2_w_re_im} and \eqref{eq:f2_t_re_im}, the dynamics of the correlation functions depend solely on the function of $f_{\hat{O}}(E_n, \omega)$; which is, in general, system- and observable-dependent. A commonly-used procedure~\cite{Mondaini:2017,Khatami:2013,Vidmar:2019} to isolate this function in generic systems involves a frequency-resolved analysis of the matrix elements of a given observable in the energy eigenbasis. As described before, one focuses on a small window of energies and extracts the off-diagonal matrix elements of an operator $\hat{O}$ in the eigenbasis of the Hamiltonian. For finite-size systems, the fluctuations present are accounted for by considering not a single eigenstate, but a collection of eigenstates around a given energy $E_n$. A coarse-grained average then leads to a smooth function $e^{-S(E_n)/2}f_{\hat{O}}(E_n, \omega)$. The entropy term, $e^{-S(E_n)/2}$, is not a function of $\omega$ and in principle needs to be evaluated. 

\begin{figure}[t]
\fontsize{13}{10}\selectfont 
\centering
\includegraphics[width=1.0\columnwidth]{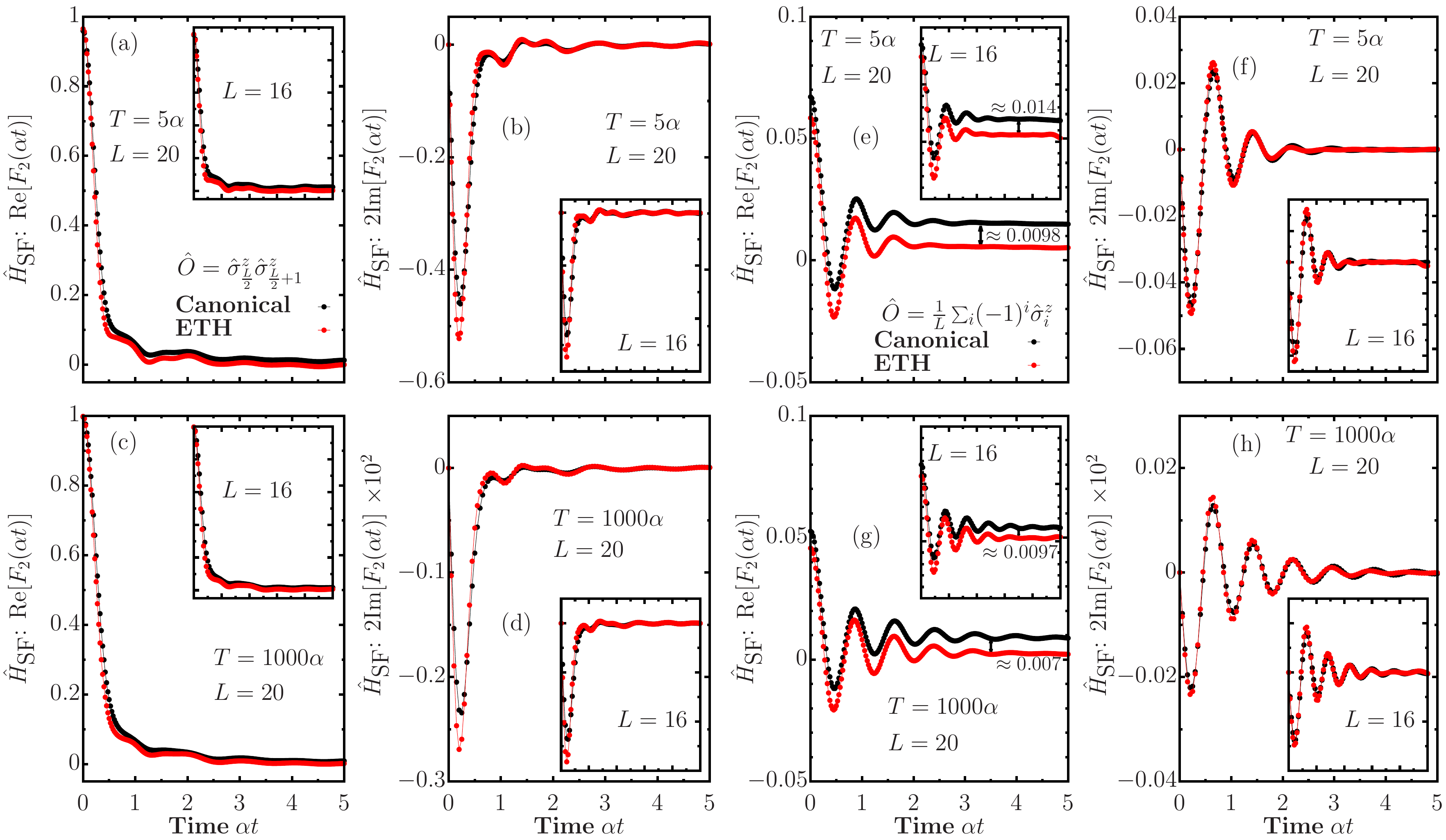}
\caption[Dynamics of the two-point correlation function evaluated in the canonical ensemble at temperature $T$ and in the ETH with a compatible energy density for the staggered field model]{Dynamics of the two-point correlation function (real and imaginary parts) evaluated in the canonical ensemble at temperature $T$ and in the ETH with a compatible energy density for [(a)-(d)] local operators and [(e)-(h)] sums of local operators for the staggered field model with $\Delta = 0.55\alpha$ and $b = \alpha$. Main panels display the correlation functions for $L = 20$, while insets show $L = 16$. Finite temperatures $T = 5\alpha$ are shown in top panels and bottom panels display very high temperature dynamics $T = 1000\alpha$.}
\label{fig:1.5.6}
\end{figure}

Instead of evaluating this term directly, we first compute the symmetric correlation function $S^{+}_{\hat{O}}(E_n, \omega)$ and normalise it by the following sum rule
\begin{align}
\label{eq:sum_rule_5}
\int_{-\infty}^{\infty} d\omega S^{+}_{\hat{O}}(E_n, \omega) = 4\pi \left[ \braket{E_n | \hat{O}^2 | E_n} - \braket{E_n | \hat{O} | E_n}^2 \right],
\end{align}
while the anti-symmetric correlation function $S^{-}_{\hat{O}}(E_n, \omega)$ follows from Eq.~\eqref{eq:f2_w_re_im}, which is the manifestation of the fluctuation-dissipation theorem. In this sense, the physical properties if the symmetric correlation function are exploited to isolate the $\omega$ dependence of $f_{\hat{O}}(E_n, \omega)$. Alternatively, one can estimate the density of states at a given temperature and from there, estimate $e^{-S(\bar{E})/2}$. In the infinite temperature regime, this term is nothing but $\mathcal{D}^{-1/2}$, where $\mathcal{D}$ is the dimension of the Hilbert space of the sub-sector under evaluation. In our calculations, however, we rely on the sum rule Eq.~\eqref{eq:sum_rule_5} to normalise the correlation functions and avoid the estimation of $e^{-S(\bar{E})/2}$.

In Fig.~\ref{fig:1.5.6}, we show the dynamics of both the real and imaginary parts of the two-time correlation function in Eq.~\eqref{eq:f2} on the staggered field model, for a local observable [panels (a)-(d)] $\hat{\sigma}^z_{L/2} \hat{\sigma}^z_{L/2 + 1}$ and for the staggered magnetisation [panels (e)-(h)], the latter being a sum of local observables at different temperatures. Correlation functions at finite temperature $T = 5\alpha$ are shown in the top panels, while infinite temperature dynamics are shown in the bottom panels. The dynamics of the two-point correlation function in the canonical ensemble were evaluated by direct diagonalisation of the propagator $e^{-\textrm{i}\hat{H}_{\textrm{SF}}t}$ acting on the density operator $\hat{\rho}$, while the dynamics from the ETH were evaluated using the procedure described above. For the latter, we computed $e^{-S(E_n)/2}f_{\hat{O}}(E_n, \omega)$ by considering a target energy $\bar{E} = \textrm{Tr}[\hat{\rho} \hat{H}]$ consistent with the canonical temperature $T$ and averaging all the off-diagonal matrix elements within an energy window of width $0.075\epsilon$, where $\epsilon \defeq E_{\textrm{max}} - E_{\textrm{min}}$ [see Refs.~\cite{Khatami:2013,Mondaini:2017} for further details on the extraction of $f_{\hat{O}}(E_n, \omega)$].  

For finite-size systems, the connected symmetric correlation function contains a time-independent term that is not present if one evaluates the same object on a single eigenstate. This term is expected to vanish in the thermodynamic limit as we described in Eq.~\eqref{eq:F2Exp}. The difference stems from the fluctuations in the canonical ensemble, a term which is already small for $\hat{A}_{\textrm{SF}}$ [Eq.~\eqref{eq:a_sf_5}] in Fig.~\ref{fig:1.5.4}[(a)-(d)], but not as much for $\hat{B}_{\textrm{SF}}$ [Eq.~\eqref{eq:b_sf_5}] in Fig.~\ref{fig:1.5.6}[(e),(g)]. Note that the finite-size difference is only present in the real part of $F_2(E_n, \omega)$, since the operators are Hermitian then the fluctuations in Eq.~\eqref{eq:F2Exp} are real-valued. The difference, however, as highlighted in Fig.~\ref{fig:1.5.6}[(e),(g)], becomes smaller as the system size is increased. The dynamics observed strongly suggest that such seemingly-constant discrepancy can be attributed to finite-size corrections.

Though the fine details and the actual form of the decay of $F_2(t)$ depend on the observable and temperature considered, it can be observed from Fig.~\ref{fig:1.5.6} that $F_2(E_n, t) \approx F_2(t)$, an approximation that becomes more accurate as the thermodynamic limit is approached. It is important to remark that such prediction is accurate not only at high temperature (bottom panels in Fig.~\ref{fig:1.5.6}), but at finite temperature (top panels in Fig.~\ref{fig:1.5.6}) as well. 

\section{A heating-relaxation numerical experiment}
\label{sec:microwave_chicken}

Having established the main concepts of eigenstate thermalisation, we are prepared to described an experimentally-relevant scenario where an isolated quantum system is allowed to relax under unitary dynamics after an initial state preparation. This {\em in silico} experiment will allow us to illustrate, to a certain extent, the degree of predictability of the eigenstate thermalisation hypothesis by studying the physical quantities we have described in this section.

Consider a quantum system which is initially prepared in a pure state $\ket{\psi_0}$ of the Hamiltonian $\hat{H}$. At time $t = 0$, the system is coupled locally to a periodic driving field which performs work onto the system. After some preparation time $t = t_{\textrm{prep}}$, the system is decoupled from the driving field. Following this procedure, the system is allowed to relax under the unitary dynamics governed by the unitary propagator $\exp(-\textrm{i}\hat{H} (t - t_{\textrm{prep}}))$. The resulting state at time $t = t_{\textrm{prep}}$ is not an eigenstate of $\hat{H}$, therefore, non-stationary. We then observe the relaxation dynamics of a local observable to understand if it follows the predictions of eigenstate thermalisation.

\begin{figure}[t]
\fontsize{13}{10}\selectfont 
\centering
\includegraphics[width=1.0\columnwidth]{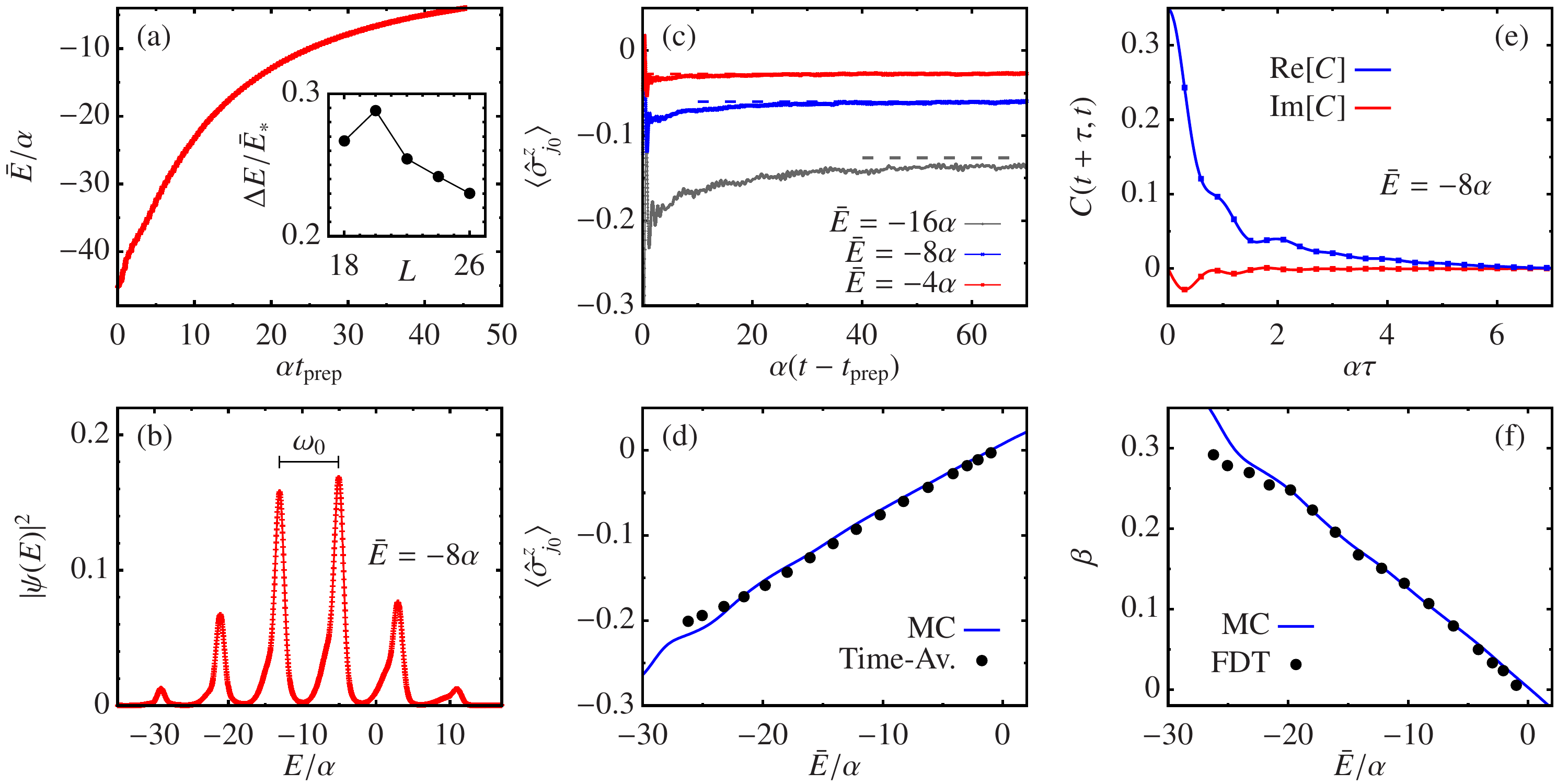}
\caption[Unitary heating of a quantum spin-$\tfrac{1}{2}$ chain]{Unitary heating of a quantum spin-$\tfrac{1}{2}$ periodic chain in the half-filling ($N = L / 2$) sector, the non-integrable staggered field model with $\Delta = 0.55\alpha$ and $b = \alpha$ from Eq.~\eqref{eq:h_sf_5}. Results are for $L=26$ sites unless otherwise indicated. (a)~Mean energy $\bar{E} =\braket{\psi(t_{\rm prep})|\hat{H}|\psi(t_{\rm prep})}$ of the chain as a function of the preparation time $t_{\rm prep}$ under local periodic driving, $\hat{H}(t) = \hat{H} + a \sin(\omega_0 t)\hat{\sigma}_{j_0}^z$, with amplitude $a=2\alpha$ and frequency $\omega_0 = 8\alpha$ applied to middle site, $j_0 = L/2$. The inset shows the energy fluctuations, $(\Delta E)^2 = \braket{\psi(t_{\rm prep})|(\hat{H}-\bar{E})^2|\psi(t_{\rm prep})}$, relative to $\bar{E}_* = \bar{E}-E_0$ with $E_0$ the ground-state energy, as a function of system size at fixed microcanonical temperature $T=10\alpha$. (b)~Energy distribution of the prepared state, $|\psi(E)|^2 = \sum_n |\braket{E_n|\psi(t_{\rm prep})}|^2\delta(E-E_n)$, where $\ket{E_n}$ is an eigenstate of $\hat{H}$ with energy $E_n$. (c)~Equilibration of the local magnetisation on site $j_0$ after the drive is switched off. Dashed lines show the corresponding microcanonical average. (d)~Time-averaged local magnetisation after equilibration (black dots, obtained by averaging over a time interval $\delta t \geq 20 \alpha^{-1}$) compared with the microcanonical average (blue line). (e)~Auto-correlation function of the local operator $\hat{A}=\sum_j u_j \hat{\sigma}^z_j$, where $u_j \propto e^{-(j-j_0)^2}$ is a Gaussian profile ($\sum_ju_j=1$). Lines show the result for $t- t_{\rm prep}=100\alpha^{-1}$, after the system has relaxed to equilibrium, while squares indicate near-identical values for $t- t_{\rm prep}=110\alpha^{-1}$. (f)~Inverse temperature estimated by fitting the low-frequency noise and response functions to the FDT $\tilde{\chi}^{\prime \prime}(\omega)/\tilde{S}(\omega)= \tanh(\beta\omega/2)$ (black dots) and the corresponding microcanonical prediction (blue line).}
\label{fig:1.5.7}
\end{figure}

Let us illustrate this physical configuration from the perspective of our archetypal staggered field model. We consider $\hat{H}_{\textrm{SF}}$ with parameters $\Delta = 0.55\alpha$ and $b = \alpha$ from Eq.~\eqref{eq:h_sf_5} with periodic boundary conditions in the zero-magnetisation sector, with $N = L/2$ excitations. To avoid the effect of global symmetries, we introduce a small perturbation $\delta \hat{\sigma}^z_1$ to the Hamiltonian, with $\delta = 0.1\alpha$. We assume that at time $t = 0$, the system is prepared in the ground state of $\hat{H}_{\textrm{SF}}$, $\ket{\psi_0}$, with energy $E_0$. We then couple the system to a periodic driving field on site $j_0$, at which point we have the total time-dependent Hamiltonian
\begin{align}
\hat{H}(t) = \hat{H}_{\textrm{SF}} + a \sin(\omega_0 t) \hat{\sigma}^z_{j_0},
\end{align}
where we are free to choose the amplitude $a = 2\alpha$, the driving frequency $\omega_0 = 8\alpha$ and $j_0 = L / 2$ for chains with even number of sites. The driving field pumps energy into the system steadily at a rate that depends on the frequency of the periodic field. At time $t = t_{\textrm{prep}}$, the system is in a state with a total energy that depends on $t_{\textrm{prep}}$, at which point the driving field is switched off. Fig.~\ref{fig:1.5.7}(a) show how, as a function of $t_{\textrm{prep}}$, the energy
\begin{align}
\bar{E}(t_{\textrm{prep}}) = \braket{\psi(t_{\textrm{prep}}) | \hat{H} | \psi(t_{\textrm{prep}})}
\end{align}  
of the chain changes as a function of the preparation time. The average energy $\bar{E}(t_{\textrm{prep}})$ can be tuned by selecting a preparation time. The energetic distribution of the class of states generated by this procedure is defined as
\begin{align}
|\psi(E)|^2 = \sum_n |\braket{n | \psi(t_{\textrm{prep}})}|^2 \delta(E - E_n),
\end{align}
where $\ket{n}$ is an eigenstate of $\hat{H}$ with energy $E_n$. It can be observed from Fig.~\ref{fig:1.5.7}(b) that the states generated by periodic driving present an energy-peaked distribution separated by the driving frequency of the driving field. Most importantly, the energy fluctuations for the class of states generated by the driving field
\begin{align}
\frac{\Delta E}{\bar{E} - E_0} \defeq \frac{\Delta E}{\bar{E}_{*}}
\end{align}
decay as the system size increases. The inset in Fig.~\ref{fig:1.5.7}(a) shows the energy fluctuations, where $(\Delta E)^2 = \braket{\psi(t_{\textrm{prep}}) | (\hat{H} - \bar{E})^2 | \psi(t_{\textrm{prep}})}$, relative to the ground state energy $\bar{E}_{*} = \bar{E} - E_0$ as a function of the system size at a fixed microcanonical temperature $T = 10\alpha$. We recall how this is a necessary condition for thermalisation as exposed in Sec.~\ref{sec:thermalisation}. The microcanonical temperature $\beta(E) = \textrm{d}S / \textrm{d}E$ can be extracted from the density of states via the Boltzmann's relation $S(E) = \ln W(E) $, where $W(E) = \Omega(E)\textrm{d}E$ corresponds to the number of microstates in small energy interval $\textrm{d}E$ from the density of states $\Omega(E) = \sum_n \delta(E - E_n)$. This procedure can be evaluated with exact diagonalisation for systems with moderate values of $L$. However, the kernel polynomial method allows one to attain larger system sizes by an appropriate approximation using Chebyshev polynomials~\cite{mitchison2021taking} .

Autonomous, unitary evolution is carried out by the out-of-equilibrium configuration obtained after the driving. Eigenstate thermalisation concerns the dynamics of a generic observable as long as it is local. After the periodic driving has been decoupled, the dynamics of the magnetisation in the $z$ direction of the central site as a function of time are shown in Fig.~\ref{fig:1.5.7}(c). The time-dependent state $\ket{\psi(t)}$ can be obtained from the numerical integration of the Schro\"odinger equation $\textrm{i} \partial_t \ket{\psi(t)} = \hat{H} \ket{\psi(t)}$ through the standard fourth-order Runge-Kutta algorithm~\cite{BrenigTypicality2015}, from which the dynamics of the observable can be computed. It can be observed that $\langle \hat{\sigma}^z_{j_0}(t) \rangle$ relaxes to an asymptotic value in the limit of long times. The saturation point depends on the average energy of the prepared initial state, as seen in Fig.~\ref{fig:1.5.7}(c).

As expected from eigenstate thermalisation, the saturation point of $\langle \hat{\sigma}^z_{j_0}(t) \rangle$ coincides with the microcanonical prediction at energy $\bar{E}$, i.e.,
\begin{align}
O(E) = \frac{1}{\Omega(E)}\sum_n O_{nn} \delta(E - E_n) \approx \lim_{t \to \infty} \langle \hat{O}(t) \rangle = \overline{O},
\end{align}
as can be observed from Fig.~\ref{fig:1.5.7}(d). The expectation value in the microcanonical ensemble can be obtained from exact diagonalisation or the kernel polynomial method described above (see Ref.~\cite{mitchison2021taking} for further details). This is the essence of eigenstate thermalisation as described in Sec.~\ref{sec:thermalisation}. 

At the level of correlation functions at thermal equilibrium, thermalisation is manifest in the fluctuation-dissipation relation. Let us consider the local operator 
\begin{align}
\hat{A} = \sum_j u_j \hat{\sigma}^z_j,
\end{align}
where $u_j \propto e^{-(j - j_0)^2}$ is a normalised Gaussian profile such that $\sum_j u_j = 1$. Correlation functions of the form
\begin{align}
C(t^{\prime}, t) = \langle \hat{A}(t^{\prime}) \hat{A}(t) \rangle - \langle \hat{A}(t^{\prime}) \rangle \langle \hat{A}(t) \rangle, 
\end{align}
become approximately stationary at long times in non-integrable systems, $C(t + \tau, t) \approx C(\tau)$, as can be observed from the numerical evaluation of $C(t + \tau, t)$ in Fig.~\ref{fig:1.5.7}(e). One can express $C(\tau)$ in terms of the symmetrised noise $S(\tau) = \textrm{Re}[C(\tau)]$ and the dissipative response function $\chi^{\prime \prime}(\tau) = \textrm{i} \textrm{Im}[C(\tau)]$. Their Fourier transforms yield the frequency-dependent signals of the correlation function, which are related by the fluctuation-dissipation theorem at thermal equilibrium
\begin{align}
S(\omega) = \coth(\beta \omega / 2) \chi^{\prime \prime}(\omega).
\end{align}
Fig.~\ref{fig:1.5.7}(f) exposes how this behaviour is manifest numerically in the staggered field model for the relaxation of the observable $\hat{A}$. We show the microcanonical temperature $\beta$, evaluated from the $\beta(E) = \textrm{d}S / \textrm{d}E$ and the microcanonical entropy from the density of states at average energy $\bar{E}$. From eigenstate thermalisation, the same temperature parameter can be evaluated from the dynamical correlation functions in the frequency domain after relaxation. In the frequency domain, this amounts to the near-zero frequency $\omega \approx 0$ behaviour of the symmetrised noise and the dissipative response function. Note that these two signals can be approximated from the ETH through $f_{\hat{O}}(\bar{E}, \omega)$, from
\begingroup
\allowdisplaybreaks
\begin{align}
\label{eq:s_chi_eth_5}
S(\omega) &= 2\pi \cosh(\beta \omega / 2) |f_{\hat{O}}(\bar{E}, \omega)|^2, \\
\chi^{\prime \prime}(\omega) &= 2\pi \sinh(\beta \omega / 2) |f_{\hat{O}}(\bar{E}, \omega)|^2,
\end{align} 
\endgroup
where the numerical estimation of $f_{\hat{O}}(\bar{E}, \omega)$ follows from the described procedure in Sec.~\ref{sec:eth_correlation_functions} through the off-diagonal matrix elements. In practice, however, to avoid full diagonalisation, one can evaluate the correlation functions from the time-evolved pure states obtained through the Runge-Kutta algorithm described before. Fig.~\ref{fig:1.5.7}(f) shows the estimation of the microcanonical temperature through 
\begin{align}
\tilde{\chi}^{\prime \prime}(\omega) / \tilde{S}(\omega) = \tanh(\beta \omega / 2).
\end{align}
It can be observed that both predictions are very close to each other for the shown values of the average energy $\bar{E}$, except for low-temperature values in which non-generic features are often found in non-integrable models.

Crucially, the experiment we have just described illustrates how thermal behaviour is manifest from a single parameter: the average energy $\bar{E}$ which then allows one to describe equilibrium quantities through a single-eigenstate thermal ensemble.   

\section[ETH in the single impurity model]{Eigenstate thermalisation in the single impurity model}
\label{sec:impurity_eth}

Integrability is believed to be unstable to perturbations~\cite{Alessio:2016}. Surprisingly, it has been shown that even a single magnetic impurity perturbation around the centre of the integrable spin-1/2 XXZ chain is enough to induce level repulsion and random matrix statistics in the spectrum~\cite{Santos:2004, santos2011domain, torres2014local, Torres_Herrera_2015, XotosIncoherentSIXXZ, Metavitsiadis2010}, as we have highlighted in Chapter~\ref{chapter:models}. However, in Chapter~\ref{chapter:kubo} we provided a study of how this model displays ballistic spin transport in linear response. This result will be solidified in Part~\ref{part:two}, when we revisit the problem from the perspective of open quantum systems. These results challenge our expectation that quantum chaotic systems (those exhibiting random matrix statistics in the spectrum) should exhibit diffusive transport.

Our intuition is that thermalisation in the model must somehow be anomalous, given that the model presents the qualities of a non-integrable model and ballistic transport at the same time. At theoretical level, non-integrability is not inconsistent with coherent transport. From Mazur's inequality in Chapter~\ref{chapter:integrability}, we can predict a vanishing spin Drude weight for any given non-integrable model. Our main point from Chapter~\ref{chapter:kubo}, however, is that a vanishing spin Drude weight does not necessarily imply incoherent transport. As we demonstrated, for the single impurity model, the vanishing Drude weight stems from lack of translation invariance. This fact prompted us to look at the finite-frequency domain in Kubo's linear response theory, which lead us to the conclusion that spin transport in the single impurity model is coherent. Hence, in this sense, non-integrability and coherent transport {\em are not mutually exclusive}. The single impurity model is the perfect example of this fact. 

Having established the fundamental aspects of eigenstate thermalisation, we are now ready to investigate thermalisation in the single impurity model. As we shall see, very interesting behaviour is observed in which the properties of the integrable XXZ model end up embedded in the non-integrable, perturbed model. We shall study the properties of the eigenstates of the unperturbed XXZ model, along with those of eigenstates of the non-integrable model obtained by perturbing it with a magnetic impurity about the centre of the chain. Let us recall the single-impurity Hamiltonian
\begin{align}
\label{eq:h_si_5}
\hat{H}_{\textrm{SI}} = \hat{H}_{\textrm{XXZ}} + h\, \hat{\sigma}^z_{L/2},
\end{align} 
where $h$ is the strength of the magnetic impurity. The interaction parameter in the XXZ model [Eq.~\eqref{eq:h_xxz}] will be considered to be $\Delta = 0.55$, for which spin transport is ballistic, but we will also show results for $\Delta = 1.1$, for which spin transport is diffusive~\cite{Bertini:2021} with $\alpha = 1$ in both cases. We henceforth set $h = 1$ so that all energy scales in our perturbed Hamiltonian are $\mathcal{O}(1)$. We focus on the zero magnetisation sector, $\sum_i \braket{\hat{\sigma}^z_i} = 0$, which is the largest sector [see Sec.~\ref{sec:global_symmetries}]. Reflection symmetry is present in $\hat{H}_{\textrm{XXZ}}$. We explicitly break it by adding a very weak magnetic field at site $i=1$, $h_1=10^{-1}$  (like open boundary conditions, this perturbation does not break integrability~\cite{Santos:2004}). We remark that individual sub-sectors stemming from the presence of a global symmetry of the Hamiltonian are independent, and cannot be coupled coherently by Hamiltonian dynamics. It is then important to consider each sub-sector individually, or, as we proceed in our analysis, explicitly remove the symmetry by a small perturbation that is expected not to change our conclusions significantly. We will use full exact diagonalisation to study matrix elements and eigenstates, and study chains with up to $L = 20$ sites, for which the Hilbert space dimension $\mathcal{D} = L! / [(L/2)!]^2 = 184\,756$.

\subsection{Diagonal ETH}
\label{sec:diagonal_eth_impurity}

Let us first study the diagonal matrix elements of two related local observables. We choose the local kinetic energy at site $i = L/4$ (far away from the boundary and the impurity), 
\begin{equation}
\label{eq:kobs}
\hat{K} \defeq \hat{K}_{\frac{L}{4}, \frac{L}{4}+1} = \left( \hat{\sigma}^x_{\frac{L}{4}}\hat{\sigma}^x_{\frac{L}{4}+1} + \hat{\sigma}^y_{\frac{L}{4}}\hat{\sigma}^y_{\frac{L}{4} + 1} \right),
\end{equation}
and the total kinetic energy per site, the average local kinetic energy, defined as
\begin{equation}
\label{eq:tobs}
\hat{T} \defeq \frac{1}{L} \sum_{i=1}^{L-1} \left( \hat{\sigma}^x_{i}\hat{\sigma}^x_{i+1} + \hat{\sigma}^y_{i}\hat{\sigma}^y_{i+1} \right).
\end{equation}
The contrast between the two shows the effect of averaging in non-translationally invariant systems. We shall also consider the local magnetisation in the $z$ direction at the position the impurity is located, i.e., $\hat{\sigma}^z_{L/4}$.

\begin{figure}[t]
\centering
\includegraphics[width=0.6\columnwidth]{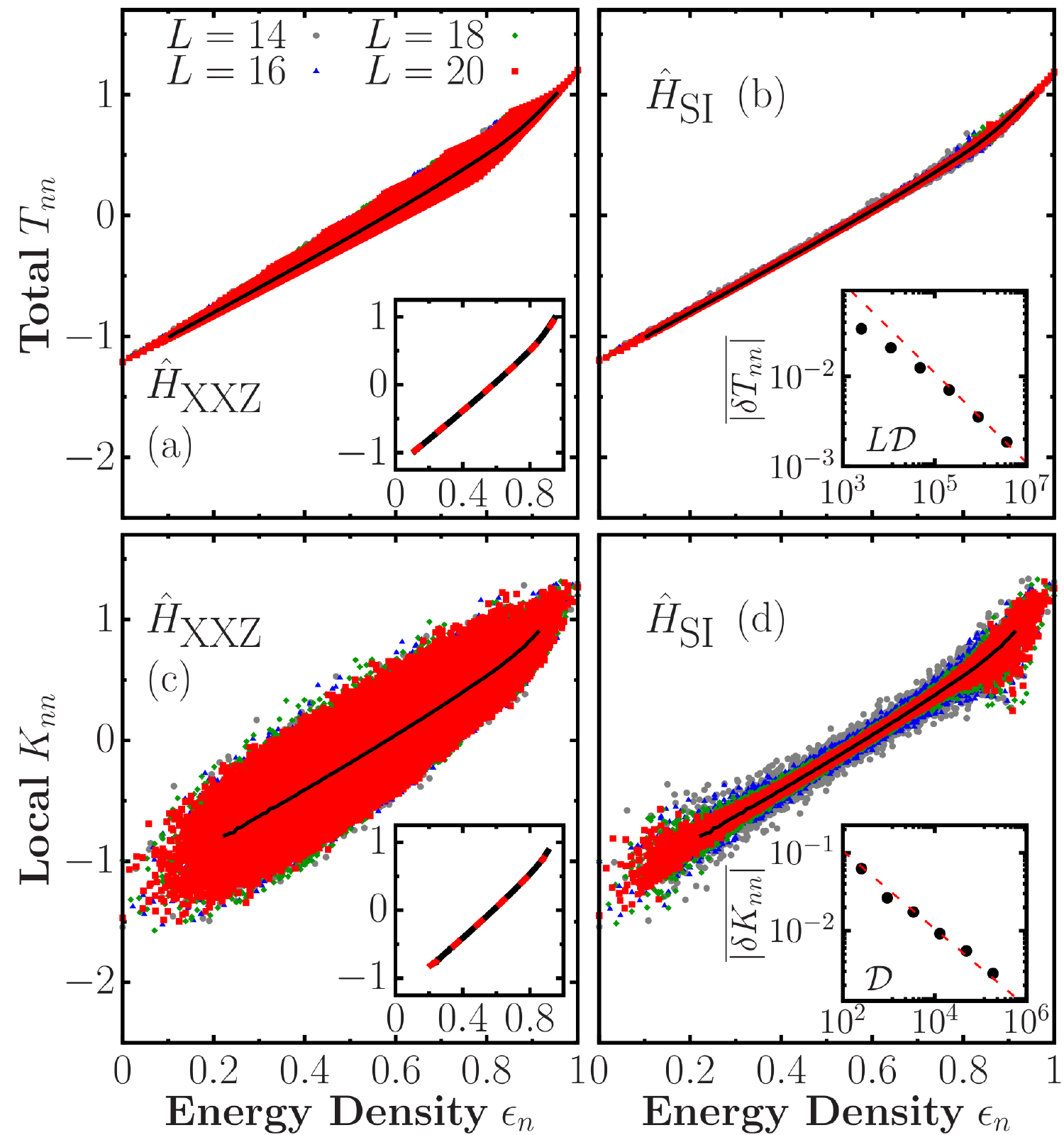}
 \caption[Diagonal matrix elements of $\hat T$ and $\hat K$ in the eigenstates of $\hat{H}_{\textrm{XXZ}}$ and $\hat{H}_{\textrm{SI}}$ for $\Delta=0.55$]{Diagonal matrix elements of $\hat T$ [(a), (b)] and $\hat K$ [(c), (d)] in the eigenstates of $\hat{H}_{\textrm{XXZ}}$ [(a), (c)] and $\hat{H}_{\textrm{SI}}$ [(b), (d)] ($\Delta=0.55$). The black lines show microcanonical averages (within windows with $\delta \epsilon_n = 0.008$) in $\hat{H}_{\textrm{XXZ}}$ for the largest chain ($L = 20$). The insets in (a) and (c) show the equivalence of the microcanonical predictions in both models for each observable, while the insets in (b) and (d) show the $(L\mathcal{D})^{-1/2}$ and $\mathcal{D}^{-1/2}$ scalings, respectively, of $\overline{|\delta O_{nn}|} \defeq \overline{|O_{nn} - O_{n+1n+1}|}$ (the dashed lines are $\propto x^{-1/2}$), where we average over the central 20\% of the eigenstates in chains with $L = 10$, 12, $\dots$, 20.}
\label{fig:1.5.8}
\end{figure}

In Fig.~\ref{fig:1.5.8}, we show the diagonal matrix elements of $\hat K$ and $\hat T$ in the eigenstates of the Hamiltonians in Eqs.~\eqref{eq:h_xxz} and \eqref{eq:h_si_5}. The results are plotted as functions of the energy density defined as $\epsilon_{n} \defeq E_{n} - E_{\textrm{min}} / E_{\textrm{max}} - E_{\textrm{min}}$, where $E_{n}$ is the $n$th energy eigenvalue, and $E_{\textrm{min}}$ ($E_{\textrm{max}}$) is the lowest (highest) energy eigenvalue. Despite the quantitative differences in the behaviour of the two observables in each model (at each energy, the spread of $T_{nn}$ is smaller than that of $K_{nn}$), they both exhibit a qualitatively similar behaviour depending on whether the model is integrable ($\hat{H}_{\textrm{XXZ}}$) or non-integrable ($\hat{H}_{\textrm{SI}}$). In the integrable model, the spread of $T_{nn}$ and $K_{nn}$ at each energy does not change with changing system size as the system does not satisfy the ETH, while in the non-integrable model it decreases exponentially fast with increasing $L$, away from the edges of the spectrum, as can be seen in the insets of Figs.~\ref{fig:1.5.8}(b) and~\ref{fig:1.5.8}(d) for a variance indicator $\overline{|\delta O_{nn}|}$. These results strongly suggest that $T_{nn}$ and $K_{nn}$ satisfy the ETH~\cite{torres2014local, Torres_Herrera_2015}.

Since the single impurity is a sub-extensive local perturbation to the XXZ chain, it does not affect the microcanonical predictions away from the edges of the spectrum for local observables away from the impurity in sufficiently large system sizes. This is confirmed in the insets in Figs.~\ref{fig:1.5.8}(a) and~\ref{fig:1.5.8}(c). Hence, a remarkable consequence of the single impurity producing eigenstate thermalisation (something that is achieved via mixing nearby unperturbed energy eigenstates) is that the smooth functions $T_{nn}$ and $K_{nn}$ are nothing but the microcanonical ensemble predictions for the integrable model. Another interesting consequence of it is that if one evolves highly excited eigenstates of $\hat{H}_{\textrm{SI}}$ under the integrable dynamics generated by $\hat{H}_{\textrm{XXZ}}$, thermalisation will occur at long times, as in the limit of vanishingly small but extensive integrability breaking perturbations~\cite{rigol_16, rigol_srednicki_12}.

Note that, as remarked in Sec.~\ref{sec:thermalisation} and Sec.~\ref{sec:thermalisation_chaotic_models}, the eigenstate-to-eigenstate fluctuations in observables composed of sums of local operators decays as $(L\mathcal{D})^{-1/2}$ due to the $1 / \sqrt{L}$ scaling of the operator Schmidt norm~\cite{Vidmar:2019,Vidmar2:2020,Leblond:2019}. This analysis was only carried out over the central eigenstates of the spectrum, related to the thermodynamical infinite-temperature regime.

\begin{figure}[t]
\centering
\includegraphics[width=0.6\columnwidth]{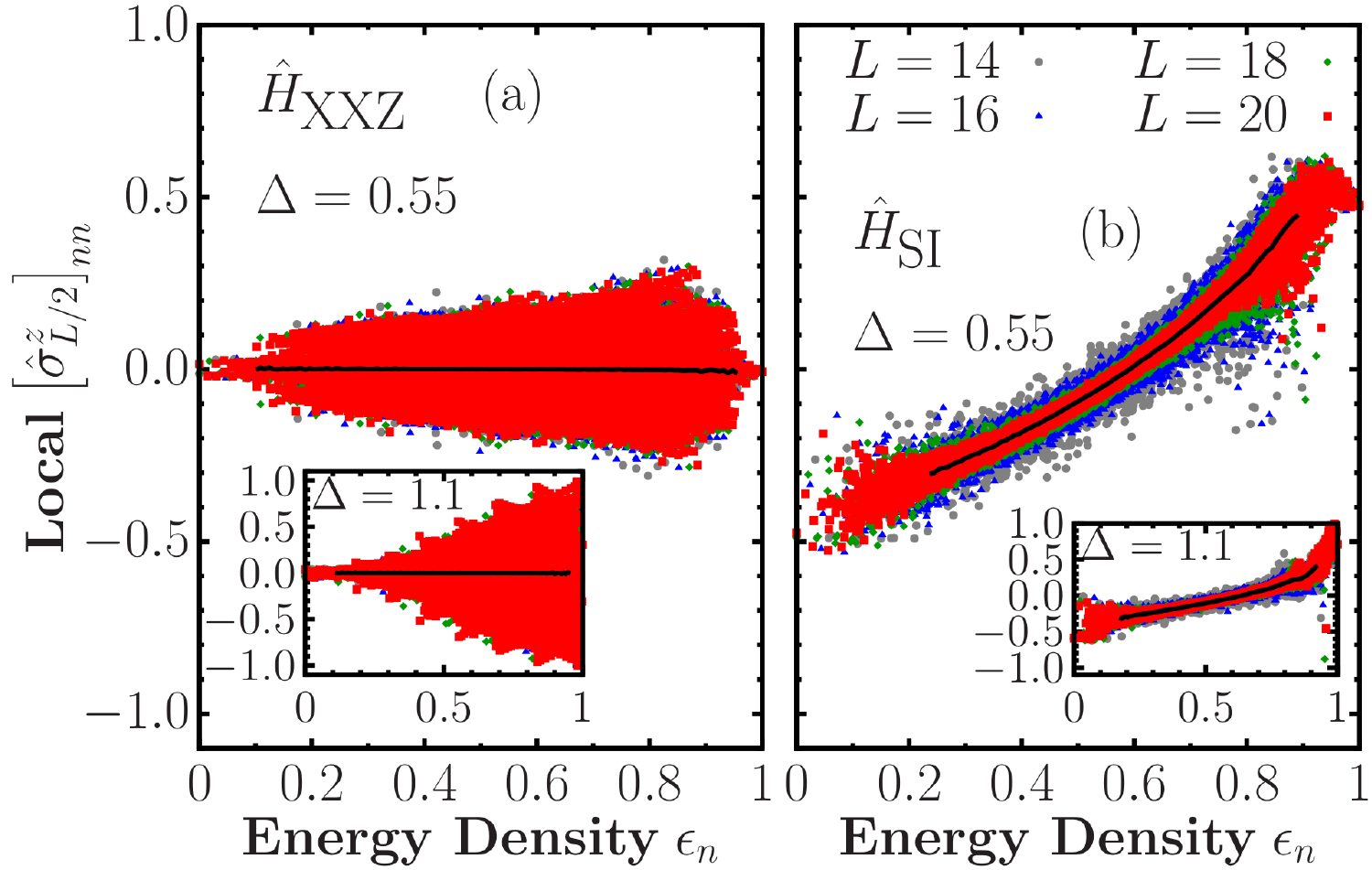}
\caption[Diagonal matrix elements of $\hat{\sigma}^z_{N/2}$ in the eigenstates of the (integrable) XXZ and the non-integrable single-impurity model for different chain sizes $L$]{Diagonal matrix elements of (a) and (b) $\hat{\sigma}^z_{N/2}$ in the eigenstates (a) of the (integrable) XXZ and (b) the non-integrable single-impurity model for $\Delta=0.55$ (main panels) and $\Delta=1.1$ (insets) for different chain sizes $L$. The black lines correspond to the microcanonical averages (within windows with $\delta \epsilon_n = 0.008$) for the largest chain ($L = 20$). We plot the matrix elements vs the energy density $\epsilon_{n}$, defined as $\epsilon_{n} \defeq E_{n} - E_{\textrm{min}} / E_{\textrm{max}} - E_{\textrm{min}}$, where $E_{n}$ is the $n$th energy eigenvalue and $E_{\textrm{min}}$ ($E_{\textrm{max}}$) is the ground-state (highest) energy eigenvalue.}
\label{fig:1.5.9}
\end{figure}

In Fig.~\ref{fig:1.5.9}, we plot the diagonal matrix elements of $\hat{\sigma}^z_{L/2}$ [Figs.~\ref{fig:1.5.9}(a) and~\ref{fig:1.5.9}(b)] in the eigenstates of $\hat{H}_{\textrm{XXZ}}$ [Figs.~\ref{fig:1.5.9}(a)] and $\hat{H}_{\textrm{SI}}$ [Figs.~\ref{fig:1.5.9}(b)] for $\Delta=0.55$ (main panels) and $\Delta = 1.1$ (insets). Fig.~\ref{fig:1.5.9}(a) shows that there is no diagonal eigenstate thermalisation for the local magnetisation in the $z$ direction in the XXZ chain, as the support of the eigenstate to eigenstate fluctuations, at any given energy, does not decrease with increasing system size. This is in contrast to the results for the single-impurity model in which the support of the eigenstate-to-eigenstate fluctuations of both observables, at any given energy away from the edges of the spectrum, decreases with increasing system size. This suggests that diagonal eigenstate thermalisation occurs for $\hat{\sigma}^z_{L/2}$ in the single-impurity model.

The results for $[\hat{\sigma}^z_{L/2}]_{nn}$ are in qualitative agreement with the results for $\hat T$ and $\hat K$, suggesting that diagonal eigenstate thermalisation occurs for local operators in the single-impurity model but not in the integrable XXZ chain. A difference to be highlighted between the diagonal ETH for $\hat T$ and $\hat K$ versus $\hat{\sigma}^z_{L/2}$ in the single-impurity model, is that for $\hat T$ and $\hat K$, the smooth functions $T(\bar{E})$ and $K(\bar{E})$ from the ETH, respectively, are the microcanonical predictions for the integrable model. This could be attributed to the fact that $\hat T$ is an average over the entire chain and the magnetic impurity is a sub-extensive perturbation, while for $\hat{K}$ the observable considered is located away from the impurity site. On the other hand, this is clearly not the case for the smooth function $\sigma^z_{L/2}(\bar{E})$ of $\hat{\sigma}^z_{L/2}$ in Fig.~\ref{fig:1.5.9}(b). The latter is expected since $\hat{\sigma}^z_{L/2}$ is the operator used to perturb the XXZ chain. The microcanonical predictions for such an observable, then, are expected to be different at any energy density away from the infinite-temperature regime.

\subsection{Off-diagonal ETH}

From the diagonal elements of local observables in the energy eigenbasis, we have found that as long as these observables are averaged over the entire system or away from the position of the impurity, the microcanonical expectation values of the unperturbed integrable model get embedded into the perturbed single impurity model. We can then ask the question if this is the case for the off-diagonal matrix elements, from which correlation functions and thus, transport regimes, can be identified.

To answer this question, we then study the off-diagonal matrix elements of the total kinetic energy per site $\hat T$ [Eq.~\eqref{eq:tobs}], and of the spin current operator per site $\hat J$,
\begin{equation}
\label{eq:jobs}
\hat{J} \defeq \frac{1}{L} \sum_{i=1}^{L-1} \left( \hat{\sigma}^x_{i}\hat{\sigma}^y_{i+1} - \hat{\sigma}^y_{i}\hat{\sigma}^x_{i+1} \right).
\end{equation}

Since $\hat T$ and $\hat J$ have Hilbert-Schmidt norms that scale as $1/\sqrt{L}$, the off-diagonal part of the ETH needs to be modified to read~\cite{Leblond:2019, Vidmar2:2020}  
\begin{equation}
\label{eq:eth_scaled}
O_{nm} = \frac{e^{-S(\bar{E})/2}}{\sqrt{L}}f_O(\bar{E}, \omega)R_{nm}.
\end{equation}
We focus on the infinite-temperature regime, in which $\bar{E}\approx 0$ and $S(\bar{E}) \approx \ln{\mathcal{D}}$. 

\begin{figure}[t]
\centering
\includegraphics[width=0.6\columnwidth]{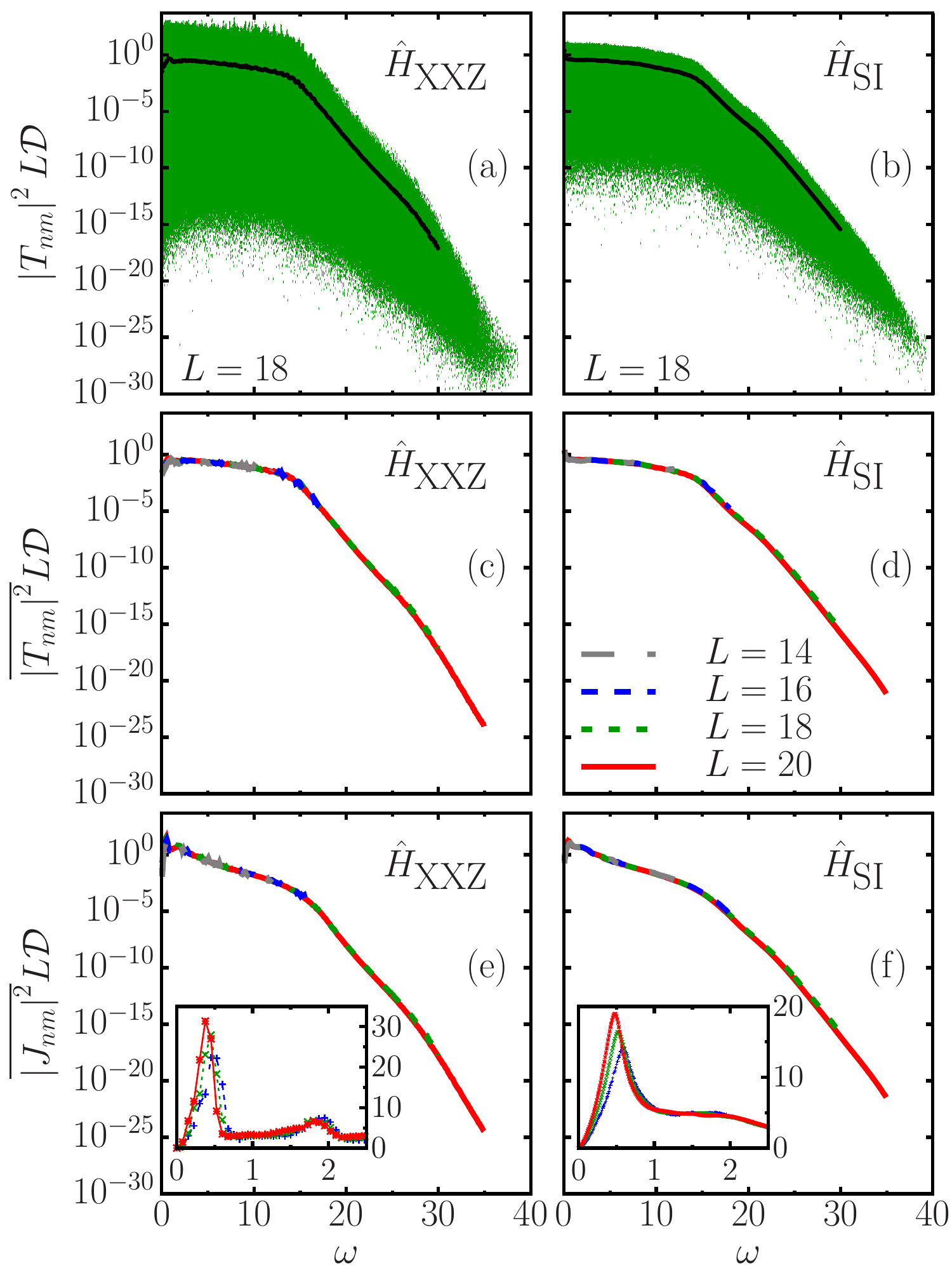}
 \caption[Off-diagonal matrix elements of $\hat T$ and $\hat J$, and the corresponding coarse-grained averages, plotted vs $\omega$ for the XXZ and single impurity models]{[(a), (b)] Off-diagonal matrix elements of $\hat T$, and the corresponding coarse-grained average [continuous (black) line], plotted vs $\omega$ for chains with $L = 18$. [(c), (d)] Coarse-grained averages of $T_{nm}$, including the ones reported in (a) and (b), for different chain sizes. [(e), (f)] Coarse-grained averages of $ J_{nm}$ for different chain sizes (the insets show results at low $\omega$). The left panels [(a), (c), and (e)] show results for $\hat{H}_{\textrm{XXZ}}$, while the right ones [(b), (d), and (f)] show results for $\hat{H}_{\textrm{SI}}$ ($\Delta=0.55$). The matrix elements were computed within a small window of energy around $\bar{E} \approx 0$ (centre of the spectrum) of width $0.05\varepsilon$ ($0.075\varepsilon$ for the insets), where $\varepsilon \defeq E_{\textrm{max}} - E_{\textrm{min}}$. The coarse-grained averages were computed using a window $\delta \omega = 0.1$ [$\delta \omega = 0.075$ and $\delta \omega = 0.01$ for the insets in (e) and (f), respectively].}
\label{fig:1.5.10}
\end{figure}

In Figs.~\ref{fig:1.5.10}(a) and~\ref{fig:1.5.10}(b), we show the off-diagonal matrix elements $|T_{nm}|^2$ in the XXZ and single-impurity models, respectively. As expected, their overall dispersion is larger in the integrable model than non-integrable one. 

Most interestingly, even in integrable models, the variance of the off-diagonal matrix elements of local observables decays as they do in non-integrable models. This decay is given by
\begin{align}
\overline{|O_{nm}|^2} \propto (L\mathcal{D})^{-1} \quad \forall n \neq m,
\end{align}
which, for the case of observables that satisfy $\overline{O_{nm}} \approx 0$, then $\overline{|O_{nm}|^2}$ is approximately the variance of the off-diagonal matrix elements. It was first conjectured that such behaviour would be present in interacting integrable and non-integrable models alike by Mallayya and Rigol in Ref.~\cite{mallayya}, and then shown numerically to be the case by Leblond {\em et al.} in Ref.~\cite{Leblond:2019} in the infinite-temperature regime. This interesting fact is crucial, as it then allows one to define a function $|f_{\hat{O}}(\bar{E}, \omega)|^2$ for both integrable and non-integrable models, even if the integrable models do not satisfy thermalisation according to the ETH. Remarkably, for integrable systems, this can only be done for the squared function $|f_{\hat{O}}(\bar{E}, \omega)|^2$ and not for $f_{\hat{O}}(\bar{E}, \omega)$. The squared function, however, is the object that enters in the expressions for correlation functions and the fluctuation-dissipation relation. It then follows that that off-diagonal matrix elements of observables encodes dynamical properties of interacting systems, irrespective of whether the system is integrable or not.

For both models, Figs.~\ref{fig:1.5.10}(a) and~\ref{fig:1.5.10}(b) show that the coarse-grained average $\overline{|T_{nm}|^2}$ (which corresponds to the variance of the off-diagonal matrix elements as $\overline{T_{nm}}=0$) is a smooth function of $\omega$~\cite{Leblond:2019}. Figures~\ref{fig:1.5.10}(c) and~\ref{fig:1.5.10}(d) for $\overline{|T_{nm}|^2}$, and Figs.~\ref{fig:1.5.10}(e) and~\ref{fig:1.5.10}(f) for $\overline{|J_{nm}|^2}$, show that such a scaling is satisfied by our observables in the XXZ and single-impurity models.

Figs.~\ref{fig:1.5.10}(c) and~\ref{fig:1.5.10}(d) [Figs.~\ref{fig:1.5.10}(e) and~\ref{fig:1.5.10}(f)] also show that the variances $\overline{|T_{nm}|^2}$ ($\overline{|J_{nm}|^2}$) are very similar in the two models and the differences are consistent within present finite-size effects. For $\overline{|J_{nm}|^2}$, see insets in Figs.~\ref{fig:1.5.10}(e) and~\ref{fig:1.5.10}(f), the similarity extends to features that occur at low frequencies. This opens the question of whether there is any difference between the off-diagonal matrix elements of observables in both models.

We find that the off-diagonal matrix elements of observables are normally distributed in the single-impurity model (qualitatively similar results have been obtained in other non-integrable models~\cite{Moessner:2015, Mondaini:2017, Leblond:2019}), while they are close to log-normally distributed in the XXZ chain~\cite{Leblond:2019}. To test the normality of the distribution in the single-impurity model for different values of $\omega$, and to contrast it to the results for the XXZ chain, we compute
\begin{equation}
\label{eq:gamma}
\Gamma_{\hat{O}}(\omega) \defeq \overline{|O_{nm}|^2} / \overline{|O_{nm}|}^2.
\end{equation}
$\Gamma_{\hat{O}} = \pi / 2$ for normally distributed matrix elements, as described in Sec.~\ref{sec:eth_correlation_functions}.

\begin{figure}[t]
\centering
\includegraphics[width=0.6\columnwidth]{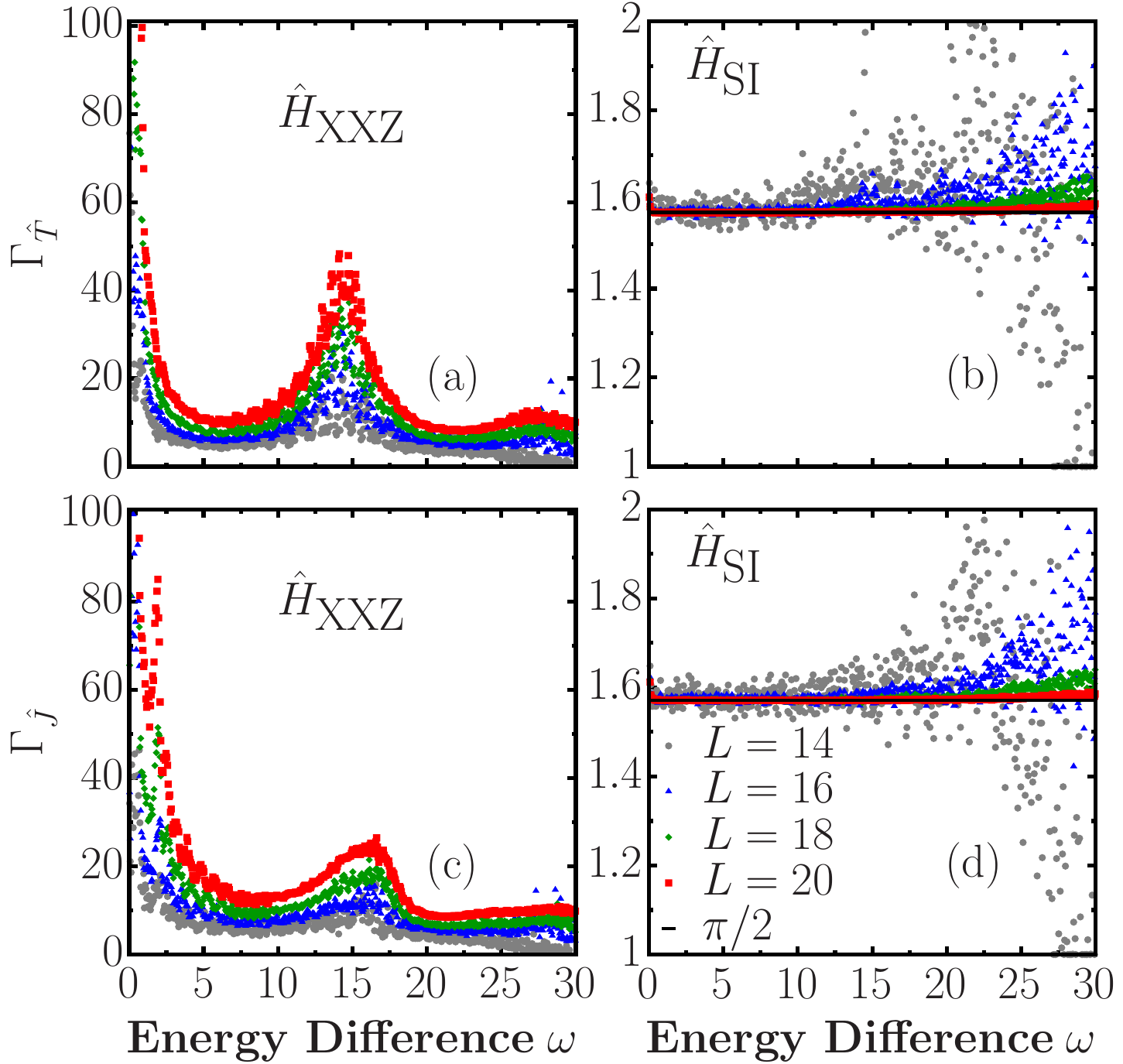}
\caption[$\Gamma_{\hat{O}}(\omega)$, see Eq.~\eqref{eq:gamma}, for the total kinetic energy per site and for the current operator, in the XXZ and single-impurity models]{$\Gamma_{\hat{O}}(\omega)$, see Eq.~\eqref{eq:gamma}, for the total kinetic energy per site [(a), (b)] and for the current operator [(c), (d)], in the XXZ [(a), (c)] and single-impurity [(b), (d)] models ($\Delta=0.55$). The horizontal line in (b) and (d) marks $\pi/2$. The matrix elements were computed using the same energy window as in Fig.~\ref{fig:1.5.10}, while the coarse-graining parameter is $\delta \omega = 0.05$.}
\label{fig:1.5.11}
\end{figure}

In Fig.~\ref{fig:1.5.11}, we show results for $\Gamma_{\hat{T}}(\omega)$ [Figs.~\ref{fig:1.5.11}(a) and~\ref{fig:1.5.11}(b)] and $\Gamma_{\hat{J}}(\omega)$ [Figs.~\ref{fig:1.5.11}(c) and~\ref{fig:1.5.11}(d)] in the XXZ [Figs.~\ref{fig:1.5.11}(a) and~\ref{fig:1.5.11}(c)] and single-impurity [Figs.~\ref{fig:1.5.11}(b) and~\ref{fig:1.5.11}(d)] models. For all values of $\omega$ shown in Figs.~\ref{fig:1.5.11}(b) and~\ref{fig:1.5.11}(d) for the single-impurity model, $\Gamma_{\hat T}(\omega)$ and $\Gamma_{\hat J}(\omega)$, respectively, approach $\pi / 2$ as $L$ increases, i.e., $T_{nm}$ and $J_{nm}$ are well described by a normal distribution. On the other hand, in Figs.~\ref{fig:1.5.11}(a) and~\ref{fig:1.5.11}(c) for the XXZ model, $\Gamma_{\hat T}(\omega)$ and $\Gamma_{\hat J}(\omega)$, respectively, depend on the system size, i.e., $T_{nm}$ and $J_{nm}$ are not normally distributed.

The results discussed so far for the matrix elements of local operators in the single-impurity model show that they are fully consistent with the ETH. The fact that the off-diagonal matrix elements are normally distributed (the variance sets all central moments) means that one can define a meaningful $f_{\hat{O}}(\bar{E}, \omega)$, while this is not the case for the XXZ chain. The question we address next is related to the ballistic spin transport in the single-impurity model, which is in stark contrast to the usual diffusive transport found in non-integrable models.

\subsubsection{Ballistic transport}

We recall that (Chapter.~\ref{chapter:kubo}) within linear response, the real part of the conductivity reads ($k_B = 1$) \cite{kubo1957statistical, kubo1957statistical2, ShastryKubo2008, RigolShastry2008, Bertini:2021}
\begin{align}
\textrm{Re}[\sigma_L(\omega)] = \pi D_L\delta(\omega) + \frac{\pi}{L}&\left(\frac{1-e^{-\beta \omega}}{\omega}\right)\sum_{\epsilon_n \neq \epsilon_m} p_n|J_{nm}|^2\delta(\epsilon_m - \epsilon_n - \omega),
\end{align}
where $D_L$ is the Drude weight, $\beta$ is the inverse temperature, $p_n = e^{-\beta E_n} / Z$ is the Boltzmann weight of eigenstate $\ket{n}$, and $Z$ is the partition function. $J_{nm}$ are the matrix elements of the spin current operator. In integrable systems with open boundary conditions (e.g., our XXZ chain), $D_L$ can be shown to be identically zero no matter the nature of the spin transport, as we demonstrated in Chapter~\ref{chapter:kubo}~\cite{RigolShastry2008}. When transport is ballistic, a peak (or peaks) appear in $\textrm{Re}[\sigma_L(\omega)]$ at a non-zero frequency (frequencies) proportional to $1/L$. When $L\rightarrow\infty$, the peak (peaks) move toward $\omega\rightarrow0$ resulting in a peak in $\textrm{Re}[\sigma_L(\omega=0)]$ that signals ballistic transport~\cite{RigolShastry2008}. Exactly the same was shown to occur in our single impurity model in Chapter~\ref{chapter:kubo}. Therefore, in our integrable and non-integrable models ballistic transport emerges because of the $\omega\rightarrow 0$ behaviour of the off-diagonal matrix elements of the current operator. 

\begin{figure}[t]
\centering
\includegraphics[width=0.6\columnwidth]{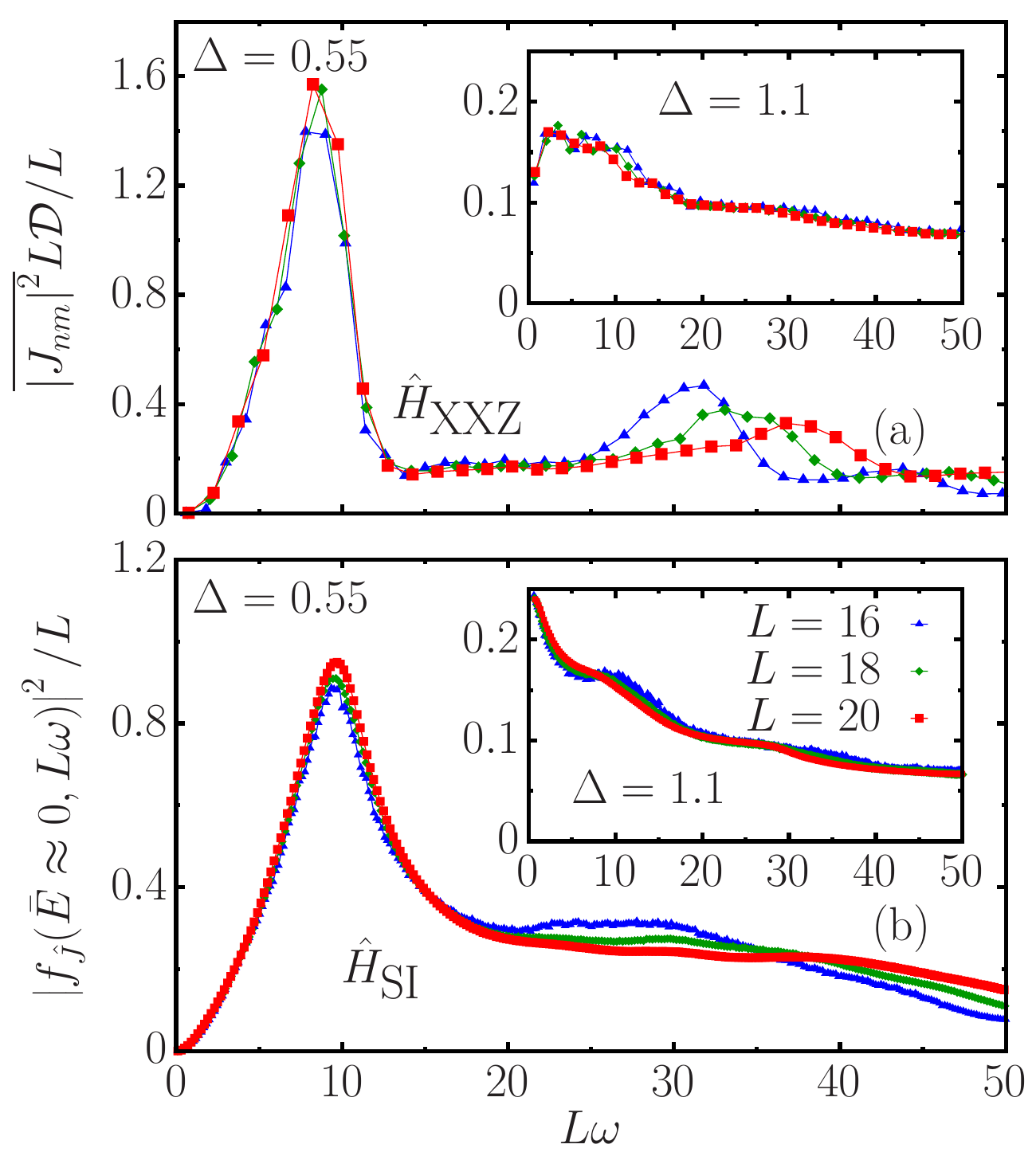}
\caption[Scaled variances of the off-diagonal matrix elements of $\hat J$ in the eigenstates of $\hat{H}_{\textrm{XXZ}}$ and $\hat{H}_{\textrm{SI}}$ plotted vs $L\omega$]{Scaled variances of the off-diagonal matrix elements of $\hat J$ in the eigenstates of $\hat{H}_{\textrm{XXZ}}$ (a) and $\hat{H}_{\textrm{SI}}$ (b) plotted vs $L\omega$. The main panels (insets) show results for $\Delta=0.55$ ($\Delta=1.1$). The matrix elements were computed within a small window of energies around $\bar{E} \approx 0$ of width $0.075\varepsilon$. For the binned averages, we used $\delta \omega = 0.075$ in (a) and $\delta \omega = 0.01$ in (b).}
\label{fig:1.5.12}
\end{figure}

In Fig.~\ref{fig:1.5.12}(a), we show the scaled variances of the matrix elements of $\hat J$ in XXZ chains with $L=16$, 18, and 20 as functions of $L\omega$ for $\Delta=0.55$. A large peak can be seen at a frequency that scales as $1/L$ whose area does not change with increasing $L$. This is consistent with the behaviour of $\textrm{Re} [\sigma_L(\omega)]$~\cite{RigolShastry2008} signalling coherent transport~\cite{Znidaric:2011}. We invite the reader to compare the behaviour of the off-diagonal matrix elements in the energy eigenbasis shown in Fig.~\ref{fig:1.5.12} with the real part of the spin conductivity shown in Fig.~\ref{fig:1.4.2}. The rather similar behaviour is indicative of two illuminating facts. First, the off-diagonal matrix elements encode dynamical quantities related to two-point correlation functions, both in interacting integrable systems and non-integrable systems. Second, in both cases, conductivity is consistent with ballistic spin transport regimes for $\Delta = 0.55$. This behaviour is expected to persist in the parameter range $0 < \Delta < 1$, for which the unperturbed XXZ model is known to display ballistic transport~\cite{Bertini:2021}. 

The position of the smaller (second) peak is nearly $L$ independent [see inset in Fig.~\ref{fig:1.5.10}(e)], appearing to mark the onset of the $L$-independent behaviour shown in Fig.~\ref{fig:1.5.10}. The variances of the matrix elements of $\hat J$ in the (non-integrable) single-impurity model, which, remarkably, define a novel $L$-independent ETH function $|f_{\hat{J}}(\bar{E}\approx 0, L\omega)|^2/L$ [Fig.~\ref{fig:1.5.12}(b)], display the same low-frequency behaviour as in the (integrable) XXZ chain. In contrast, as shown in the insets in Fig.~\ref{fig:1.5.12}, the scaled variances of the matrix elements of $\hat J$ behave completely differently for $\Delta=1.1$, for which spin transport is diffusive. The nature of the spin transport in the absence and presence of the single magnetic defect, for $\Delta$ in the easy-plane ($0 < \Delta < 1$) and easy-axis ($\Delta > 1$) regimes, is something that can readily be probed in ultracold gases experiments~\cite{Jepsen:2020}.

\subsection{Discussion and outlook}

We have demonstrated in Chapter~\ref{chapter:kubo} that spin transport in the easy-plane regime of the XXZ model perturbed with a single magnetic impurity located around the centre of the chain is ballistic, just as the unperturbed model. In this section, we have shown that the matrix elements of observables in such a model are fully consistent with the ETH. Unique to breaking integrability with local perturbations, we argued that statistical mechanics and transport properties of the unperturbed integrable model can end up embedded in properties of the eigenstates of the perturbed (quantum chaotic) one. Thermalisation, in this sense, is manifest from the ergodic behaviour observed in the single impurity model, yet the system thermalises to the microcanonical predictions of the unperturbed model. We have shown this to be the case as long as the observables considered are either averaged over the entire chain or locally constraint in system sites located away from the impurity site.

We showed that the ETH is fully fulfilled when breaking integrability with a local perturbation and that, in such setups, it can inherit statistical mechanics and transport properties of the integrable model. Specifically, we showed that the diagonal matrix elements of observables in the perturbed energy eigenstates can follow the microcanonical predictions for the integrable model, and that ballistic transport in the integrable model can result in a novel $L$-independent ETH function $|f_{\hat{J}}(\bar{E}\approx 0, L\omega)|^2/L$ that characterises the off-diagonal matrix elements of the current operator in the perturbed energy eigenstates at low frequencies. It is quite peculiar that a system that fulfils the ETH is consistent with ballistic transport, since the expectation on physical grounds related to ergodic systems is that scattering processes and complexity will lead to incoherent transport. We have shown the single impurity model to be a counterexample of this intuition. 

An open question is connected to the concept of pre-thermalisation~\cite{Mori:2018}. This phenomenon is characterised by a two-step process in which there exists relaxation to a non-thermal state, that could be described the GGE or another ensemble, followed by a long-time relaxation to a thermal state. This could be the thermalisation mechanism for the single impurity model and we leave it to future work to determine whether this is the case. 
\chapter{Fine structure of eigenstate thermalisation}
\label{chapter:fine_eth}

We have introduced thermalisation in Chapter~\ref{chapter:eth}, through the framework of the eigenstate thermalisation hypothesis. The premise is based on the ansatz that describes the structure of the matrix elements of local observables in the eigenbasis of a non-integrable Hamiltonian, such that ergodicity is satisfied, in the sense that the long-time expectation value of a local observable coincides with the predictions from statistical mechanics. Most importantly, the dynamics of temporal correlation functions related to linear response at thermal equilibrium which, in turn, describe transport, noise and response fall within the range of applicability of the eigenstate thermalisation hypothesis (ETH). Though no formal proof exists for the ETH, overwhelming numerical and experimental evidence attest to its predictive power~\cite{Alessio:2016}. In this sense, the ETH synthesises the conditions to be satisfied by the matrix elements of an operator $\hat{O}$ in the energy eigenbasis
\begin{equation}
\label{eq:eth_6}
O_{n m} = O(\bar{E}) \delta_{n m} + e^{-S(\bar{E}) / 2}f_{\hat{O}}(\bar{E}, \omega)R_{n m},
\end{equation}
to have expectation values and correlation functions indistinguishable from their corresponding finite-temperature counterparts. Eq.~\eqref{eq:eth_6} has been described in Chapter~\ref{chapter:eth} [see Eq.~\eqref{eq:eth}].

Whenever the ETH is satisfied, it is difficult to contrast the coherence of a pure state with that of a statistical mixture by means of standard measurements. Therefore, a question that naturally comes to mind is: will pure state dynamics possess detectable features beyond thermal noise? This question, posed recently by Kitaev~\cite{kitaevTalk} in the context of black-hole physics, lead him to suggest the study of
a peculiar type of out-of-time-order correlations (OTOCs), originally introduced by Larkin and Ovchinikov~\cite{larkin1969quasiclassical}. This object, as a result of a nested time structure, detects quantum chaos and correlations beyond thermal ones. It was recently shown~\cite{kurchan,Chan2019} that OTOCs are controlled by correlations beyond ETH. We must remark, however, that the interpretation of the connection between the OTOC and the underlying quantum state dynamics is, in general, complex.

In Sec.~\ref{sec:entanglement_eth} we shall show that the task of discriminating a pure state that {\em looks} thermal from a true, thermal Gibbs density matrix might be achieved by a different physical quantity: the quantum Fisher information (QFI)~\cite{helstrom1969quantum, Tth2014, pezze2014quantum} with respect to the pure state and the thermal state, respectively. The QFI is a quantity of central importance in metrology~\cite{giovannetti2011advances,Pezze:2018} and entanglement theory~\cite{Hyllus2012,Tth2012}. The first observation in this chapter is that the QFI computed in the eigenstates of the Hamiltonian  ${\cal F}_{\textrm{ETH}}$ (or in the asymptotic state of a quenched dynamics), and the one computed in the Gibbs state at the corresponding inverse temperature $\beta$, ${\cal F}_{\textrm{Gibbs}}$  \cite{Hauke2016, Gabbrielli2018}, satisfy the inequality ${\cal F}_{\textrm{ETH}} \geq {\cal F}_{\textrm{Gibbs}}$, where the equality holds at zero temperature. By computing both terms, we quantify the difference. The corresponding multipartite entanglement structure, as obtained from the Fisher information densities $f_Q={\cal F}/L$ is in stark contrast. For example, in systems possessing finite temperature phase transitions, we argue that $\mathcal F_{\textrm{ETH}}$ diverges with system size at critical points (implying extensive multipartiteness of entanglement in the pure state), while it is only finite in the corresponding Gibbs ensemble \cite{Hauke2016,Gabbrielli2018,Frrot2018}.

The second part of this chapter is intended to provide a connection between the statistical correlations in the matrix elements of local operators in the energy eigenbasis [Eq.~\eqref{eq:eth_6}] and the OTOCs. In isolated classical systems, thermalisation relies on the emergence of chaos and ergodicity, which together lead phase-space trajectories starting from the same energy to become indistinguishable when averaged over time~\cite{lebowitz1973modern}. The equivalent notion of indistinguishability in quantum many-body systems is provided by the ETH~\cite{Deutsch:1991,Srednicki:1994,Alessio:2016}, which states that nearby energy eigenstates cannot be distinguished by local observations. 

More recently, thermalisation has been explored from a new quantum information perspective, with emphasis on the notion of scrambling~\cite{swingle2018}. Information scrambling is a more primordial feature of quantum dynamical systems where information, initially stored locally, gets dynamically distributed in global degrees of freedom \cite{Hosur2016Chaos}. This process is explained as a consequence of the growth of operator complexity under time evolution~\cite{Parker19}. Although traditional tools can hardly be of any help in studying this phenomenon, a variety of ideas have emerged recently for this task. Among them, the OTOCs~\cite{larkin1969quasiclassical}, suggested to characterise synthetic analogues of black-holes~\cite{kitaevTalk,shenker2014black,maldacena2016}, has arisen as an important figure of merit for scrambling, ergodicity and quantum chaos in complex many-body quantum systems. Several experimental studies with a variety of platforms have demonstrated that OTOCs indeed characterise scrambling following the operation of a unitary circuit~\cite{garttner2017,li2020,Nie20,mi2021}. 

Recently, Foini and Kurchan~\cite{Foini2019} argued that correlations between the matrix elements of operators in the energy eigenbasis must exist in the ETH to account for the positive exponential growth rate of OTOCs in chaotic models~\cite{maldacena2016}. Based on this result, Murthy and Srednicki~\cite{Murthy19} were able to derive known bounds on the growth rate from the ETH. Chan et al.~\cite{Chan2019} showed that in locally interacting systems the butterfly effect for OTOCs implies a universal form for these correlations. The existence of frequency-dependent correlations has recently been confirmed by Richter et al.~\cite{Richter2020} and a distinction with a regime in which these correlations vanish was identified, by a numerical investigation of the statistical distributions of matrix elements in non-integrable systems.

It remains an open question to establish if these frequency-dependent correlations can be observed in the dynamics of OTOCs and if the timescales associated with late-time chaos can be connected to the presence, or lack thereof, of matrix-element correlations. It is still not clear if temperature plays a role and, furthermore, the scaling as a function of system size of the frequency scale that divides correlated and uncorrelated regimes has yet to be estimated.

In the Sec.~\ref{sec:otoc_eth}, we carry out a thorough study of the frequency and energy dependence of the statistics of off-diagonal matrix elements and of the OTOCs of extensive observables in two experimentally relevant models: hardcore bosons with dipolar interactions in a harmonic trap~\cite{Khatami:2013} and an Ising chain with longitudinal and transverse fields~\cite{Kim2013}. In all instances, the statistical matrix appears to have some common features. The matrix elements $R_{nm}$ at a given energy $\bar{E}$ and frequency $\omega$ obey Gaussian statistics~\cite{Beugeling2015, Leblond:2019, Khaymovich2019, Richter2020, Leblond:2020, Santos2020}, in contrast with non-ergodic systems~\cite{Luitz2016Anomalous, Luitz2016Long, Foini2019Eigenstate}. We demonstrate that this feature persists well-beyond the infinite-temperature limit. We also further characterise the statistical correlations between $R_{nm}$ at well-separated frequencies. However, these correlations disappear between matrix elements close to the diagonal, indicating the emergence of random-matrix-like behaviour at small frequencies $|E_n-E_m|<\omega_{\rm GOE}$, where $\omega_{\rm GOE}$ is a model- and operator-dependent energy scale, as first demonstrated for non-extensive observables in Ref.~\cite{Richter2020}. We show explicitly that this rich structure is naturally reflected in the dynamics of OTOCs. A comparison between the OTOCs computed on a thermal ensemble and those computed assuming the ETH with a random uncorrelated Gaussian statistical matrix shows convergence of the two on time scales that appear to be related to $\omega_{\rm GOE}^{-1}$. We use this observation to provide an estimation of the scaling as a function of the system size of $\omega_{\rm GOE}^{-1}$ in the infinite-temperature regime. This suggests that an experimental study of OTOCs could be an efficient way to probe the energy scales beyond which a complex, interacting system displays Gaussian random-matrix behaviour in local observables.   

\section[Multipartite entanglement]{Multipartite entanglement in the eigenstate thermalisation hypothesis}
\label{sec:entanglement_eth}

In this section, the eigenstate thermalisation hypothesis will be shown to to be intimately related to the notion of multipartite entanglement through the quantum Fisher information. In Sec.~\ref{sec:bipartite_entanglement} we introduce the concept of entanglement and entanglement measures, while Sec.~\ref{sec:multipartite_entanglement} briefly introduces mutipartite entanglement and its connection to the quantum Fisher information. We then provide the main result of the section in Sec.~\ref{sec:qfi_eth}, in which we show the multipartite entanglement structure within the ETH. Sec.~\ref{sec:evaluation_qfi} provides a numerical example using the staggered field model and we finalise with a short summary in Sec.~\ref{sec:summary_qfi_eth}.

\subsection{Bipartite entanglement and correlations}
\label{sec:bipartite_entanglement}

Entanglement is at the core of quantum mechanics. The inception of the Einstein, Podolsky and Rosen paradoxical thought experiment~\cite{Einstein:1935}, later reformulated by Bohm~\cite{Bohm:1951} in its most commonly-referred form, incited the notion that the so-called {\em orthodox} view of physical reality\footnote{A view in which pre-existing physical quantities are {\em created} rather than {\em revealed} at the moment of measurement} was not a complete description. It was not until Bell proposed that quantum mechanics is incompatible with a local hidden variable theory~\cite{Bell:1964}, solidified experimentally by the work of Aspect, Dalibard and Roger~\cite{Aspect:1982}, that the notion of entanglement as part of the physical reality became widespread. Today, quantum entanglement is a concept of much interest in many-body quantum physics~\cite{Amico:2008} and lies at the forefront of developing quantum technologies~\cite{Preskill2018}.

There exists a plethora of approaches to the problem of measuring entanglement~\cite{Amico:2008}. Let us first consider the the case of bipartitions. Most generally, in this sense, a system is divided in two bipartite sections $A$ and $B$. We then start from the fact that a pure bipartite quantum state $\ket{\psi_A}$ is not entangled with its counterpart $\ket{\psi_B}$, referred to as its {\em complement}, if, and only if, the quantum state $\ket{\psi_{AB}}$ pertaining to the system $A+B$ can be written down as a tensor product of the states of the partitions. If this is the case, $\ket{\psi_{AB}} = \ket{\psi_{A}} \otimes \ket{\psi_{B}}$ and we say that the state $\ket{\psi_{AB}}$ is {\em separable}.

For the specific case of bipartitions, the global Hilbert space is a tensor product of the Hilbert space of the two sub-systems $\mathcal{H} = \mathcal{H}_A \otimes \mathcal{H}_B$ and defines the Schmidt decomposition
\begin{align}
\ket{\psi_{AB}} = \sum_{i = 1}^{n} \sqrt{\lambda_i} \ket{i_A} \otimes \ket{i_B},
\end{align}
where $\ket{i_A}$ $(\ket{i_B})$ is an orthonormal basis pertaining to the local Hilbert space $\mathcal{H}_A$ $(\mathcal{H}_B)$ and $n \leq \textrm{min}\{ \textrm{dim}[\mathcal{H}_A], \textrm{dim}[\mathcal{H}_B] \}$. The $\lambda_i$ are the Schmidt coefficients and satisfy $0 < \lambda_i \leq 1$. It follows from $\braket{\psi_{AB} | \psi_{AB}} = 1$ that $\lambda_i \geq 0$ and $\sum_{i=1}^n \lambda_i = 1$. The Schmidt bases coincide with the eigenbases of the corresponding reduced density operators
\begin{align}
\hat{\rho}_{B/A} = \textrm{Tr}_{A/B}[\ket{\psi_{AB}}] = \sum_{i=1}^{n} \lambda_i \ket{\psi_{B/A, i}} \bra{\psi_{B/A, i}},
\end{align}
where $\textrm{Tr}_{A}[\cdot]$ ($\textrm{Tr}_{B}[\cdot]$) denotes the partial trace over the basis $\ket{i_A}$ ($\ket{i_B}$), yielding a state which does not depend on the degrees of freedom of its complement. An interesting point is that the density operators $\hat{\rho}_A$ and $\hat{\rho}_B$ have a common spectrum and they are equally mixed. Furthermore, only product states $\ket{\psi_{AB}} = \ket{\psi_{A}} \otimes \ket{\psi_{B}}$ lead to pure reduced density matrices, so a measure of their mixedness can be envisaged to quantify entanglement. In particular, from the Schmidt decomposition, the state $\ket{\psi_{AB}}$ is only separable using the bipartition that defines the regions $A$ and $B$ if there exists only one non-zero $\lambda_i$~\cite{Amico:2008}. As it turns out, the only measure of entanglement that satisfies invariance under unitary operations, continuity and additivity, is the von Neumann entropy of reduced density matrices, defined as
\begin{align}
S(\hat{\rho}_A) = S(\hat{\rho}_B) = -\textrm{Tr}(\hat{\rho}_A \log \hat{\rho}_A) = -\textrm{Tr}(\hat{\rho}_B \log \hat{\rho}_B).
\end{align}
Note that $S(\hat{\rho}_A) = S(\hat{\rho}_B)$ since its corresponding density operators share the same spectrum. It then follows that the entanglement entropy is a property of the bipartition and it vanishes for separable states. This object has been the subject of much study, as its finite-size properties allude at information regarding quantum phase transitions in many-body systems~\cite{Laflorencie:2016, Eisert:2010, KimThesis:2014, Amico:2008}.

\begin{figure}[t]
\centering
\includegraphics[width=0.7\columnwidth]{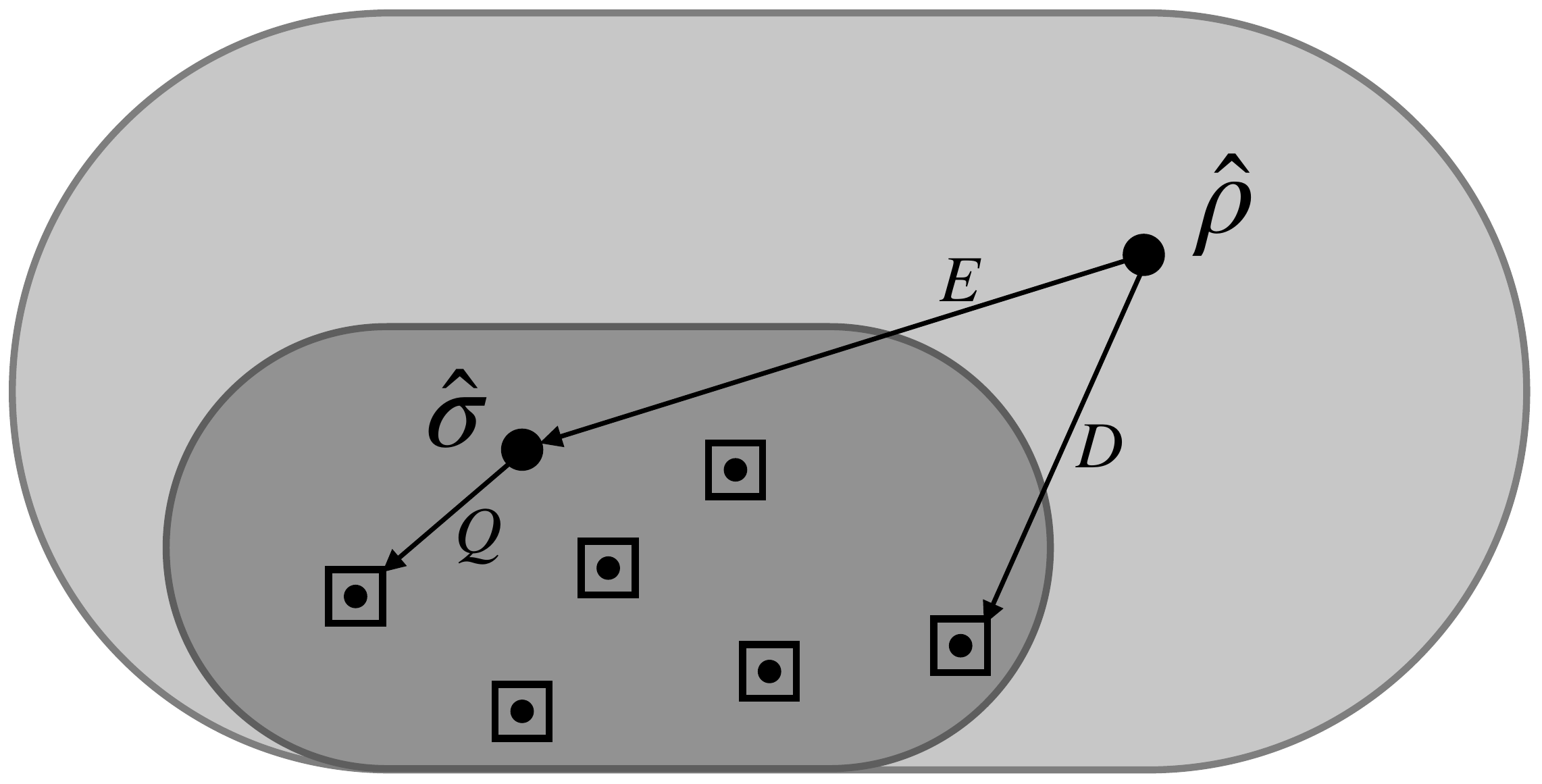}
\caption[A depiction of correlations as distance]{A depiction of correlations as distance from Ref.~\cite{Modi:2010}. The large ellipse represents the set of all possible states, while the smaller one contains the set of the separable states. The squares inside the set of separable states denote the set of classical states while the dots within constitute the set of product states. The state $\hat{\rho}^{\textrm{P}}_{AB}$ from Eq.~\eqref{eq:rho_2l_p}, for example, is a separable state. $E$ denotes the entanglement, $D$ the discord and $Q$ the dissonance.}
\label{fig:1.6.1}
\end{figure}

The generalisation of the classical mutual information for probability densities leads to the quantum mutual information, defined as 
\begin{align}
I(A:B) = S(\hat{\rho}_A) + S(\hat{\rho}_B) - S(\hat{\rho}_{AB}),
\end{align}
which vanishes for completely uncorrelated separable states and it is maximised for entangled states. It is important to remark that this quantity can be evaluated on mixed states as well as pure states, in which case it is a measure of both quantum and classical correlations. Certain states can be separable, i.e., unentangled, but still display classical correlations which $I(A:B)$ quantifies. Consider the following mixed state written down as a separable state between two different two-level systems
\begin{align}
\label{eq:rho_2l_p}
\hat{\rho}^{\textrm{P}}_{AB} &= \frac{1}{2} \left[ \ket{0} \bra{0}_A \otimes \ket{0} \bra{0}_B + \ket{1} \bra{1}_A \otimes \ket{1} \bra{1}_B \right] \defeq  \frac{1}{2} \left[ \ket{00} \bra{00} + \ket{11} \bra{11} \right].
\end{align} 
It follows that $\hat{\rho}^{\textrm{P}}_A = \hat{\rho}^{\textrm{P}}_B = \frac{1}{2} \left[ \ket{0} \bra{0} + \ket{1} \bra{1} \right]$ from the partial traces. This state is diagonal in the $\{ \ket{0}_{A/B}, \ket{1}_{A/B} \}$ bases. Since the state is written as a mixture of product states, by definition it constitutes a separable state. However, for this particular state, it trivially follows that $S(\hat{\rho}_A) = S(\hat{\rho}_B) = S(\hat{\rho}_{AB}) = \log(2)$, which then leads to $I(A:B) = \log(2)$.

More generally, the nature of total correlations, both quantum and classical, can be unified from the perspective of the relative entropy as described by Modi {\em et al} in Ref.~\cite{Modi:2010}, see Fig.~\ref{fig:1.6.1}.

The relative entropy 
\begin{align}
S(\hat{x} || \hat{y}) \defeq -\textrm{Tr}[\hat{x} \log \hat{y}] - \textrm{Tr}[\hat{x} \log \hat{x}]
\end{align}
is non-negative quantity related to the geometric distance between the states $\hat{x}$ and $\hat{y}$ in Hilbert space. Note that this quantity is not symmetric, so in the technical sense it does not constitute an appropriate distance measure. The relative entropy provides a quantifier of all possible state correlations, i.e., the entanglement $E$, the discord $D$ and the dissonance $Q$. As defined in Ref.~\cite{Modi:2010},
\begin{align}
E = \textrm{min}_{\hat{\sigma} \in \mathcal{S}} S(\hat{\rho} || \hat{\sigma}), D = \textrm{min}_{\hat{\chi} \in \mathcal{C}} S(\hat{\rho} || \hat{\chi}), Q = \textrm{min}_{\hat{\chi} \in \mathcal{C}} S(\hat{\rho} || \hat{\chi}),
\end{align}
where $\mathcal{S}$ is the set of separable states and $\mathcal{C}$ is the set of classical states. We remark that this mathematical picture formally presents a complete and unified view of quantum and classical correlations. However, in general, its actual evaluation could turn out to be very complicated.

\subsection{Multipartite entanglement and quantum Fisher information}
\label{sec:multipartite_entanglement}

The quantification of many-party entanglement is a difficult task~\cite{Amico:2008}. It is of great utility, however, to determine whether a state subdivided in $L$ parties is entangled or not. Certain quantifiers exist to determine the presence, or lack thereof, of entanglement. Such quantifiers could be employed to ascertain whether a state is entangled.

A pure quantum state is separable in a system of $L$ particles or qubits, for instance, if it can be written as a product state
\begin{align}
\ket{\psi_{\textrm{sep}}} = \ket{\psi^{(1)}} \otimes \ket{\psi^{(2)}} \otimes \cdots \otimes \ket{\psi^{(L)}},
\end{align}
where $\ket{\psi^{(l)}}$ is the state of the $l$th individual element. On the other hand, a mixed state is separable if it can be written as a mixture of pure states~\cite{Pezze:2018}
\begin{align}
\hat{\rho}_{\textrm{sep}} = \sum_q p_q \ket{\psi_{\textrm{sep}, q}}\bra{\psi_{\textrm{sep}, q}},
\end{align}
with $p_q \leq 0$ and $\sum_q p_q = 1$. For the case where $L > 2$, multipartite entanglement is characterised by the number of individual elements in the {\em largest non-separable subset}. We can state that a pure state of $L$ individual elements is $k$-separable if it can be written as the product
\begin{align}
\ket{\psi_{k-\textrm{sep}}} = \ket{\psi^{(L_1)}} \otimes \ket{\psi^{(L_2)}} \otimes \cdots \otimes \ket{\psi^{(L_N)}},
\end{align}
where $\ket{\psi^{(L_l)}}$ is a state of $L_l \leq k$ individual elements and $\sum_{l=1}^N L_l = L$. Just as before, a mixed state is $k$-separable if it can be written down as a statistical mixture of $k$-separable pure states. In particular, a state that is $k$-separable but not $(k - 1)-$separable is dubbed $k-$particle entangled, and it contains at least one state of $k$ particles that does not factorise. In maximally-multipartite-entangled states, $k = L$ and each individual element is entangled with all the others~\cite{Pezze:2018}.

The quantum Fisher information (QFI) $\mathcal{F}$ can be used to ascertain the previous structure in the multipartite entanglement. This quantity was introduced to bound the precision of the estimation of a parameter $\phi$, conjugated to an observable $\hat{O}$ using a quantum state $\hat{\rho}$. 

If one is interested in the estimation of a parameter $\phi$, unknown {\em a-priori}, the phase estimation protocol begins by preparing a probe state $\hat{\rho}_0$ and letting it interact within the system via a generic transformation~\cite{Pezze:2018}
\begin{align}
\label{eq:transformation_phi}
\hat{T}(\phi): \hat{\rho}_0 \rightarrow \hat{\rho}_{\phi}.
\end{align}
In principle, the transformation can be given by that generated by unitary dynamics from the Hamiltonian in isolated systems, $\hat{\rho}_{\phi} = e^{\textrm{i}\phi\hat{H}} \hat{\rho}_0 e^{-\textrm{i}\phi\hat{H}}$. The parameter $\phi$ cannot be measured directly and the estimation is done via the results of the measurements performed on identical copies of the output state $\hat{\rho}_{\phi}$. The measurements can correspond to expectation values, denoted by $\mu$. We define $P(\mu | \phi)$ as the conditional probability of a result $\mu$ given the parameter $\phi$. A sequence of {\em independent} $M$ measurements allows one to factorise the probability of observing the sequence $\vec{\mu} = {\mu_1, \cdots, \mu_M}$ as
\begin{align}
P(\vec{\mu} | \phi) = \prod_{i=1}^{M} P(\mu_i | \phi).
\end{align}  
One then defines an estimator function $\Phi(\vec{\mu})$ as a generic function that associates the measurement outcomes $\vec{\mu}$ with the estimation of $\phi$. Given that this function is the result of random outcomes, it constitutes a random variable itself and, therefore, characterised by a mean
\begin{align}
\overline{\Phi(\vec{\mu})} = \sum_{\mu} P(\vec{\mu} | \phi) \Phi(\vec{\mu})
\end{align}
and a variance
\begin{align}
(\Delta \phi)^2 = \sum_{\mu} P(\vec{\mu} | \phi) \left[ \Phi(\vec{\mu}) - \overline{\Phi(\vec{\mu})} \right]^2.
\end{align}
At this point we should mention that not all choices of measurement observables are optimal. The optimal measurements are the ones which maximise the sensitivity to changes in $\phi$~\cite{Pezze:2018}. Furthermore, the estimators themselves could in principle be different and yield different values when applied to the same outcomes $\vec{\mu}$. It is common to consider {\em unbiased} estimators, for which $ \overline{\Phi(\vec{\mu})} = \phi$ and $\partial \overline{\Phi(\vec{\mu})} / \partial \phi = 1$~\cite{Pezze:2018,Hyllus2012}.

The Cramer-Rao bound can then be derived for an unbiased estimator~\cite{Pezze:2018} 
\begin{align}
\Delta \phi \geq \Delta \phi_{\textrm{CR}} = \frac{1}{\sqrt{M F(\phi)}},
\end{align}
which connects the fundamental limit of precision attainable from the estimation protocol to the Fisher information
\begin{align}
F(\phi) = \sum_{\mu} \frac{1}{P(\mu | \phi)} \left( \frac{\partial P(\mu | \phi)}{\partial \phi} \right)^2.
\end{align}
Its quantum version is obtained by maximising over all possible generalised measurements $\mathcal{F}(\hat{\rho}_{\phi}) = \textrm{max}_{\hat{E}} F(\phi)$ and it {\em upper bounds} the classical Fisher information $F(\phi) \leq \mathcal{F}(\hat{\rho}_{\phi})$. In this sense, the classical Cramer-Rao relation upper bounds its quantum counterpart
\begin{align}
\Delta \phi_{\textrm{CR}} \geq \Delta \phi_{\textrm{QCR}} = \frac{1}{\sqrt{M \mathcal{F}(\hat{\rho})}}.
\end{align}
It then follows that quantum protocols provide a lower fundamental limit in the uncertainty in parameter estimation. For a composite system of $L$ individual entities (particles, spins), there exists a fundamental difference in the scaling of the QFI as a function of $L$~\cite{Hyllus2012} for separable states
\begin{align}
\mathcal{F}(\hat{\rho}_{\textrm{sep}}) \leq L
\end{align} 
and entangled states,
\begin{align}
\mathcal{F}(\hat{\rho}) \leq L^2.
\end{align} 
It is this fact that allows one to establish a connection between multipartite entanglement and the QFI\footnote{In the above expressions we considered the dependance of the QFI as a function of the states $\hat{\rho}$ as opposed to the parameter $\phi$, assuming that both $\hat{\rho}$ and $\phi$ are connected via the transformation $\hat{T}(\phi)$ from Eq.~\eqref{eq:transformation_phi}.}. In passing, we mention that it is this difference in scaling that allows one to surpass the classical limit on the uncertainty in phase estimation. We remark that the scaling bounds above have only been shown to apply to a specific class of phase-shift generators $\hat{T}(\phi)$~\cite{Hyllus2012}.

Most importantly, the QFI has key mathematical properties~\cite{Braunstein1994,PETZ2011,Tth2014,Pezze:2018}, such as convexity, additivity, monotonicity and it can, following our discussion above, be used to probe the multipartite entanglement structure of a quantum state~\cite{Hyllus2012,Tth2012}. If, for a certain Hermitian operator $\hat{O}$ that generates $\hat{\rho}_{\phi} = e^{\textrm{i}\phi\hat{O}} \hat{\rho}_0 e^{-\textrm{i}\phi\hat{O}}$, the QFI density satisfies
\begin{align}
 f_Q(\hat O)=\frac{\mathcal{F}(\hat{O})}{L} > m,
\end{align}
then at least $(m+1)$ parties in the system are entangled (with $1\leq m \leq L-1$ a divisor of $L$). Hence, $m$ represents the size of the biggest entangled block of the quantum state. In particular, if 
\begin{align}
L-1 \leq f_Q(\hat O) \leq L,
\end{align} 
then the state is called {\em genuinely} $L$-partite entangled. In this sense, the quantum Fisher information density can be used to ascertain the multipartite entanglement structure.

In general, different operators $\hat{O}$ lead to different bounds and there is no systematic method to choose the optimal one without some knowledge on the physical system~\cite{Hauke2016,Pezz2017Gabri}. The operator, however, will typically be an extensive sum of local operators. An educated guess, however, based on some knowledge of the system allows the detection of multipartite entanglement in physically relevant situations, e.g., choosing the order parameter in the proximity of quantum phase transitions~\cite{Hauke2016,Pezz2017Gabri}.

\subsection{Connection between quantum Fisher information, eigenstate thermalisation and linear response}
\label{sec:qfi_eth}

There exists a connection between the QFI evaluated for a given observable in the regime of linear response for a generic quantum system and the response functions, as first noted by Hauke {\em et al.} in Ref.~\cite{Hauke2016}. The relation between the QFI and the response functions will allow us to study the consequences of eigenstate thermalisation on the QFI and hence, on the multipartite entanglement structure present in systems that satisfy the ETH.

It was shown by Braunstein and Caves in Ref.~\cite{Braunstein1994} that the QFI can be explicitly written in terms of the decomposition of a general mixed state in the energy eigenbasis, i.e., $\hat{\rho} = \sum_n p_n \ket{n}\bra{n}$, where the $\ket{n}$ are the eigenstates of the Hamiltonian. In their work, Braunstein and Caves showed that the QFI
\begin{align}
\label{QFI}
{\mathcal F}(\hat O)=2\sum_{n,n^{\prime}} \frac{(p_n-p_{n^{\prime}})^2}{p_n+p_{n^{\prime}}} |\langle n| \hat O|n^{\prime}\rangle|^2 {\leq 4 \, \langle \Delta^2 \hat{O} \rangle },
\end{align}
with $\langle  \Delta^2 \hat{O} \rangle =\textrm{Tr}(\hat{\rho} \hat{O}^2 ) -  [\textrm{Tr}(\hat{\rho}\hat{O} )]^2$. 

It is crucial that, as shown in Ref.~\cite{Braunstein1994}, the equality holds in the case of pure states $\hat \rho = \ket{\psi}\bra{\psi}$. 

Let us now contrast the QFI computed on a thermodynamic ensemble with the one of a single energy eigenstate for an operator satisfying ETH. When computed on a canonical Gibbs state with $p_n=e^{-\beta E_n}/Z$, with $Z = \textrm{Tr}[e^{-\beta \hat{H}}]$, in Eq.~\eqref{QFI}, it was shown by Hauke {\em et al.} in Ref.~\cite{Hauke2016} that
\begin{align}
\label{eq:f_gibbs}
{\cal F}_{\textrm{Gibbs}}(\hat{O})= \frac{2}{\pi}\int_{-\infty}^{+\infty} d\omega  \, \tanh \left(\frac{\beta \omega}{2}\right)  \chi_{\hat{O}}^{\prime\prime}(\omega),
\end{align}
where $\chi_{\hat{O}}^{\prime\prime}(\omega) = \textrm{Im}[\chi_{\hat{O}}(\omega)]$ is the susceptibility obtained from the linear response function
\begin{align}
\label{eq:chi_def}
\chi_{\hat O}(t_1, t_2) = - \textrm{i} \theta(t_1-t_2)\, \langle [\hat O(t_1), \hat O(t_2)]\rangle,
\end{align}
which is connected to the symmetrised noise
\begin{align}
\label{eq:S_def}
S_{\hat O}(t_1, t_2) = \langle \{\hat O(t_1), \hat O(t_2)\}\rangle - 2 \langle \hat O(t_1) \rangle \, \langle \hat O(t_2) \rangle
\end{align}
via the fluctuation-dissipation relation $S_{\hat{O}}(\omega)=2\coth(\frac{\beta \omega}{2}) \chi_{\hat{O}}^{\prime \prime}(\omega)$ in the frequency domain. The same result holds in the microcanonical ensemble where ensemble equivalence is expected to hold.

If in contrast one considers a pure eigenstate $\ket{n}$ at the same temperature, i.e., one with energy $\bar{E}=E_n=\textrm{Tr}(\hat{H}e^{-\beta \hat{H}}/Z)$ compatible with the average energy of a canonical state in the system, the QFI is
\begin{align}
\begin{split}
\label{eq:qfi_eth}
{\cal F}_{\textrm{ETH}}(\hat{O}) &= 4\, \langle n | \Delta^2 \hat{O} | n \rangle =  \int_{-\infty}^{+\infty} \frac{d\omega}{\pi} S_{\hat{O}}(E_n, \omega) \\
& = \frac{2}{\pi}\int_{-\infty}^{+\infty} d\omega  \, \coth \left(\frac{\beta \omega}{2}\right)  \chi_{\hat{O}}^{\prime\prime}(E_n, \omega)\ ,
\end{split}
\end{align}
where $S_{\hat{O}}(E_n, \omega)$ in the previous equation is determined by the function $f_{\hat{O}}(\bar{E}, \omega)$ appearing in Eq.~\eqref{eq:eth_6} as described in Chapter~\ref{chapter:eth}. Since $S_{\hat{O}}(E_n, \omega)$ evaluated explicitly from ETH is equivalent to its canonical counterpart as presented in Chapter~\ref{chapter:eth}, then the following result holds 
\begin{equation}
    \label{eq:bound}
    {\cal F}_{\textrm{ETH}}(\hat{O})\ge {\cal F}_{\textrm{Gibbs}}(\hat{O}).
\end{equation}
This is one of the main results of this chapter. This analysis has immediate consequences for the QFI and the entanglement structure of asymptotic states in out-of-equilibrium unitary dynamics.

In this framework, the expectation value of operators $\hat{O}(t)=\bra{\psi} \hat{O}(t) \ket{\psi}$ (or of the correlation functions defined above) are taken with respect to an initial pure state $\ket{\psi}$, which is not an eigenstate of the Hamiltonian $\hat{H}$. Provided that the QFI attains an asymptotic value at long times $\mathcal F_{\infty}$, taking the long-time, whenever there are no degeneracies or only a sub-extensive number of them, we have that ${{\overline{\mathcal F(\hat O)} = \mathcal F_{\infty}}(\hat O)}= 4 \langle \Delta^2 \hat O \rangle_{\text {DE}}$ with ${\langle \,\cdot\, \rangle_{\text{DE}} = \text{Tr}(\hat\rho_{\text{DE}}\,\cdot\,)}$~\cite{Rossini2014, pappalardi2017multipartite}, and the diagonal ensemble defined as $\hat{\rho}_{\text{DE}}=\sum|c_{n}|^2\ket{n}\bra{n}$ with $c_n = \bra{\psi}E_n\rangle$. We remark that, since the out-of-equilibrium global state is pure, $\mathcal {F}_{\infty}(\hat{O})$ is given by the variance of $\hat O$ over the diagonal ensemble which is different from the QFI computed on the state $\hat \rho_{\text{DE}}$ using Eq.~\eqref{QFI}.

For sufficiently chaotic Hamiltonians, the initial state $\ket{\psi}$ considered is usually a microcanonical superposition around an average energy $\langle E \rangle  = \bra{\psi} \hat{H} \ket{\psi}$ with variance $(\delta E)^2 = \bra{\psi} \hat{H}^2 \ket{\psi} - \bra{\psi}\hat{H}\ket{\psi}^2$, i.e. $|c_n|^2$ has a narrow distribution around $\langle E \rangle$ with small fluctuations $(\delta E)^2 / \langle E \rangle ^2 \sim 1/L$~\cite{Alessio:2016}. 
It follows, then, that 
\begin{align}
\langle\Delta^2 \hat{O} \rangle_{\text{DE}} = \bra{n}\Delta^2 \hat{O} \ket{n} +  \left[ \frac{\partial O}{\partial{\bar{E}}} \right]_{E_n}^2 (\delta E)^2,
\end{align}
where the first term represents fluctuations inside each eigenstate --computed before in Eq.~\eqref{eq:qfi_eth} -- and the second is related to energy fluctuations. This observation, together with the bound Eq.~\eqref{eq:bound}, leads to 
\begin{align}
\label{eq:final_bound}
{\cal F}_{\infty}(\hat{O}) \ge {\cal F}_{\textrm{ETH}}(\hat{O})\ge {\cal F}_{\textrm{Gibbs}}(\hat{O}) \ ,
\end{align}
where the first equality holds in the thermodynamic limit and the second in the low temperature limit $T\to 0$. This also implies that $4 \langle \Delta^2 \hat  O\rangle_{\text{Gibbs}}  \geq \mathcal F_{\text{ETH}}(\hat{O})$, which follows from employing the same procedure described before for the diagonal ensemble on the canonical ensemble.
    
These expressions set a hierarchy in the entanglement content of {\em thermal states} at the same temperature, yet of different nature (mixed/pure). Furthermore, via Eqs.~\eqref{eq:f_gibbs}-\eqref{eq:qfi_eth}, one can quantify this difference via 
\begin{align}
\Delta \mathcal F = \mathcal F_{\text{ETH}}-\mathcal F_{\text{Gibbs}}= 1/\pi \int d\omega S_{\hat{O}}(\omega) / \cosh^2(\beta\omega/2).
\end{align}

\subsubsection{Multipartite entanglement at thermal criticality}

The major difference between the ETH and Gibbs multipartite entanglement
can be appreciated at critical points of thermal phase transitions, where $\hat O$ in (\ref{QFI}) is the order parameter of the theory.

While it is well known that the QFI does not ascertain divergence of multipartiteness at thermal criticality, i.e. $\mathcal F_{\textrm{Gibbs}}/L \sim \text{const.}$ \cite{Hauke2016,Gabbrielli2018}, on the other hand, the ETH result obeys the following critical scaling with the system size $L$
\begin{equation}
  f_Q^{\textrm{ETH}} \sim \frac{ \mathcal F_{\textrm{ETH}}}L \sim L^{\gamma / (\nu\, d)} \ ,
\end{equation}
where $\gamma $ and $\nu$ are the critical exponents of susceptibility and correlation length of the thermal phase transition, respectively, and $d$ is the dimensionality of the system \cite{cardy2012finite}. 

\subsection{Evaluation}
\label{sec:evaluation_qfi}

We now turn to the evaluation of Eq.~\eqref{eq:bound} in the context of a physical system with a microscopic Hamiltonian description. For this evaluation we return to our staggered field model
\begin{align}
\label{eq:h_sf_6}
\hat{H}_{\textrm{SF}} = \hat{H}_{\textrm{XXZ}} + b\, \sum_{i\, \textrm{even}} \hat{\sigma}^z_i,
\end{align} 
where
\begin{align}
\label{eq:h_xxz_6}
\hat{H}_{\textrm{XXZ}} = \sum_{i=1}^{L-1}\left[\left(\hat{\sigma}^x_{i}\hat{\sigma}^x_{i+1} + \hat{\sigma}^y_{i}\hat{\sigma}^y_{i+1}\right) + \Delta\,\hat{\sigma}^z_{i}\hat{\sigma}^z_{i+1}\right],
\end{align} 
as we have defined throughout this thesis. Note that for this particular case, we are considering the open-boundary chain with $L$ sites. This model is quantum chaotic with Wigner-Dyson level spacing statistics and diffusive transport as described in previous chapter, it then constitutes yet once more a valuable testbed for ergodic properties in an experimentally-relevant model. Recall that these models commute with the total magnetisation operator in the $z$ direction, $[\hat{H}_{\textrm{XXZ}},  \sum_i\hat{\sigma}^z_i]=[\hat{H}_{\textrm{SF}}, \sum_i\hat{\sigma}^z_i]=0$ and are, therefore, $U(1)$-symmetric. Even with OBCs, parity symmetry is present in the system. We break this symmetry by adding a small perturbation $\delta\hat{\sigma}^z_1$ on the first site, with $\delta = 0.1\alpha$. For our calculations, we set $\Delta = 0.5\alpha$ and $b = \alpha$, while all energy variables are given in terms of $\alpha$. The operator we will consider for our evaluation will be the total staggered magnetisation
\begin{align}
\hat{O} = \sum_i (-1)^i \hat{\sigma}^z_i.
\end{align}
Note that this observable differs from $\hat{B}_{\textrm{SF}}$ in Eq.~\eqref{eq:b_sf_5} considered in our previous analyses by a factor of $1 / L$. The {\em extensive} observable $\hat{O}$ is more relevant for this particular study of multipartite entanglement.

To evaluate our results in the canonical ensemble and in the context of ETH, we proceed with the full diagonalisation of $\hat{H}_{\textrm{SF}}$ in the largest $U(1)$ sector, in which $\sum_i \langle \hat{\sigma}^z_i \rangle = 0$, and compute all the matrix elements of $\hat{O}$ in the eigenbasis of the Hamiltonian $\hat{H}_{\textrm{SF}}$. 

\begin{figure}[t]
\fontsize{13}{10}\selectfont 
\centering
\includegraphics[width=0.8\columnwidth]{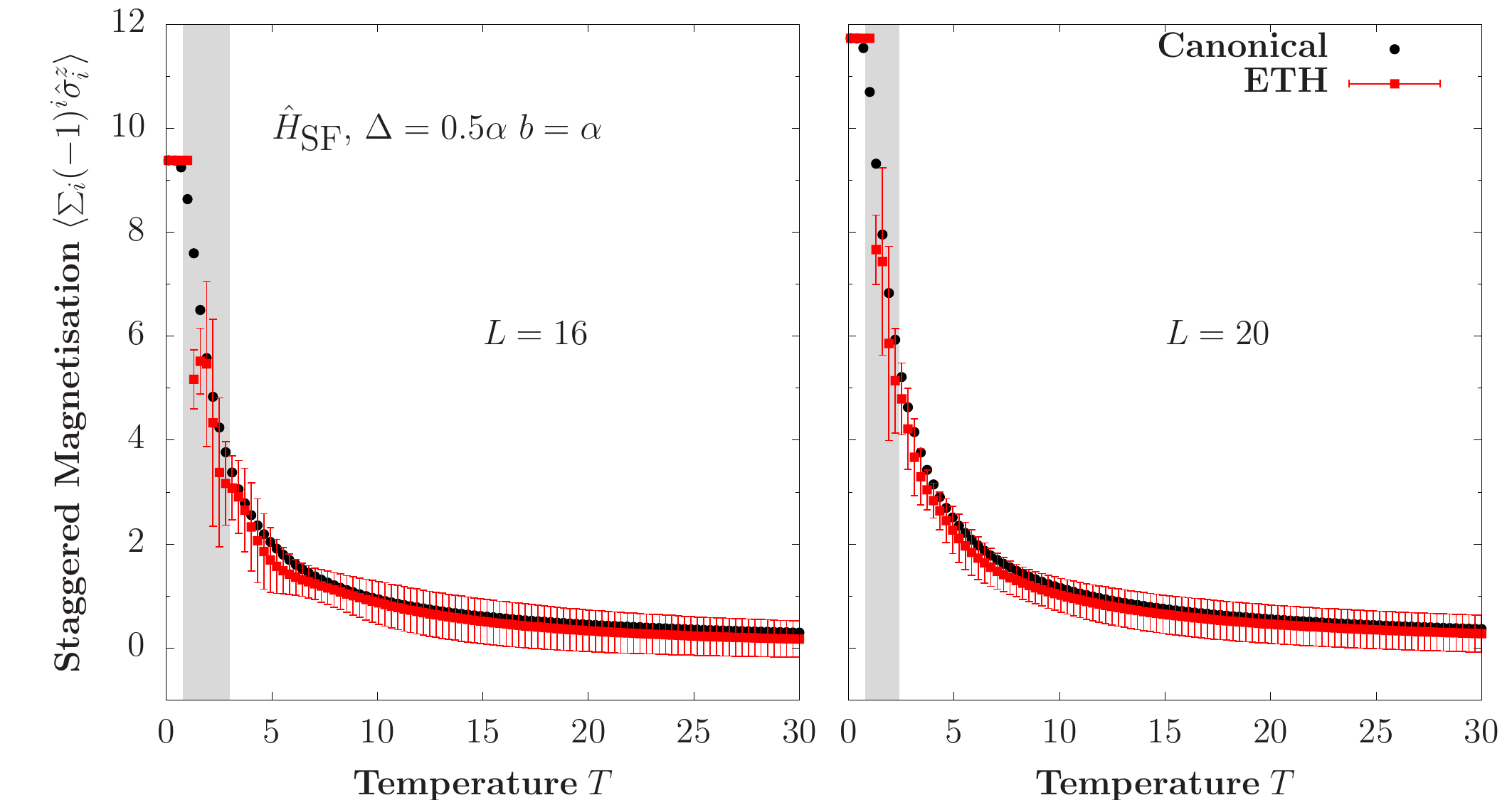}
\caption[Expectation value of the staggered magnetisation as a function of temperature in both the canonical ensemble and the ETH prediction]{Expectation value of the staggered magnetisation as a function of temperature in both the canonical ensemble and the corresponding ETH prediction for and $L = 16$ (left) $L = 20$ (right). Grey area highlights the low temperature regime, close to the edges of the spectrum where the ETH prediction gives the largest fluctuations.}
\label{fig:1.6.2}
\end{figure}

Our starting point is to evaluate the expectation value of $\hat{O}$ in the canonical ensemble and compare it with the ETH prediction. In the thermodynamic limit, a single eigenstate $\ket{n}$ with energy $E_n$ suffices to obtain the canonical prediction: $\langle \hat{O} \rangle =\bra{n} \hat{O} \ket{n} {=} \text{Tr}(\hat {O}\, e^{-\beta \hat H})/Z$, with an inverse temperature $\beta$ that yields an average energy $\langle E \rangle = E_n$. For finite-size systems, we instead focus on a small energy window centred around $E_n$ of width $0.1\epsilon$ in order to average eigenstate fluctuations, where $\epsilon$ is the bandwidth of the Hamiltonian for a given $L$. Fig.~\ref{fig:1.6.2} shows $\langle \hat{O} \rangle$ as a function of temperature for two different system sizes, including $L=20$, the largest system we have access to (Hilbert space dimension $\mathcal{D} = L!/[(L/2)!(L/2)!] = 184\;756$). The results exhibit the expected behaviour predicted from ETH for finite-size systems: the thermal expectation value is well approximated away from the edges of the spectrum (low temperature, section highlighted in grey on Fig.~\ref{fig:1.6.2}), and, moreover, the canonical expectation value is better approximated as the system size increases. These results are strongly suggestive that the ETH is fulfilled for the staggered field model and the operator considered. We recall that the same observation holds for the local magnetisation near the centre of the chain in Fig.~\ref{fig:1.5.2} and eigenstate-to-eigenstate fluctuations decay as expected for a system and observables that satisfy the ETH in Fig.~\ref{fig:1.5.1}.

We now turn to the evaluation of $\mathcal{F}_{\textrm{ETH}}$ and $\mathcal{F}_{\textrm{Gibbs}}$. The task requires to either compute $S_{\hat{O}}(E_n, \omega)$ or $\chi^{\prime \prime}_{\hat{O}}(E_n, \omega)$ in each respective framework. For the former, in the context of ETH, we can employ Eq.~\eqref{eq:s_chi_eth_5} which depends only on $f_{\hat{O}}(E_n, \omega)$. As before, we focus on a small window of energies and extract all the relevant off-diagonal elements of $\hat{O}$ in the eigenbasis of $\hat{H}_{\textrm{SF}}$. Fluctuations are then accounted for by computing a bin average over small windows $\delta \omega$, chosen such that the resulting average produces a smooth curve that is not sensitive to the particular choice of $\delta \omega$, such value, changes depending on dimension $\mathcal{D}$ of the sector studied~\cite{Mondaini:2017,Khatami:2013}, as described in Chapter~\ref{chapter:eth}. 

\begin{figure}[t]
\fontsize{13}{10}\selectfont 
\centering
\includegraphics[width=0.55\columnwidth]{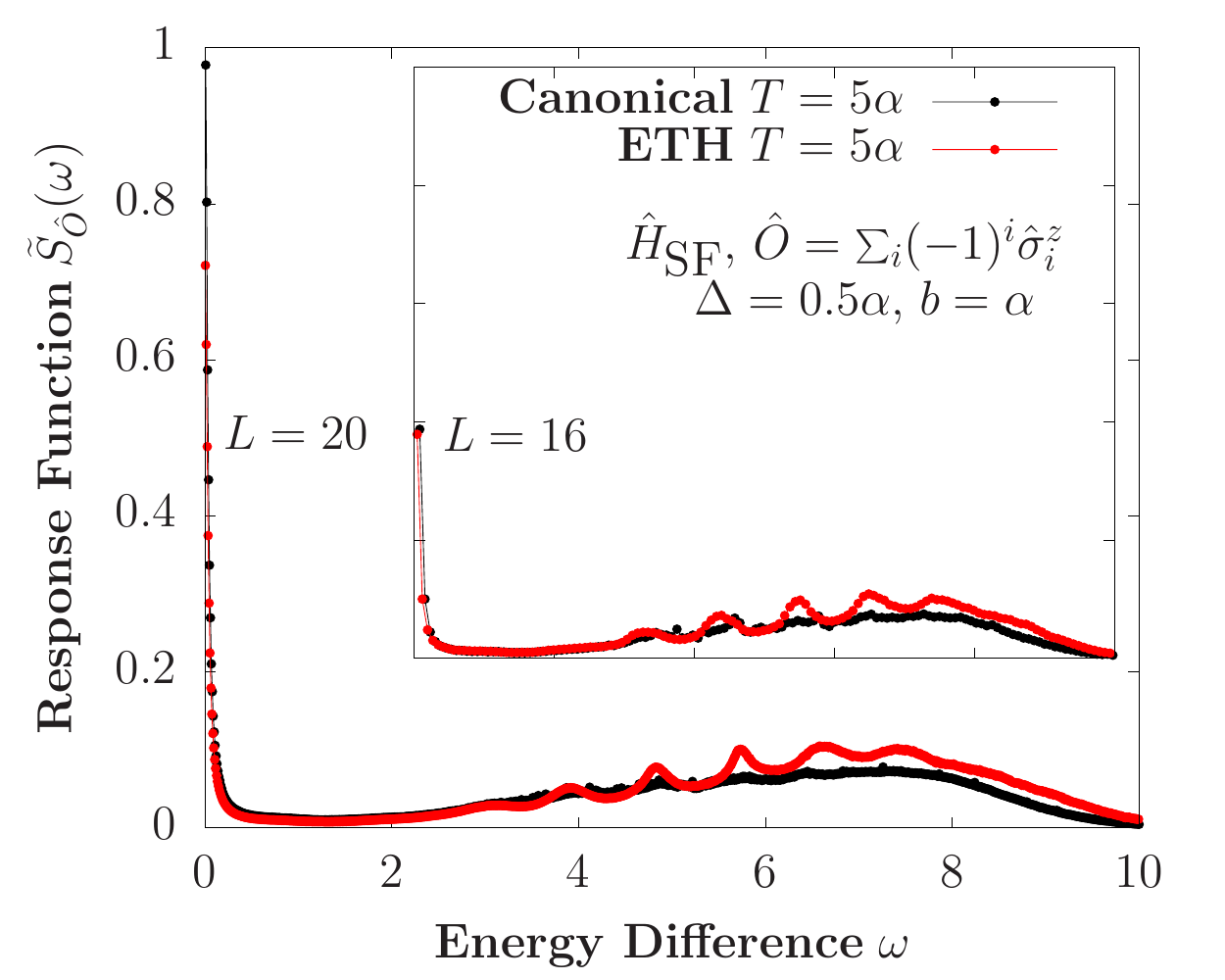}
\caption{Response function $S_{\hat{O}}(\omega)$ computed directly from ETH and in the canonical ensemble for $L=16$ (inset) and $L=20$ (main) for $T=5\alpha$.}
\label{fig:1.6.3}
\end{figure}

The procedure leads to a smooth function $e^{-S(E_n)/2}f_{\hat{O}}(E_n, \omega)$, in which the first factor is a constant value with respect to $\omega$. The entropy factor can be left undetermined in our calculations if we normalise the curve by the sum rule shown in Eq.~\eqref{eq:qfi_eth}, computed in this case from the ETH prediction of the expectation value of $\langle \Delta^2\hat{O} \rangle$. In the context of the canonical ensemble, $S_{\hat{O}}(\omega)$ can be explicitly evaluated by computing the thermal expectation value of the non-equal correlation function in the frequency domain given by
\begin{align}
\label{eq:s_canonical}
S_{\hat{O}}(\omega) = 2\pi \coth{\left( \frac{\beta \omega}{2} \right)} \sum_{n, n^{\prime}} (p_n - p_{n^{\prime}}) |\braket{n | \hat{O} | n^{\prime}}| \delta(\omega + E_{n} - E_{n^{\prime}}).
\end{align}

In Fig.~\ref{fig:1.6.3} we show $S_{\hat{O}}(\omega)$ for both the canonical ensemble for $T = 5\alpha$ and the corresponding ETH prediction normalised by the sum rule mentioned before. The sum rule is evaluated from the expectation values computed within both the canonical ensemble and ETH, correspondingly. It can be observed that the main features of the response function can be well approximated from the corresponding ETH calculation. For this particular case, however, the approximation is only marginally improved by increasing the system size. This behaviour is expected given that overall fluctuations for extensive observables carry an extensive energy fluctuation contribution, as mentioned before~\cite{Alessio:2016}.

\begin{figure}[t]
\fontsize{13}{10}\selectfont 
\centering
\includegraphics[width=0.5\columnwidth]{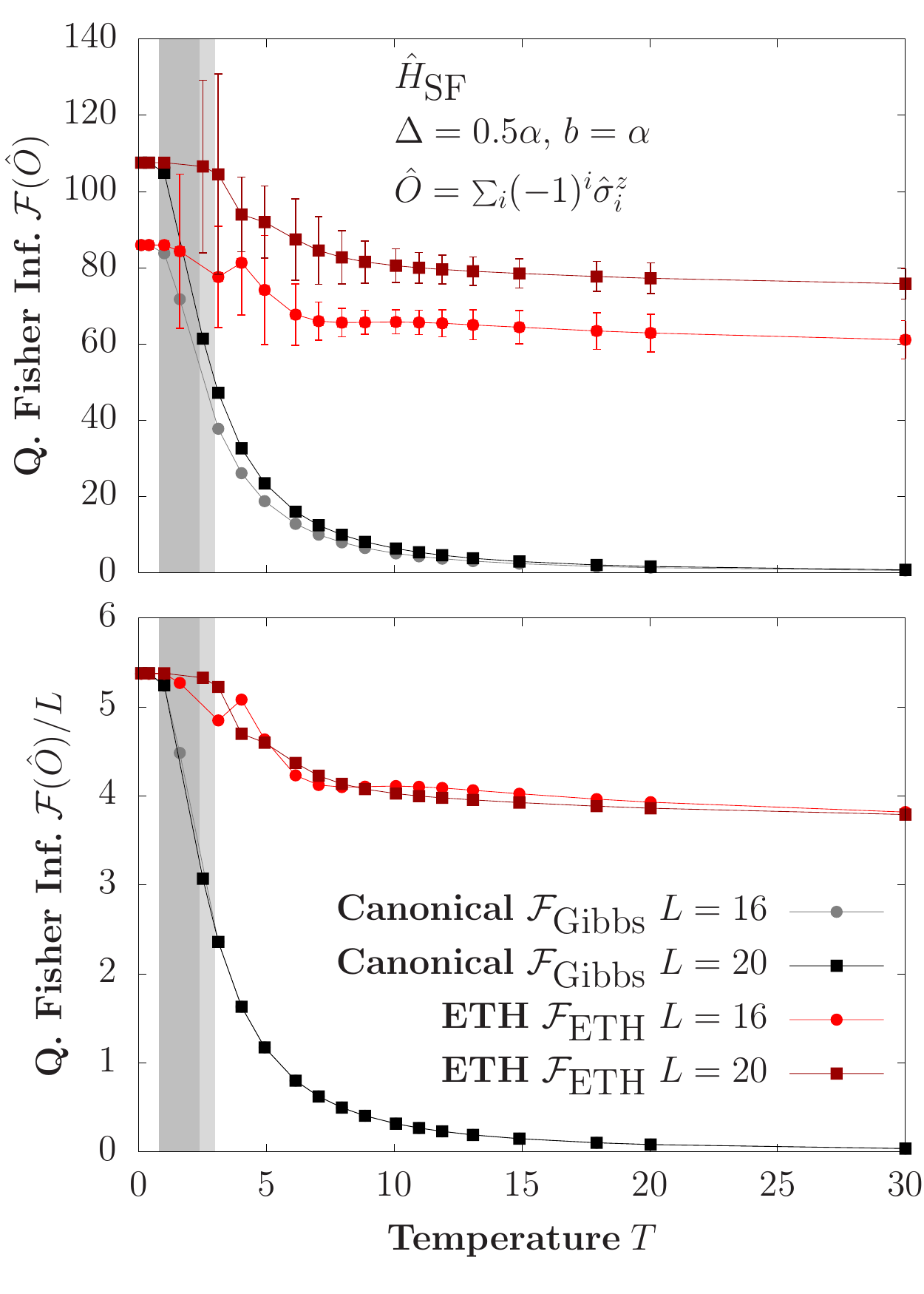}
\caption[The quantum Fisher information and the corresponding density for different system sizes as a function of temperature in both the canonical ensemble and corresponding ETH prediction]{The quantum Fisher information and the corresponding density for different system sizes as a function of temperature in both the canonical ensemble ($\mathcal{F}_{\textrm{Gibbs}}$) and corresponding ETH prediction ($\mathcal{F}_{\textrm{ETH}}$). At infinite temperature ETH predicts the presence of multipartite entanglement while there is none in the canonical ensemble.}
\label{fig:1.6.4}
\end{figure}

The previous analysis unravels the agreement between the thermal expectation values of non-equal correlation functions in time and those predicted by ETH. From these results, as $S_{\hat{O}}(\omega)$ (and, consequently, $\chi^{\prime \prime}_{\hat{O}}(\omega)$ from the fluctuation-dissipation relation) is well approximated by means of ETH, the inequality in Eq.~\eqref{eq:bound} is satisfied. 

Finally, we compute the QFI for $\hat{O}$ in our model within both contexts: $\mathcal{F}_{\textrm{ETH}}$ and $\mathcal{F}_{\textrm{Gibbs}}$. The results are shown in Fig.~\ref{fig:1.6.4}. The fluctuations in the ETH calculation of $\mathcal{F}_{\textrm{ETH}}$ are inherited from the fluctuations of the predicted expectation value of $\langle \Delta^2\hat{O} \rangle$, which, as expected for finite-size systems, decrease away from the edges of the spectrum. Both predictions for the QFI, canonical and ETH, are equivalent at vanishing temperatures. Remarkably, the QFI predicted from ETH is finite at infinite temperature, while the QFI from the canonical ensemble in this regime vanishes. We emphasise that although the QFI can be used in order to infer the structure of multipartite entanglement i.e. the number of sub-systems entangled, it is not a measure of these correlations in the mathematical sense of the formal theory of entanglement~\cite{Horodecki:2009}.

\subsection{Summary}
\label{sec:summary_qfi_eth}

We have shown that the QFI detects the difference between a pure state satisfying ETH and the Gibbs ensemble at the corresponding temperature. We note that it is possible to extend these results to integrable systems, by considering the quantum Fisher information in the context of the generalised Gibbs ensemble. This topic is left for future work.

Even though it can be anticipated that the expectation values of global observables might be sensitive to the ensemble in which they are evaluated, i.e., the difference between the evaluation over pure states and over the canonical Gibbs ensemble~\cite{Garrison2018}, several operators including sum of local ones and the non-local entanglement entropy appear to coincide at the leading order with the thermodynamic values when ETH is considered~\cite{Garrison2018, Polkovnikov2011, Santos2011entropy, Gurarie2013,  Alba2017}.

In this section, the difference between ETH/Gibbs multipartite entanglement, which can be macroscopic in the proximity of a thermal phase transition, is observed numerically in a XXZ chain with integrability breaking term, when the temperature grows toward infinity. The consequences of this could be observed in ion trap and cold-atom experiments via phase estimation protocols on pure state preparations evolved beyond the coherence time. Our result suggests that although at a local level all thermal states look the same, a quantum information perspective indicates that there are many ways to be thermal. 

\section{High-order correlation functions in time}
\label{sec:otoc_eth}

In many-body quantum systems, particularly from the perspective of condensed-matter theory, it is common to evaluate the dynamics of entanglement spreading to characterise a given system from the perspective of unitary evolution. Such an evaluation can lead to interesting insights with respect to how a quantum system behaves, e.g., close to a phase transition~\cite{Amico:2008}. On the other hand, {\em operator scrambling} relates to the growth of operator complexity through unitary evolution of an operator $\hat{O}$ in the Heisenberg picture $\hat{O}(t) = e^{\textrm{i}\hat{H}t} \hat{O}(0) e^{-\textrm{i}\hat{H}t}$. The out-of-time order correlators mentioned in the introduction to this section have recently been suggested to characterise quantum chaos~\cite{Cotler:2017, Cotler:2018} and even studied in random unitary circuit platforms~\cite{Nahum:2018}. 

Due to its nested structure, the square commutator
\begin{align}
\label{eq_sc}
c(t) \defeq - \left ( \langle [ \hat O(t), \hat O]^2 \rangle - \langle [ \hat O(t), \hat O]\rangle^2 \right ),
\end{align}
where the expectation value is considered over an ensemble of statistical mechanics, has been introduced as an object to study operator complexity growth~\cite{larkin1969quasiclassical,Parker19}. It is of particular interest to consider this object in quantum systems with a well-defined semi-classical limit, in which some connections can be established between operator complexity growth and the classical Lyapunov exponents, which characterise the exponential departure of nearby trajectories in the classical domain.

As we stated in the introduction to this section, it has been argued that the OTOC dynamics in chaotic systems attests to the fact that there exist probabilistic correlations in the matrix elements of the observables in the energy eigenbasis of chaotic Hamiltonians~\cite{kurchan}. The presence of these correlations has been studied numerically for local observables in experimentally-relevant physical systems~\cite{Richter2020}. 

In this section, we will further characterise the nature of these correlations numerically for a more general class of observables in two different physical models in Secs.~\ref{sec:models_otoc}, \ref{sec:gaussian_otoc} and \ref{sec:correlations_otoc}, following our analysis of the probability distributions introduced in Chapter~\ref{chapter:eth} and the approach taken in Ref.~\cite{Richter2020} for the numerical evaluation of the probabilistic correlations. We shall then evaluate the dynamics of the OTOC in Sec.~\ref{sec:dynamics_otoc}, which will establish a connection between matrix-element correlations and the dynamics of the OTOC. We then use this connection to provide an estimation of the scaling with respect to the system size frequency parameter that characterises the onset of matrix-element correlations in Sec.~\ref{sec:scaling_otoc}. We finalise with a short summary in Sec.~\ref{sec:summary_otoc}.

\subsection{Models and observables}
\label{sec:models_otoc}

To address the generic behaviour of thermalising systems that is independent of microscopic details, we consider two different non-integrable models: the first describing hard-core bosons with dipolar interactions in a harmonic trap~\cite{Khatami:2013}, while the second is a quantum Ising chain with both transverse and longitudinal fields~\cite{Kim2013}. The Hamiltonian of the first model is ($\hbar \defeq 1$)
\begingroup
\allowdisplaybreaks
\begin{align}
\label{eq:h_hb}
\hat{H}_{\textrm{HB}} = -J\sum_{i=1}^{L-1} \left( \hat{b}^{\dagger}_i \hat{b}^{\phantom{\dagger}}_{i + 1} + \textrm{H.c.} \right) + \sum_{i < l} \frac{V \hat{n}_{i} \hat{n}_{l}}{|i - l|^3} + \sum_i g x_i^2 \hat{n}_i
\end{align}
\endgroup
for a one-dimensional chain with $L$ sites where $\hat{b}^{\dagger}_i$ and $\hat{b}^{\phantom{\dagger}}_i$ are hard-core bosonic creation and annihilation operators, respectively, at site $i$, $\hat{n}_i = \hat{b}^{\dagger}_i \hat{b}^{\phantom{\dagger}}_i$ and $x_i=|i-L/2|$. Hereafter, all energies are given in units of the hopping amplitude $J$ and we set the strength of the dipolar interaction and confining potential to be $V = 2J$ and $g = 16J / (L - 1)^2$, respectively (parameters selected from Ref.~\cite{Khatami:2013}). The system conserves the total number of bosons, which is guaranteed from $[\hat{H}_{\textrm{HB}}, \sum_i\hat{n}_i]=0$. This symmetry is resolved throughout this work. We focus on the half-filled sub-sector, in which the Hilbert space dimension is $\mathcal{D} = L!/[(L/2)!(L/2)!]$. To avoid parity (spatial inversion) or reflection (spin inversion) symmetries, we add a small perturbation  $\delta\hat{n}_1$ to the Hamiltonian ($\delta = 0.1J$). 

The second model has the following Hamiltonian:
\begingroup
\allowdisplaybreaks
\begin{align}
\label{eq:h_is}
\hat{H}_{\textrm{IS}} = \sum_{i=1}^{L} w \hat{\sigma}^x_i + \sum_{i=1}^{L} h \hat{\sigma}^z_i + \sum_{i = 1}^{L - 1} J \hat{\sigma}^z_{i} \hat{\sigma}^z_{i+1}\ .
\end{align}
\endgroup
We measure energies in units $J$ and set $w = J\sqrt{5} / 2$, $h = J(\sqrt{5} + 5) / 8$ (see Ref.~\cite{Kim2013}). The only known symmetry associated to this model is parity. We consider the even parity sub-sector for chains with an even number of sites, with a corresponding Hilbert space dimension $\mathcal{D} = 2^L - [(2^L - 2^{L/2})/2]$. 

We consider extensive observables, composed of sums of local operators spanning the entire system
\begin{align}
\label{eq:obs_ext}
\hat{B}_{\textrm{HB}} = \frac{1}{L}\sum_i [ 1 + (-1)^i ]\hat{n}_{i} \ , \quad 
\hat{B}_{\textrm{IS}} = \frac{1}{L}\sum_i \hat{\sigma}^z_{i}\ .
\end{align}

\begin{figure}[t]
\fontsize{13}{10}\selectfont
\centering
\includegraphics[width=0.5\columnwidth]{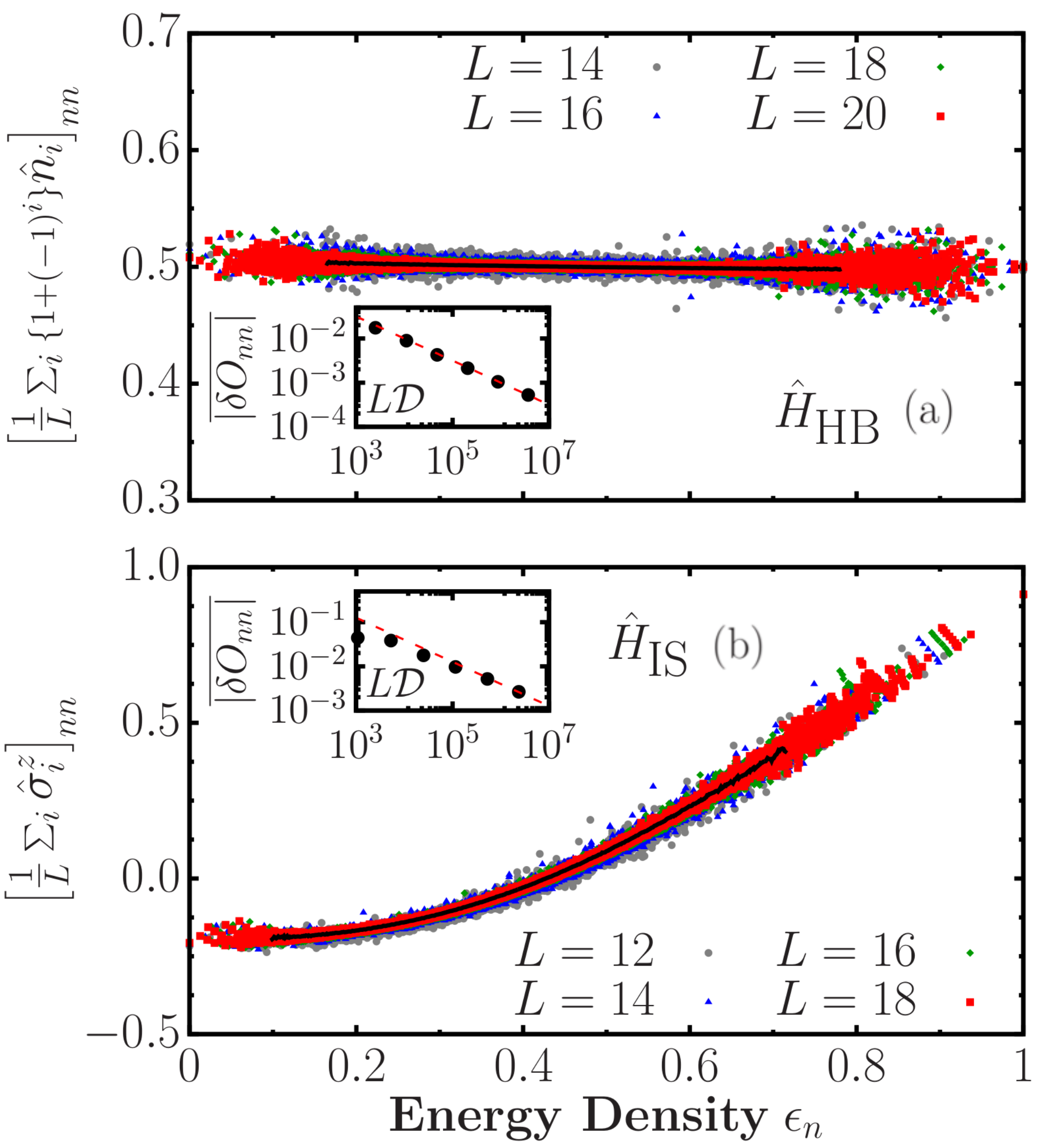}
\caption[Diagonal matrix elements of $\hat{B}_{\textrm{HB}}$ and $\hat{B}_{\textrm{IS}}$ as a function of the energy density $\epsilon_n \defeq (E_n - E_{\textrm{min}}) / (E_{\textrm{max}} - E_{\textrm{min}})$ and of the system size $L$]{Diagonal matrix elements of $\hat{B}_{\textrm{HB}}$ (a) and $\hat{B}_{\textrm{IS}}$ (b) as a function of the energy density $\epsilon_n \defeq (E_n - E_{\textrm{min}}) / (E_{\textrm{max}} - E_{\textrm{min}})$ and of the system size $L$. The black lines depict an approximation of the smooth function $O(\bar{E})$ obtained from a coarse-grained average of the data for the largest system size. The insets show the eigenstate-to-eigenstate fluctuations for different systems sizes, obtained from the eigenvalues in the central region. The dashed lines on the insets show the $(L \mathcal{D})^{-1/2}$ scaling.}
\label{fig:1.6.5}
\end{figure}

These observables satisfy the ETH from the perspective of the diagonal matrix elements in the energy eigenbasis of each corresponding Hamiltonian. As introduced in Chapter~\ref{chapter:eth}, we show in Fig.~\ref{fig:1.6.5} the diagonal matrix elements of of $\hat{B}_{\textrm{HB}}$ [panel (a)] and of $\hat{B}_{\textrm{IS}}$ [panel (b)] in the eigenbasis of their corresponding Hamiltonian. The support over which the matrix elements exist shrink suggesting that the behaviour can be described by a smooth function $O(\bar{E})$ corresponding to the microcanonical prediction and, furthermore, the eigenstate-to-eigenstate fluctuations [see Eq.~\eqref{eq:ete_fluctuations}] decay exponentially for the 20\% of the total eigenvalues in the centre of the spectrum, as shown in the insets in Fig.~\ref{fig:1.6.5}. We remark that for this class of observables, the eigenstate-to-eigenstate fluctuations decay as $L\mathcal{D}^{-1/2}$, which can be traced back to the $1 / \sqrt{L}$ scaling of the Schmidt norm~\cite{Vidmar:2019,Vidmar2:2020,Leblond:2019}. The results shown in Fig.~\ref{fig:1.6.5} indicate that the ETH is obeyed by the models and observables considered in the main text for the parameters selected, away from non-generic features observed at the edges of the spectrum. 

It is crucial to recognise that, within the ETH, the one- and two-point correlation functions in time do not depend on the details of the matrix elements of the statistical matrix $R_{nm}$. In particular, two-point correlators are determined by the smooth function $f_{\hat{O}}(\bar{E},\omega)$ entering Eq.~\eqref{eq:eth_6}, which itself depends on the variance of matrix elements $O_{nm}$ near a given energy $\bar{E}$ and frequency $\omega$~\cite{Khatami:2013, Beugeling2015, Mondaini:2017, Leblond:2020}. 

\begin{figure}[t]
\fontsize{13}{10}\selectfont 
\centering
\includegraphics[width=1.0\columnwidth]{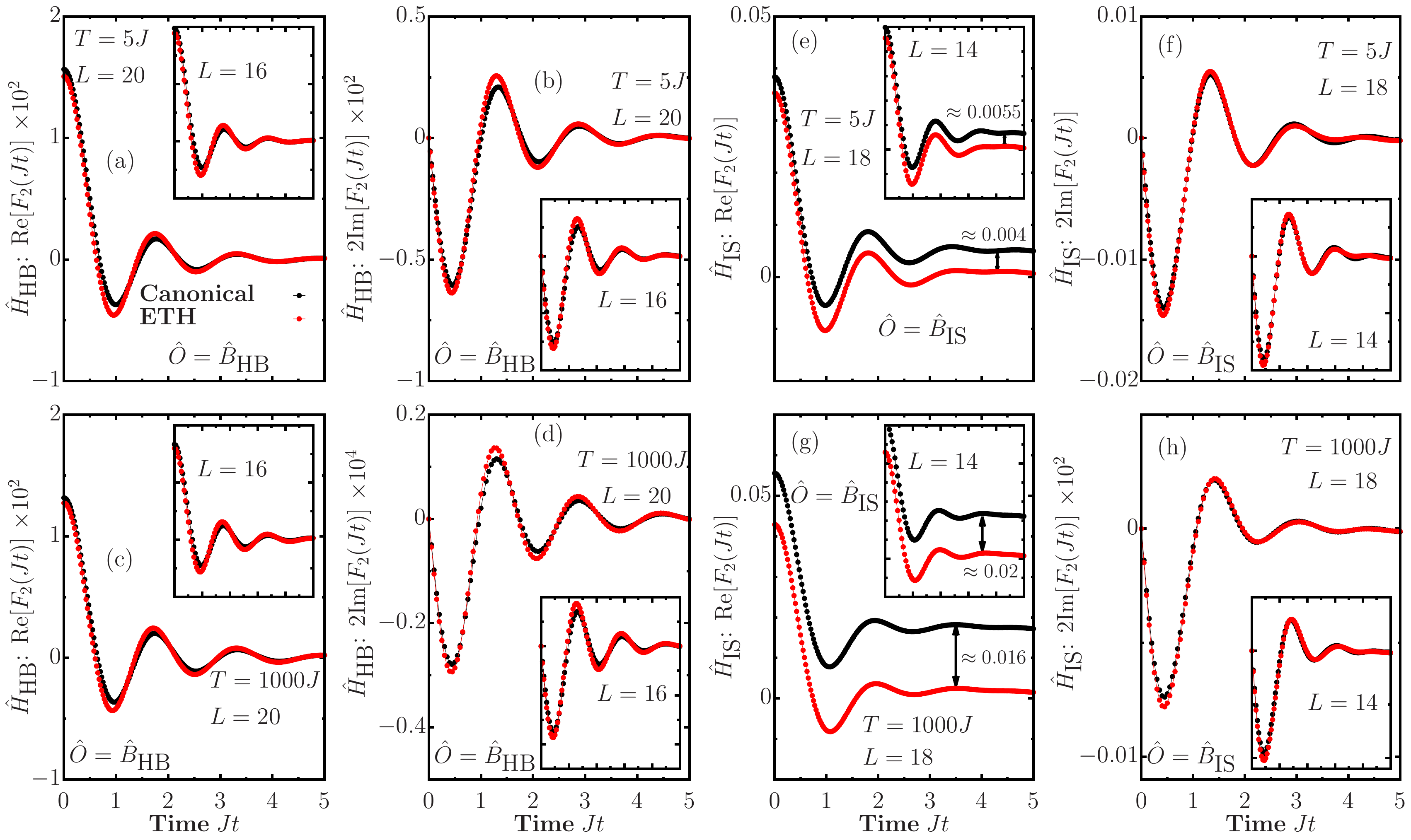}
\caption[Dynamics of the two-point correlation function evaluated in the canonical ensemble at temperature $T$ and in the ETH with a compatible energy density for sums of local operators]{Dynamics of the two-point correlation function evaluated in the canonical ensemble at temperature $T$ and in the ETH with a compatible energy density for sums of local operators. In [(a)-(d)] we show the results for $\hat{H}_{\textrm{HB}}$ and in [(e)-(h)] the corresponding results for $\hat{H}_{\textrm{IS}}$ at different temperatures for different system sizes as highlighted in the figure.}
\label{fig:1.6.6}
\end{figure}

This fact can be observed numerically in the physical models considered in this section. Just as we proceeded in Chapter~\ref{chapter:eth} for the staggered field model, the connected symmetric and anti-symmetric two-point correlation functions (see Sec.~\ref{sec:eth_correlation_functions})
\begin{align}
S^{+}_{\hat{O}}(t) &\defeq \langle \{ \hat{O}(t), \hat{O}(0) \} \rangle_c = 2\,\textrm{Re}[F_2(t)]\nonumber \\
S^{-}_{\hat{O}}(t) &\defeq \langle \,[ \hat{O}(t), \hat{O}(0) \,] \rangle_c = 2\textrm{i}\, \textrm{Im}[F_2(t)],
\end{align}
evaluated in the ensembles of statistical mechanics are equivalent to the same objects evaluated for a single eigenstate as the thermodynamic limit is approached in ergodic systems. As noted in Sec.~\ref{sec:eth_correlation_functions}, correlations that may be present in the $R_{nm}$ are not relevant for two-point functions. We recall that, within the ETH, the two-point correlation functions depend solely on the features of $f_{\hat{O}}(E_n, \omega)$, given that
\begin{align}
S^{+}_{\hat{O}}(E_n, \omega) &\approx 4\pi \cosh(\beta \omega /2) |f_{\hat{O}}(E_n, \omega)|^2, \nonumber \\
S^{-}_{\hat{O}}(E_n, \omega) &\approx 4\pi \sinh(\beta \omega /2) |f_{\hat{O}}(E_n, \omega)|^2.
\end{align}
Using the procedure described in Sec.~\ref{sec:eth_correlation_functions}, we proceed to extract $f_{\hat{O}}(E_n, \omega)$ for $\hat{B}_{\textrm{HB}}$ and $\hat{B}_{\textrm{IS}}$ from the off-diagonal elements in the eigenbasis of their corresponding Hamiltonian. The results are shown in Fig.~\ref{fig:1.6.6}, where we observe excellent agreement between the dynamics of two-point functions computed with respect to a single eigenstate and the canonical ensemble at the same average energy, without any particular considerations about the statistical matrix $R_{nm}$, other than its mean and variance. We recall that there exists a finite-size term in the symmetric correlation function that decreases as the system size is increased, as observed in Fig~\ref{fig:1.6.6}[(e),(g)]. This term is much smaller for $\hat{B}_{\textrm{HB}}$ in Fig~\ref{fig:1.6.6}[(a),(c)] due to distribution of diagonal matrix elements as can be seen in Fig.~\ref{fig:1.6.5}(a).

The precise distribution of these elements, as well as correlations between matrix elements at different frequencies, thus encode the fine structure of the ETH beyond linear-response theory~\cite{Foini2019}. In the following, we investigate how this structure influences the dynamics of higher-order correlators such as the OTOC.

\subsection{Gaussian statistics}
\label{sec:gaussian_otoc}

As we first noted in Chapter~\ref{chapter:eth}, the probability distribution of off-diagonal matrix elements of local observables in the energy eigenbasis of chaotic Hamiltonians is Gaussian, while their distribution might be very different in integrable systems as we observed in Sec.~\ref{sec:impurity_eth}. Let us present the probability distributions for the models and observables considered in this section.

\begin{figure}[t]
\fontsize{13}{10}\selectfont 
\centering
\includegraphics[width=0.7\columnwidth]{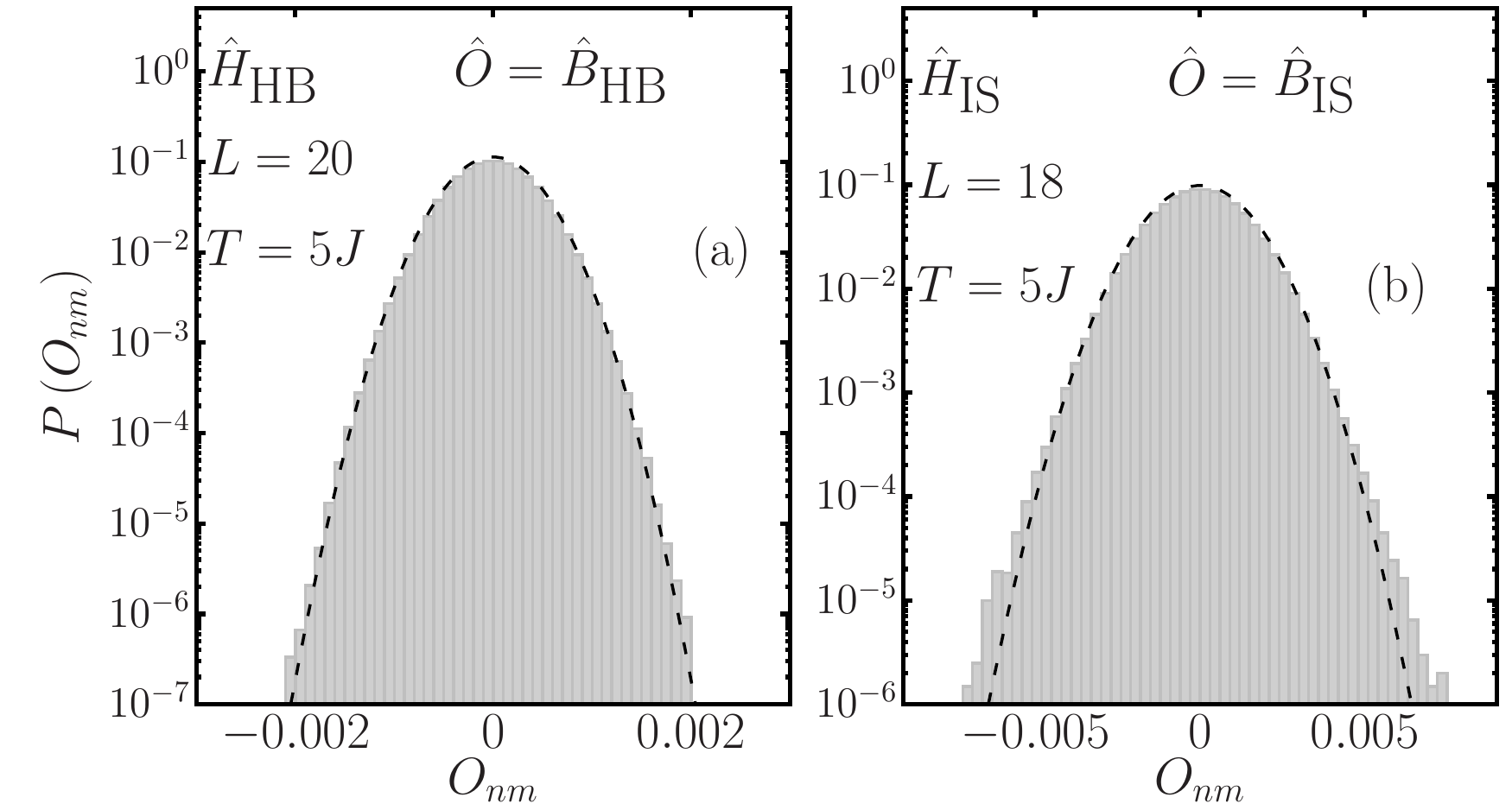}
\caption[Probability distributions of off-diagonal matrix elements in a small frequency range $\omega \lesssim 0.05$ for $\hat{B}_{\textrm{HB}}$ and in (b) for $\hat{B}_{\textrm{IS}}$]{Probability distributions of off-diagonal matrix elements in a small frequency range $\omega \lesssim 0.05$. The average energy $\bar{E}$ selected is consistent with a finite canonical temperature $T = 5J$. The distributions are shown in (a) for $\hat{B}_{\textrm{HB}}$ and in (b) for $\hat{B}_{\textrm{IS}}$. Results obtained for finite-sized systems of $L = 20$ for $\hat{H}_{\textrm{HB}}$ and $L = 18$ for $\hat{H}_{\textrm{IS}}$. Dashed lines depict a Gaussian distribution with the same mean and variance.}
\label{fig:1.6.7}
\end{figure}

The number of matrix elements is very large even at small system sizes, so we shall start by considering the off-diagonal matrix elements near the zero-frequency regime first, to evaluate their distribution. We then consider $O_{nm} = \braket{E_n | \hat{O} | E_m}$ in a small frequency-resolved window $\omega \lesssim 0.05$ and a finite temperature $T = \beta^{-1} = 5J$ ($k_B\defeq 1$). We recall that the temperature can be estimated by associating the average energy $\bar{E}$ with a canonical density matrix $\hat{\rho} = e^{-\beta\hat{H}}/Z$ as $\bar{E} = \textrm{Tr}[\hat{\rho} \hat{H}]$, with $Z = \textrm{Tr}[e^{-\beta\hat{H}}]$. The probability distribution can then be studied from the histogram of all the possible off-diagonal matrix elements satisfying these conditions. As observed in Fig.~\ref{fig:1.6.7}, the matrix elements are Gaussian-distributed for the extensive operators in both of the models we have studied, even away from the infinite-temperature regime.

We can proceed to evaluate the frequency-dependent ratio~\cite{Leblond:2019}
\begin{align}
\label{eq:gamma_6}
\Gamma_{\hat{O}}(\omega) \defeq \overline{|O_{nm}|^2} / \overline{|O_{nm}|}^2,
\end{align}
to understand if these probability distributions can be seen over small portions of the entire spectrum away from zero frequency at all temperatures where the ETH is expected to hold. We perform the averages over small frequency windows $\delta\omega = 0.05$, as described in Sec.~\ref{sec:eth_correlation_functions}. We recall that $\Gamma_{\hat{O}}(\omega) = \pi / 2$ for normally-distributed probability distributions. In this analysis, we consider $\omega = E_m - E_n$ over the entire spectrum, while the average energy $\bar{E} = (E_n + E_m) / 2$ is chosen to be compatible with a corresponding canonical temperature. To account for finite-size eigenstate-to-eigenstate fluctuations, the quantity is computed within a small energy window $0.05\epsilon$, where $\epsilon \defeq E_{\textrm{max}} - E_{\textrm{min}}$ is the bandwidth of the Hamiltonian.

\begin{figure}
\fontsize{13}{10}\selectfont 
\centering
\includegraphics[width=0.56\columnwidth]{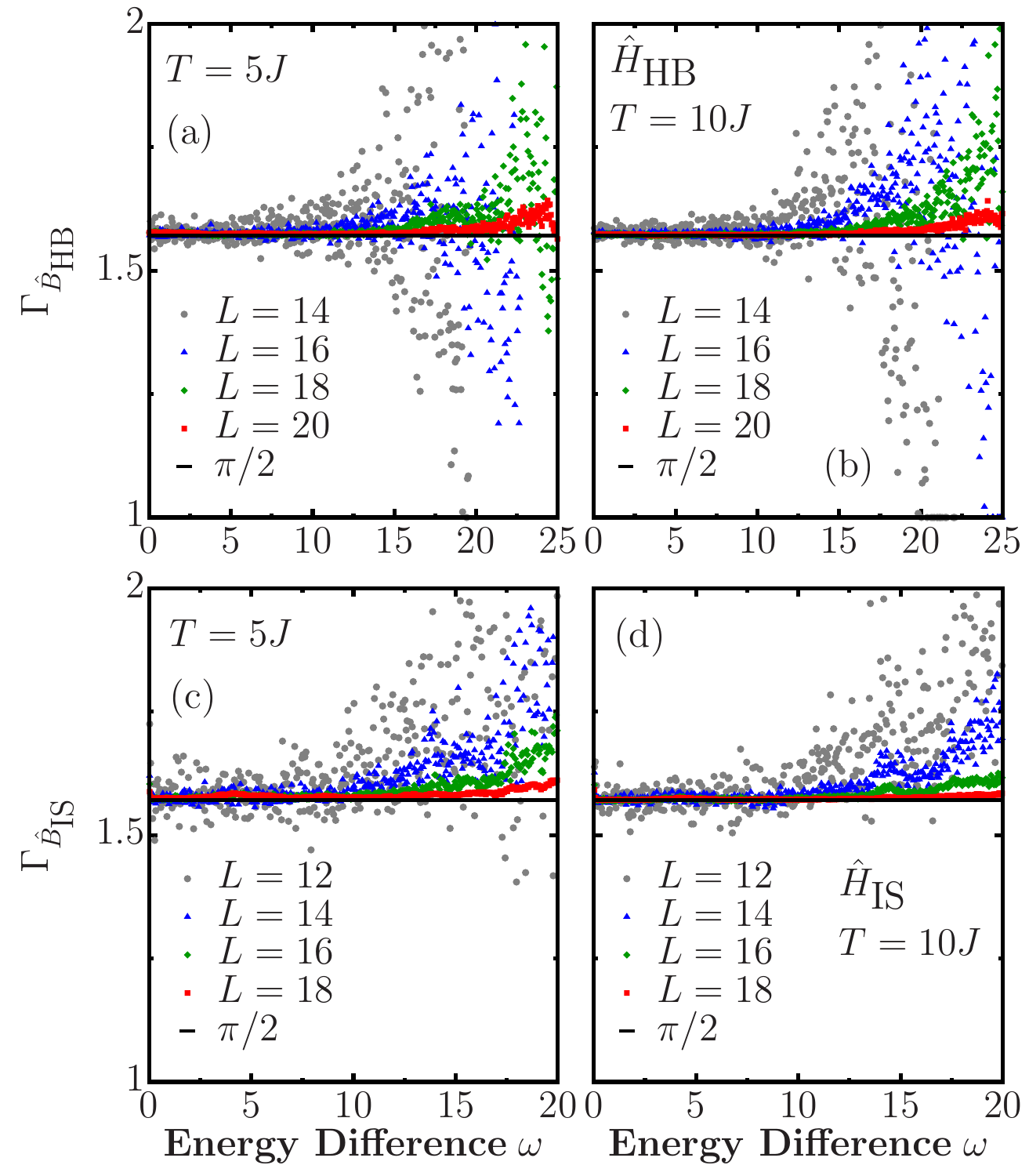}
\caption[$\Gamma_{\hat{O}}(\omega)$, for operators $\hat{B}_{\textrm{HB}}$ and $\hat{B}_{\textrm{IS}}$ in the eigenbasis of $\hat{H}_{\textrm{HB}}$ and $\hat{H}_{\textrm{IS}}$]{$\Gamma_{\hat{O}}(\omega)$, from Eq.~\eqref{eq:gamma_6}, for operators $\hat{B}_{\textrm{HB}}$ [(a) and (b)] and $\hat{B}_{\textrm{IS}}$ [(c) and (d)] in the eigenbasis of $\hat{H}_{\textrm{HB}}$ and $\hat{H}_{\textrm{IS}}$, respectively. Two different finite temperatures were chosen, $T = 5J$ [(a) and (c)] and $T = 10J$ [(b) and (d)]. The black horizontal line shows the value $\Gamma_{\hat{O}}(\omega) = \pi / 2$. The matrix elements were computed in a small energy window $0.05\epsilon$ where $\epsilon \defeq E_{\textrm{max}} - E_{\textrm{min}}$, and a frequency window $\delta \omega = 0.05$.}
\label{fig:1.6.8}
\end{figure}

In Fig.~\ref{fig:1.6.8} we show the $\Gamma_{\hat{O}}(\omega)$ ratio as a function of $\omega$ and of the system size $L$ for both $\hat{H}_{\textrm{HB}}$ [panels (a) and (b)] and $\hat{H}_{\textrm{IS}}$ [panels (c) and (d)], evaluated for the operators $\hat{B}_{\textrm{HB}}$ and $\hat{B}_{\textrm{IS}}$ from Eq.~\eqref{eq:obs_ext} and for two different temperatures $T = 5J$ and $T = 10J$. We have chosen to display our results for values of temperature away from the infinite-temperature regime. Gaussian statistics emerge at all frequencies, i.e.~ $\Gamma_{\hat{O}} \approx \pi / 2$ for increasing values of $\omega$ as the system size increases.

\subsection{Correlations between matrix elements}
\label{sec:correlations_otoc}

We have shown how Gaussian statistics emerge in the probability distributions of off-diagonal matrix elements of local observables in chaotic non-integrable models, this does not imply that these matrix elements can be considered independently- and identically-distributed random variables. In this section we address the presence of correlations between matrix elements. 

Let us now examine the overall structure of the statistical matrix $R_{nm}$ as a function of the mean energy $\bar{E}$ and frequency $\omega$. In particular, we are interested in correlations between matrix elements at different frequencies, which are encoded in the eigenvalue distribution of the matrix $O_{nm}$. In the absence of correlations, the eigenvalue distribution should coincide with that of the Gaussian orthogonal ensemble (GOE), where each matrix element is an independent, identically distributed random variable~\cite{Mehta:2004}. Therefore, any deviation from the GOE prediction heralds the presence of correlations between matrix elements.

In order to investigate the temperature- and frequency-dependence of such correlations, we consider sub-matrices of $\hat{O}$ restricted to a finite frequency window and construct the corresponding eigenvalue distributions, following Ref.~\cite{Richter2020}. To fix the temperature, we first extract a $\mathcal{D}^{\prime} \times \mathcal{D}^{\prime}$ sub-matrix from $\hat{O}$, centred around the diagonal matrix element $O_{nn}$ such that $E_n = \textrm{Tr}[\hat{\rho} \hat{H}]$. The size $\mathcal{D}'$ of this sub-matrix is selected to encompass an energy range of width $0.125\epsilon$, where $\epsilon \defeq E_{\textrm{max}} - E_{\textrm{min}}$ is the bandwidth. We then further restrict our attention to frequencies $|\omega| < \omega_c$ by setting
\begin{align}
\label{eq:o_wc}
O^{\omega_c}_{nm} \defeq \begin{cases} O_{nm}, \, &\textrm{if}\;  |E_m - E_n| < \omega_c \\ 0, &\textrm{otherwise.} \end{cases}
\end{align}
To test for correlations between these matrix elements, we follow the procedure introduced in Ref.~\cite{Richter2020}: we generate a sign-randomised matrix from the original sub-matrix
\begin{align}
\label{eq:o_tilde}
\widetilde{O}^{\omega_c}_{nm} \defeq \begin{cases} O^{\omega_c}_{nm}, \, &\textrm{probability} = 1/2, \\ -O^{\omega_c}_{nm}, \, &\textrm{probability} = 1/2, \end{cases}
\end{align}
where we apply the sign randomisation on the elements $n \neq m$ to retain the mean and support of the original sub-matrix.
The random sign destroys correlations between matrix elements, leading to the semi-elliptical eigenvalue distribution that is characteristic of the GOE~\cite{Mehta:2004,Livan2018introduction}. Comparing the eigenvalue distributions of $O^{\omega_c}_{nm}$ and $\widetilde{O}^{\omega_c}_{nm}$ thus probes correlations between the matrix elements of $\hat{O}$ within a frequency range controlled by the cutoff $\omega_c$. 

\begin{figure}[t]
\fontsize{13}{10}\selectfont 
\centering
\includegraphics[width=0.7\columnwidth]{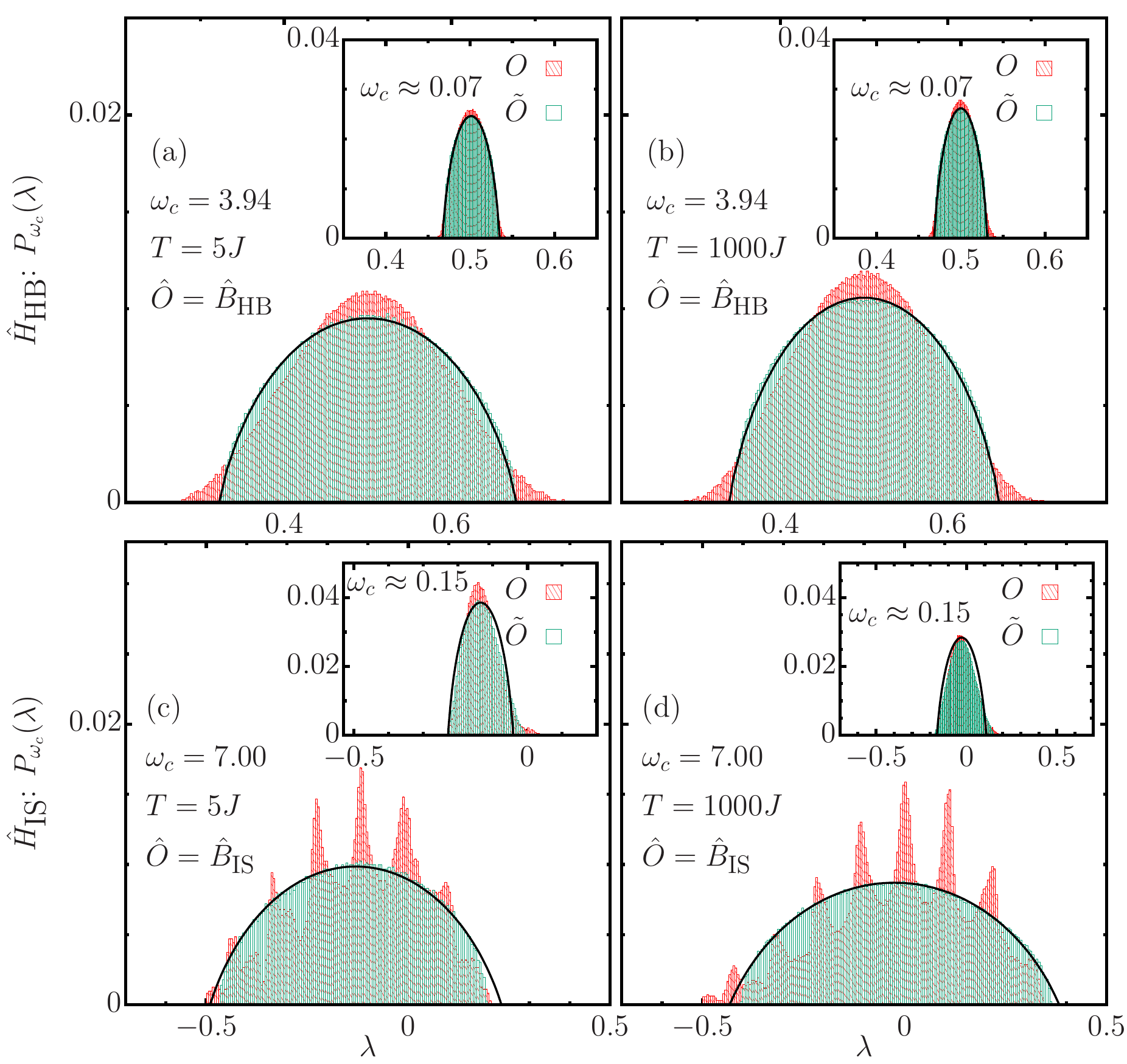}
\caption[Probability distributions $P_{\omega_c}(\lambda)$ of the eigenvalues of the full and randomised sub-matrices (Eqs.~\eqref{eq:o_wc} and~\eqref{eq:o_tilde}) in an energy window $0.125\epsilon$ for $\hat{B}_{\textrm{HB}}$ and $\hat{B}_{\textrm{IS}}$]{Probability distributions $P_{\omega_c}(\lambda)$ of the eigenvalues of the full and randomised sub-matrices [Eqs.~\eqref{eq:o_wc} and~\eqref{eq:o_tilde}] in an energy window $0.125\epsilon$ for $\hat{B}_{\textrm{HB}}$ [(a), (b)] and $\hat{B}_{\textrm{IS}}$ [(c), (d)]. In the insets of [(a), (b)] and [(c), (d)] we show $P_{\omega_c}(\lambda)$ for the banded operators $\hat{B}_{\textrm{HB}}$ and $\hat{B}_{\textrm{IS}}$, respectively, with the smallest cutoff frequency $\omega_c$ such that the eigenvalue distribution follows the GOE (see section below on localisation effects). All the panels on the left correspond to $T = 5J$, while $T = 1000J$ for the panels on the right. We show the results for $L = 18$ in $\hat{H}_{\textrm{HB}}$ and $L = 16$ in $\hat{H}_{\textrm{IS}}$.}
\label{fig:1.6.9}
\end{figure}

The distribution of all the $\mathcal{D}^{\prime}$ eigenvalues $\lambda^{\omega_c}_{\alpha}$ of $\hat{O}^{\omega_c}$ is expressed as
\begin{align}
P_{\omega_c}(\lambda) = \frac{1}{\mathcal{D}^{\prime}} \sum_{\alpha=1}^{\mathcal{D}^{\prime}} \delta \left( \lambda - \lambda^{\omega_c}_{\alpha} \right),
\end{align}
where all the individual $\delta(\cdot)$ peaks are collected in small bins to describe a given probability distribution. The function $P_{\omega_c}(\lambda)$ can be studied as a function of $\omega_c$ and yields a semi-circular distribution if the eigenvalues are uncorrelated. If correlations are to arise, deviations from a semi-circle distribution are observed.

The eigenvalues $\{\lambda\}$ of the sub-matrices in Eq.~\eqref{eq:o_wc} and Eq.~\eqref{eq:o_tilde} are evaluated numerically and the corresponding distributions, $P_{\omega_c}(\lambda)$, are shown in Fig.~\ref{fig:1.6.9} for extensive operators. The eigenvalues of the entire sub-matrix within the chosen energy window show a departure from the semi-elliptical distribution (Fig.~\ref{fig:1.6.9}[(a),(b)] for $\hat{B}_{\textrm{HB}}$ and Fig.~\ref{fig:1.6.9}[(c),(d)] for $\hat{B}_{\textrm{IS}}$), signalling substantial correlations between matrix elements at significantly different frequencies. These correlations are seen for high ($T = 1000J$) [Fig.~\ref{fig:1.6.9}(b,d)] and low  ($T = 5J$) [Fig.~\ref{fig:1.6.9}(a,c)] temperatures alike. For smaller values of the cutoff $\omega_c$, however, the eigenvalue distributions begin to resemble the GOE prediction (Fig.~\ref{fig:1.6.9}[(a),(b)] insets for $\hat{B}_{\textrm{HB}}$ and Fig.~\ref{fig:1.6.9}[(c),(d)] insets for $\hat{B}_{\textrm{IS}}$). Our data are therefore consistent with a crossover to Gaussian random-matrix-like behaviour at low frequencies~\cite{Richter2020}. The frequency scale of the crossover can be estimated from the value of $\omega_c$ at which the distributions appear to coincide with the GOE prediction, $\omega_c = \omega_{\rm GOE}$. Note that, for even smaller values of $\omega_c$, the eigenvalue statistics eventually become Poissonian due to well-known localisation effects~\cite{Alessio:2016}. The insets in Fig.~\ref{fig:1.6.9} display the eigenvalue distributions for smallest frequency values which are still above the localised regime. This indicates that $\omega_{\textrm{GOE}}$ refers to a different frequency scale, as first denoted in Ref.~\cite{Richter2020}. 

\subsubsection{Localisation effects}

We have considered banded sub-matrices to determine the degree of correlations within the statistical matrix $R_{nm}$. From Eq.~\eqref{eq:o_wc}, $\omega_c$ determines the frequency value associated to a given banded sub-matrix. We considered $\omega_c$ only above a given threshold, due to the fact that below this threshold the eigenvalues of $\hat{O}^{\omega_c}$ become uncorrelated because of localisation effects~\cite{Richter2020}. In this frequency regime, the adjacent eigenvalue level spacings of $\hat{O}^{\omega_c}$ are Poisson-distributed. 

The relevant frequency regime in our work is the one dictated by the largest $\omega_c$ where the eigenvalues $\lambda^{\omega_c}_{\alpha}$ are still uncorrelated. This implies that there are resonant timescales $t \sim 2\pi / \omega_c$ for which the dynamics are dictated by uncorrelated energy modes. In Ref.~\cite{Richter2020}, it was shown that the eigenvalues of $\hat{O}^{\omega_c}$ are uncorrelated even in the regime where the distribution of level spacings follows the GOE. For these reasons, it is important to restrict ourselves to values of $\omega_c$ for which the eigenvalues of $\hat{O}^{\omega_c}$ are uncorrelated and the level spacings follow the GOE. These regimes can be probed by studying the mean ratio of adjacent level spacings, defined as
\begin{align}
\braket{r_{\omega_c}} \defeq \frac{1}{M}\sum_{\alpha} \frac{\textrm{min}\{ \Delta_{\alpha}, \Delta_{\alpha + 1} \}}{\textrm{max}\{ \Delta_{\alpha}, \Delta_{\alpha + 1} \}},
\end{align}
where $\Delta_{\alpha} = | \lambda_{\alpha + 1}^{\omega_c} - \lambda_{\alpha}^{\omega_c}|$. We performed the average over all adjacent level spacings, i.e., $M \approx D^{\prime}$. We have that $\braket{r_{\omega_c}} \approx 0.53$ for a distribution following the GOE and $\braket{r_{\omega_c}} \approx 0.39$ for Poisson-distributed random variables.  

\begin{figure}[t]
\fontsize{13}{10}\selectfont 
\centering
\includegraphics[width=0.65\columnwidth]{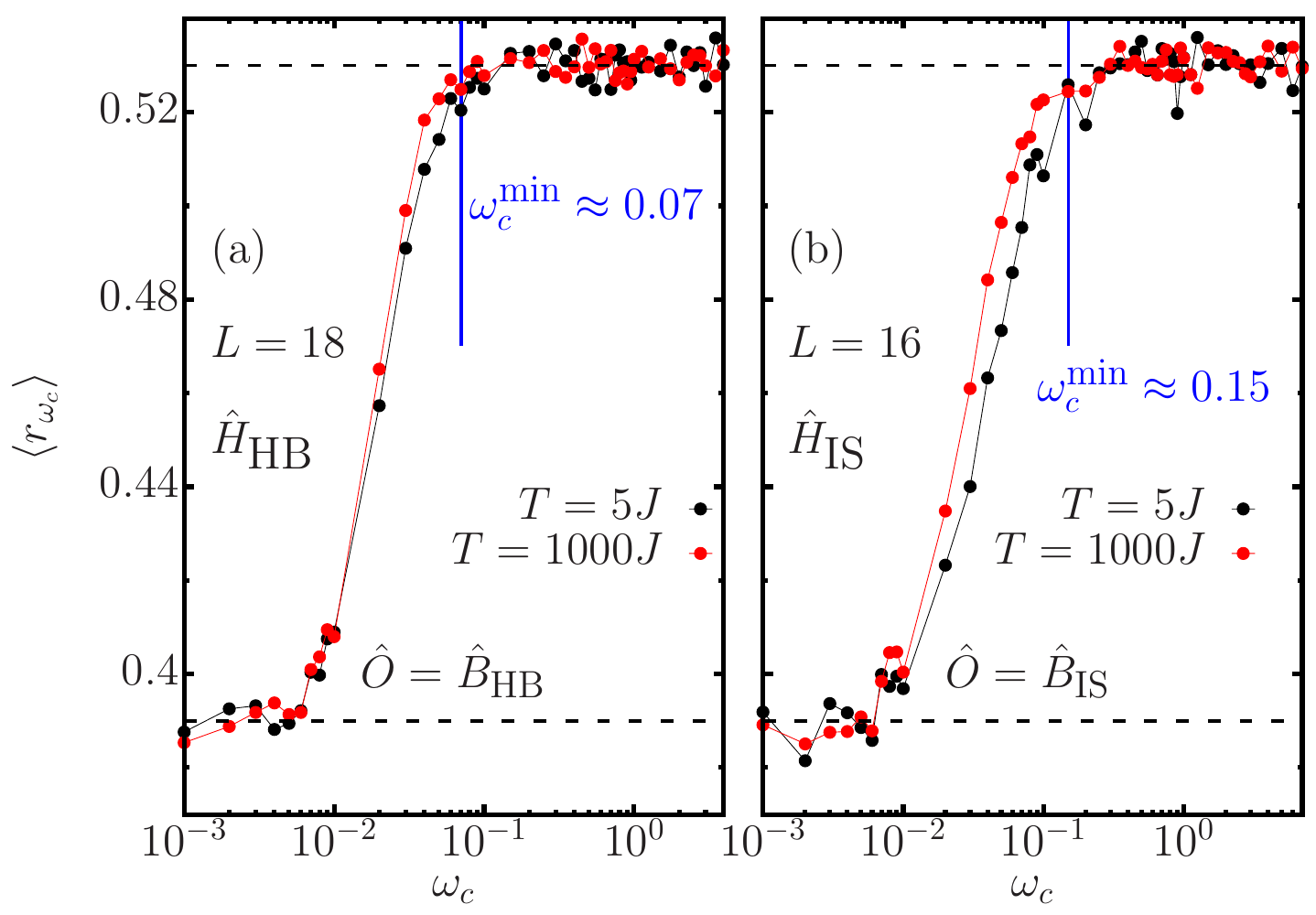}
\caption[Mean ratio of adjacent level spacings $\langle r_{\omega_c} \rangle$ as a function of the cutoff frequency $\omega_c$ for the (a) $\hat{H}_{\textrm{HB}}$ and (b) $\hat{H}_{\textrm{IS}}$ models]{Mean ratio of adjacent level spacings $\langle r_{\omega_c} \rangle$ as a function of the cutoff frequency $\omega_c$ for the (a) $\hat{H}_{\textrm{HB}}$ and (b) $\hat{H}_{\textrm{IS}}$ models, at fixed system size for two different values of temperature $T = 5J$ and $T = 1000J$.}
\label{fig:1.6.10}
\end{figure}

In Fig.~\ref{fig:1.6.10}(a) we show the mean ratio of adjacent level spacings for the $\hat{H}_{\textrm{HB}}$ model as a function of $\omega_c$. It can be observed that the value of $\omega_c$ for which the onset of the GOE is observed is rather similar between the values of temperature $T = 5J$ and $T = 1000J$ chosen. We do not expect this behaviour to be generic. On the contrary, the onset of the GOE is typically observed at different values of $\omega_c$ for different system sizes $L$ and observables. To avoid the aforementioned localisation effects, we restricted our analyses to the values of $\omega_c$ above the $\omega_c^{\textrm{min}}$ denoted in Fig.~\ref{fig:1.6.10}. For all $\omega_c > \omega_c^{\textrm{min}}$, the associated sub-matrices exhibit a value of $\langle r_{\omega_c} \rangle \approx 0.53$. This, however, does not imply that the matrix elements display correlations or lack thereof. As we have shown, there exist a regime for which correlations build up as $\omega_c$ is increased above $\omega_c^{\textrm{min}}$. Similar results are observed for the $\hat{H}_{\textrm{IS}}$ model in Fig.~\ref{fig:1.6.10}(b), with the only difference noticed at the specific $\omega_c^{\textrm{min}}$ values for which the onset of the GOE mean ratio of level spacings is obtained. 

\subsection{Dynamics of the OTOC}
\label{sec:dynamics_otoc}

We now study the implications of the nature of correlations in the off-diagonal matrix elements for the observable dynamics. As discussed above, two-point correlation functions are independent of the statistical correlations between matrix elements. It is thus crucial to examine higher-order correlators and the OTOC is a natural example. We focus in particular on the squared commutator
\begin{align}
c(t) \defeq - \left ( \langle [ \hat O(t), \hat O]^2 \rangle - \langle [ \hat O(t), \hat O]\rangle^2 \right ).
\end{align}
To detect the dynamical effect of matrix-element correlations, we compute $c(t)$ in two different ways: : {\it i)}~by a thermal average in the canonical ensemble at temperature $T$, and {\it ii)}~using a single eigenstate $\ket{E_n}$ and assuming independent, identically distributed (IID) Gaussian statistics for $R_{nm}$ in the ETH Eq.~\eqref{eq:eth}~\cite{Cotler2017Chaos, Foini2019, Murthy19}. 
To evaluate the square commutator within the ETH assuming IID statistics, let us consider the following four point connected correlator
\begin{align}
\label{eq:defF4c}
F_c(t_1, t_2, t_3, t_4) \defeq \braket{ \hat O(t_1)\, \hat O(t_2)\,\hat O(t_3)\,\hat O(t_4) } - \braket{ \hat O(t_1) \,\hat O(t_2)}\,\braket{\hat O(t_3)\,\hat O(t_4) }
\end{align}
where all operators are written in the Heisenberg representation, $\hat O(t)=e^{i\hat Ht}\hat Oe^{-i\hat Ht}$. All time-ordered and out-of-time-ordered correlation functions can be constructed from $F_c(t_1, t_2, t_3, t_4)$ with a suitable choice of arguments. In particular, we focus on the standard OTOC
\begin{equation}
\label{eq:Fotoc}
F_{\text{OTO}}(t)  \defeq \braket{ \hat O(t) \hat O  \hat O(t) \hat O }  - \braket{ \hat O(t) \hat O }^2 =  F_c( t, 0, t, 0),
\end{equation}
and the square commutator
\begin{align}
\label{eq_scApp}
c(t) \defeq - \left ( \langle [ \hat O(t), \hat O]^2 \rangle - \langle [ \hat O(t), \hat O]\rangle^2 \right ) = F_c( t, 0, 0, t)+F_c( 0, t, t, 0)- 2 \text{Re} F_c( t, 0, t, 0).
\end{align}
We now restrict our analysis assuming that the matrix elements in the ETH are uncorrelated Gaussian variables, i.e. 
\begin{align}
\label{eq:Rexp}
\overline{R_{\alpha\beta}\;R_{\gamma\delta}} &= \delta_{\alpha\delta}\;\delta_{\beta\gamma}
\end{align}
and
\begin{align}
\overline{R_{\alpha\beta}\;R_{\beta\gamma}\;R_{\gamma\delta}\;R_{\delta\alpha}} &= \overline{R_{\alpha\beta}\;R_{\beta\gamma}} \; \overline{R_{\gamma\delta}\;R_{\delta\alpha}} +  \overline{R_{\alpha\beta}\;R_{\gamma\delta}} \; \overline{R_{\beta\gamma}\;R_{\delta\alpha}} + \overline{R_{\alpha\beta}\;R_{\delta\alpha}} \; \overline{R_{\beta\gamma}\;R_{\gamma\delta}}.
\end{align}
This allows us to re-write the four point function Eq.~\eqref{eq:defF4c} evaluated over $\hat \rho=\sum_n p_n \ket{n}\bra{n}$ as
\begin{equation}
F_c(t_1, t_2, t_3, t_4) = \sum_n p_n \, F_c(E_n, t_1, t_2, t_3, t_4) \ ,
\end{equation}
where $F_c(E_n, t_1, t_2, t_3, t_4)$ is the micro-canonical expectation value, i.e., Eq.~\eqref{eq:defF4c} computed over a single eigenstate $\ket{n}$. One can observe that the same result holds from a purely out-of-equilibrium calculation, where the expectation value in Eq.~\eqref{eq:defF4c} is taken over an initial pure state $\ket{\psi}$. In this case, the distribution of the $p_n$ is given by the overlaps with the initial state $p_n = |\langle n|\psi\rangle|^2$.
One can then show that if $p_n$ is compatible with statistical mechanics, with is guaranteed when the average energy $\bar{E}$ is defined and the variance $(\delta E)^2/\langle E \rangle^2\sim 1/L$ decays with the system size, then the leading term of Eq.~\eqref{eq:defF4c} is given by the single-eigenstate expectation value. One can, however, expect a constant divergence at finite sizes that decreases with increasing $L$. In the following, we consider only the microcanonical single eigenstate four-point functions and omit $E_n$ in the notations.

Using Eq.~\eqref{eq:Rexp}, one can re-write the four-point function in Eq.~\eqref{eq:defF4c} directly in terms of the two-point functions from Eq.~\eqref{eq:f2} as
\begin{align}
\label{eq:result}
F_c(t_1, t_2, t_3, t_4) = f_1(t_1, t_2, t_3, t_4) + c_1(t_2-t_3) + c_2(t_1-t_4, t_2-t_3).
\end{align}
We then see that $F_c(t_1, t_2, t_3, t_4)$ is composed of 
\begin{align}
\label{eq:c1}
c_1(t_2-t_3) \defeq O'^2 F_2''(t_2-t_3),
\end{align}
with
\begin{align}
\label{eq:c2}
c_2(t_1-t_4, t_3-t_2) \defeq F_2(t_1-t_4) \;F_2(t_2-t_3)  + F'_2(t_1-t_4) \;\frac{\partial F_2(t_2-t_3) }{\partial E},
\end{align}
and
\begin{align}
\label{eq:f1}
f_1(t_1, t_2, t_3, t_4)  \defeq O^2 &\left [F_2(t_1-t_3) + F_2(t_2-t_4) + F_2(t_1-t_4) + F_2(t_2-t_3) \right ] \nonumber \\
&+ O O' \left [F'_2(t_1-t_3) + F'_2(t_2-t_4) + 2F'_2(t_2-t_3) \right ] \nonumber \\
&+ \frac 12 O O'' \left [F''_2(t_1-t_3) + F''_2(t_2-t_4) + 2F''_2(t_2-t_3) \right ],
\end{align}
where $F_2(t)$ is written as defined in Eq.~\eqref{eq:f2} $O\,$, $O\,'$ and $O\,''$ are obtained from the diagonal expectation value of the operator $\hat{O}$ and its first and second derivative with respect to energy evaluated at $E_n$. The functions $F'_2(t)$  and $F''_2(t)$ can be written as
\begingroup
\allowdisplaybreaks
\begin{subequations}
\label{eq:F22Def}
\begin{align}
\label{f2primo}
2 \text {Re} F'_2(t) = -  \sum_{\beta\neq n} \omega_{n \beta} (e^{\textrm{i} \omega_{n \beta} t} +e^{-\textrm{i} \omega_{n \beta} t} )\, |f_{n \beta}|^2\, e^{-S_{n \beta}} = -  \frac 1{2\pi}\, \int d\omega \; \omega\; S_{\hat O}(E_n, \omega) e^{\textrm{i}\omega t} \ ,\\
\label{f2secondo}
2 \text {Re}F''_2(t) = \sum_{\beta\neq n} \omega_{n \beta}^2 (e^{\textrm{i} \omega_{n \beta} t} +e^{-\textrm{i} \omega_{n \beta} t} )\, |f_{n \beta}|^2\, e^{-S_{n \beta}} =  \frac 1{2\pi}\, \int d\omega \;\omega^2\; S_{\hat O}(E_n, \omega) e^{\textrm{i}\omega t}\ ,
\end{align}
\end{subequations}
\endgroup
where in the second line of Eq.~\eqref{eq:F22Def} we have identified sums with integrals and expanded the entropy terms around energy $E_n$. 

The resulting $ S_{\hat O}(E_n, \omega)$ is the symmetric response function which can be computed within ETH as stated in Eq.~\eqref{eq:f2_w_re_im}. It is very important to note that $f_1$ and $c_1$ vanish whenenver $O(\bar{E})=0$, and neither of these functions contain information about the distribution of the $R_{n \beta}$ in the ETH. We can use this to our advantage to simplify the calculations. Finally, the OTOC and square-commutator can be directly written as combinations of $f_1$, $c_1$ and $c_2$ [Eqs.~\eqref{eq:c1}, \eqref{eq:c2} and \eqref{eq:f1}] as
\begin{align}
F_{\text{OTO}}(t) & =  \; f_1(t, 0, t, 0) +  \; c_1(t) + \; c_2(t, -t) \\
\label{eq:ct_eth}
c(t)  & = 2\left[ c_1(0) + c_2(0, 0) - c_1(t) - c_2(t, -t) \right] \ .
\end{align}
Within this approximation, the leading terms in the system size of the square-commutator and of the OTOC read
\begin{align}
\label{eq:cteth}
    F_{\text{OTO}} & = |F_2(t)|^2 + 2\, O(E_n)^2\,  
    \textrm{Re}{\left [ F_2(t)+F_2(0)\right ]} \ , \\
    c(t) & = 2 |F_2(0)|^2 - 2 |F_2(t)|^2 \ .
\end{align}
In fact, all the terms containing derivatives with respect to energy are proportional to $1/L$ and are therefore sub-leading in $L$ with respect to $|F_2(t)|^2$, both for local or sums of local operators.
Note, however, that these terms could be relevant in large-$L$ chaotic models at intermediate times, where the square-commutator is expected to grow exponentially as $\sim e^{2\lambda t}/L^2$, with $\lambda$ the classical Lyapunov exponent \cite{Pappalardi2020Quantum}.

The square commutator as defined in Eq.~\eqref{eq_scApp} can be directly evaluated from the ETH assuming uncorrelated $R_{nm}$ from Eq.~\eqref{eq:eth} using the procedure described above. The resulting expression depends only on the dynamics of two-point functions, which can be evaluated using the procedure described in Sec.~\ref{sec:eth_correlation_functions}. 

\subsubsection{Square commutator dynamics in non-integrable systems}

Under the approximation that the $R_{nm}$ in the ETH are IID Gaussian random variables, we have 
\begin{align}\label{ETH_Uncorrelated}
[c(t)]_{\textrm{ETH Unc.}} \approx 2 |F_2(0)|^2 - 2 |F_2(t)|^2,
\end{align}
where
\begin{align}
F_2(t) \defeq \langle \hat{O}(t) \hat{O}(0) \rangle_c \defeq \langle \hat{O}(t) \hat{O}(0) \rangle - \langle \hat{O}(t) \rangle \langle \hat{O}(0)\rangle.
\end{align}

We can then proceed to evaluate numerically these results in our non-integrable systems $\hat{H}_{\textrm{HB}}$ and $\hat{H}_{\textrm{IS}}$. The OTOC dynamics in the canonical ensemble are computed by exact diagonalisation, representing $\hat{O}(t)$ as a time-dependent matrix in the Heisenberg picture following the computation of the commutators in Eq.~\eqref{eq_sc}. On the other hand, the dynamical evaluation of $c(t)$ assuming IID Gaussian statistics in the ETH is done by obtaining the two-point functions from $f_{\hat{O}}(\bar{E}, \omega)$, as described in Sec.~\ref{sec:models_otoc}.

\begin{figure}[t]
\fontsize{13}{10}\selectfont 
\centering
\includegraphics[width=0.65\columnwidth]{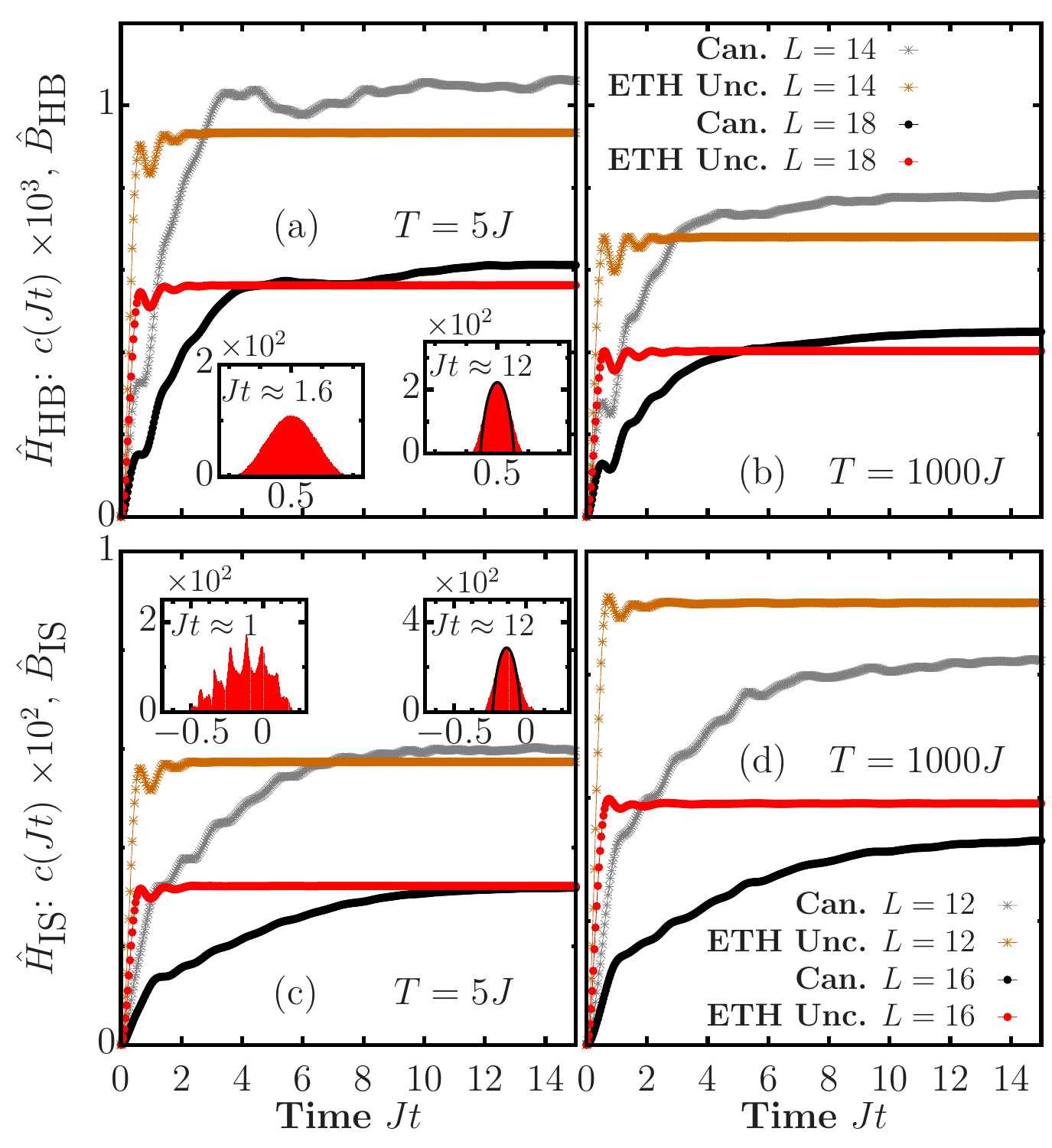}
\caption[Time-dependent square-commutator \eqref{eq_sc} for the operators $\hat{B}_{\textrm{HB}}$ and $\hat{B}_{\textrm{IS}}$ for $\hat{H}_{\textrm{HB}}$ and $\hat{H}_{\textrm{IS}}$]{Time-dependent square-commutator \eqref{eq_sc} for the operators $\hat{B}_{\textrm{HB}}$ [(a),(b)] and $\hat{B}_{\textrm{IS}}$ [(c),(d)] for $\hat{H}_{\textrm{HB}}$ and $\hat{H}_{\textrm{IS}}$ at temperatures $T = 5J$ [(a) and (c)] and $T = 1000J$ [(b) and (d)]. The expectation value obtained for a canonical state is compared with the one obtained assuming the ETH and uncorrelated $R_{nm}$ for increasing system size $L$. Insets show the distribution of eigenvalues of the matrix $\hat{O}^{\omega_c=2\pi/t}$ \eqref{eq:o_wc} for the largest system size displayed in each case.}
\label{fig:1.6.11}
\end{figure}

The result of this comparison is shown in Fig.~\ref{fig:1.6.11} for sums of local operators. A discrepancy between the two predictions at short times signals that this regime is indeed governed by correlations between the matrix elements. However, the curves saturate to a similar value at longer times, differing in some cases by a small correction that we attribute to energy fluctuations in the canonical ensemble at finite size, given that these deviations are less prominent for larger system sizes. Fig.~\ref{fig:1.6.11} shows that the time $t_s$ at which saturation occurs qualitatively increases with system size. 

Interestingly, our data suggest that this saturation time is related to the frequency $\omega_{\rm GOE}$ by $t_s\approx 2\pi/\omega_{\rm GOE}$. The procedure employed so far only allows one to study system sizes available to exact diagonalisation techniques. Regardless, we can proceed visually by showing the distribution of eigenvalues of the matrix $\hat{O}^{\omega_c=2\pi/t}$ at different times (insets of Fig.~\ref{fig:1.6.11}).  While at short times the distribution deviates from the GOE prediction, these deviations are strongly reduced when the OTOC nears saturation, approximately leading to semi-circular distributions. This behaviour indicates that the OTOC's long-time dynamics encodes the statistical properties of $R_{nm}$ and the emergence of random-matrix behaviour at low frequencies.

\subsection{Estimation of the scaling of $\omega_{\rm GOE}$ with system size in the infinite-temperature regime}
\label{sec:scaling_otoc}

Our previous results strongly suggests that the frequency scales divided by $\omega_{\rm GOE}$, studied from the spectrum of banded matrices, have a connection to the saturation timescales of the OTOCs, denoted by $t_s$. This connection could be used to estimate the behaviour of $\omega_{\rm GOE}$ as a function of the system size $L$ from the saturation point of the dynamics of the OTOCs. Notice that the saturation time $t_s$ is not related to the relaxation time of two-point correlations, the dephasing time $t_{\varphi}$, which is expected to be an intensive quantity on general grounds~\cite{sachdev2011}. The saturation time $t_s$ is also generically larger than $t_{\varphi}$. This observation is verified by Fig.~\ref{fig:1.6.11}, since $t_{\varphi}$ determines the fast saturation of the OTOC computed according to the uncorrelated approximation in Eq.~\eqref{ETH_Uncorrelated}. 

In our previous calculations, establishing the connection between $t_s$ and $\omega_{\rm GOE}$ entailed the computation of the unitary operator $\hat{U}$ that renders the Hamiltonian diagonal, i.e., $\tilde{H} = \hat{U}^{\dagger} \hat{H} \hat{U}$, where $\tilde{H}$ is a diagonal matrix with the eigenvalues in its entries. This exact diagonalisation procedure is computationally costly due to the rapid increase of $\mathcal{D}$ as a function of the system size. Having established the relation between $\omega_{\rm GOE}$ and $t_s$, we could evaluate the saturation point in the dynamics of the OTOCs using a different approach to provide and estimation of the scaling of $\omega_{\rm GOE}$.

For this purpose, we employ the concept of dynamical quantum typicality~\cite{Goldstein:2006,Popescu:2006,Luitz:2017,Chiaracane:2021,Richter:2018}. In this framework, it is possible to approximate the unitary dynamics of a given system within an equilibrium ensemble from a single pure state $\ket{\psi}$, which is drawn at random from the Haar measure~\cite{Luitz:2017} on an arbitrary basis $\{ \ket{\phi}_k \}_{k = 1}^{\mathcal{D}}$. We start with 
\begin{align}
\ket{\psi} = \hat{R}\sum_{k=1}^{\mathcal{D}}c_{k} \ket{\phi_k}, \quad c_k \defeq a_k + \textrm{i}b_k,
\end{align}
where $\hat{R}$ is an arbitrary operator on the Hilbert space and $a_k$ and $b_k$ are independent random variables drawn from a normal distribution. The averaged expectation value of an operator $\hat{O}$ in the typical state is equivalent to the expectation value computed with respect to a density matrix $\hat{\rho}$, such that $\overline{O} \defeq \overline{\braket{\psi | \hat{O} | \psi}} \approx \textrm{Tr}[\hat{\rho} \hat{O}]$. In this particular case, $\hat{\rho} = \hat{R} \hat{R}^{\dagger}$. It can be shown that the approximation is more accurate as $\mathcal{D}$ increases~\cite{Chiaracane:2021}. 

We now focus on the infinite-temperature regime, in which we can write
\begin{align}
\hat{\rho} = \frac{\mathds{1}}{\mathcal{D}}\quad \textrm{and} \quad \ket{\psi} = \frac{1}{\sqrt{\mathcal{D}}}\sum_{k=1}^{\mathcal{D}}c_k \ket{\phi_k}.
\end{align}
With this procedure, we may approximate the dynamics of $c(t)$ from Eq.~\eqref{eq_sc} in the infinite-temperature regime by
\begin{align}
\label{eq:c_t_s}
c(t) \approx - \left ( \langle \psi | [ \hat O(t), \hat O]^2 | \psi \rangle - \langle \psi | [ \hat O(t), \hat O] | \psi \rangle^2 \right ),
\end{align}
where the approximation becomes more accurate as $L$ is increased. To provide a better approximation for smaller values of $L$ we carry out an averaging procedure using several different random states $\ket{\psi}$. The dynamics is evaluated in the Schr\"odinger picture, using the method of Krylov subspaces to evaluate time-evolved states. The idea is to evaluate the action of the propagator onto a pure state to obtain a time-evolved state, i.e., $\ket{\psi(t)} = e^{-\textrm{i}\hat{H}t}\ket{\psi(0)}$. With this method, we evaluate $\ket{\psi(t)}$ by computing the action of $e^{-\textrm{i}\hat{H}t}$ onto $\ket{\psi(0)}$. This is done by a polynomial approximation to $\ket{\psi(t)}$ from within the Krylov subspace
\begin{align}
\mathcal{K}_m = \textrm{span}\left\{\ket{\psi(0)}, \hat{H}\ket{\psi(0)}, \hat{H}^2\ket{\psi(0)}, \dots, \hat{H}^{m-1}\ket{\psi(0)}\right\}.
\end{align} 
The optimal approximation is obtained by an Arnoldi decomposition procedure of the upper Hessenberg matrix $A_m$, defined as $A_m \defeq V^T_mHV_m$, where $V_m$ corresponds to the orthonormal basis resulting from the decomposition. $A_m$ can be seen as the projection of $\hat{H}$ onto $\mathcal{K}_m$ with respect to the basis $V_m$. In the previous description $m$ is the dimension of the Krylov subspace. In principle, the Arnoldi decomposition procedure can be replaced by a three-term Lanczos recursion for the specific case of Hermitian matrices. The latter amounts to a more efficient algorithm, yet to one that may suffer from numerical instabilities for ill-conditioned matrices.

The desired solution is then approximated by
\begin{equation}
\ket{\psi(t)}\approx V_{m}\exp(-\textrm{i}tA_m)\ket{e_1},
\end{equation}
where $\ket{e_1}$ is the first unit vector of the Krylov subspace. The approximation becomes an exact solution when $m \geq \mathcal{D}$, however, the method has been proven to be accurate even if $m \ll \mathcal{D}$ for short enough time-steps~\cite{expokit,brenes2019massively}. For the particular case when $m \ll \mathcal{D}$, the much smaller matrix exponential $\exp(-\textrm{i}tA_m)$ can be evaluated using standard numerical techniques, such as a Pad\`e approximation with a scaling-and-squaring algorithm. The error in the method behaves like $\mathcal{O}(e^{m-t||A||_{2}}(t||A||_{2}/m)^m)$ when $m \leq 2t||A||_{2}$, which indicates that the technique can be applied successfully if a time-stepping strategy is implemented along with error estimations~\cite{expokiterror}. In practice, the dimension of the Krylov subspace $m$ is a free parameter of the simulation, while the time-step is estimated such that the above {\em a priori} error estimation is kept under control.  

We remark that for the case of $c(t)$, evaluating terms of the form $\braket{\psi | [\hat{O}(t), \hat{O}]^2 | \psi}$ is more complicated, since it requires running both forwards and backwards time evolution of operators $\hat{O}$ acting on pure states $\ket{\psi}$, where the $\ket{\psi}$ are typical states introduced above. Yet, this procedure can be carried out efficiently dividing the entire time evolution into several time-steps~\cite{Luitz:2017}. 

\subsubsection{Evaluation in non-integrable models}

\begin{figure}[t]
\fontsize{13}{10}\selectfont 
\centering
\includegraphics[width=0.55\columnwidth]{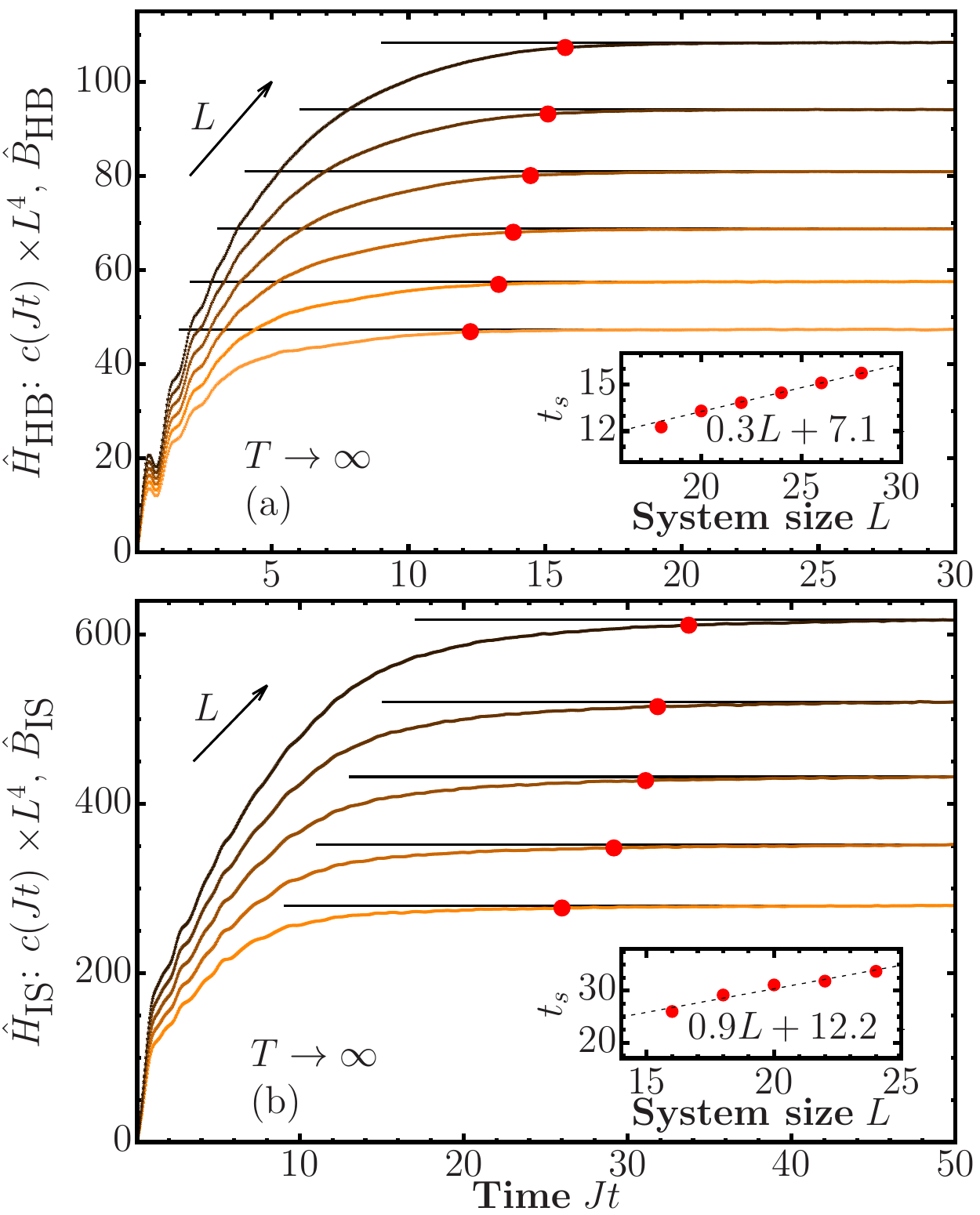}
\caption[Dynamics of the OTOC averaged over many typical states in the infinite-temperature regime for the (a) $\hat{H}_{\textrm{HB}}$ ($L = 18, \cdots, 28$ increasing for even $L$) and (b) $\hat{H}_{\textrm{IS}}$ ($L = 14, \cdots, 24$ increasing for even $L$) models and their corresponding observables composed of sums of local operators.]{Dynamics of the OTOC averaged over many typical states in the infinite-temperature regime for the (a) $\hat{H}_{\textrm{HB}}$ ($L = 18, \cdots, 28$ increasing for even $L$) and (b) $\hat{H}_{\textrm{IS}}$ ($L = 14, \cdots, 24$ increasing for even $L$) models and their corresponding observables composed of sums of local operators. The circles denote our estimation of the saturation time $t_s$. The insets exhibit the finite-size scaling analysis of the saturation time $t_s$ as a function of $L$, where the dashed lines depict a fit to a linear function $aL+b$.}
\label{fig:1.6.12}
\end{figure}

In Fig.~\ref{fig:1.6.12} we show the results of $c(t)$ averaged over several typical states using the procedure above for the $\hat{H}_{\textrm{HB}}$ [Fig.~\ref{fig:1.6.12}(a)] and the $\hat{H}_{\textrm{IS}}$ [Fig.~\ref{fig:1.6.12}(b)] models for their respective extensive observables. The numerical approach described before allows us to study much larger system sizes (up to $L = 28$ for $\hat{H}_{\textrm{HB}}$ and $L = 24$ for $\hat{H}_{\textrm{IS}}$). The dynamics displayed correspond to the average between many different typical states, which range from $1000$ for the smallest values of $L$, to $2$ for the largest values. The number of realisations is chosen such that the standard deviation along each point in the time trajectory does not surpass $1\%$ of the mean value.

The circles marked in the main panels correspond to the saturation points of the OTOCs. The values of $c(Jt)$ have been scaled by a factor of $L^4$ for visualisation purposes and do not affect the saturation times. To evaluate these saturation values, we first estimate the long-time value of the OTOCs $c(Jt \to \infty)$ from the average of the late dynamics (we use $Jt \in [28,30]$ for $\hat{H}_{\textrm{HB}}$ and $Jt \in [48,50]$ for $\hat{H}_{\textrm{IS}}$). The saturation time is then selected at the value for which $c(Jt_s) = \varepsilon c(Jt \to \infty)$ is reached, where $\varepsilon$ is a certain threshold parameter. The saturation times are highlighted by the circles in the main panels of Fig.~\ref{fig:1.6.12}. There exist some very small finite-size oscillations throughout the dynamics, which introduce a level of uncertainty into the estimation of $t_s$. To account for these, we compare $c(Jt \to \infty)$ against a running-average value in the vicinity of $Jt$. Explicitly so, we compare the long-time estimation of $c(Jt \to \infty)$ against the average within the set $[Jt - 0.5, Jt + 0.5]$ to approximate $t_s$ more accurately. The insets in Fig.~\ref{fig:1.6.12} display the scaling of the saturation time $t_s$ as a function of the system size $L$. In both cases, the saturation time appears to scale linearly with $L$. This behaviour is robust to changes on the parameter $\varepsilon$, as long as $\varepsilon \approx 1$. The displayed $t_s$ results in Fig.~\ref{fig:1.6.12} were obtained with $\varepsilon = 0.99$. 

\subsection{Summary and outlook}
\label{sec:summary_otoc}

The observed linear scaling of the saturation time $t_s$, reminiscent of ballistic transport, is qualitatively consistent with the linear front propagation at the butterfly velocity expected in chaotic systems~\cite{shenker2014black,Aleiner16}. This observation and the results obtained for both models would imply that, consequently, $\omega_{\textrm{GOE}} \propto 1/L$ for extensive operators in chaotic many-body systems. The connection of this energy scale to the Thouless energy $\omega_{Th}$, characterising random matrix behaviour in the spectrum of the Hamiltonian, depends on the details of the system. For standard diffusive scaling, $\omega_{\textrm{Th}} \propto 1/L^2 $, one observes $\omega_{\textrm{GOE}} > \omega_{\textrm{Th}}$ (in apparent contrast with Ref.~\cite{Dymarsky2018}, in which strictly local operators were considered). The careful numerical study of the scaling of $\omega_{\textrm{Th}}$ with system size required to clarify this issue is beyond the scope of this paper and it is left for future studies. 

We have performed a systematic analysis of statistical correlations within the ETH and explored their consequences for the dynamics of quantum information scrambling. Remarkably, we find that correlations between off-diagonal matrix elements indicate the timescale for the onset of random-matrix dynamics in the corresponding OTOC, an experimentally observable quantity. This operator- and temperature-dependent timescale is not apparently connected to hydrodynamic behaviour of linear-response functions, given that the dynamics of stationary two-point correlation functions are independent of statistical matrix-element correlations. Moreover we have provided an estimation of the scaling of the timescale $t_s$ as a function of the system size $L$, which appears to behave linearly with $L$, consistently with the expected ballistic propagation of combustion-like waves associated to the butterfly effect~\cite{shenker2014black,Aleiner16}. This timescale appears to be connected to the frequency scale $\omega_{\rm GOE} \sim t_s^{-1}$ where random-matrix behaviour is observed from the analysis of banded sub-matrices. The estimation of the scaling as a function of the system size was possible by employing the concept of dynamical typicality in conjunction with computationally optimised Krylov subspace techniques for time evolution. Our results lie at the limit of system size that can be achieved with this numerical approach using parallel algorithms in supercomputers, due to the long timescales required to study saturation of the OTOCs. Finite system size thus remains a limitation to our estimations, despite the fact that the exposed technique allows us to access much larger systems than possible with exact diagonalisation techniques. 
\part{Open quantum systems}
\label{part:two}

\chapter{Introduction}

The miniaturisation of technologies in combination with the exquisite control now available over nanoscale systems has motivated increasing interest in thermal machines that operate in the quantum regime~\cite{Kosloff2014, Goold:2016, Benenti:2017, Binder2018, Mitchison2019}. While recent demonstrations with trapped ions~\cite{Rossnagel2016,Maslennikov2019,Horne2020, Lindenfels2019}, nanomechanical oscillators~\cite{Klaers2017} and diamond colour centres~\cite{Klatzkow2019} serve as impressive proofs of principle, practical applications such as thermoelectric power generation call for electronic devices. To that end, the focus of experiments in mesoscopic physics has expanded beyond traditional questions of charge transport to include the manipulation of heat currents in platforms such as semiconductor quantum dots~\cite{Linke2018}, superconducting circuits~\cite{Ronzani2018} and molecular junctions~\cite{Mosso2019}. Understanding the non-equilibrium thermodynamics of these systems is a formidable theoretical challenge, due to the simultaneous presence of strong system-reservoir coupling, inter-particle interactions and finite temperatures. 

Existing approaches to modelling energy transport in complex quantum systems typically depend on perturbative arguments, which require a clear separation of energy or time scales. For example, a quantum master equation can be derived under the assumption of weak system-reservoir coupling~\cite{BreuerPetruccione}. However, the approximations needed to ensure positivity of the density matrix may fail to capture quantum coherences far from equilibrium~\cite{Wichterich2007,Purkayastha2016,Kirsanskas2018,Mitchison2018}, while a first-principles derivation requires full diagonalisation of the system Hamiltonian and thus becomes infeasible for large open systems. 

A more tractable approach for many-body problems is a local master equation, where incoherent sinks and sources create and remove excitations at the boundaries of the system. This method has been successfully applied to study infinite-temperature transport in strongly interacting systems~\cite{Prosen_2015}, but its finite-temperature predictions may violate basic thermodynamic laws~\cite{Levy2014,Stockburger_2016,Gonzalez2017,Hofer2017} unless a specific kind of periodically modulated system-bath interaction is assumed~\cite{karevski2009quantum,Clark_2010,barra2015thermodynamic,Strasberg2017,De_Chiara_2018}. Alternatively, non-equilibrium Green functions~\cite{Stefanucci_2009} can be used to model energy transport under strong system-reservoir coupling, but at the cost of treating many-body interactions within the system perturbatively~\cite{Wang_2013}. Another possibility is the numerical renormalisation group, which can handle strong interactions but is typically limited to near-equilibrium transport properties~\cite{Bulla2008}. The related chain representation of unitary system-bath dynamics~\cite{Prior2010} is also capable of non-perturbative transport calculations~\cite{Nuesseler2019} at finite temperatures~\cite{Tamascelli2019} but its scalability to large system size remains unclear.

In Part~\ref{part:two}, we start by providing the background theory behind local master equations, in a configuration known as boundary driving. Particularly, in Chapter~\ref{chapter:lindblad_high}, we introduce the concept of local master equations, mathematical expressions for the spin current, scaling theory and a numerical framework based on matrix product states and operators to solve for non-equilibrium steady states. Our motivation to introduce this topic is twofold. First, in Part~\ref{part:one}, we demonstrated from the perspective of linear response that spin transport in single impurity model is ballistic in the gapless phase $0 < \Delta < 1$, for $\alpha = 1$ in Eq.~\eqref{eq:h_si} of the XXZ model. We were able to demonstrate this from the finite-frequency structure of the spin conductivity. However, such calculations are limited by the available system size reachable with exact diagonalisation techniques. The method of boundary driving will allow us to overcome this limitation to solidify these results, and to demonstrate the diffusive regime of the staggered field model in Eq.~\eqref{eq:h_sf}, implying that the nature of integrability-breaking terms results in different transport regimes. The second reason is related to the generalisation of thermodynamics and transport in the finite-temperature regime. In Chapter~\ref{chapter:finite_temperature}, we shall put forward a general and efficiently scalable numerical approach to quantum thermodynamics that can deal with simultaneously strong intra-system and system-bath interactions and which works arbitrarily far from equilibrium. We focus on autonomous thermal machines, where macroscopic fermion reservoirs held at different temperatures and chemical potentials drive currents through a complex quantum working medium.
 
For the approach introduced Chapter~\ref{chapter:finite_temperature}, we model the macroscopic reservoirs by a finite collection of fermionic modes that are continuously damped towards thermal equilibrium by an appropriate Lindblad master equation. We use a purification scheme based on auxiliary {\em superfermion} modes~\cite{Dzhioev2011} to compute the non-equilibrium steady states of both non-interacting and interacting working media. For interacting systems, we develop a tensor-network algorithm to efficiently simulate the real-time dynamics of the entire configuration, working directly in the energy eigenbasis of the reservoirs. Our approach is well suited to far-from-equilibrium problems in which all energy scales are comparable, such that perturbative or linear-response theories fail. To exemplify this, we demonstrate that the efficiency of a three-site quantum heat engine is enhanced by repulsive interactions and is further improved when the system-reservoir coupling is increased.

The concept of modelling infinite baths by a finite set of damped modes has been widely adopted and adapted since the seminal work of Imamoglu~\cite{Imamoglu1994} and Garraway~\cite{Garraway1997a,Garraway1997b}. In the context of open quantum systems coupled to bosonic reservoirs, this representation has been placed on a mathematically rigorous footing~\cite{Tamascelli2018,Mascherpa2019}, while its amenability to tensor-network simulations has been demonstrated~\cite{Somoza2019}. Related approaches have been used to study quantum heat engines~\cite{Strasberg_2016,Newman2017} and thermalisation in few-level~\cite{IlesSmith2014} and many-particle systems~\cite{Uzdin2018,Reichental2018}. In the fermionic setting, conditions under which continuum baths can be modelled by mesoscopic reservoirs have been recently discussed in Refs.~\cite{Gruss2016,Elenewski2017,Chen2019}. Such mesoscopic reservoirs have been used quite extensively over the last several years for studying transport in non-interacting systems~\cite{Dzhioev2011, Ajisaka2012, Ajisaka2013, Zelovich2014, Landi:2016,Gruss2016, Elenewski2017}, including under time-dependent driving fields~\cite{Oz2020}. For interacting systems, a mesoscopic-reservoir description was recently applied to study particle transport and Kondo phenomena in impurity models~\cite{Schwarz2016,Schwarz2018}, while a related approach to simulating non-equilibrium many-body problems via an auxiliary master equation has been reported~\cite{Dorda2015, Titvinidze2015}. 

A key feature of our work that differs from previous approaches is a novel tensor-network algorithm that exploits the superfermion representation to simulate Lindblad dynamics directly in the energy eigenbasis of the baths (the so-called star geometry). This configuration is particularly favourable in fermionic systems, where only a limited energy window participates in the dynamics at finite temperature due to Pauli exclusion effects at low energies. Although we focus here on steady states of autonomous machines, our methods can be adapted to study transient dynamics or time-dependent Hamiltonians. Moreover, our tensor-network algorithm is inherently scalable to many-body problems, as we demonstrate by first extracting the super-diffusive transport exponents of the isotropic Heisenberg model at high temperature, and then by studying finite-temperature regimes in the gapless phase of the anisotropic Heisenberg model beyond the predictions of single-site boundary driving configurations. Our work thus paves the way for simulations of heat transport in strongly correlated systems that probe heretofore inaccessible regimes of temperature and system size. 
\chapter{Local master equation for high-temperature transport}
\label{chapter:lindblad_high}

From the theoretical perspective, studying transport in non-integrable models represents a significant computational challenge, as both large system sizes and long-time limits are required~\cite{Heidrich-Meisner_Honecker_review_07}. This requirement is even more prevalent at high energies where effective low-energy field theories fail~\cite{prelovvsek2002transport}. A relatively modern approach for extracting high-temperature transport properties of non-integrable one-dimensional quantum systems is known as boundary driving~\cite{karevski2009quantum, Prosen:2009, Benenti:2009, Znidaric:2010, Znidaric:2010b, Prosen:2011, Znidaric:2011, Mendoza:2013a, Mendoza:2013b, karevski2013exact, Landi:2015}. Boundary driving is a setup which stems from the theory of open quantum systems, in which Lindblad jump operators are applied at the boundaries of the chain in order to model spin sources and sinks that drive the chain into a non-equilibrium steady state (NESS). In some cases, it may be combined with the power of matrix product operator techniques~\cite{Schollwock2011} to reach system sizes beyond those accessible via full exact diagonalisation or Lanczos based techniques. Transport properties can be determined by means of finite-size scaling of the current operator in the NESS. This approach has been successful in providing an accurate numerical characterisation of high temperature transport properties of the XXZ model~\cite{Znidaric:2010, Znidaric:2011} and of the ergodic regime of spin chains that exhibit many-body localisation~\cite{vznidarivc2016diffusive, vznidarivc2017dephasing, mendoza2018asymmetry, schulz2018energy}. These works have shown that strong integrability breaking need not result in diffusive transport in the steady state, and that anomalous diffusion could exist in a variety of circumstances.

In this chapter, we shall describe and employ the approach of boundary driving to address the high-temperature transport in the spin-$\frac{1}{2}$ XXZ chain in the presence of integrability breaking in the form of a single (static) magnetic defect. This model is known in the literature to lead to quantum chaos~\cite{Santos:2004, santos2011domain, torres2014local, XotosIncoherentSIXXZ}, and, as we have described in Part~\ref{part:one}, presents ballistic spin transport. We will contrast the results for that model with those from a model in which the (global) integrability breaking perturbation applied to the XXZ chain is a staggered magnetic field. The latter perturbation is known to render transport fully diffusive~\cite{Prosen:2009, Huang2013, JuanThesis:2014, BrenigTypicality2015}. We recall the Hamiltonian of the anisotropic Heisenberg model for an open chain of length $D$\footnote{In Part~\ref{part:two} we shall use $D$ to denote the number of sites/spins of a given system, while other symbols such as $N$ and $L$ shall be used for reservoirs' degrees of freedom in Chapter~\ref{chapter:finite_temperature}.},
\begin{align}
\label{eq:h_xxz_22}
\hat{H}_{\textrm{XXZ}} = \sum_{i=1}^{D-1}\left[\alpha\left(\hat{\sigma}^x_{i}\hat{\sigma}^x_{i+1} + \hat{\sigma}^y_{i}\hat{\sigma}^y_{i+1}\right) + \Delta\,\hat{\sigma}^z_{i}\hat{\sigma}^z_{i+1}\right],
\end{align} 
described in detail in Part~\ref{part:one}. Integrability breaking perturbations can be introduced in the form of a single impurity
\begin{align}
\label{eq:h_si_22}
\hat{H}_{\textrm{SI}} = \hat{H}_{\textrm{XXZ}} + h\, \hat{\sigma}^z_{D/2},
\end{align}
or a staggered magnetic field
\begin{align}
\label{eq:h_sf_22}
\hat{H}_{\textrm{SF}} = \hat{H}_{\textrm{XXZ}} + b\,\sum_{i\,odd} \hat{\sigma}^z_{i}.
\end{align}
By coupling the edges of the spin chain to jump operators that induce excitations, we study the linear response spin transport in the models above. In Sec.~\ref{sec:boundary_driving} we introduce the open-system configuration, provide a derivation to the resulting Lindblad master equation and describe spin current operators in non-equilibrium steady states. Sec.~\ref{sec:mpos_ness} describes the approach we employ to address the solution of the master equation in the long-time limit, by means of a tensor network approach. We then proceed to evaluate spin transport in Sec.~\ref{sec:transport_ness} and finalise with a summary in Sec.~\ref{sec:summary_ness}.

\section{Boundary driving for interacting systems}
\label{sec:boundary_driving}

\begin{figure}[t]
\centering
\includegraphics[width=0.65\columnwidth]{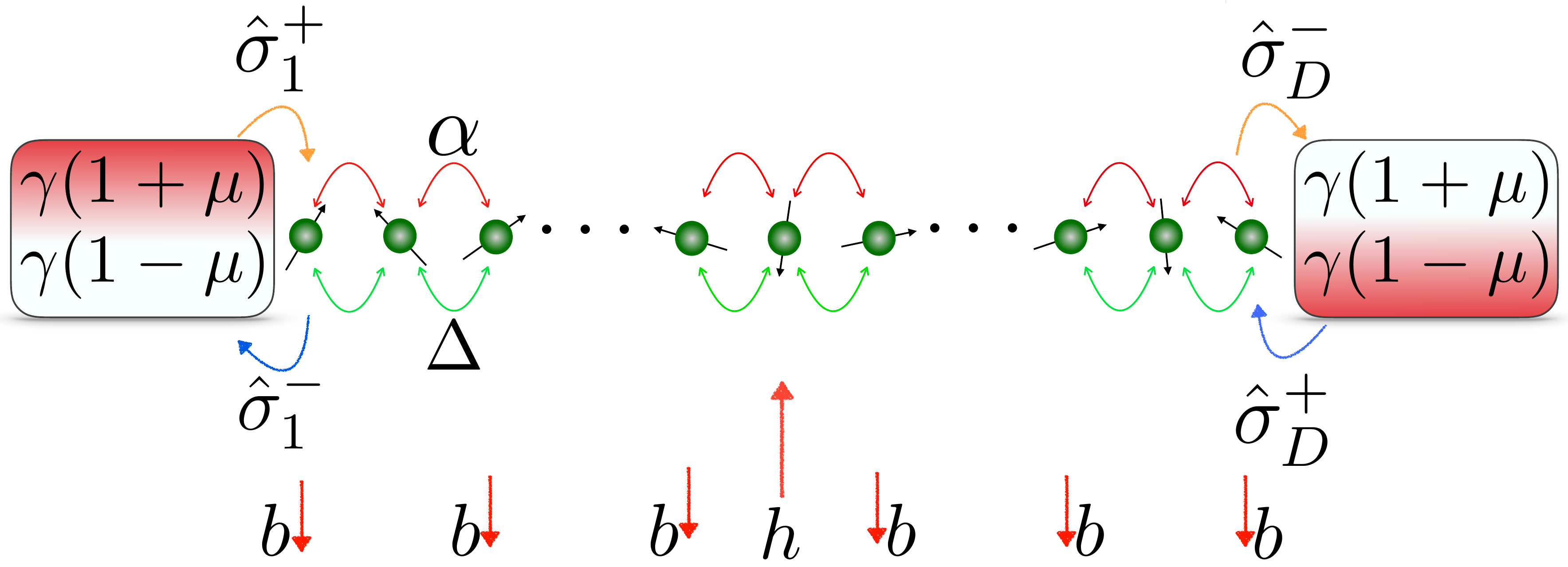}
\caption[Diagrammatic depiction of the non-equilibrium configuration used to study transport in the boundary-driven scheme]{Diagrammatic depiction of the non-equilibrium configuration used to study transport in the boundary-driven scheme. Excitations induced by the baths can propagate through the system (top red arrows) while interactions occur because of the third term in Eq.~\eqref{eq:h_xxz_22} (bottom green arrows). The system-bath coupling strength is given by $\gamma$, while $\mu$ represents the driving strength. A sufficiently strong (but finite) field in either configuration (a single magnetic impurity of strength $h$ or a staggered field of strength $b$) renders the system non-integrable.}
\label{fig:2.2.1}
\end{figure}

In order to study transport in a genuinely non-equilibrium steady-state in a long chain, we couple the latter to two Markovian baths that create and remove excitations at the boundaries. The dynamics of such a setup can be analysed by means of the Lindblad master equation 
\begin{align}
\label{eq:lme}
\frac{d\hat{\rho}}{dt} &= -\textrm{i}[\hat{H},\hat{\rho}] + \mathcal{L}\{\hat{\rho}\} \nonumber \\ 
&= -\textrm{i}[\hat{H},\hat{\rho}] + \mathcal{L}_{\tt L}\{\hat{\rho}\} + \mathcal{L}_{\tt R}\{\hat{\rho}\},
\end{align}
where $\hat{\rho}$ is the density matrix of the system and $\mathcal{L}_{\tt L, \tt R}$ are dissipative super-operators that act on $\hat{\rho}$ inducing excitations in terms of spin creation and annihilation operators given by $\hat{\sigma}^{\pm}_j = (\hat{\sigma}^x_j \pm i\hat{\sigma}^y_j) / 2$ for site at position $j$. Specifically, we have
\begin{align}
\label{eq:lind1}
\mathcal{L}_{m}\{\hat{\rho}\} = \sum_{s=\pm} 2\hat{L}_{s,m}\,\hat{\rho}\, \hat{L}_{s,m}^{\dag} - \{\hat{L}_{s,m}^{\dag}\hat{L}_{s,m},\hat{\rho}\},
\end{align}
where $m = \tt L, \tt R$ (left and right, respectively) and $\{\cdot\,,\cdot\}$ is the anticommutator. The operators in Eq.~\eqref{eq:lind1} are defined as follows:
\begin{align}
\label{eq:lind2}
\hat{L}_{+,\tt L} = \sqrt{\gamma(1 + \mu)}\,\hat{\sigma}_1^{+}, \quad \hat{L}_{-, \tt L} = \sqrt{\gamma(1 - \mu)}\,\hat{\sigma}_1^{-}, \nonumber \\
\hat{L}_{+,\tt R} = \sqrt{\gamma(1 - \mu)}\,\hat{\sigma}_D^{+}, \quad \hat{L}_{-,\tt R} = \sqrt{\gamma(1 + \mu)}\,\hat{\sigma}_D^{-},
\end{align}
where $\gamma$ is the bath coupling parameter and $\mu$ is a parameter that dictates the strength of the boundary driving. A diagrammatic depiction of the non-equilibrium configuration is presented in Fig.~\ref{fig:2.2.1}. The Lindblad master equation [Eq.~\eqref{eq:lme}] can be obtained from a microscopic derivation, such as the one used in the repeated interactions scheme, which allows one to obtain expressions for thermodynamic quantities such as heat and work~\cite{barra2015thermodynamic, PereiraHeatXXZ}.

\subsection{Repeated interactions derivation of the Lindblad master equation for boundary-driven spin chains}
\label{sec:repeated_interactions}

The Lindblad master equation (Eq.~\ref{eq:lme}) can be derived from first principles using the repeated interactions scheme. We employ the approach used in Ref.~\cite{PereiraHeatXXZ} and references therein.

Consider a quantum system with a time-independent Hamiltonian $H_{\tt S}$, coupled to two thermal baths on each boundary, with Hamiltonian $H_{\tt R}$ and $H_{\tt L}$ for right and left bath, respectively. At time $t=0$, we assume the system to be decoupled from the baths. We can then write the density matrix of system + environment as
\begin{align}
\hat{\rho}_{\tt T}(0) = \hat{\rho}_{\tt S}(0) \otimes \hat{\rho}_{\tt E}(0),
\end{align}
where we used subscripts $\tt T$ for total,  $\tt S$ for system and $\tt E$ for environment. The repeated interactions scheme then goes as follows: We couple the system to the left and right baths and allow the configuration to evolve to time $\tau$. Subsequently, we take a partial trace over the baths, couple the system to a new copy of baths and allow the configuration to evolve up to time $2\tau$, before taking another partial trace of the new set of baths. This procedure is done iteratively up to time $t$. We thus express the baths as an infinite collection of copies acting in different intervals of time. We write
\begin{align}
\hat{H}_{m} = \sum_n \hat{H}_{m}^n,
\end{align} 
where $m = {\tt L}, {\tt R}$ for left and right, respectively. The Hamiltonian $\hat{H}_{m}^n$ interacts with the system in the time interval $t \in \left[ (n-1)\tau, n\tau \right]$. The interaction between system and the baths is, crucially, time dependent in this scheme and given by $\hat{V}(t) = \hat{V}^n$, where $\hat{V}^{n} = \hat{V}^{n}_{\tt L} + \hat{V}^{n}_{\tt R}$. We remark that this interaction at each timestep is of the same form, however, the interactions occur with different copies of $\hat{H}_{\tt L}^n + \hat{H}_{\tt R}^n$ in different time intervals.

We write the density matrix of the baths as a product state between all the copies, i.e., $\hat{\rho}_{\tt E} = \otimes_n \hat{\rho}_n$ where we have
\begin{align}
\hat{\rho}_n = \hat{\omega}_{\beta_{\tt L}}\left( \hat{H}^n_{\tt L} \right) \otimes \hat{\omega}_{\beta_{\tt R}}\left( \hat{H}^n_{\tt R}\right),
\end{align} 
where $\hat{\omega}_{\beta_{m}}$ are the initial density operators of the $m$-th bath, which may be expressed as Gibbs states $\hat{\omega}_{\beta_{m}}(\hat{H}_{m}) = e^{-\beta_{m} \hat{H}_{m}} / Z_{m}$, with $Z_{m} = \textrm{Tr}[e^{-\beta_{m} \hat{H}_{m}}]$. According to the scheme described before, we have that the state of the system after the $n$-th step is
\begin{align}
\label{eq:rhos1}
\hat{\rho}_{\tt S}(n\tau) = \textrm{Tr}_n \left( \hat{U}_n \{ \hat{\rho}_{\tt S} [ (n-1)\tau ] \otimes \hat{\rho}_n \} \hat{U}^{\dagger}_n \right),
\end{align}
where $\textrm{Tr}_n$ denotes the partial trace over the $n$-th bath. $\hat{U}_n$ is the global unitary using the total Hamiltonian and given by $\hat{U}_n = \textrm{exp}\left[ -\textrm{i}\tau \left( \hat{H}_{\tt S} + \hat{H}_{\tt L}^n + \hat{H}^n_{\tt R} + \hat{V}^n \right) \right]$ with $\hbar = 1$. 

Let us now consider, for the specific case at hand, the Hamiltonian of the system to be the anisotropic Heisenberg model for a one-dimensional spin chain of size $D$
\begin{align}
\hat{H}_{\tt S} = \sum_i \left[ \alpha \left( \hat{\sigma}^x_i \hat{\sigma}^x_{i+1} + \hat{\sigma}^y_i \hat{\sigma}^y_{i+1} \right) + \Delta \hat{\sigma}^z_i \hat{\sigma}^z_{i+1} \right].
\end{align}
For the Hamiltonian of the baths we have 
\begin{align}
\hat{H}_{m} = \frac{h_{m}}{2}\hat{\sigma}^z_m,
\end{align}
where $h_{m}$ is a constant factor. Note that this is a very specific driving configuration for the Heisenberg model. The interaction Hamiltonian is given by
\begin{align}
\hat{V}_{\tt L} &= g \left(  \hat{\sigma}^x_{\tt L} \hat{\sigma}^x_1 + \hat{\sigma}^y_{\tt L} \hat{\sigma}^y_1 \right), \\
\hat{V}_{\tt R} &= g \left(  \hat{\sigma}^x_D \hat{\sigma}^x_{\tt R} + \hat{\sigma}^y_D \hat{\sigma}^y_{\tt R} \right),
\end{align}
for left and right baths, respectively. With this type of interaction, there exists transfer of spin excitations (or, equivalently, transfer of particles in the form of spinless fermions) between the  degrees of freedom of the system and the baths. We can express Eq.~\ref{eq:rhos1} in power series of $\tau$ as
\begin{align}
\label{eq:expansion1}
\hat{\rho}_{\tt S} (n\tau) &=  \textrm{Tr}_n \left( \hat{U}_n \{ \hat{\rho}_{\tt S} [ (n-1)\tau ] \otimes \hat{\rho}_n \} \hat{U}^{\dagger}_n \right) \nonumber \\
&= \textrm{Tr}_n \left( e^{-\textrm{i}\hat{H}_{\tt T} \tau} \hat{\rho}_{\tt S} \left[ (n-1) \tau \right] \otimes \hat{\rho}_n e^{\textrm{i} \hat{H}_{\tt T} \tau} \right) \nonumber \\
&= \textrm{Tr}_n \left( \hat{\rho} - \textrm{i}\tau [\hat{H}_{\tt T}, \hat{\rho}] - \frac{\tau ^2}{2}[\hat{H}_{\tt T}, [\hat{H}_{\tt T}, \hat{\rho}]]+ \cdots \right),
\end{align}
where we have used the notation $\hat{\rho} \defeq \hat{\rho}_{\tt S} \otimes \hat{\rho}_n$. 

We now proceed to take the partial traces. The first term yields $\textrm{Tr}_n ( \hat{\rho}_{\tt S} \otimes \hat{\rho}_n ) = \hat{\rho}_{\tt S}$. The second term yields
\begin{align}
\textrm{Tr}_n \left( [\hat{H}_{\tt T}, \hat{\rho}_{\tt S} \otimes \hat{\rho}_n ] \right) = [\hat{H}_{\tt S}, \hat{\rho}_{\tt S}],
\end{align}
given that our interaction Hamiltonian does not involve $\hat{\sigma}^z$ terms. There is a subtle point to be considered now: if we take the limit $\tau \to 0$, we find that the interaction Hamiltonian vanishes as well. In order to consider a proper interaction, we need $\hat{V}$ properly scaled with the interaction time. We can achieve this by defining $g \defeq \sqrt{\lambda_{m} / \tau}$ such that
\begin{align}
\hat{V}_{\tt L} = \sqrt{\frac{\lambda_{\tt L}}{\tau}} \left(  \hat{\sigma}^x_{\tt L} \hat{\sigma}^x_1 + \hat{\sigma}^y_{\tt L} \hat{\sigma}^y_1 \right), \quad \hat{V}_{\tt R} = \sqrt{\frac{\lambda_{\tt R}}{\tau}} \left(  \hat{\sigma}^x_D \hat{\sigma}^x_{\tt R} + \hat{\sigma}^y_D \hat{\sigma}^y_{\tt R} \right).
\end{align}
Now, from Eq.~\ref{eq:expansion1}, after some algebra we find that
\begin{align}
\hat{\rho}_{\tt S} (n\tau) = \hat{\rho}_{\tt S} \left( \{n-1\}\tau \right) - &\textrm{i}\tau \left[ \hat{H}_{\tt S}, \hat{\rho}_{\tt S} \left(\{n-1\}\tau \right) \right] \nonumber \\
&+ \tau \left\{ \mathcal{L}_{\tt L}\{\hat{\rho}_{\tt S}([n-1]\tau) \} + \mathcal{L}_{\tt R}\{\hat{\rho}_{\tt S}([n-1]\tau) \}  \right\} + O(\tau^{>1}),
\end{align}
where
\begin{align}
\tau \mathcal{L}_{m} \{ \hat{\rho}_{\tt S} \} = -\frac{\tau ^2}{2}\textrm{Tr}_n \left[ \hat{V}_{m}, [\hat{V}_{m}, \hat{\rho}_{\tt S}] \right].
\end{align}
We can then write an equation for $\hat{\rho}_{\tt S} (n\tau) - \hat{\rho}_{\tt S} \left( \{n-1\}\tau \right)$, divide by $\tau$ and take the limit $\tau \to 0$ to obtain
\begin{align}
\frac{d\hat{\rho}}{dt} = -\textrm{i}[\hat{H},\hat{\rho}] + \mathcal{L}\{\hat{\rho}\} = -\textrm{i}[\hat{H},\hat{\rho}] + \mathcal{L}_{\tt L}\{\hat{\rho}\} + \mathcal{L}_{\tt R}\{\hat{\rho}\},
\end{align}
where
\begin{align}
\mathcal{L}_{m}\{\hat{\rho}\} = \sum_{s=\pm} 2\hat{L}_{s,m}\,\hat{\rho}\, \hat{L}_{s,m}^{\dag} - \{\hat{L}_{s,m}^{\dag}\hat{L}_{s,m},\hat{\rho}\},
\end{align}
and, by setting $\lambda_{m} = \gamma / 2$, we finally identify
\begin{align}
\hat{L}_{+,\tt L} = \sqrt{\gamma(1 + \mu)}\,\hat{\sigma}_1^{+},\quad \hat{L}_{-,\tt L} = \sqrt{\gamma(1 - \mu)}\,\hat{\sigma}_1^{-}, \nonumber \\
\hat{L}_{+,\tt R} = \sqrt{\gamma(1 - \mu)}\,\hat{\sigma}_D^{+},\quad \hat{L}_{-,\tt R} = \sqrt{\gamma(1 + \mu)}\,\hat{\sigma}_D^{-}.
\end{align}

\subsection{Spin current and steady state}
\label{sec:current_steady_state}

The configuration described previously drives the system towards a non-equilibrium steady state, denoted by $\hat{\rho}_{\textrm{NESS}}$, given by
\begin{align}
\label{eq:supop}
\mathcal{W}\{\hat{\rho}_{\textrm{NESS}}\} = -\textrm{i}[\hat{H},\hat{\rho}_{\textrm{NESS}}] +\mathcal{L}_{\tt L}\{\hat{\rho}_{\textrm{NESS}}\} + \mathcal{L}_{\tt R}\{\hat{\rho}_{\textrm{NESS}}\} = 0,
\end{align}
which implies that the steady state is the one that spans the null space of the super-operator $\mathcal{W}$. It can be proven that this state exists and is unique if and only if the set of operators $\{\hat{H}, \hat{L}_{+,\tt L}, \hat{L}_{+, \tt R}, \hat{L}_{-, \tt L}, \hat{L}_{-, \tt R}\}$ generate, under multiplication and addition, the entire Pauli algebra. This condition is fulfilled in our case~\cite{ProsenUnique}. Another property of the NESS is related to the time evolution of the system. Given the mathematical existence and uniqueness of this particular state, any initial state will converge to the NESS in the long time limit
\begin{align}
\label{eq:ness}
\lim_{t\rightarrow\infty}\hat{\rho}(t) = \hat{\rho}_{\textrm{NESS}}.
\end{align}
Since, by construction, we introduce an imbalance in the strength of the boundary driving $\mu$, the NESS is characterised by a constant flow of magnetisation in the $z$ direction from one boundary to the other. The boundary driving parameter establishes the degree of imbalance between the Markovian baths and thus affects transport in the bulk of the spin chain. We focus on the regime $0 \leq \mu \leq 1$. For $\mu = 0$ there is no imbalance and the state in the bulk is given by an infinite temperature steady state, $\hat{\rho} = \mathds{1}/2^{D}$. For any nonzero $\mu$, effective spin excitations are introduced and removed from the system. For $\mu = 1$ the system is at maximum driving, i.e., maximum bias.    

We can determine the flux of magnetisation by means of the equation dictating the dynamics of the expectation value of $\hat{\sigma}^z_i$. We then turn to Eq.~\eqref{eq:lme} to obtain, in the bulk of the chain
\begin{align}
\label{eq:magevo}
\frac{\textrm{d}\langle \hat{\sigma}^z_i \rangle}{\textrm{d}t} &= \frac{\textrm{d}}{\textrm{d}t}\textrm{Tr}\left(\hat{\rho}\hat{\sigma}^z_i\right) 
= \textrm{Tr}\left(\hat{\sigma}^z_i\frac{\textrm{d}\hat{\rho}}{\textrm{d}t}\right) 
= -\textrm{i}\,\textrm{Tr}\left(\hat{\sigma}^z_i[\hat{H},\hat{\rho}]\right) \nonumber \\
&= \textrm{i}\,\textrm{Tr}\left([\hat{H},\hat{\sigma}^z_i]\hat{\rho}\right);\quad \forall i = 2, \cdots, D-1\,.
\end{align}
Using Pauli matrix commutation relations, one obtains for Eq.~\eqref{eq:magevo}:
\begin{align}
\label{eq:magevocurrent}
\frac{\textrm{d}\langle \hat{\sigma}^z_i \rangle}{\textrm{d}t} = \langle \hat{j}_{i-1} \rangle - \langle \hat{j}_i \rangle;\quad \forall i = 2, \cdots, D-1,
\end{align}
where
\begin{align}
\label{eq:spincurrent_22}
\hat{j}_{i} \defeq 2\alpha\left(\hat{\sigma}^x_i\hat{\sigma}^y_{i+1} - \hat{\sigma}^y_i\hat{\sigma}^x_{i+1}\right).
\end{align}

We call this object the {\em spin current} operator. Up to this point, Eq.~\eqref{eq:magevocurrent} is ill-defined for the leftmost and the rightmost sites of the chain. However, we can obtain the dynamics of the magnetisation in these sites by interpreting $\mu$ as the average magnetisation of the Markovian baths, where we therefore identify 
\begin{align}
\label{eq:boundcurrent}
\frac{\textrm{d}\langle \hat{\sigma}^z_1 \rangle}{\textrm{d}t} &= \langle \hat{j}_{\tt L} \rangle - \langle \hat{j}_1 \rangle, \\
\frac{\textrm{d}\langle \hat{\sigma}^z_D \rangle}{\textrm{d}t} &= \langle \hat{j}_{D-1} \rangle - \langle \hat{j}_{\tt R} \rangle,
\end{align}
with the corresponding values of the current on the boundaries given by 
\begin{align}
\label{eq:boundarycurrent}
\langle \hat{j}_{\tt L} \rangle &= \textrm{Tr}\left(\hat{\sigma}^z_1\mathcal{L}_{\tt L}\{\hat{\rho}\}\right) = 4\gamma\,(\mu - \langle \hat{\sigma}^z_1 \rangle), \\
\langle \hat{j}_{\tt R} \rangle &=  \textrm{Tr}\left(\hat{\sigma}^z_D\mathcal{L}_{\tt R}\{\hat{\rho}\}\right) = 4\gamma\,(\mu + \langle \hat{\sigma}^z_D \rangle).
\end{align}

With these definitions, the continuity equation of the magnetisation in the $z$ direction is consistent. In the NESS, the relation $\textrm{d}\langle \hat{\sigma}^z_i \rangle / \textrm{d}t= 0$ holds for all sites, which means that the spin current is homogeneous across the chain (in one dimension): 
\begin{align}
\label{eq:homogeneous}
\langle \hat{j}_{\tt L} \rangle = \langle \hat{j}_1 \rangle = \cdots = \langle \hat{j}_D \rangle = \langle \hat{j}_{\tt R} \rangle \defeq \langle \hat{j} \rangle.  
\end{align}

\section{Solution to the non-equilibrium steady state}
\label{sec:mpos_ness}

The mathematical properties of the NESS can be obtained from properties of the Liouville super-operator. In order to visualise them, it is convenient to use a vectorisation procedure on the density matrix~\cite{LandiFluxRectificationXXZ, EduLinearW}. The procedure consists in concatenating the columns of the density matrix onto a vector. This allows the factorisation of the Liouville super-operator in matrix form that acts on a vector form of the density matrix. Using a matrix representation of the super-operator, we can write the Lindblad master equation as
\begin{align}
\label{eq:linmaster}
\frac{d|\hat{\rho}\rangle\rangle}{dt} = \hat{W}|\hat{\rho}\rangle\rangle,
\end{align}
where $|\cdot\rangle\rangle$ is a vectorised matrix built by concatenating its columns, and $\hat{W}$ is the matrix representation of the super-operator in Eq.~\eqref{eq:supop}. The master equation (Eq.~\eqref{eq:lme}) can be expressed in such a way because the vectorisation procedure is a linear operation, and all the terms in Eq.~\eqref{eq:lind1} are of the form $\hat{A}\hat{B}\hat{C}$, where $\hat{A}$, $\hat{B}$, and $\hat{C}$ are matrices. In light of this, the following relation can be used to obtain Eq.~\eqref{eq:linmaster} \cite{LandiFluxRectificationXXZ}:
\begin{align}
\label{eq:vectorized}
|\hat{A}\hat{B}\hat{C}\rangle\rangle = (\hat{C}^{T} \otimes \hat{A})|\hat{B}\rangle\rangle.
\end{align}
From Eq.~\eqref{eq:lind1}, this relation is the only one needed to reduce the Lindblad master equation to Eq.~\eqref{eq:linmaster} in terms of the density matrix and the Pauli spin matrices. 
In this chapter, we shall employ two different methods to solve for the NESS. In the first one, we solve a system of linear equations using a matrix representation of the super-operator $\mathcal{W}$ from Eq.~\eqref{eq:supop}, limited only by the accessible system sizes; while the second one is based on time-dependent Matrix Product States (tMPS) \cite{Schollwock2011, Verstraete2008} in combination with a fourth-order Suzuki-Trotter decomposition of the Liouville propagator. Let us provide a brief description about both of these approaches.

\subsection{Exact numerical approach to the solution of the non-equilibrium steady state}
\label{sec:directmethod}

Using the vectorised form of the density matrix described in Sec.~\ref{sec:mpos_ness}, one can write a matrix representation of the Liouville super-operator, and combine operations of the form in Eq.~\eqref{eq:vectorized} in order to factorise this operator from the density matrix. In this picture, Eq.~\eqref{eq:supop} transforms to 
\begin{align}
\label{eq:vecformsupop}
\hat{W}|\hat{\rho}_{\textrm{NESS}}\rangle\rangle = 0,
\end{align}
where $\hat{W}$ is a non-Hermitian matrix of dimension $d_{\mathcal{H}}^2$ and $|\hat{\rho}_{\textrm{NESS}}\rangle\rangle$ is the vector form of the density matrix representing the NESS, with the same dimension. At this point it is clear that, given that the Hilbert space dimension is effectively increased by a power of 2, the computational cost of studying interacting open quantum systems is immensely higher than in closed quantum systems.

The solution of Eq.~\eqref{eq:vecformsupop} is found by directly solving the system of linear equations constrained to the trace preserving property of the density matrix
\begin{align}
\label{eq:trace1}
\langle\langle\mathds{1}|\hat{\rho}\rangle\rangle = \textrm{Tr}(\hat{\rho}) = 1,
\end{align}
where $|\mathds{1}\rangle\rangle$ is the vectorised identity. 

One can then define~\cite{EduLinearW}
\begin{align}
\label{eq:rescaleW}
\widetilde{W} = \hat{W} + |0\rangle\rangle\langle\langle\mathds{1}|,
\end{align}
such that
\begin{align}
\label{eq:linearsolve}
\widetilde{W}|\hat{\rho}_{\textrm{NESS}}\rangle\rangle &= \hat{W}|\hat{\rho}_{\textrm{NESS}}\rangle\rangle + |0\rangle\rangle\langle\langle\mathds{1}|\hat{\rho}_{\textrm{NESS}}\rangle\rangle, \nonumber \\
\widetilde{W}|\hat{\rho}_{\textrm{NESS}}\rangle\rangle &= |0\rangle\rangle, \nonumber \\
\implies |\hat{\rho}_{\textrm{NESS}}\rangle\rangle &= \widetilde{W}^{-1}|0\rangle\rangle,
\end{align}
where $|0\rangle\rangle$ is the vectorised form of the first state in the Hilbert space. The choice of the matrix $|0\rangle\rangle\langle\langle\mathds{1}|$ is in principle arbitrary, with the only condition that the trace of the density matrix is preserved. In the present case, $|0\rangle\rangle\langle\langle\mathds{1}|$ is a matrix of zeroes, with ones only in the first row in the columns corresponding to the diagonal elements of $\hat{\rho}$.

It is impractical to evaluate $\widetilde{W}^{-1}$ given that, even if $\widetilde{W}$ is sparse, $\widetilde{W}^{-1}$ will not be sparse in general. Therefore, the solution to the linear system is normally tackled by means of direct or indirect methods. In general, direct methods are more expensive in both computational and memory terms. However, indirect methods such as Krylov subspace techniques normally require preconditioning or other additional techniques to attain acceptable numerical convergence with a low number of operations. 

The main drawback of the exact numerical approach is intractability, in light of the $d_{\mathcal{H}}^2$ scaling of the Hilbert space. In our work, we used this method only for small system sizes $D \sim 10$. These system sizes are generally too small to identify the transport regime in boundary driven spin chains. We resort to the tMPS technique, briefly described in Sec.~\ref{sec:mpos}, and use the exact approach to evaluate the numerical fidelity of the results obtained with tMPS.

\subsection{Matrix product states-operators approach to the solution of the non-equilibrium steady state}  
\label{sec:mpos}

In order to appreciate and properly quantify the transport properties of boundary driven systems, it is usually required to analyse large system sizes to overcome finite-size effects. To this end, we can apply the time-dependent Matrix Product States algorithm to study the evolution of any initial state under Eq.~\ref{eq:lme}, which allows the study of large system sizes.

We start by writing the density matrix of the system in the form
\begin{align}
\label{eq:rhomps}
|\rho \rangle = \sum_{\sigma_1,\cdots,\sigma_D} c_{\sigma_1\cdots\sigma_D}| \sigma_1, \cdots, \sigma_D \rangle,
\end{align}
where there are $d^D$ coefficients $c_{\sigma_1\cdots\sigma_D}$ that describe the state of the system and $\sigma_{i}$ is the local basis at site $i$ for a system with $D$ sites. The Pauli basis is a natural and commonly used choice to represent the local basis, such that at site $i$ the local basis is given by
\begin{align}
\label{eq:basis}
\{\sigma_i\} = \left\{ \frac{1}{2}\mathds{1}, \frac{1}{2}\sigma^x, \frac{1}{2}\sigma^y, \frac{1}{2}\sigma^z \right\}.
\end{align}
In the following, we use the vectorised form of this local basis, i.e., $\textrm{vec}(\sigma^{\nu})$ such that the density matrix operator can be represented as an MPS in the extended Hilbert space. 
The power of the MPS representation of the density matrix resides on the fact that it provides a mathematical sense of locality to the state, while preserving the inherent quantum non-locality features. In order to achieve this, we write Eq.~\ref{eq:rhomps} in MPS form by first reshaping the $d^D$ dimensional vector $c_{\sigma_1\cdots\sigma_D}$ into a matrix $\Psi$ of dimension $d \times d^{D-1}$
\begin{align}
\label{eq:reshapec1}
c_{\sigma_1\cdots\sigma_D} \rightarrow \Psi_{\sigma_1, (\sigma_2\cdots\sigma_D)},
\end{align}
and apply the singular value decomposition (SVD) to the resulting matrix:
\begin{align}
\label{eq:firstsvd}
\Psi_{\sigma_1, (\sigma_2\cdots\sigma_D)} = \sum_{a_1}^{r_1}U_{\sigma_1,a_1}S_{a_1,a_1}(V^{\dagger})_{a_1,(\sigma_2\cdots\sigma_D)} \defeq \sum_{a_1}^{r_1}U_{\sigma_1,a_1}c_{a_1\sigma_2\cdots\sigma_D},
\end{align}
where $U$, $S$ and $V$ are the matrices resulting from the SVD. In the last definition $S$ and $V^{\dagger}$ have been multiplied together and the result shaped back into a vector. At this stage, there can only be $r_1 \leq d$ finite Schmidt coefficients from the SVD procedure. 

We now proceed to represent $U_{\sigma_1,a_1} \rightarrow A_{a_1}^{\sigma_1}$ as a collection of $d$ row vectors and $c_{a_1\sigma_2\cdots\sigma_D}$ as a matrix of dimension $r_1d \times d^{D-2}$ to obtain
\begin{align}
\label{eq:reshapec2}
c_{\sigma_1\cdots\sigma_D} \rightarrow \sum_{a_1}^{r_1}A_{a_1}^{\sigma_1}\Psi_{(a_1\sigma_2),(\sigma_3\cdots\sigma_D)}.
\end{align}
If we proceed exactly as before we can write:
\begin{align}
\label{eq:secondsvd}
c_{\sigma_1\cdots\sigma_D} \rightarrow &\sum_{a_1}^{r_1}\sum_{a_2}^{r_2}A_{a_1}^{\sigma_1}U_{(a_1\sigma_2),a_2}S_{a_2,a_2}(V^{\dagger})_{a_2,(\sigma_3\cdots\sigma_D)} \nonumber \\
=&\sum_{a_1}^{r_1}\sum_{a_2}^{r_2}A_{a_1}^{\sigma_1}A_{a_1,a_2}^{\sigma_2}\Psi_{(a_2\sigma_3),(\sigma_4\cdots...\sigma_D)},
\end{align}
where $U$ has been replaced by a collection of $d$ matrices $A^{\sigma_2}$ of dimension $r_1 \times r_2$ with entries $A_{a_1,a_2}^{\sigma_2} = U_{(a_1\sigma_2),a_2}$. Just like before, $S$ and $V^{\dagger}$ have been multiplied together and reshaped into a matrix $\Psi$ of dimension $r_2d \times d^{D-3}$, with $r_2 \leq r_1d \leq d^2$. 

Proceeding iteratively, we obtain
\begingroup
\allowdisplaybreaks
\begin{align}
\label{eq:mpsform}
c_{\sigma_1\cdots\sigma_D} &=  \sum_{a_1,\cdots,a_{D-1}}A_{a_1}^{\sigma_1}A_{a_1,a_2}^{\sigma_2} \cdots A_{a_{D-2},a_{D-1}}^{\sigma_{D-1}}A_{D-1}^{\sigma_D} \nonumber \\
&= A^{\sigma_1}A^{\sigma_2} \cdots A^{\sigma_{D-1}}A^{\sigma_D},
\end{align}
\endgroup
where the last $A^{\sigma_D}$ corresponds to a collection of $d$ column vectors and the sums over $a_{i}$ have been represented as matrix multiplications. Finally,
\begin{align}
\label{eq:rhofinalmps}
| \rho \rangle = \sum_{\sigma_1,\cdots,\sigma_D} A^{\sigma_1}A^{\sigma_2} \cdots A^{\sigma_{D-1}}A^{\sigma_D} | \sigma_1,\cdots,\sigma_D \rangle 
\end{align}
is the MPS form of the density matrix in the vectorised Pauli basis described above. The SVD procedure possesses a normalisation condition on the $U$ matrices, given by $U^{\dagger}U = I$. This condition translates to the $A^{\sigma_n}$ for the $n$th site such that
\begin{align}
\label{leftnormalised}
\sum_{\sigma_n} \left( A^{\sigma_n} \right)^{\dagger} A^{\sigma_n} = I.
\end{align}
The matrices pertaining to an MPS that satisfy this condition are known as left-normalised, while the MPS itself is known as left-canonical. In our description the MPS form of the density matrix was built from left to right so this condition is satisfied. The MPS can be built from right to left and from the edges to the centre as well, respectively these MPS forms are known as right-canonical and mixed-canonical. We, however, restrict our description to left-canonical MPS. Eq.~\ref{eq:rhofinalmps} is valid only for {\em open} boundary conditions.

\begin{figure}[t]
\centering
\includegraphics[width=1.0\textwidth]{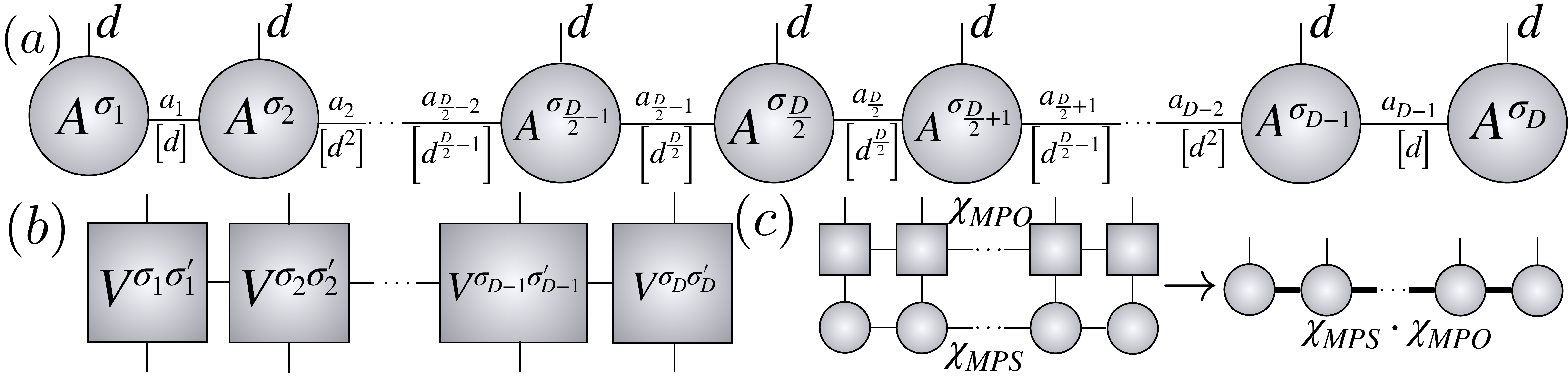}
\caption{Diagrammatic depiction of (a) an MPS, (b) an MPO and (c) a contraction of an MPS with an MPO that returns a new MPS.}
\label{fig:2.2.2}
\end{figure}

The representation used in MPS form can be depicted graphically in a neat way using nodes and lines connecting the $A^{\sigma_n}$ matrices. From this graphical depiction one can relate each collection of $A^{\sigma_n}$ matrices to the lattice site they belong to. It is a common practice to use vertical lines to represent the {\em physical} degrees of freedom and horizontal lines to represent the {\em auxiliary} degrees of freedom, connected lines mean degrees of freedom being summed over. We show how an MPS is represented using this graphical depiction in Fig.~\ref{fig:2.2.2}(a), below the horizontal lines we have written down the {\em biggest} possible dimension that the auxiliary degrees of freedom can take (in square braces). This value is commonly known as {\em bond dimension}.

From Fig.~\ref{fig:2.2.2}(a) we can see that the use of an MPS representation does not readily solve the intractability problem, one is still left to face exponentially increasing matrix sizes. In this form, however, the SVD procedure exposes the degree of both quantum and classical correlations present in the state and {\em truncation} can be made on the sizes of the matrices used for the MPS based on this observation. We denote the maximum value used for bond dimension with $\chi$, as the maximum amount of auxiliary degrees of freedom used for the $A^{\sigma_n}$ matrices. 

For the specific case at hand, one can keep the degree of correlations under control by using an initial product state, say for instance, the identity state; and evolving the system under dynamics that keep the state close to an identity state throughout the evolution as the NESS is reached. From Eq.~\ref{eq:lind2}, this can be achieved for {\em small} values of $\mu$. Increasing this parameter (e.g., $\mu \sim 1$) induces a state into the system much further away from the identity in terms of quantum correlations, i.e, states that require a large bond dimension to be represented with fidelity; particularly for large system sizes.  

Just like states, operators can be written down in MPS form in a representation known as Matrix Product Operators (MPOs). Given that any quantum operator can be expressed as
\begin{align}
\label{eq:mpo1}
O = \sum_{\sigma_1,\cdots,\sigma_D,\sigma^{\prime}_1,\cdots,\sigma^{\prime}_D} c_{(\sigma_1,\cdots,\sigma_D), (\sigma^{\prime}_1,\cdots,\sigma^{\prime}_D)} | \sigma_1,\cdots,\sigma_D \rangle \langle \sigma^{\prime}_1,\cdots,\sigma^{\prime}_D |,
\end{align}
we can decompose $O$ the same way we did for an MPS with the double index $\sigma_i \sigma^{\prime}_i$ taking the role of the single index $\sigma_i$ to give
\begin{align}
\label{eq:mpo2}
O = \sum_{\sigma_1,\cdots,\sigma_D,\sigma^{\prime}_1,\cdots,\sigma^{\prime}_D} V^{\sigma_1,\sigma^{\prime}_1}V^{\sigma_2,\sigma^{\prime}_2} \cdots V^{\sigma_{D-1},\sigma^{\prime}_{D-1}} V^{\sigma_D,\sigma^{\prime}_D} | \sigma_1,\cdots,\sigma_D \rangle \langle \sigma^{\prime}_1,\cdots,\sigma^{\prime}_D | ,
\end{align}
where we have omitted the sums auxiliary indices as they can be recognised as matrix multiplications.
 
Graphically, an MPO is represented just like an MPS with an extra physical set of degrees of freedom vertically coming out of each node, as depicted in Fig.~\ref{fig:2.2.2}(b). At this point we note that technically a density matrix should be represented as an MPO instead of an MPS, however, the vectorisation procedure allows the density matrix to be represented as an MPS and, as we shall see, the Liouvillian propagator to be represented as an MPO.

One of the most relevant operations involving MPSs and MPOs for us is the contraction of an MPS with an MPO that returns a new MPS. Analytically, we proceed with this operation as follows:
\begingroup
\allowdisplaybreaks
\begin{align}
\label{eq:mpomps}
O |\rho \rangle &= \sum_{\boldsymbol{\sigma},\boldsymbol{\sigma^{\prime}}} \left(V^{\sigma_1,\sigma^{\prime}_1}V^{\sigma_2,\sigma^{\prime}_2} \cdots \right) \left( A^{\sigma_1^{\prime}}A^{\sigma_2^{\prime}} \cdots \right)  | \boldsymbol{\sigma} \rangle \nonumber \\
&= \sum_{\boldsymbol{\sigma},\boldsymbol{\sigma^{\prime}}} \sum_{ \boldsymbol{a}, \boldsymbol{b}} \left(V^{\sigma_1,\sigma^{\prime}_1}_{1,b_1}V^{\sigma_2,\sigma^{\prime}_2}_{b_1,b_2} \cdots \right) \left( A^{\sigma_1^{\prime}}_{1,a_1}A^{\sigma_2^{\prime}}_{a_1,a_2} \cdots \right)  | \boldsymbol{\sigma} \rangle \nonumber \\
&= \sum_{\boldsymbol{\sigma},\boldsymbol{\sigma^{\prime}}} \sum_{ \boldsymbol{a}, \boldsymbol{b}} \left(V^{\sigma_1,\sigma^{\prime}_1}_{1,b_1} A^{\sigma_1^{\prime}}_{1,a_1} \right) \left( V^{\sigma_2,\sigma^{\prime}_2}_{b_1,b_2} A^{\sigma_2^{\prime}}_{a_1,a_2} \right) \cdots | \boldsymbol{\sigma} \rangle \nonumber \\
&= \sum_{\boldsymbol{\sigma}} \sum_{ \boldsymbol{a}, \boldsymbol{b}} N^{\sigma_1}_{(1,1),(b_1,a_1)} N^{\sigma_2}_{(b_1,a_1),(b_2,a_2)} \cdots | \boldsymbol{\sigma} \rangle = \sum_{\boldsymbol{\sigma}} N^{\sigma_1} N^{\sigma_2} \cdots | \boldsymbol{\sigma} \rangle,
\end{align}
\endgroup
where $\boldsymbol{\sigma} \defeq | \sigma_1,\cdots,\sigma_D \rangle$, $\boldsymbol{a} \defeq a_1,\cdots,a_{D-1}$ and $\boldsymbol{b} \defeq b_1,\cdots,b_{D-1}$. From this procedure we note two important observations. First, the contraction of an MPS with an MPO has to be done in a certain order to avoid an exponentially-complex operation. Particularly, the tensors corresponding to a certain site are contracted together {\em vertically} and not {\em horizontally} by grouping them in relation to the site they belong to, as shown above. Second, the bond dimension of the MPS resulting from the MPS-MPO contraction is bigger than the one from its predecessors, namely, an MPS contracted with an MPO containing bond dimensions $\chi_{\textrm{MPS}}$ and $\chi_{\textrm{MPO}}$ respectively, corresponds to an MPS with bond dimension $\chi_{\textrm{MPS}} \cdot \chi_{\textrm{MPO}}$. This operation is depicted in Fig.~\ref{fig:2.2.2}(c). The contraction, when operated in this fashion, has an overall complexity of $\mathcal{O}(Dd^2\chi_{\textrm{MPS}}^2\chi_{\textrm{MPO}}^2)$~\cite{Schollwock2011}.

\subsubsection{Real time evolution}

To obtain the NESS, we target the solution to the master equation numerically given by
\begin{align}
\label{eq:solutionmps}
| \rho(\tau) \rangle = e^{W\tau} | \rho(0) \rangle,
\end{align}
in the limit $\tau \rightarrow +\infty$, with $| \rho(\tau) \rangle$ being the density matrix of the state at time $t = \tau$, $| \rho(0) \rangle$ describing the density matrix of the initial state and $W$ a linearised form of the super-operator $\mathcal{W}$ written down in Eq.~\ref{eq:supop}. As described before, in this form, $W$ corresponds to a square non-Hermitian matrix while the density operators correspond to vectors in an extended Hilbert space. The MPS formalism allows us to describe in a tractable way the states of the system, so now we devote the following to the MPO description of the Liouville propagator $e^{W\tau}$.

\begin{figure}[t]
\centering
\includegraphics[width=1.0\textwidth]{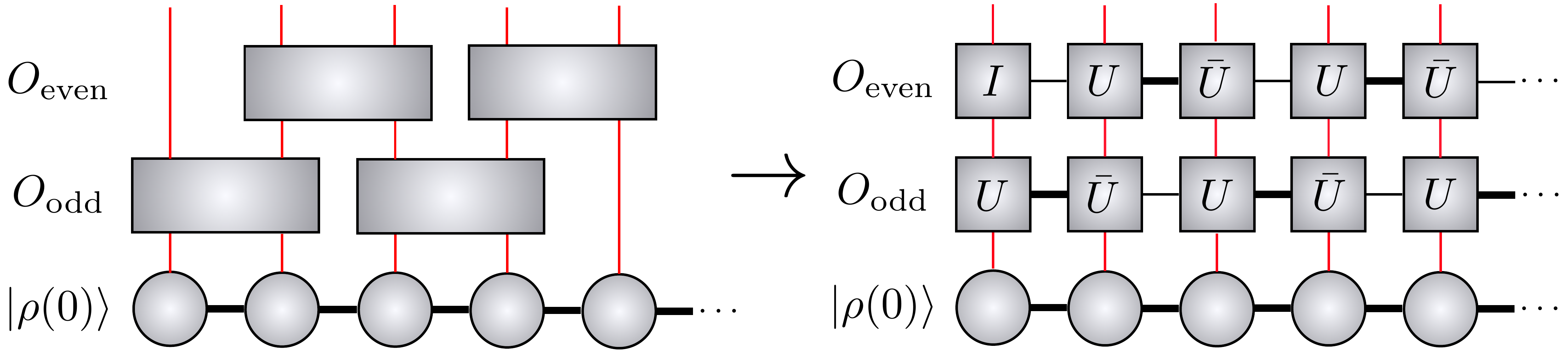}
\caption{First order even-odd Trotter decomposition}
\label{fig:2.2.3}
\end{figure} 

We first start by noting that the Liouville superoperator can be written as a sum of terms involving only two sites
\begin{align}
\label{twositeterms}
W = \sum_{i = 1}^{D-1}W_{i, i+1},
\end{align}  
given that the Hamiltonian involves only two-site terms and the Lindblad operators act locally\footnote{This is the case when one considers local Hamiltonians, of the form $\hat{H} = \sum_i \hat{h}_{i,i+1}$, such as $\hat{H}_{\textrm{XXZ}}$ from Eq.~\eqref{eq:h_xxz_22}}. This structure allows for the so-called Trotter decomposition of the Liouville propagator. Let us first describe the first-order decomposition, given by 
\begin{align}
\label{eq:firstordertrotter1}
e^{W\tau} = \prod_{i=1}^{D-1} e^{W_{i,i+1}\tau} + \mathcal{O}(\tau ^2).
\end{align}
The error introduced in this decomposition is due to the fact that nearest-site local Hamiltonians do not commute. However, next nearest-site local Hamiltonians do commute and this allows for an even-odd decomposition of the Liouville propagator that can be carried out at the same time. In such a way, we can define 
\begin{align}
\label{eq:firstordertrotter2}
O_{\textrm{odd}} &\defeq e^{W_{1,2}\tau} \otimes \mathds{1} \otimes e^{W_{3,4}\tau} \otimes \mathds{1} \otimes \cdots \\
O_{\textrm{even}} &\defeq \mathds{1} \otimes e^{W_{2,3}\tau} \otimes \mathds{1} \otimes e^{W_{4,5}\tau} \otimes \cdots,
\end{align}
such that $O_{\textrm{odd}}$ and $O_{\textrm{even}}$ can be applied at the same time $\tau$. It can be noticed that each of the $e^{W_{i,i+1}\tau}$ acts on two sites so in this form the MPO structure is no longer present as shown in Fig.~\ref{fig:2.2.3}. To recover MPO form, we need decompose the operators in a way that preserves the locality attributed to the MPS; to do that, we can reshape the operators and apply the SVD procedure as follows: 
\begingroup
\allowdisplaybreaks
\begin{align}
\label{eq:firstordertrotter3}
e^{W_{i,i+i}\tau} &\rightarrow \sum_{\sigma_i \sigma_{i+1},\sigma_i^{\prime} \sigma_{i+1}^{\prime}} O^{\sigma_i \sigma_{i+1},\sigma_i^{\prime} \sigma_{i+1}^{\prime}} | \sigma_i \sigma_{i+1} \rangle \langle \sigma_i^{\prime} \sigma_{i+1}^{\prime} | \nonumber \\
&\rightarrow \sum_{\sigma_i \sigma_{i+1},\sigma_i^{\prime} \sigma_{i+1}^{\prime}} O_{(\sigma_i \sigma_{i}^{\prime}),(\sigma_{i+1} \sigma_{i+1}^{\prime})} | \sigma_i \sigma_{i+1} \rangle \langle \sigma_i^{\prime} \sigma_{i+1}^{\prime} | \nonumber \\
&= \sum_{\sigma_i \sigma_{i+1},\sigma_i^{\prime} \sigma_{i+1}^{\prime}} \sum_k U_{(\sigma_i \sigma_{i}^{\prime}),k}S_{k,k}\left(V^{\dagger}\right)_{k,(\sigma_{i+1} \sigma_{i+1}^{\prime})} | \sigma_i \sigma_{i+1} \rangle \langle \sigma_i^{\prime} \sigma_{i+1}^{\prime} | \nonumber \\
&= \sum_{\sigma_i \sigma_{i+1},\sigma_i^{\prime} \sigma_{i+1}^{\prime}} \sum_k U^{\sigma_i \sigma_{i}^{\prime}}_k \bar{U}^{\sigma_{i+1} \sigma_{i+1}^{\prime}}_k | \sigma_i \sigma_{i+1} \rangle \langle \sigma_i^{\prime} \sigma_{i+1}^{\prime} | \nonumber \\
&= \sum_{\sigma_i \sigma_{i+1},\sigma_i^{\prime} \sigma_{i+1}^{\prime}} \sum_k U^{\sigma_i \sigma_{i}^{\prime}}_{1,k} \bar{U}^{\sigma_{i+1} \sigma_{i+1}^{\prime}}_{k,1} | \sigma_i \sigma_{i+1} \rangle \langle \sigma_i^{\prime} \sigma_{i+1}^{\prime} |,
\end{align}
\endgroup
with $U^{\sigma_i \sigma_{i}^{\prime}}_k = U_{(\sigma_i \sigma_{i}^{\prime}),k}\sqrt{S_{k,k}}$ and $\bar{U}^{\sigma_{i+1} \sigma_{i+1}^{\prime}}_k = \sqrt{S_{k,k}}\left(V^{\dagger}\right)_{k,(\sigma_{i+1} \sigma_{i+1}^{\prime})}$. The last step is just an extension of the tensor rank by a dummy index to obtain the MPO form. In Fig.~\ref{fig:2.2.3} we show in the pictorial depiction the previously described decomposition. With this procedure, the Liouville propagator is brought into MPO form and can be operated with an MPS to yield a time-evolved state.

To attain higher accuracy, instead of implementing the first-order decomposition described above we use a higher order approximation, namely, the fourth-order Trotter-Suzuki decomposition given by
\begin{align}
\label{eq:fourthordertrotter1}
e^{W\tau} = \mathcal{U}(\tau_1)\mathcal{U}(\tau_2)\mathcal{U}(\tau_3)\mathcal{U}(\tau_2)\mathcal{U}(\tau_1) + \mathcal{O}(\tau^5),
\end{align}
with
\begin{align}
\label{eq:fourthordertrotter2}
\mathcal{U}(\tau_i) = e^{W_{\textrm{odd}}\tau_i / 2}e^{W_{\textrm{even}}\tau_i }e^{W_{\textrm{odd}}\tau_i / 2}
\end{align}
and
\begin{align}
\label{eq:fourthordertrotter3}
\tau_1 = \tau_2 = \frac{\tau}{4-4^{1/3}};\; \tau_3 = \tau - 2\tau_1 - 2\tau_2.
\end{align}

The even-odd propagators from Eq.~\ref{eq:fourthordertrotter2} are decomposed as described above in terms of $U$ and $\bar{U}$, we then construct two different MPO representations: one given by the contraction of the MPOs for each term on the right of Eq.~\ref{eq:fourthordertrotter2} for $\tau_1 = \tau_2$ and another for $\tau_3$. Once these two MPOs are operated in the sequence shown in Eq.~\ref{eq:fourthordertrotter1} on an initial state $| \rho(0) \rangle$, the MPS for $| \rho(\tau) \rangle$ is obtained. This procedure is done iteratively until the NESS is reached in light of Eq.~\ref{eq:ness}, evaluating expectation values of observables after each time step. To contract the Liouville propagator in MPO form and the MPS at time $t$, we combine both methods presented in Ref.~\cite{Schollwock2011} to contract an MPS: SVD truncation and the variational approach. We find that convergence is achieved by providing the SVD-truncated state as an initial guess for the variational algorithm with only a few variational sweeps ($\approx$ 3-5); this approach provides better numerical results than using one of the two contraction methods on its own for a fixed value of $\chi$, albeit at a higher computational cost. We refer the reader to Ref.~\cite{Schollwock2011,Verstraete2008} for details on both contraction techniques.

The described method presents two main sources of error. One of them is a truncation error due to the maximum value of bond dimension $\chi$ used, in the specific case of simulations to reach non-equilibrium steady states, this error strongly depends on the system size $D$, the strength of the driving $\mu$ and the interaction parameter $\Delta$ at fixed $\alpha$ from Eqs.~\ref{eq:h_xxz_22}, \ref{eq:h_si_22} and \ref{eq:h_sf_22}. This error can be analysed by studying the expectation value of the current operator (Eq.~\ref{eq:spincurrent_22}) for the largest system size desired at fixed $\mu$ for different values of $\chi$. A specific value of $\chi$ that introduces a small tolerable error in simulations is then selected. 

The second main source of error is related to the Trotter-Suzuki decomposition from Eq.~\ref{eq:fourthordertrotter1}, which introduces an error of order $\mathcal{O}(M\tau^5)$ for the M-th time step. This error has also been found to linearly depend on the system size $D$~\cite{GobertDMRG2005}, though in the particular case of NESS simulations, this error is not as important as the truncation error, given that the state does not change after the NESS is reached. For practical applications, in light of Eq.~\ref{eq:homogeneous}, enough time steps can be applied such that the standard deviation of the expectation value of the current operator averaged over all sites becomes very small ($\approx 0.5\%$ in the calculations presented in Sec.~\ref{sec:transport_ness}).

\section{Transport from non-equilibrium steady states in non-integrable models}
\label{sec:transport_ness}

The procedure described in Sec.~\ref{sec:boundary_driving} shall now be used to describe spin transport in the single impurity model $\hat{H}_{\textrm{SI}}$ [Eq.~\eqref{eq:h_si_22}] and the staggered field model $\hat{H}_{\textrm{SF}}$ [Eq.~\eqref{eq:h_sf_22}]. As we have described in Part~\ref{part:one}, the single impurity model displays anomalous thermalisation and ballistic transport, despite its quantum-chaotic properties. We shall now confirm those results from the perspective of boundary-driven configurations, which allow much larger system sizes to be reached. We shall also revisit spin transport in the staggered field model as a testbed for our calculations.

\subsection{Transport and scaling theory}
\label{sec:scaling_ness}

The behaviour of $\langle \hat{j} \rangle$ [Eq.~\eqref{eq:spincurrent_22}] changes depending on the transport regime of the system, and can be analysed using scaling theory. From basic microscopic transport theory, the variance of a local inhomogeneity $\langle \Delta x^2 \rangle$ grows in space as a function of time $t$ as
\begin{align}
\label{eq:variance}
\langle \Delta x^2 \rangle = 2\mathcal{D}\,t^{2\delta}\,,
\end{align} 
where $\delta$ ($0 < \delta \leq 1$) is the transport coefficient, and $\mathcal{D}$ as the diffusion coefficient. The value of $\delta$ is set by how perturbations propagate across the system. This parameter can also be extracted by studying the scaling of the expectation value of the current in the NESS (from here on, unless otherwise specified, all expectation values are taken in the NESS) as a function of chain size as 
\begin{align}
\label{eq:scaling}
\langle \hat{j} \rangle \propto \frac{1}{D^{\nu}}
\end{align}
where $\nu \geq 0$ is the transport exponent. The parameters $\delta$ and $\nu$ are related by $\delta = 1 / (1 + \nu)$~\cite{LiScaling2003}. 

Different transport regimes are identified based on the value of $\nu$ as follows: $\nu = 0$ implies no dependence on system size and occurs when excitations in the system propagate without scattering, i.e., the system behaves as a perfect conductor and transport is {\em ballistic} (also known as {\em coherent}). This regime is expected for integrable systems~\cite{Ilievski2013}. A known exception is the XXZ model for $\Delta \geq 1$ for $\alpha = 1$, which is integrable yet exhibits non-ballistic spin transport \cite{Znidaric:2011}. $\nu = 1$ implies a regular {\em diffusive} regime and spin transport in the system obeys Fick's law, so the current across the system is proportional to the gradient of the driving field. The cases $0 < \nu < 1$ and $\nu > 1$ are referred to as {\em anomalous diffusion}, specifically, super-diffusion and sub-diffusion, respectively. In these cases, the constant of proportionality (the diffusion coefficient $\mathcal{D}$) in Eq.~\eqref{eq:scaling} picks up a dependence on the system size given by $\mathcal{D} \propto D^{1 - \nu}$~\cite{Ilievski2013}.

In Sec.~\ref{sec:results_ness}, we use finite-size scaling of the expectation values of the current in the NESS to probe the effect of integrability breaking in Eqs.~\eqref{eq:h_si_22} and~\eqref{eq:h_sf_22}.

\subsection{High-temperature transport in non-integrable systems}
\label{sec:results_ness}

In this section, transport will be investigated by studying the expectation value of the current operator $\langle \hat{j} \rangle$ in the non-equilibrium steady state. Following our discussion above, transport regimes can be identified from the finite-size scaling of $\langle \hat{j} \rangle$.

\begin{figure}[t]
\fontsize{13}{10}\selectfont 
\centering
\includegraphics[width=0.6\columnwidth]{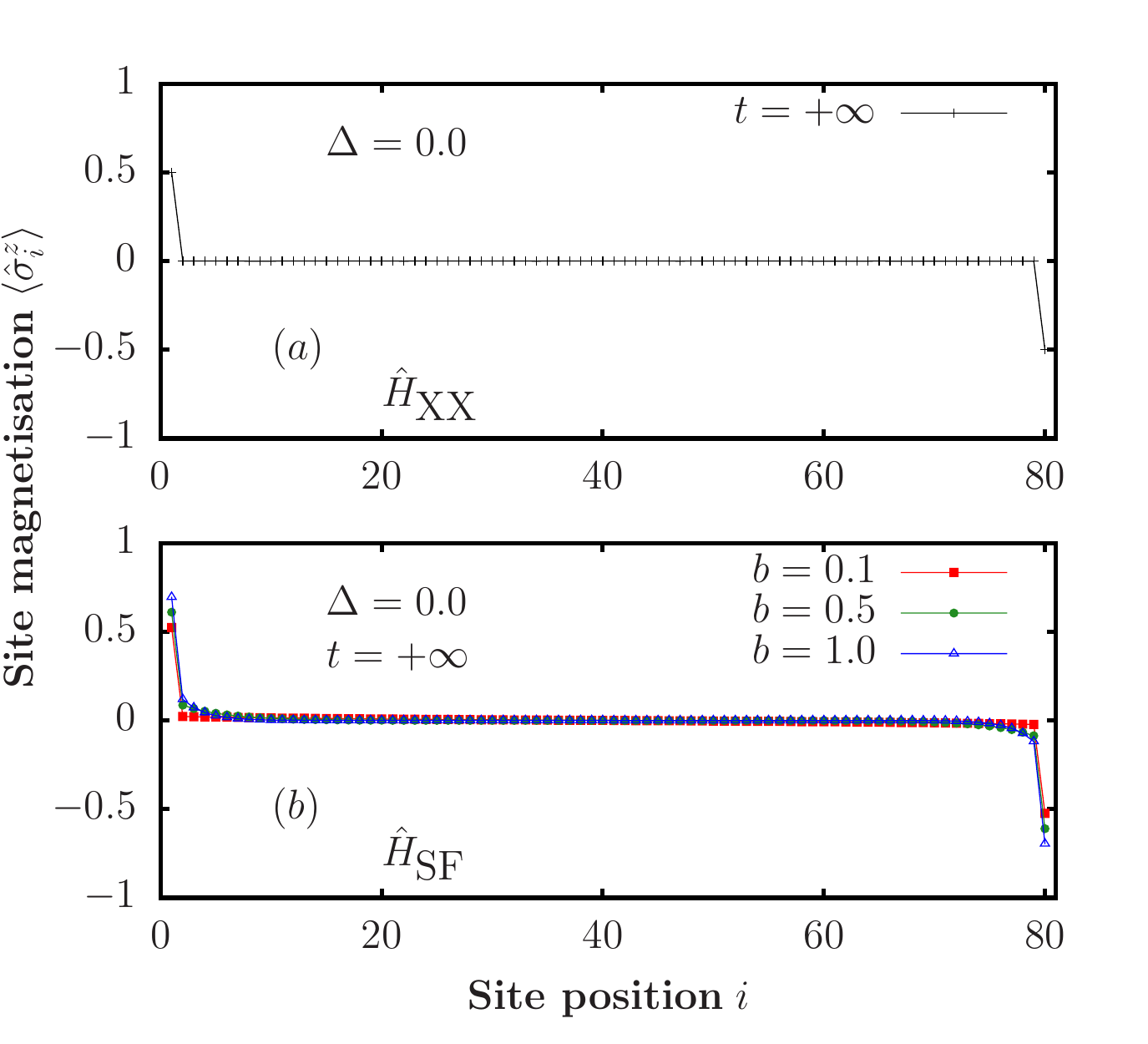}
\caption[Magnetisation profile of the non-equilibrium steady state for the non-interacting XXZ model and for the non-interacting model XXZ in the presence of a staggered magnetic field with different values of $b$]{(a) Magnetisation profile of the non-equilibrium steady state for the non-interacting XXZ model and (b) for the non-interacting model XXZ in the presence of a staggered magnetic field with different values of $b$. The driving parameters are $\gamma = 1.0$ and $\mu = 1.0$.}
\label{fig:2.2.4}
\end{figure}

\subsubsection{The non-interacting regime: $\Delta = 0$}

Let us begin with the non-interacting regime $\Delta = 0$ in the single impurity and staggered field models from Eqs~\eqref{eq:h_si_22} and \eqref{eq:h_sf_22}, respectively. In this limit, all these models are (trivially) integrable, and one can use the approach proposed in Ref.~\cite{DephasingMarko2013} to solve large system sizes at a low computational cost. Within this approach, a perturbative expansion is used to obtain the exact form of the non-equilibrium steady state by solving an equation of the Lyapunov type for any value of the boundary driving strength $\mu$ (we use $\mu = 1$). In the non-interacting limit, the dependence of the expectation value of the local magnetisation and the spin current on $\mu$ is always linear. This is in contrast with the interacting case, which shows a non-linear dependence for sufficiently strong boundary driving~\cite{NegDiffCondBenenti2009}. 

\begin{figure}[t]
\fontsize{13}{10}\selectfont 
\centering
\includegraphics[width=0.6\columnwidth]{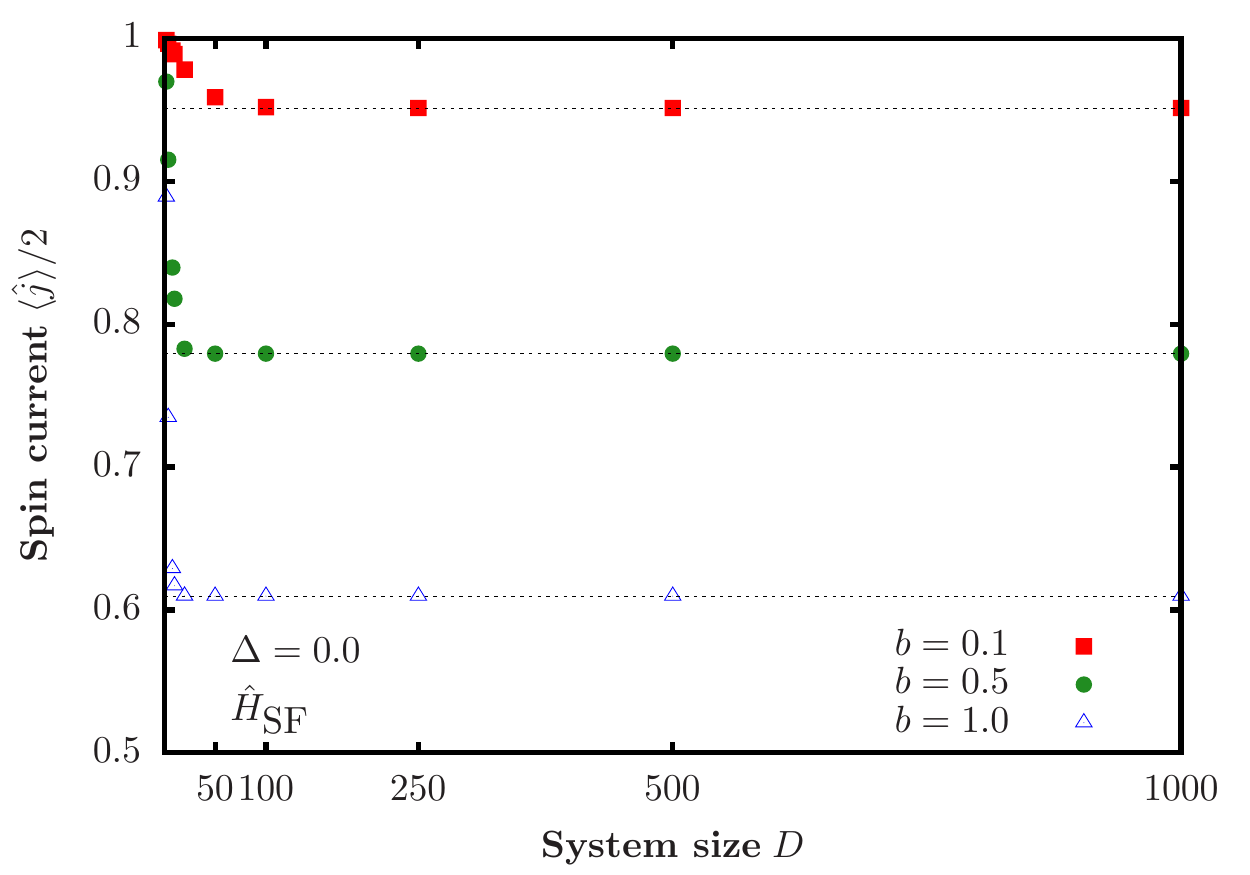}
\caption[Scaling of the expectation value of the current operator in the non-equilibrium steady state as a function of system size ($D=4,\cdots,1000$) for the non-interacting model in the presence of a staggered magnetic field]{Scaling of the expectation value of the current operator in the non-equilibrium steady state as a function of system size ($D=4,\cdots,1000$) for the non-interacting model in the presence of a staggered magnetic field. The driving parameters are $\gamma = 1.0$ and $\mu = 1.0$.}
\label{fig:2.2.5}
\end{figure}

In Fig.~\ref{fig:2.2.4}, we show the magnetisation profile of the NESS for the $\hat{H}_{\textrm{XXZ}}$ model with $\Delta = 0$, dubbed $\hat{H}_{\textrm{XX}}$, and for the staggered field model with different values of $b$ in the non-interacting regime. The magnetisation profile across the chain in the non-equilibrium steady state is a signal of the transport regime being investigated. This follows from the fact that hydrodynamical behaviour is characterised by Fick's law and a gradient has to be developed across the chain for the system to display incoherent transport~\cite{Prosen:2009}. Therefore, it is interesting to observe the expectation value of the local magnetisation in the $z$ direction $\hat{\sigma}^z_i$ for all the sites across the chain. As can be observed in Fig.~\ref{fig:2.2.4}, the magnetisation across the chain is flat and there exist no gradient of magnetisation, signalling ballistic transport. The magnetisation at the boundaries is different due to the imposed magnetisation by the driving scheme.

Indeed, if one considers the expectation value of the current operator in the non-equilibrium steady state, it can be observed that as the system size is increased the current reaches an asymptotic value in the thermodynamic limit. This is clear from Fig.~\ref{fig:2.2.5} for the non-interacting Heisenberg model and the non-interacting staggered field model. As a function of the strength of the staggered field, the absolute value of the current decreases when $b$ is increased. The regime of spin transport, however, is clearly ballistic. From Eq.~\eqref{eq:scaling}, we have that $\nu = 0$ for the non-interacting models, as detailed in Chapter~\ref{chapter:kubo}.

\begin{figure}[t]
\fontsize{13}{10}\selectfont 
\centering
\includegraphics[width=0.5\columnwidth]{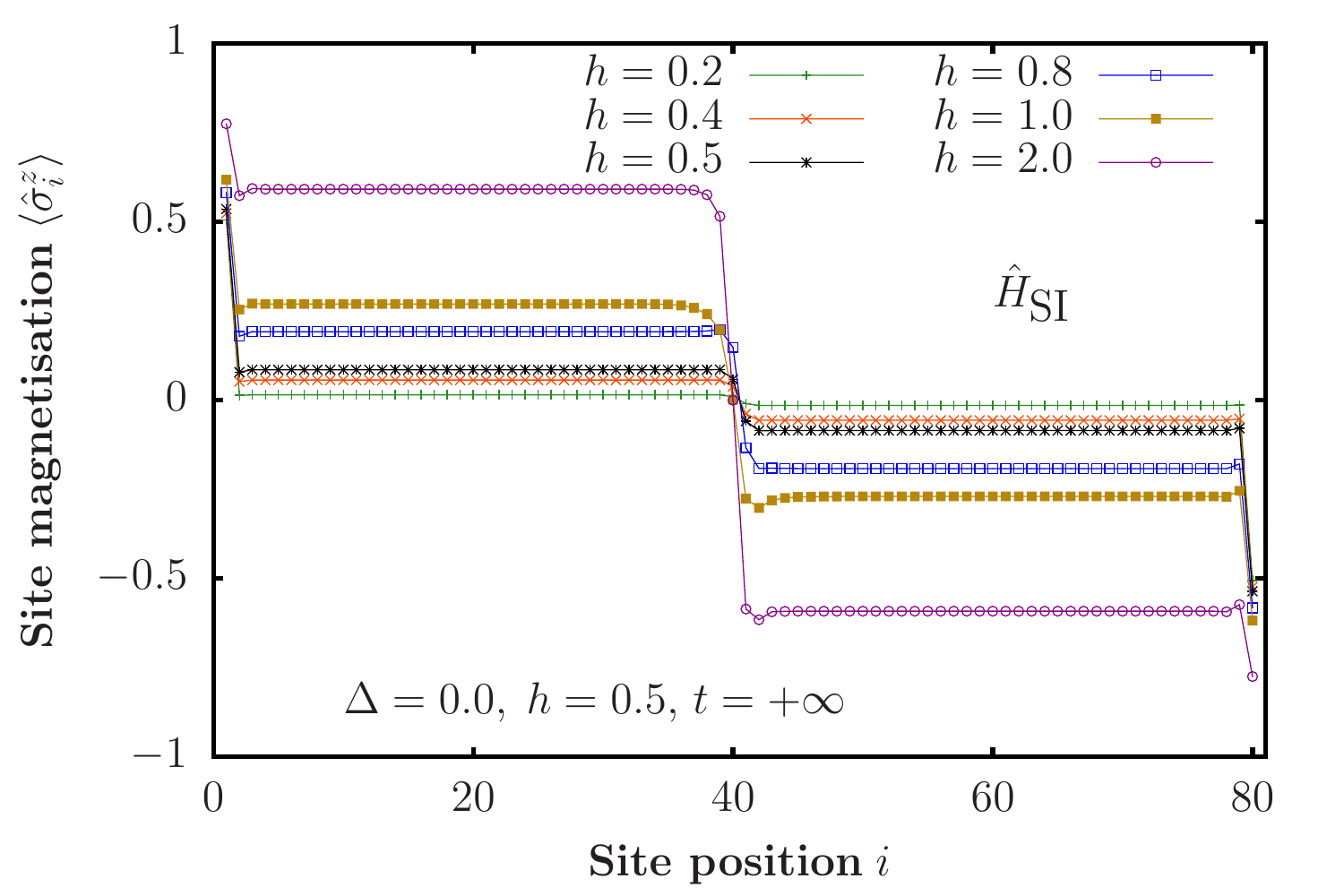}\includegraphics[width=0.5\columnwidth]{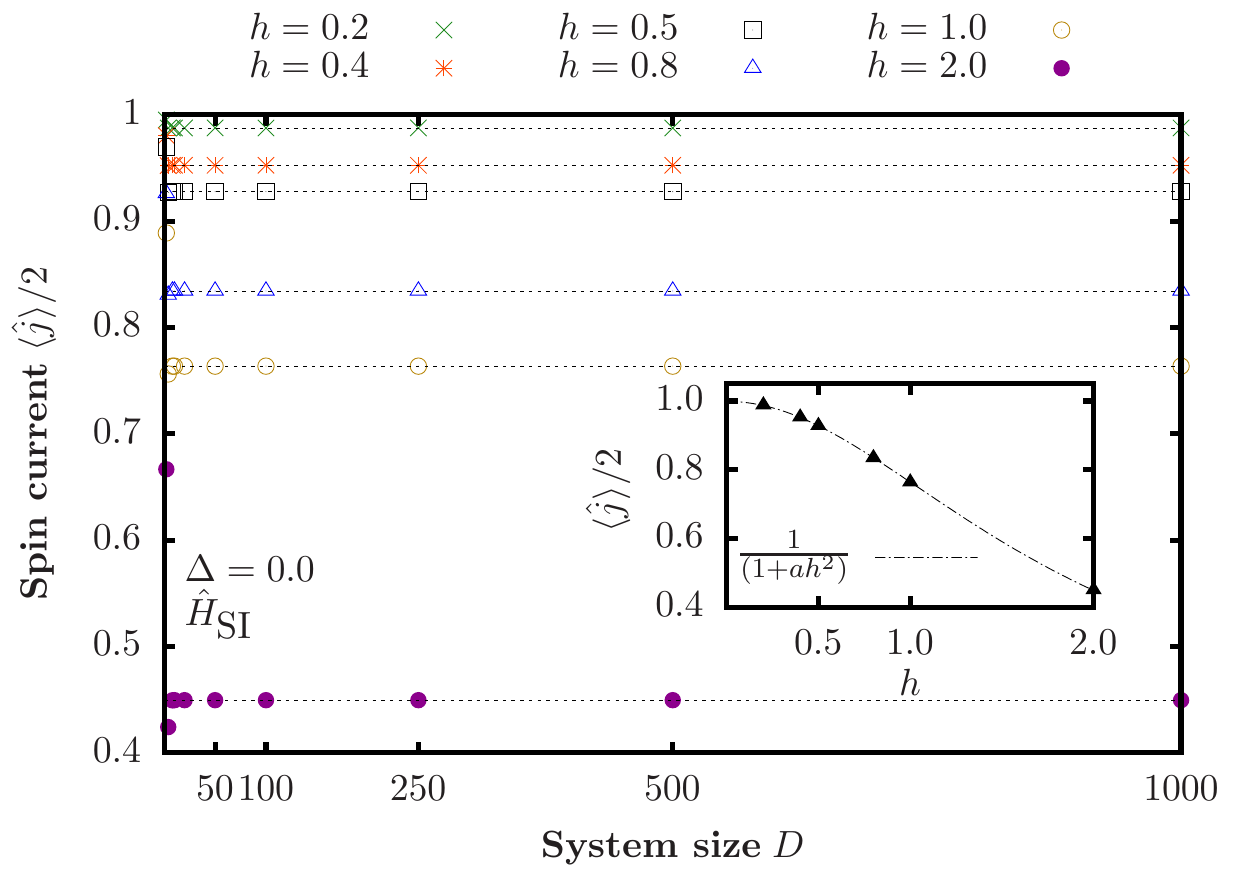}
\caption[Magnetisation profiles and expectation values of the current in the NESS of the non-interacting limit of the $\hat{H}_{\textrm{SI}}$ model ($\Delta = 0$) in the presence of a single magnetic impurity with different strengths $h$]{(Left) Magnetisation profiles in the NESS of the non-interacting limit of the $\hat{H}_{\textrm{SI}}$ model ($\Delta = 0$) in the presence of a single magnetic impurity with different strengths $h$. The profiles were obtained with $D = 80$, $\gamma = 1.0$, and $\mu = 1.0$. (Right) Scaling of the expectation value of the current operator in the non-equilibrium steady state of the $\hat{H}_{\textrm{SI}}$ model with $\Delta = 0$ as a function of system size ($D=4,\cdots,1000$), for different values of $h$. The driving parameters are $\gamma = 1.0$ and $\mu = 1.0$.}
\label{fig:2.2.6}
\end{figure}

The same procedure can be applied to the single impurity model $\hat{H}_{\textrm{SI}}$ as a function of the impurity strength $h$ in the non-interacting regime. In Fig.~\ref{fig:2.2.6}(left panel) we present the magnetisation profile in the $z$ direction for this particular case. These profiles of magnetisation, as before, are an indication of ballistic transport since no gradient of magnetisation is present across the chain, with the exception of the site of the impurity where the perturbation is located. The strength of the perturbation dictates the magnetisation slant, yet the profile in the bulk of the chain remains flat, as expected for ballistic systems.

In Fig.~\ref{fig:2.2.6}(right panel) we show the expectation value of the spin current operator in the NESS $\langle \hat{j} \rangle$ as a function of the chain sizes. One can see that, for sufficiently large system sizes, $\langle \hat{j} \rangle$ becomes independent of $D$, in analogy to the results for the XXZ model and the staggered field model in the non-interacting regime $\Delta = 0$. The absolute value of $\langle \hat{j} \rangle$ decreases with increasing the strength of the impurity as $1 / (1 + ah^2)$ [see the inset in Fig.~\ref{fig:2.2.6}(right)]. This functional form is obtained from the transmission probability of free particles through a barrier at high temperatures \cite{ryndyk2016nano}. We can conclude then, that the effect of the single impurity in the non-interacting regime is to act as a barrier for the spin transport, without changing its transport regime but merely the absolute value of the expectation value of the current. These results are consistent with the prediction that the non-interacting models remain ballistic for any values of the perturbation strength in the thermodynamic limit $D \to \infty$.

\subsubsection{The interacting regime: $\Delta \neq 0$}

In the interacting regime $\Delta \neq 0$, the solution to the non-equilibrium steady state is non-trivial. The tensor network method presented in this chapter can be used to approximate the long-time state $\ket{\rho(t \to \infty)}$ from Eq.~\eqref{eq:lme}, via a time-stepped time evolution of the density matrix with a fourth-order Trotter decomposition. At each time step, the expectation values of observables such as the current of the local magnetisation can be evaluated by expressing these operators in MPO form and contracting the networks with the time-evolved state~\cite{Schollwock2011}. Given Eq.~\eqref{eq:homogeneous}, we apply enough time steps such that the current across the chain is homogeneous up to numerical tolerance.

\begin{figure}[t]
\fontsize{13}{10}\selectfont 
\centering
\includegraphics[width=0.65\columnwidth]{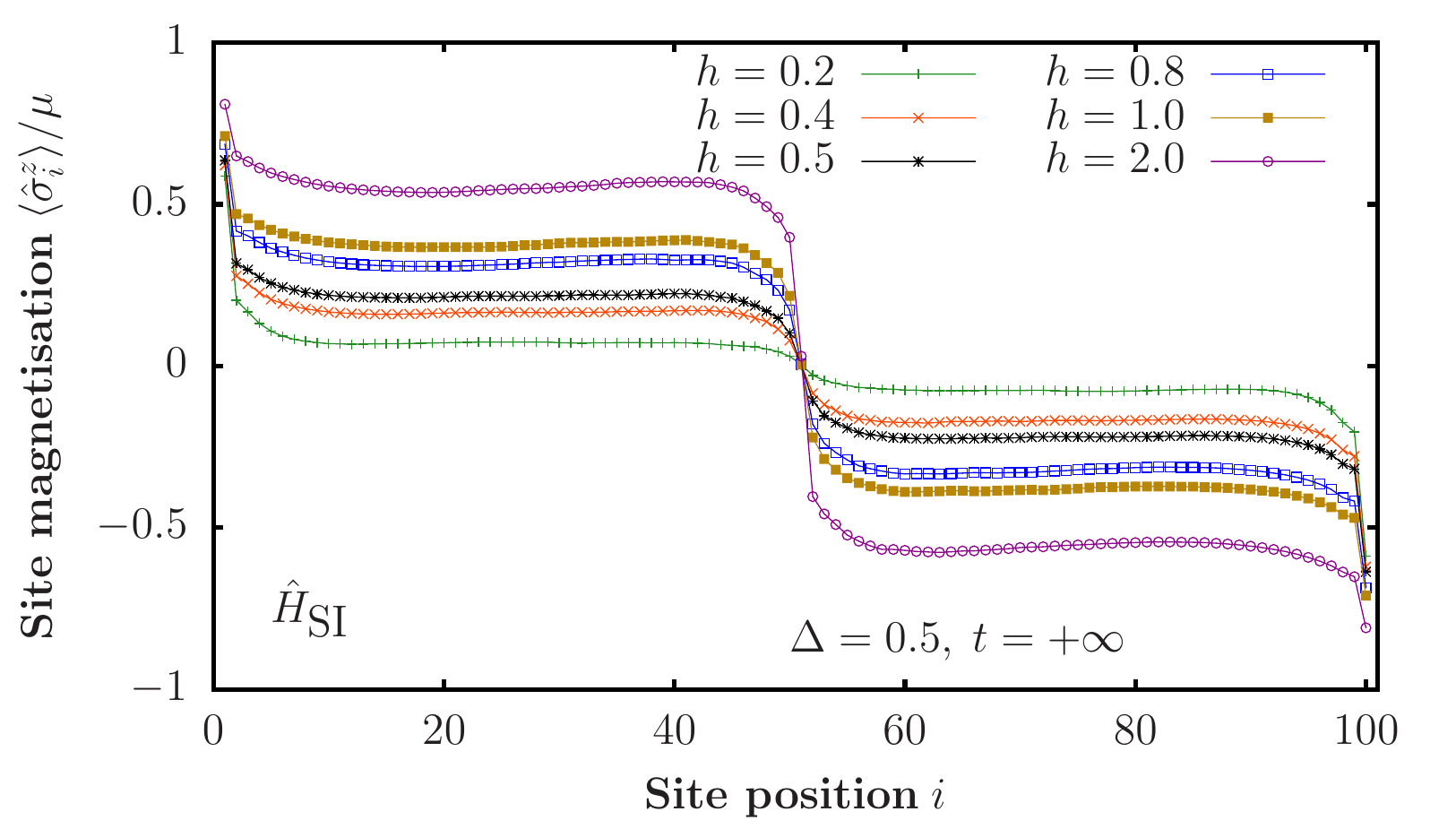}
\caption[Magnetisation profile in the non-equilibrium steady state of the anisotropic Heisenberg model in the presence of a single magnetic impurity with different values of $h$]{Magnetisation profile in the non-equilibrium steady state of the anisotropic Heisenberg model in the presence of a single magnetic impurity with different values of $h$ [see Eq.~\eqref{eq:h_si_22}]. The profiles were obtained for chains with $D = 100$, $\Delta = 0.5$, $\gamma = 1.0$, and $\mu = 0.005$.}
\label{fig:2.2.7}
\end{figure}

It is enlightening to first look at the magnetisation profile across the chain in the NESS in the presence of the impurity perturbation. In fact, the profile itself is determined by the transport regime of the system. In Fig.~\ref{fig:2.2.7}, we show the expectation value of $\hat{\sigma}_{z}$ in the NESS, as a function of site positions, for different values of the impurity strength $h$. The profiles reveal strong boundary effects induced by the driving at the edges of the chain, and are nearly {\em flat} in the bulk of the chain, with the exception of the site where the impurity is located. The ``kink'' at the latter point is larger the stronger the impurity field. The {\em flat} profiles in Fig.~\ref{fig:2.2.7} are a first indicator that transport is ballistic, as seen in integrable models such as the unperturbed XXZ chain~\cite{Znidaric:2011,Prosen:2009}.

\begin{figure}[t]
\fontsize{13}{10}\selectfont 
\centering
\includegraphics[width=0.65\columnwidth]{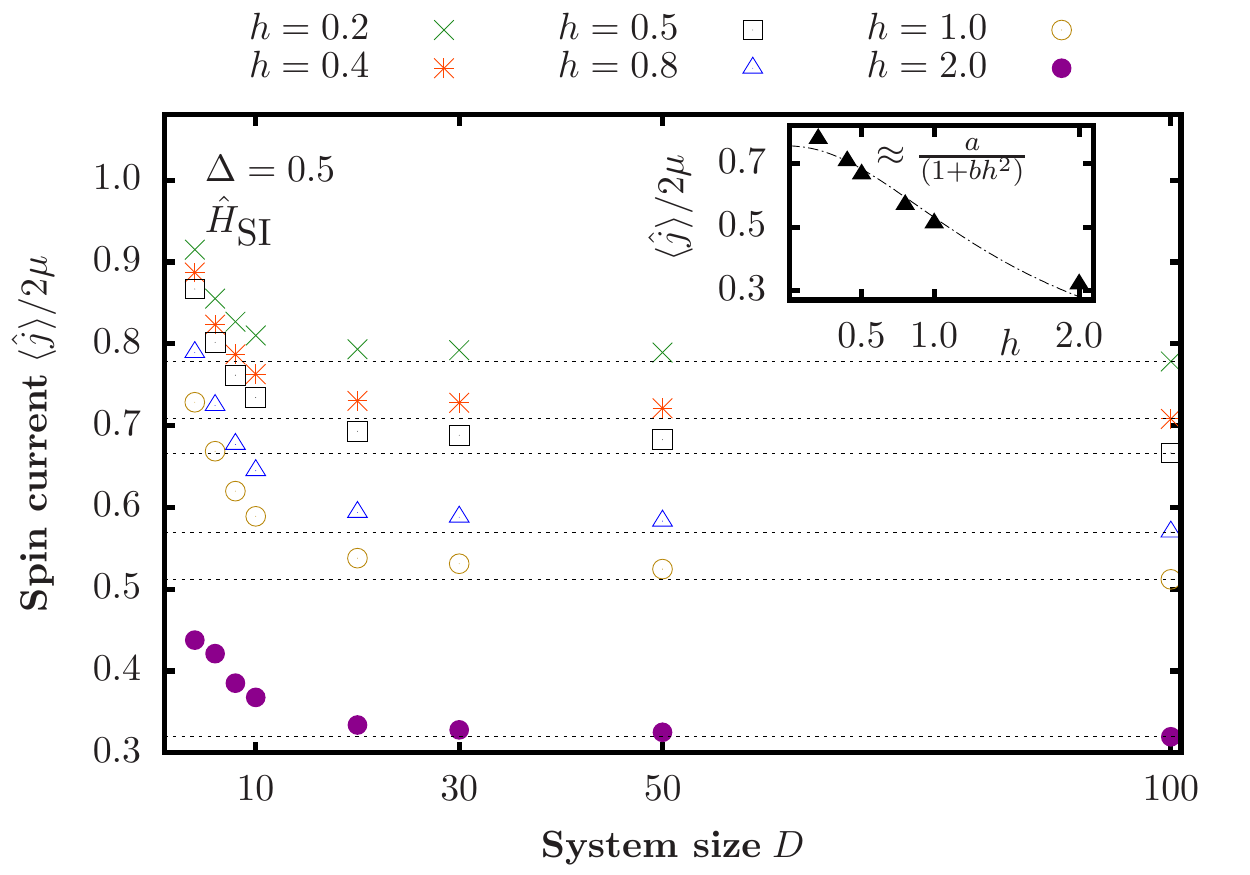}
\caption[Scaling of the expectation value of the current operator in the non-equilibrium steady state of the anisotropic Heisenberg model in the presence of a single magnetic impurity, plotted as a function of system size]{Scaling of the expectation value of the current operator in the non-equilibrium steady state of the anisotropic Heisenberg model in the presence of a single magnetic impurity [see Eq.~\eqref{eq:h_si_22}], plotted as a function of system size ($D=4,\cdots,100$), for $\Delta=0.5$ and different values of $h$. The driving parameters are $\gamma = 1.0$ and $\mu = 0.005$.}
\label{fig:2.2.8}
\end{figure}

Next, we quantify how the current in the NESS scales with increasing system size. In Fig.~\ref{fig:2.2.8}, we plot $\langle \hat{j} \rangle$ vs $D$ for $\Delta = 0.5$ and different values of $h$. Transport in the XXZ model is ballistic for any $0 < \Delta < 1$, a regime that is expected to change to incoherent, either diffusive or anomalous, when integrability is broken. We chose $\Delta=0.5$ because the system is in the strongly-interacting regime, and obtaining the NESS numerically is not as difficult as for $\Delta \approx 1$. The main observation in Fig.~\ref{fig:2.2.8} is that, for sufficiently large system sizes, $\langle \hat{j} \rangle$ becomes independent of $D$, a property of systems that exhibit coherent/ballistic transport. 

Our high-temperature non-equilibrium calculations indicate that, even though a single magnetic impurity breaks the integrability of the XXZ chain as seen from the probability distribution of energy level spacings (Fig.~\ref{fig:1.3.3}), transport remains ballistic and the system behaves as a perfect conductor. This becomes apparent in the scaling of the spin current only for sufficiently large system sizes, see Fig.~\ref{fig:2.2.8}, in analogy with the integrable XXZ case~\cite{Znidaric:2011}. We stress that this behaviour persists for all the values of $h$ studied, and that we expect it to persists for any finite non-vanishing magnetic impurity strength (for $h=0$ one has an integrable XXZ chain, and for $h=\infty$ one has two disconnected integrable XXZ chains). This is the first example known to us in which a quantum many-body system exhibits a Wigner-Dyson level spacing distribution and displays coherent transport. The latter can be understood to be the result of excitations traveling in a ballistic fashion on either side of the integrability breaking defect and scattering only at the impurity site. 

The inset in Fig.~\ref{fig:2.2.8} shows the scaling of the steady-state spin current, for $D = 100$, with the impurity strength. $\langle \hat{j} \rangle$ vs $h$ can be well fitted with the function $a / (1 + bh^2)$, an ansatz that follows from results for the non-interacting case discussed in this section. The main effect of increasing the magnitude of $h$ is to decrease the magnitude of  $\langle \hat{j} \rangle$, while transport remains ballistic.

The results reported here suggest that a single impurity is not sufficient to render transport incoherent, despite the fact that it is enough to render the system quantum chaotic, as indicated by the distribution of energy levels.

\begin{figure}[t]
\centering
\includegraphics[width=0.6\columnwidth]{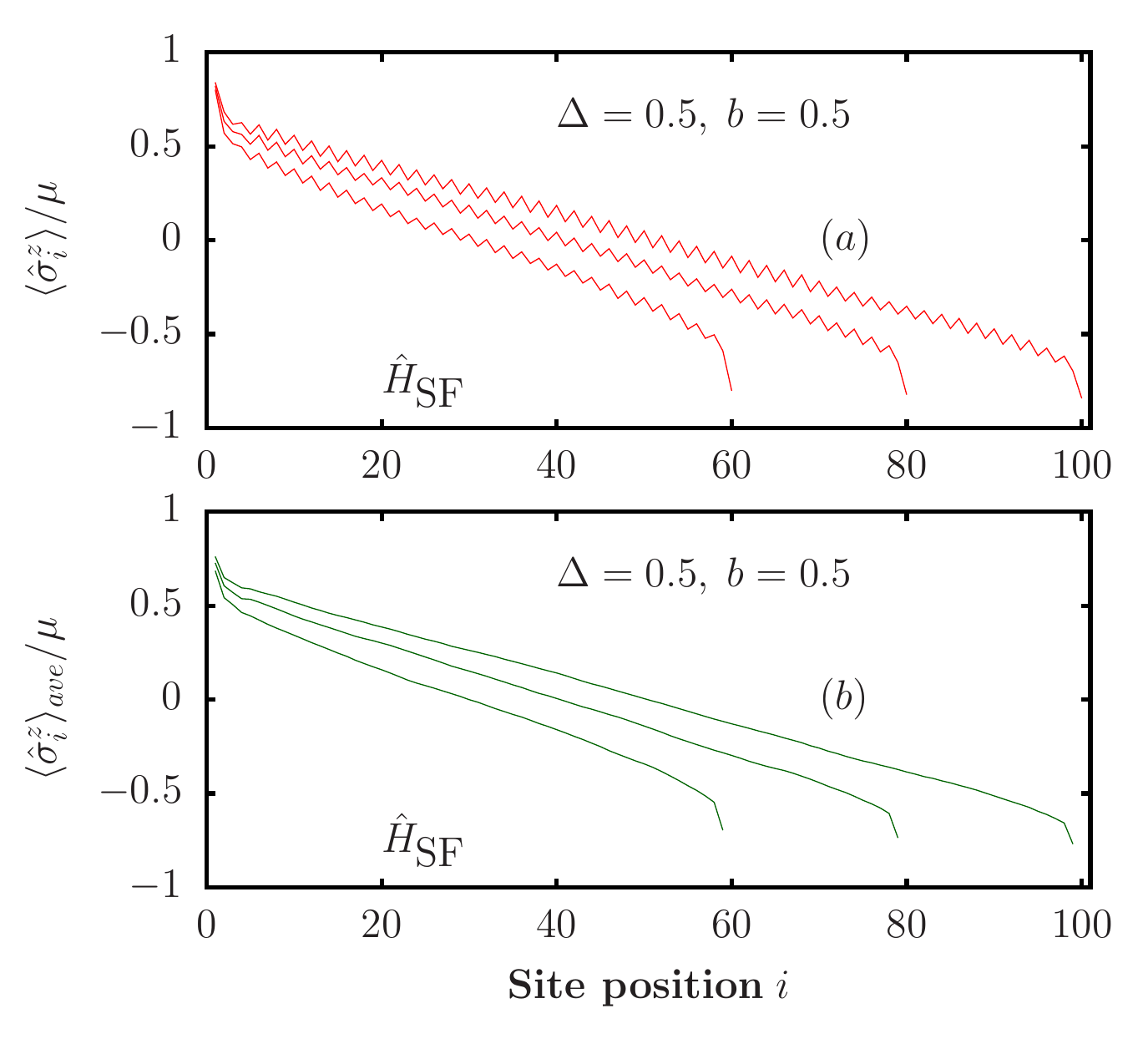}
\caption[Magnetisation profile of the non-equilibrium steady state of the anisotropic Heisenberg model in the presence of a staggered magnetic field]{Magnetisation profile of the non-equilibrium steady state of the anisotropic Heisenberg model in the presence of a staggered magnetic field with $b = 0.5$. The results were obtained for $\Delta = 0.5$, $\gamma = 1.0$, and $\mu = 0.001$. (a) Magnetisation, and (b) average magnetisation [see Eq.~\eqref{eq:avemag}].}
\label{fig:2.2.9}
\end{figure}

While it is known that the gapless XXZ model ($0 < \Delta < 1$) exhibits ballistic spin transport, and it is therefore an ideal conductor~\cite{Znidaric:2011, JuanThesis:2014}, breaking integrability by means of a staggered magnetic field renders the system chaotic and spin transport  becomes diffusive~\cite{Prosen:2009}. We shall revisit spin transport in the $\hat{H}_{\textrm{SF}}$ model [see Eq.~\eqref{eq:h_sf_22}] to contrast it with that in the $\hat{H}_{\textrm{SI}}$ model [see Eq.~\eqref{eq:h_si_22}].

Figure~\ref{fig:2.2.9}(a) shows the magnetisation profile in the NESS of the $\hat{H}_{\textrm{SF}}$ model for $\Delta=0.5$, $b=0.5$, and different chain sizes. Unlike the magnetisation profile in the NESS for the $\hat{H}_{\textrm{SI}}$ model, the staggered field induces a ramp-like linear profile in the magnetisation across the chain. The small oscillations of the magnetisation are due to the presence of the staggered field. In Fig.~\ref{fig:2.2.9}(b), we show the average
\begin{align}
\label{eq:avemag}
\langle \hat{\sigma}^z_i \rangle_{ave} = \left(\langle \hat{\sigma}^z_i \rangle + \langle \hat{\sigma}^z_{i+1} \rangle\right) / 2.
\end{align}
Figure~\ref{fig:2.2.9}(b) makes apparent that, aside from boundary effects, the magnetisation profile is linear.

\begin{figure}[t]
\centering
\includegraphics[width=0.6\columnwidth]{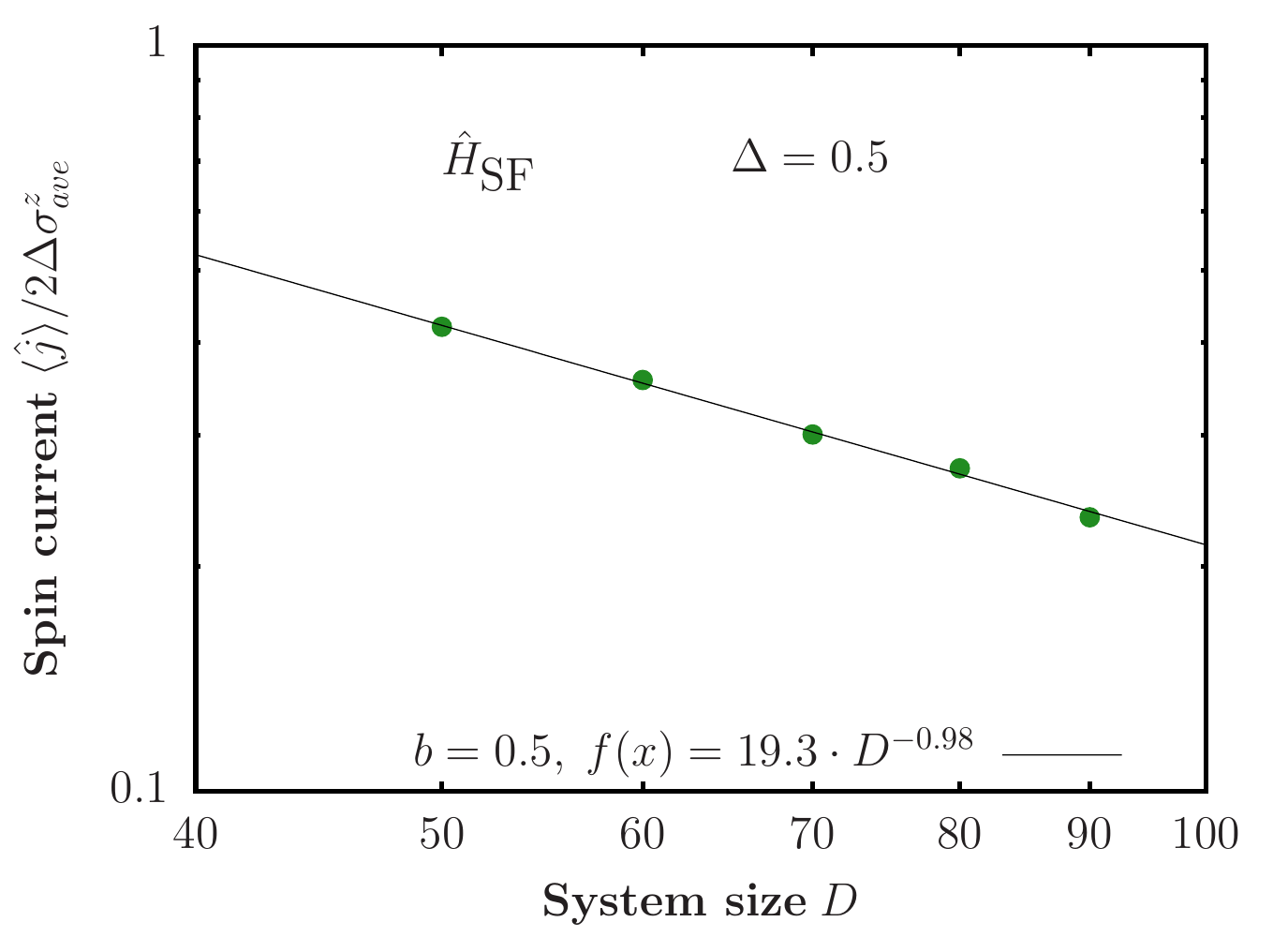}
\caption[Scaling of the spin current in the NESS of the staggered field model as a function of system size]{Scaling of the spin current in the NESS of the staggered field model as a function of system size ($D = 60, 70, 80, 90, 100$), for $\Delta=0.5$ and $b = 0.5$ (same parameters as in Fig.~\ref{fig:2.2.9}). The driving parameters are $\gamma = 1.0$ and $\mu = 0.001$. To reduce finite-size effects, in our calculations we discard the five leftmost and the five rightmost sites of the chains.}
\label{fig:2.2.10}
\end{figure}

Figure~\ref{fig:2.2.10} shows results for the finite-size scaling of the spin current in the NESS of the $\hat{H}_{\textrm{SI}}$ model, for the same parameters used in Fig.~\ref{fig:2.2.9}. We obtain the diffusion parameters, $\mathcal{D} = 19.3$ (the diffusion coefficient) and $\nu = 0.98$, from
\begin{align}
\label{scaling_sf}
\frac{\langle \hat{j} \rangle}{2\Delta \hat{\sigma}^z_{ave}} = \frac{\mathcal{D}}{(D-10)^{\nu}}.
\end{align}
Our results show that the current obeys the diffusion equation (Fick's law). They are in agreement with the results in Ref.~\cite{Prosen:2009}. Note that we have truncated the boundaries by five sites on each side, to reduce boundary effects in the estimation of the transport exponent.

\subsubsection{Linear response regime and error analysis}

We have now fully characterised spin transport in the single impurity model and have found it to be ballistic for any strength of the impurity, as long as $0 < \Delta < 1$, just as in the unperturbed model. In Chapter~\ref{chapter:kubo}, however, we estimated transport regimes from the theory of linear response. It is therefore crucial to ascertain that our boundary-driven configurations follow under the same predictions of linear response theory. 

\begin{figure}[t]
\fontsize{13}{10}\selectfont 
\centering
\includegraphics[width=0.6\columnwidth]{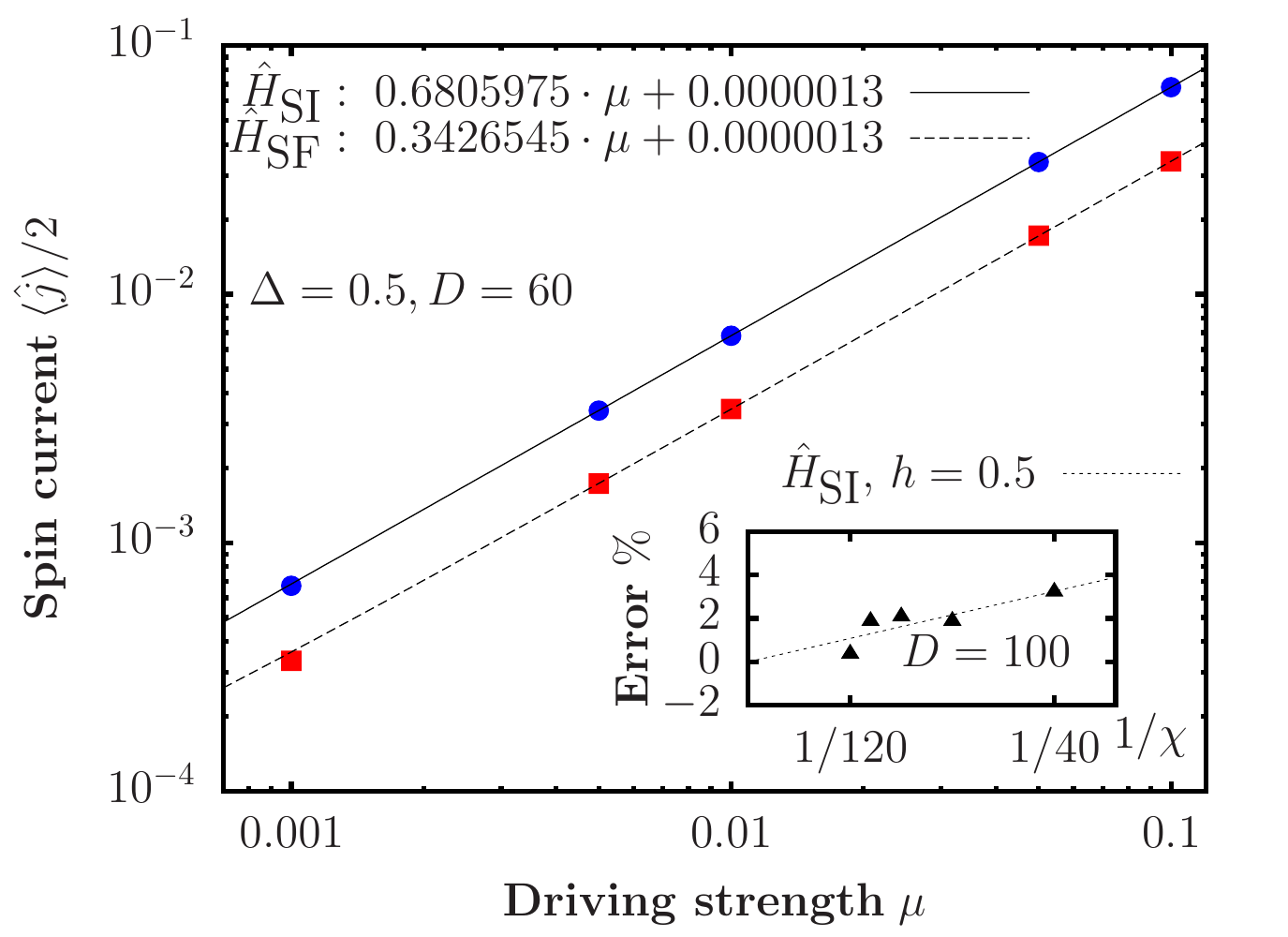}
\caption[Expectation value of the current in the NESS as a function of driving]{Expectation value of the current in the NESS as a function of driving strength. (inset) Truncation error in the tMPS method versus the bond dimension $\chi$ for the largest system size we simulated for the $\hat{H}_{\textrm{SI}}$ model with $h = 0.5$. }
\label{fig:2.2.11}
\end{figure}

We have compared transport properties of different models using non-equilibrium configurations and linear response theory, we shall now demonstrate that the non-equilibrium transport calculations are within linear response regime.  Figure~\ref{fig:2.2.11} shows that the magnitude of the spin current depends linearly on the driving strength for values well above those used in our simulations. This implies that the transport properties in our systems depend linearly on $\mu$,
\begin{align}
\langle \hat{j} \rangle \propto \mu,
\end{align}
and can be well-captured by linear response theory, as linear dependence on the driving field is the hallmark of linear response. For $\mu = 0$, the fit shown in Fig.~\ref{fig:2.2.11} is very close to zero, as no boundary driving implies no excitations propagating through the chain. The small value in the intercept is a witness of the numerical fidelity of calculations we have carried out.

We analysed the truncation error (induced by using a finite value of $\chi$) by studying the expectation value of the current operator [Eq.~\eqref{eq:spincurrent_22}] for the largest system size we simulated at fixed $\mu$ for different values of $\chi$. We then selected a value of $\chi$ that introduces a small tolerable error in our simulations. In the inset of Fig.~\ref{fig:2.2.11}, we show the error defined as $|\langle \hat{j}(\chi) \rangle - \langle \hat{j}(\infty) \rangle | / \langle \hat{j}(\infty) \rangle \times 100$, where $\langle \hat{j}(\infty) \rangle$ is an extrapolated value of the current, as a function of the bond dimension $\chi$, i.e., $\hat{j} = \hat{j}(\chi)$. The scaling of the bond dimension suggests convergence for $\chi \rightarrow \infty$, as expected. In our calculations, we used $\chi = 100$ which results in an error due to the truncation that is $\lesssim 2\%$ for the specific case of the current in the single impurity model.

\section{Summary and outlook}
\label{sec:summary_ness}

In this chapter, we have introduced the approach known as boundary driving. Combined with the theory of tensor networks, it constitutes an extremely powerful approach to address the transport in both integrable and non-integrable systems alike. We have derived the {\em local} master equation from first principles and described a numerical technique to address non-equilibrium steady states, from which transport can be inferred. As a numerical problem, identifying transport regimes in interacting quantum systems is a computationally-difficult problem since both the long-time and thermodynamic limits need to be described. 

Complementing the results exposed in Chapter~\ref{chapter:kubo}, we have found from the theory of open systems that the single impurity model is indeed a special case of a non-integrable system with interesting thermalisation properties (Chapter~\ref{chapter:eth}) that displays coherent transport in the thermodynamic limit. Global perturbations, such as the introduction of a staggered magnetic field to the XXZ model renders the model diffusive. We remark that traditional linear response theory points towards the same conclusion and our results are, therefore, fully consistent. As stated in Chapter~\ref{chapter:kubo}, the notion of integrability breaking and ballistic transport from Mazur's inequality can be made to be fully consistent. For the particular case of the single impurity model, we found that translational invariance plays a pivotal role in transport regimes. 

The local Lindblad approach introduced here in the context of transport in the single impurity model, opens further interesting questions. 

We bring our attention, in particular, to the question of thermalisation and finite-temperature transport in quantum thermal machines. Akin to the configuration we employed in this chapter, quantum thermal machines are composed of a central quantum system, which may be strongly interacting, coupled to thermal reservoirs that may be kept away from equilibrium. One can then study the properties of a quantum system as the working medium for a thermal machine. These particular configurations fall outside of the range of applicability of the local Lindblad approach employed in this chapter, where only the degrees of freedom of the edges are coupled to a driving that enforces a local magnetisation. One may only attempt to study infinite-temperature regimes from such an approach. The reason behind the correspondence between the transport observed at infinite temperature and the one obtained from boundary-driven configurations is subtle. We shall explore this in further detail in Chapter~\ref{chapter:finite_temperature}.

For the description of a more general scenario where a quantum system is coupled to true thermal reservoirs, one needs to go beyond boundary-driven schemes. These considerations drive our motivation for the following chapter.
 
\chapter{Finite temperature transport and autonomous quantum thermal machines}
\label{chapter:finite_temperature}

Having established the fundamentals of boundary driving configurations and non-equilibrium steady states, we now proceed to develop a framework to address finite-temperature transport in autonomous quantum thermal machines.

In this chapter, we build our methodology step by step. We begin by providing a description of the pitfalls of {\em local} and {\em global} master equations in Sec.~\ref{sec:local_v_global}. This insight provides the urgency to address finite-temperature transport, from which thermodynamic properties of interacting systems connected to thermal reservoirs follow. We then proceed to introduce autonomous thermal machines in Sec.~\ref{sec:autonomous_thermal_machines}, where the problem to be solved is precisely defined. Following the statement of the problem, we outline the mesoscopic-reservoir approach and demonstrate its connection to the infinite-bath scenario in Sec.~\ref{sec:mesoscopic_leads}. Subsequently, in Sec.~\ref{sec:landauer_buttiker}, we introduce Landauer-B\"uttiker theory as a fundamental framework to study finite-temperature transport in systems where incoherent effects are not present. This theory will serve as a benchmark to our mesoscopic reservoir construction. The superfermion representation description then follows in Sec.~\ref{sec:superfermion}, and we employ this description to find an analytical expression for the non-equilibrium steady state of a non-interacting (quadratic) system. 

In Sec.~\ref{sec:thermo_meso} we explain how to compute particle and energy currents within our framework.  Equipped with the exact solution for quadratic systems, we study a non-interacting quantum-dot heat engine and compare the results with Landauer-B\"uttiker theory in order to identify the number and distribution of modes in the mesoscopic reservoirs needed to accurately reproduce the continuum limit. Next, in Sec.~\ref{sec:tensor} we detail our tensor-network algorithm for studying interacting problems. We then apply this algorithm in Sec.~\ref{sec:interacting} to study a three-site interacting heat engine and a many-body Heisenberg spin model at infinite and finite temperatures. Finally, we summarise and conclude in Sec.~\ref{sec:conclusions}.

\section[Local vs global master equations]{Local vs global master equations and their pitfalls}
\label{sec:local_v_global}

\subsection{Local master equations}

In Chapter~\ref{chapter:lindblad_high}, we introduced boundary driving as an open-systems approach to address transport in interacting spin chains. We derived a {\em local} master equation from first principles using the repeated interactions scheme. Such master equation has the form
\begin{align}
\label{eq:lme_23}
\frac{d\hat{\rho}}{dt} &= -\textrm{i}[\hat{H},\hat{\rho}] + \mathcal{L}\{\hat{\rho}\} \nonumber \\ 
&= -\textrm{i}[\hat{H},\hat{\rho}] + \mathcal{L}_{\tt L}\{\hat{\rho}\} + \mathcal{L}_{\tt R}\{\hat{\rho}\},
\end{align}
for the specific configuration in which only the degrees of freedom pertaining to the boundaries are locally driven to a given magnetisation parameter. This follows from the Lindblad super-operators $\mathcal{L}_{\alpha}$ which act only on the left and right boundaries of the chain $\alpha = \tt L, \tt R$, and  
\begin{align}
\label{eq:lind1_23}
\mathcal{L}_{\alpha}\{\hat{\rho}\} = \sum_{s=\pm} \gamma^s_{\alpha} \left[ 2\hat{L}_{s,{\alpha}}\,\hat{\rho}\, \hat{L}_{s,\alpha}^{\dag} - \{\hat{L}_{s,\alpha}^{\dag}\hat{L}_{s,\alpha},\hat{\rho}\} \right],
\end{align}
where
\begin{align}
\label{eq:lind2_23}
\hat{L}_{+,\tt L} = \hat{\sigma}_1^{+}, \quad \hat{L}_{-,\tt L} = \hat{\sigma}_1^{-}, \nonumber \\
\hat{L}_{+,\tt R} = \hat{\sigma}_D^{+}, \quad \hat{L}_{-,\tt R} = \hat{\sigma}_D^{-},
\end{align}
for a spin chain of length $D$. Note that instead of absorbing the coupling $\gamma^s_{\alpha}$ into the definitions of $\hat{L}_{s,\alpha}$ as we did in Chapter~\ref{chapter:lindblad_high}, we have instead kept it as a pre-factor in Eq.~\eqref{eq:lind1_23}. These local Lindblad operators act as incoherent sinks and sources that create and remove excitations from the boundaries, yielding a true and parameter-dependent out-of-equilibrium configuration. This approach allowed us to overcome the achievable system size limitations inherent to other techniques, such as exact diagonalisation, when applied in conjunction with tensor network techniques.  

If one is interested, however, in modelling true thermal reservoirs connected to a generic and interacting many-body quantum system, the local Lindblad description would fail to capture basic thermodynamic principles. Specifically, a many-body quantum system coupled to Lindblad jump operators acting only on its boundaries, would fail to thermalise to the state imposed by the macroscopic parameters of the reservoir.

Let us describe this in further detail~\cite{barra2015thermodynamic}. Within the approach of boundary driving, the canonical density matrices $\hat{\omega}_{\beta_{\alpha}}(\hat{H}_{\alpha}) = e^{-\beta_{\alpha} \hat{H}_{\alpha}} / Z_{\alpha}$, with $Z_{\alpha} = \textrm{Tr}[e^{-\beta_{\alpha} \hat{H}_{\alpha}}]$ are only imposed on the spin degrees of freedom pertaining locally to the boundaries. As described from the repeated interactions scheme in Sec.~\ref{sec:repeated_interactions}, a single copy of the Hamiltonian of the reservoirs is $\hat{H}_{\alpha} = (h_{\alpha}/2) \hat{\sigma}^z_{\alpha}$ suffices to characterise the density matrices $\hat{\omega}_{\beta_{\alpha}}(\hat{H}_{\alpha})$ of the boundaries via its magnetisation
\begingroup
\allowdisplaybreaks
\begin{align}
\label{eq:tanh_beta}
\langle \hat{\sigma}^z_{\alpha} \rangle &= \textrm{Tr}[\hat{\sigma}^z_{\alpha} \hat{\omega}_{\beta_{\alpha}}] \nonumber \\
&= \textrm{Tr}[\hat{\sigma}^z_{\alpha} e^{-\beta_{\alpha} \hat{H}_{\alpha}}] / Z_{\alpha} \nonumber \\
&= \frac{e^{-\beta_{\alpha} h_{\alpha} / 2} - e^{\beta_{\alpha} h_{\alpha} / 2}}{e^{-\beta_{\alpha} h_{\alpha} / 2} + e^{\beta_{\alpha} h_{\alpha} / 2}} \\
&= -\tanh(\beta_{\alpha} h_{\alpha} / 2).
\end{align}
\endgroup
As described in Sec.~\ref{sec:repeated_interactions}, we find that the couplings are connected to the magnetisation of the boundaries via
\begin{align}
\gamma^s_{\alpha} = \lambda_{\alpha} (1 \pm \langle \hat{\sigma}^z_{\alpha} \rangle),
\end{align}
which then implies, from Eq.~\eqref{eq:tanh_beta},
\begin{align}
\label{eq:plus_d_minus_bd}
\frac{\gamma^+_{\alpha}}{\gamma^-_{\alpha}} = e^{-\beta_{\alpha} h_{\alpha}}.
\end{align}
If we disconnect the spin degree of freedom being driven from the rest of the chain, a local detailed-balance condition is respected with respect to the canonical state $\hat{\omega}_{\beta_{\alpha}}(\hat{H}_{\alpha})$. This follows from the fact that after evaluating Eq.~\eqref{eq:lind1_23} with $\hat{\omega}_{\beta_{\alpha}}(\hat{H}_{\alpha})$, we find\footnote{Take, for instance, $\mathcal{L}_{\tt L}\{ \hat{\rho} \} = \gamma^+_{\tt L} \left( [ 2\hat{\sigma}^+_1 \hat{\rho} \hat{\sigma}^-_1 - \left\{ \hat{\sigma}^-_1\hat{\sigma}^+_1, \hat{\rho} \right\} ] + e^{\beta_l h_1} [2\hat{\sigma}^-_1 \hat{\rho} \hat{\sigma}^+_1 - \left\{ \hat{\sigma}^+_1\hat{\sigma}^-_1, \hat{\rho} \right\}] \right)$, where we have used Eq.~\eqref{eq:plus_d_minus_bd} for the second coupling factor. It is straightforward to then show $\mathcal{L}_{\tt L}\{ \hat{\omega}_{\beta_{\tt L}}(\hat{H}_{\tt L}) \} = 0$} 
\begin{align}
\mathcal{L}_{\alpha}\{ \hat{\omega}_{\beta_{\alpha}}(\hat{H}_{\alpha}) \} = 0,
\end{align}
which then implies that 
\begin{align}
0 = -\textrm{i}[\hat{H}_{\tt S}, \hat{\rho}] + \mathcal{L}_{\alpha}\{\hat{\rho}\}
\end{align}
is satisfied for $\alpha = \tt L, \tt R$ if $\hat{\rho} = \hat{\omega}_{\beta_{\alpha}}(\hat{H}_{\alpha})$, since the single spin degree of freedom of the boundary has been disconnected from the rest of the chain. However, if the full many-body system is only driven from its boundary, we find that the above condition is not satisfied, given that in such case $\mathcal{L}_{\alpha}\{\hat{\rho}\} \neq 0$. 

We have then shown that, in boundary driving, local detailed-balance is only satisfied for the single spin degree of freedom being driven. This immediately implies that appropriate thermalisation is not guaranteed from such a configuration if the full many-body system is connected only from its boundary. 

In fact, the question of the necessary requirements for appropriate thermalisation to be achieved using {\em local} couplings between the reservoirs and the system has been addressed independently by Guimaraes {\em et al.} in Ref.~\cite{Landi:2016} and Reichental {\em et al.} in Ref.~\cite{Reichental2018}. It has been shown numerically, particularly in Ref.~\cite{Reichental2018}, that a single set of dissipators acting locally on the boundaries of the system fails to thermalise the system to the parameters dictated by the reservoirs. As we have discussed, this is expected from the lack of local detailed balance. Crucially, however, thermalisation can be achieved by coupling the system not just from its boundaries, but by employing a set of multiple dissipators that enforce local detail balance in the sites of the chain that compose the reservoir. For thermalisation to be achieved, weak system-reservoir coupling is also required.

Most interestingly, the local Lindblad master equation Eq.~\eqref{eq:lme_23} for boundary-driven configurations (Fig.~\ref{fig:1.2.1}), can be shown to satisfy appropriate thermodynamic laws (first law of thermodynamics and positivity of the entropy production) when a time-dependent interaction is introduced in the form of the repeated interactions discussed in Sec.~\ref{sec:repeated_interactions}. For the specific case of the boundary-driven XXZ chain, however, it was shown by Pereira in Ref.~\cite{PereiraHeatXXZ} that the spin current and the energy currents are connected by a constant factor, i.e., these thermodynamic quantities are always proportional to each other. This behaviour is not expected to be generic at finite temperatures, but a restriction imposed by the driving scheme.

Finally, we remark that even though the local Lindblad master equation Eq.~\eqref{eq:lme_23} introduces parameter-dependent rates $\gamma_{\alpha}^{\pm}$, there exists no energy dependence of the excitations being introduced and removed from the boundaries. The rates $\gamma_{\alpha}^{\pm}$ depend on the temperature through Eq.~\eqref{eq:plus_d_minus_bd}, however, this only provides a given probability to induce or remove an excitation via $\hat{\sigma}^{\pm}_{\alpha}$. The excitation is created locally in configuration space, which implies that in energy-momentum space, the excitation is fully delocalised. This heuristic argument brings us to the following illuminating conclusion: boundary driving configurations induce and remove excitations with {\em equal probability for every possible energy density}. In a true reservoir, with an infinite collection of degrees of freedom, this can only happen at infinite temperature. It follows that the properties of a given quantum system deduced from boundary driven configurations can only represent infinite-temperature behaviour with fidelity.  

\subsection{Global master equations}

The discussion above brings us to the topic of {\em global master equations}. A global approach is also characterised by the dynamical evolution of a quantum master equation of the the form 
\begin{align}
\label{eq:lme_global}
\frac{d\hat{\rho}}{dt} = -\textrm{i}[\hat{H},\hat{\rho}] + \mathcal{L}\{\hat{\rho}\},
\end{align}
where $\hat{H}$ and $\hat{\rho}$ are the Hamiltonian and density operators associated to the system alone, while Eq.~\eqref{eq:lme_global} describes the evolution under dissipation. As before, $\mathcal{L}\{\hat{\rho}\}$ needs to be expressed in Lindblad form [Eq.~\eqref{eq:lind1_23}] for the evolution of the density operator to completely positive and trace-preserving~\cite{BreuerPetruccione}. It is common to assume in multi-reservoir configurations that each reservoir is independent, which then implies that generators $\mathcal{L}_{\alpha}\{\hat{\rho}\}$ for the $\alpha$th reservoir are additive, in the sense that $\mathcal{L}\{\hat{\rho}\} = \sum_{\alpha} \mathcal{L}_{\alpha}\{\hat{\rho}\}$~\cite{Mitchison2018}.

The global approach to the master equation rests on fixing the state $\hat{r}_{\alpha}$ such that
\begin{align}
\mathcal{L}_{\alpha} \{ \hat{r}_{\alpha} \} = 0.
\end{align}
For the specific case of the local approach from before, the generators that satisfy the fixed-point condition can be obtained microscopically under certain approximations, as we did in Sec.~\ref{sec:repeated_interactions}. For the global approach, the following fixed-point condition defines the Lindblad generators~\cite{Mitchison2018}:
\begin{align}
\mathcal{L}_{\alpha} \{ \hat{r}_{\alpha} \} = 0 \Longleftrightarrow \hat{r}_{\alpha} = \frac{e^{-\beta_{\alpha} (\hat{H}_{\tt S} - \mu_{\alpha} \hat{N}_{\tt S})}}{Z},
\end{align} 
where $\hat{r}_{\alpha}$ is the grand-canonical ensemble state, $Z = \textrm{Tr}[e^{-\beta_{\alpha} (\hat{H}_{\tt S} - \mu_{\alpha} \hat{N}_{\tt S})}]$, $\hat{H}_{\tt S}$ and $\hat{N}_{\tt S}$ are system Hamiltonian and total number operators, respectively. The generators that satisfy this condition can be derived using a microscopic model and applying the Born-Markov and secular approximations~\cite{BreuerPetruccione}, which can usually be justified under certain conditions of the system Hamiltonian and system-reservoir couplings. 

Most importantly, even in the case of a system coupled to two thermal reservoirs kept away from equilibrium, the global approach yields a stationary state 
\begin{align}
\hat{\rho}_{\textrm{NESS}} = \sum_{n = 1}^{d} p_n \ket{n}\bra{n},
\end{align}
where $d$ is the dimension of the Hilbert space being considered and $\ket{n}$ are eigenstates of the system Hamiltonian $\hat{H}_{\tt S} \ket{n} = E_n \ket{n}$~\cite{Wichterich2007}. This immediately implies that that the non-equilibrium steady state is diagonal in the energy eigenbasis and the expectation value of current operators can be expressed as
\begin{align}
\langle \hat{J}(t \to \infty) \rangle = \textrm{Tr}[\hat{\rho}_{\textrm{NESS}} \hat{J}] = \sum_{n = 1}^{d} p_n \braket{n | \hat{J} | n}.
\end{align}
Crucially, it is expected that $\braket{n | \hat{J} | n} = 0$ as a very plausible condition, since energy eigenstates $\ket{n}$ are stationary and do not carry currents for systems with {\em open boundary conditions}. This implies 
\begin{align}
\langle \hat{J}(t \to \infty) \rangle = 0
\end{align}
in the global approach. In other words, the global approach {\em fails to capture the steady state coherences required for the existence of currents of conserved quantities}. This is an un-physical consequence, which occurs even in the case of two reservoirs kept out of equilibrium coupled to the system~\cite{Wichterich2007}. Furthermore, it has been shown that such an approach may violate the additivity condition required in order to write a master equation in the Lindblad form~\cite{Mitchison2018}.

The considerations introduced in this section pose a problem to understand quantum thermal machines, in which a quantum system is coupled to thermal reservoirs kept away from equilibrium. The local approach violates thermalisation and basic thermodynamic laws. Although the first and second laws of thermodynamics can be conciliated from the repeated interactions scheme, it yields a non-generic proportionality between particle (spin) and energy currents. Thermalisation is not satisfied and, furthermore, only infinite temperature properties can be understood from such a description. The global approach conciliates thermalisation, but does not capture the required coherences required for a current to be induced in the steady state. It follows that another description is required, which is the driving motivation for this chapter. We shall introduce a fully thermodynamically-consistent approach, to address and understand quantum thermal machines in which the working medium is a generic interacting quantum system.

\section{Autonomous thermal machines}
\label{sec:autonomous_thermal_machines}

This chapter is concerned with autonomous thermal machines whose working medium is a quantum system $\tt S$, which may be a complex entity comprising many interacting subsystems. The working medium is connected to multiple fermionic reservoirs labelled by the index $\alpha$. These reservoirs are macroscopic systems described by equilibrium temperatures $T_\alpha = 1/\beta_\alpha$ and chemical potentials $\mu_\alpha$ (we set $k_{\rm B} = 1 = \hbar$). The total Hamiltonian of such a setup takes the form 
\begin{equation}
    \label{eq:global_Hamitonian}
    \hat{H}_{\rm tot} = \hat{H}_{\tt S} + \sum_{\alpha}\left( \hat{H}_\alpha + \hat{H}_{\tt S \alpha}\right),
\end{equation}
where $\hat{H}_{\tt S}$ is the system Hamiltonian, $\hat{H}_\alpha$ is the Hamiltonian of bath $\alpha$ and $\hat{H}_{\tt S \alpha}$ describes its coupling to the system. We will consider exclusively Hamiltonians $\hat{H}_{\rm tot}$ that conserve fermion number $\hat{N} = \hat{N}_{\tt S} + \sum_\alpha \hat{N}_\alpha$, where $\hat{N}_{\tt S}$ and $\hat{N}_\alpha$ are the total particle number operators for the system and each bath $\alpha$, respectively.

Crucially, the baths are taken to have an infinite volume and heat capacity, implying a diverging number of degrees of freedom, $N\to\infty$. Moreover, it is typical to assume a factorised initial state of the form 
\begin{equation}
\label{eq:product_state}
    \hat{\rho}_{\rm tot}(0) = \hat{\rho}(0) \hat{\rho}_{\tt B},
\end{equation}
where $\hat{\rho}(0)$ is the initial system state and $\hat{\rho}_{\tt B}  = \prod_\alpha \hat{\rho}_\alpha$, with $\hat{\rho}_\alpha  = \ee^{-\beta_\alpha(\hat{H}_\alpha - \mu_\alpha \hat{N}_\alpha)}/Z_\alpha$ a thermal state and $Z_\alpha$ the partition function of each reservoir. Evolving into the long-time limit the system ${\tt S}$ will generically relax to a steady state given by 
\begin{equation}
    \label{eq:NESS_unitary}
    \hat{\rho}(\infty) = \lim_{t\to \infty} \lim_{N\to\infty} \Tr_{\tt B}\left[ \ee^{-\ii \hat{H}_{\rm tot}t} \hat{\rho}_{\rm tot}(0) \ee^{\ii \hat{H}_{\rm tot}t} \right],
\end{equation}
where $\Tr_{\tt B}$ denotes the trace over all bath degrees of freedom. If the temperatures or chemical potentials of the reservoirs differ, $\hat{\rho}(\infty)$ will be a non-equilibrium steady state (NESS) possessing currents of particles and energy.

\begin{figure}[t]
\fontsize{13}{10}\selectfont 
\centering
\includegraphics[width=0.5\columnwidth]{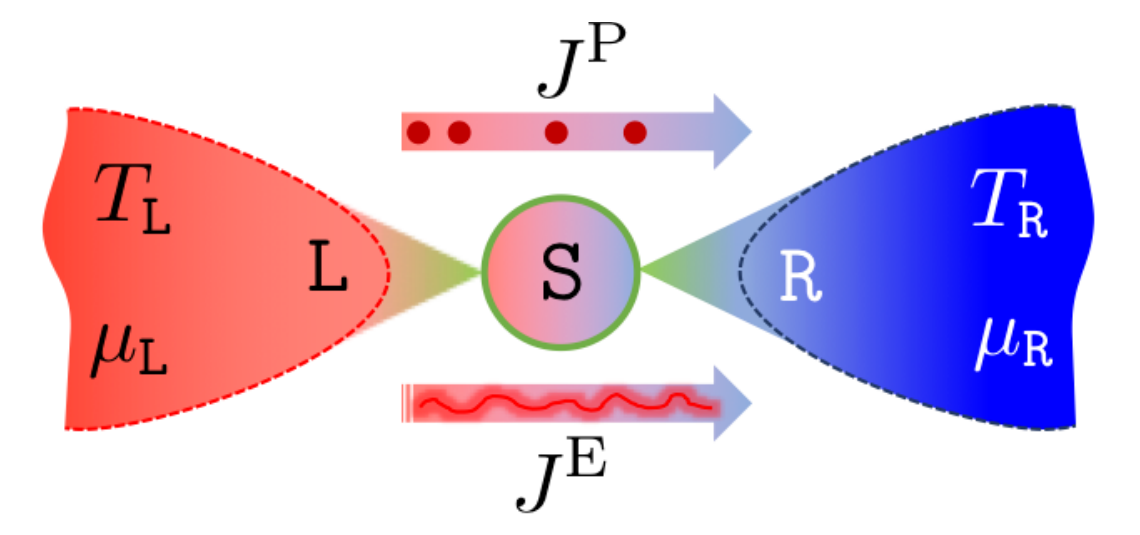}
\caption{A simple thermal machine scenario in which the system $\tt S$ is coupled to two reservoirs $\tt L$ and $\tt R$ at temperatures $T_{\tt L}> T_{\tt R}$ and possessing a chemical-potential difference $\mu_{\tt R} - \mu_{\tt L}> 0$. A particle $J^{\rm P}$ and energy $J^{\rm E}$ current is thus sustained through $\tt S$.}
\label{fig:2.3.1}
\end{figure}

We focus especially on the simplest scenario depicted in Fig.~\ref{fig:2.3.1}, with two reservoirs labelled by $\alpha = {\tt L},{\tt R}$. The sustained fluxes of particles and energy in this setup can be exploited, for example by operating the device as an autonomous heat engine. In this case a temperature gradient, $T_{\tt L}> T_{\tt R}$, drives a current that performs work by moving fermions against a chemical-potential difference $V = \mu_{\tt R} - \mu_{\tt L} > 0$. The power developed per unit time is given by
\begin{equation}
    \label{eq:power_def}
    P = V J^{\rm P},
\end{equation}
where $J^{\rm P}$ is the particle current, defined to be positive when flowing from left to right. The concomitant energy current $J^{\rm E}$ (also from left to right) transfers heat out of the left lead and into the right lead at a rate~\cite{Benenti2017}
\begin{equation}
    \label{eq:heat_current_def}
    \dot{Q}_\alpha = J^{\rm E} - \mu_\alpha J^{\rm P},
\end{equation}
so that the first law of thermodynamics can be written as $P = \dot{Q}_{\tt L} - \dot{Q}_{\tt R}$. The second law of thermodynamics imposes the relation $\beta_{\tt R} \dot{Q}_{\tt R} \geq \beta_{\tt L} \dot{Q}_{\tt L}$. The efficiency of heat-to-work conversion is thus given by 
\begin{equation}
    \label{eq:efficiency_def}
    \eta = \frac{P}{\dot{Q}_{\tt L}}  = 1 - \frac{\dot{Q}_{\tt R}}{\dot{Q}_{\tt L}}\leq \eta_{\rm C},
\end{equation}
where $\eta_{\rm C} = 1 - T_{\tt R}/T_{\tt L}$ is the Carnot efficiency. Thus, the performance of an autonomous thermal machine depends on the currents and their relationship to the thermodynamic properties of the reservoirs. 

Evaluating the currents requires finding the NESS of the quantum system. In general, however, the computation of Eq.~\eqref{eq:NESS_unitary} is a difficult task. Analytical solutions are available only if the global Hamiltonian is non-interacting, while a direct numerical approximation with finite baths may require prohibitively large values of $N$ in order to avoid Poincar\'e recurrences within the timescale of relaxation. On the other hand, perturbative schemes are limited to cases where either the internal interactions within $\tt S$ or its couplings to the reservoirs are weak. 

It is thus desirable to consider an alternative approach that is not bounded by these restrictions. Part~\ref{part:two} of this thesis proposes an alternate approach, in which the macroscopic reservoirs are replaced with mesoscopic leads comprising $L$ sites, which are continuously damped towards thermal equilibrium by dissipative processes. As a consequence, as we shall demonstrate, convergence can be obtained with only moderate values of $L$, bringing the non-equilibrium thermodynamics of complex many-body quantum systems within reach.

\subsection{From macroscopic reservoirs to mesoscopic leads}
\label{sec:mesoscopic_leads}

The thermal reservoirs described before are composed of an infinite amount of degrees of freedom. Our purpose now is to describe the reservoirs in a more tractable way, while attempting to retain the properties inherent to infinite, macroscopic reservoirs. 

The main details of our approach to studying the problem described in Sec.~\ref{sec:autonomous_thermal_machines} are outlined in this section, where an infinite bath is replaced by a finite collection of damped modes. For the formal mathematical description, we refer the reader to Appendix~\ref{app:meso_equivalence}.

\begin{figure}[t]
\fontsize{13}{10}\selectfont 
\centering
\includegraphics[width=0.5\columnwidth]{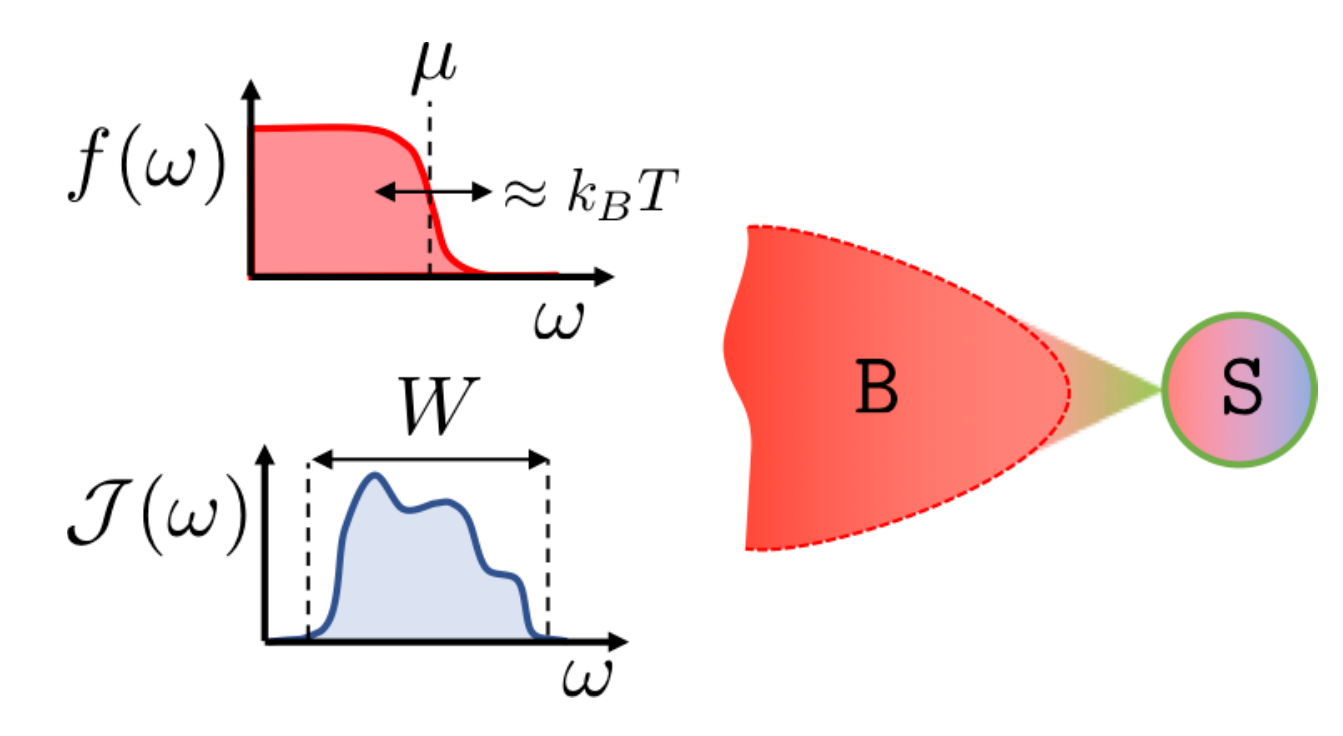}
\caption{The dynamics of a system coupled to a single thermal bath is determined by the bath's spectral density $\mathcal{J}(\omega)$, with a bandwidth $W$, and the Fermi-Dirac distribution $f(\omega)$ corresponding to its chemical potential $\mu$ and temperature $T$.}
\label{fig:2.3.2}
\end{figure}

The system $\tt S$ is assumed to be a lattice of $D$ sites, with arbitrary geometry and interactions, while the baths are modelled by infinite collections of non-interacting spinless fermionic modes. To illustrate the approach, we consider first the case of a single bath $\tt B$, as shown in Fig.~\ref{fig:2.3.2}, described by the Hamiltonian
\begin{align}
\label{eq:H_B_infinite}
\hat{H}_{\tt B} = \sum_{m=1}^\infty \omega_m \hat{b}^{\dagger}_m \hat{b}_m,
\end{align}
where $\hat{b}^\dagger_m$ creates a fermion with energy $\omega_m$. Each site $j$ of the system is described by a fermionic operator $\hat{c}_j$. A particular site $p$ of the system exchanges particles and energy with the bath via a tunnelling interaction
\begin{align}
\label{eq:H_SB_infinite}
\hat{H}_{\tt SB} = \sum_{m=1}^{\infty} \left( \lambda_{m} \hat{c}^{\dagger}_p \hat{b}_m + \lambda^*_{m} \hat{b}^{\dagger}_m \hat{c}_p \right),
\end{align}
where $\lambda_m$ is its coupling to bath mode $m$.

The Heisenberg equation for the system operators reads as
\begin{equation}
    \label{eq:Heisenberg_cj}
    \frac{{\rm d}}{{\rm d} t}\hat{c}_j(t) = \ii [\hat{H}_{\tt S}, \hat{c}_j(t)] + \delta_{jp} \left[\hat{\xi}(t) - \int_0^t {\rm d}t'\, \chi(t-t') \hat{c}_p(t')\right].
\end{equation}
Here, we have defined the noise operator
\begin{equation}
\label{eq:noise_operator}
\hat{\xi}(t) = -{\rm i}\sum_m \lambda_{m}\ee^{-\ii\omega_m t}\hat{b}_m,
\end{equation}
and the memory kernel $\chi(t-t') = \langle \{\hat{\xi}(t),\hat{\xi}^\dagger(t')\}\rangle $. The Gaussian statistics of the noise operator with respect to the initial product state Eq.~\eqref{eq:product_state} are defined by $\langle \hat{\xi}(t)\rangle  = 0$ and
\begin{align}
    \label{eq:memory_kernel}
    \langle \{\hat{\xi}(t),\hat{\xi}^\dagger(t')\}\rangle & = \int \frac{{\rm d}\omega}{2\pi} \mathcal{J}(\omega) \ee^{-\ii \omega(t-t')},\\
    \label{eq:noise_correlations}
    \langle \hat{\xi}^\dagger(t)\hat{\xi}(t')\rangle & = \int \frac{{\rm d}\omega}{2\pi} \mathcal{J}(\omega) f(\omega) \ee^{\ii \omega(t-t')},
\end{align}
where we have defined the spectral density as
\begin{equation}
    \label{eq:spectral_density_def}
    \mathcal{J}(\omega) = 2\pi\sum_{m=1}^\infty |\lambda_{m}|^2 \delta(\omega - \omega_m),
\end{equation}
and introduced the Fermi-Dirac distribution
$f(\omega) = (\ee^{\beta(\omega-\mu)} + 1)^{-1}$. The average system-bath coupling strength is typically quantified as
\begin{equation}
    \label{eq:coup_spectral_density}
    \Gamma = \frac{1}{2W}\int_{-\infty}^\infty {\rm d}\omega\, \mathcal{J}(\omega),
\end{equation}
where $2W$ denotes the reservoir bandwidth, namely the size of the energy range over which $\mathcal{J}(\omega)$ has support. The state of $\tt S$ is completely determined by $f(\omega)$ and $\mathcal{J}(\omega)$ via the noise statistics, since for an overall closed system the solution of Eq.~\eqref{eq:Heisenberg_cj} is sufficient to reconstruct all $n$-point correlation functions.

Our approach is based on a key insight. Namely, that the open-system dynamics in Eq.~\eqref{eq:Heisenberg_cj}, induced by an infinite bath with spectral function $\mathcal{J}(\omega)$, can be accurately approximated by instead coupling the system to a finite collection of damped modes. Indeed, let us consider a {\em lead} of size $L$ coupled to site $p$ of the system, described by the Hamiltonian
\begin{align}
\label{eq:H_lead}
    \hat{H}_{\tt L} & = \sum_{k=1}^L \varepsilon_k \hat{a}^\dagger_k \hat{a}_k,\\
    \label{eq:H_lead_sys}
    \hat{H}_{\tt SL} & = \sum_{k=1}^{L} \left( \kappa_{kp} \hat{c}^{\dagger}_p \hat{a}_k + \kappa^*_{kp} \hat{a}^{\dagger}_k\hat{c}_p \right),
\end{align}
where $\hat{a}^\dagger_k$ creates a fermion in the lead with energy $\varepsilon_k$, and $\kappa_{kp}$ is the coupling strength. Each energy eigenmode $k$ of the lead is coupled to an independent thermal bath modelled by an infinite non-interacting fermion reservoir ${\tt B}_k$, as illustrated in Fig.~\ref{fig:2.3.3} (see Appendix~\ref{app:meso_equivalence} for details). These baths have identical temperatures and chemical potentials, but crucially they are characterised by a structureless frequency-independent spectral density $\mathcal{J}_k(\omega) = \gamma_k$, where $\gamma_k$ is a characteristic damping rate whose value may be different for each bath.

\begin{figure}[t]
\fontsize{13}{10}\selectfont 
\centering
\includegraphics[width=0.825\columnwidth]{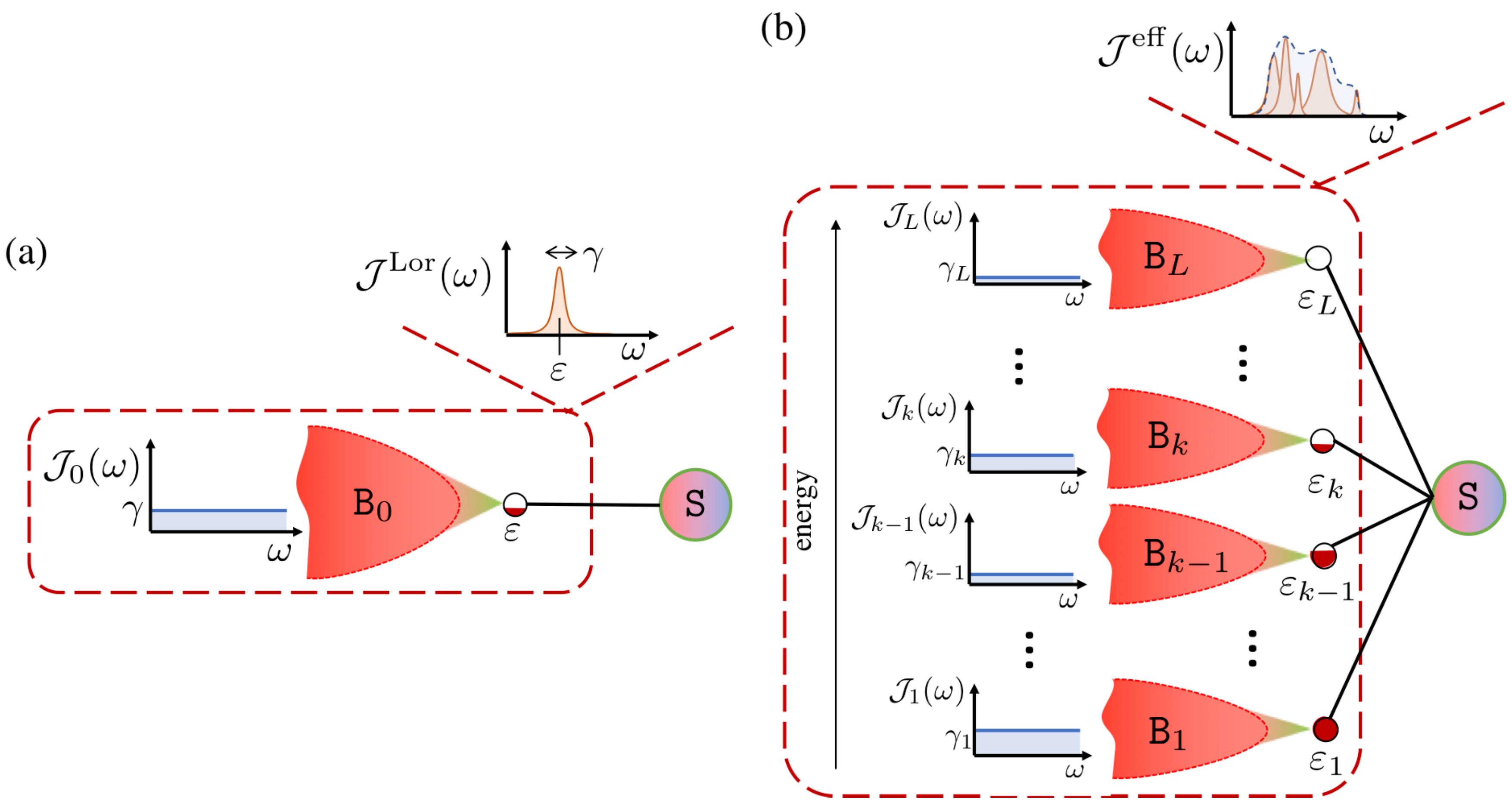}
\caption[Spectral densities from coupling structureless auxiliary modes and effective spectral densities]{(a)~A Lorentzian spectral density $\mathcal{J}^{\rm Lor}(\omega)$ is equivalent to coupling the system to a single auxiliary mode damped by a structureless reservoir. (b)~A mesoscopic reservoir comprising many damped modes gives rise to an effective spectral density $\mathcal{J}^{\rm eff}(\omega)$ that is a sum of Lorentzians. By tuning the damping of each mode and its coupling to the system $\mathcal{J}^{\rm eff}(\omega)$ can approximate $\mathcal{J}(\omega)$ of the infinite bath depicted in Fig.~\ref{fig:2.3.2}.}
\label{fig:2.3.3}
\end{figure}

\begin{figure}[t]
\fontsize{13}{10}\selectfont 
\centering
\includegraphics[width=0.45\columnwidth]{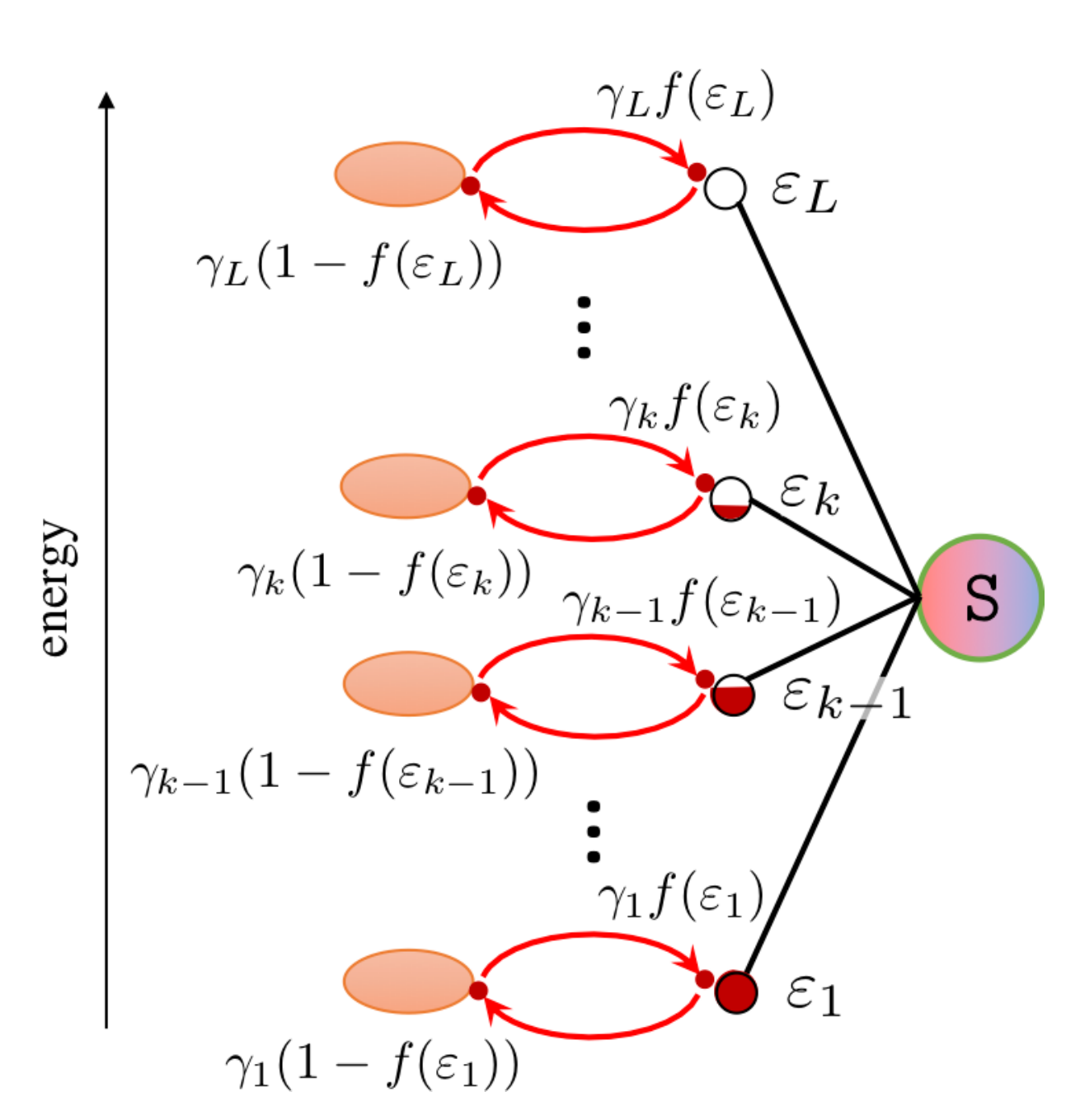}
\caption{In the limit $L\gg 1$ modes in the lead each bath ${\tt B}_k$ is sufficiently weakly coupled its corresponding lead mode that it can be accurately modelled by a Lindblad dissipator. The dissipator on a lead mode then that injects and ejects fermions at rates which in isolation damp the mode into a thermal state.}
\label{fig:2.3.4}
\end{figure}

To analyse the steady-state physics it is sufficient to focus on long times, such that $t \gg \gamma_k^{-1},\tau_{\rm rel}$. Here $\tau_{\rm rel}$ represents the characteristic relaxation timescale of ${\tt S}$ due to its coupling with the bath\footnote{Note that some systems, such as glassy systems, may never relax when coupled to a bath. In such cases, our arguments regarding the equivalence of mesoscopic and infinite reservoirs do not hold. Indeed, one expects that for such systems the effect of a bath must be highly dependent on the microscopic details of the bath and its coupling to the system.}. In this limit, we find that the Heisenberg equations for the system variables in this configuration are identical to Eq.~\eqref{eq:Heisenberg_cj}, but the statistics of the noise operator are now determined by an effective spectral density
\begin{equation}
    \label{eq:spectral_density_effective}
    \mathcal{J}^{\rm eff}(\omega)  = \sum_{k=1}^L \frac{|\kappa_{kp}|^2 \gamma_k}{(\omega-\varepsilon_k)^2 + (\gamma_k/2)^2}.
\end{equation}
It follows that this damped mesoscopic lead configuration reproduces the correct steady state of $\tt S$, so long as the true spectral density $\mathcal{J}(\omega)$ can be well approximated by a sum of Lorentzians as above. This is depicted in Fig.~\ref{fig:2.3.3}. In particular, consider a given set of lead energies $\varepsilon_k$ that sample the spectral density and are arranged in ascending order, with energy spacing $e_k = \varepsilon_{k+1} - \varepsilon_k$. By taking $\kappa_{kp} = \sqrt{\mathcal{J}(\varepsilon_k)e_k/2\pi}$ and $\gamma_k = e_k$, we have $\gamma_k \sim L^{-1}$ so that Eq.~\eqref{eq:spectral_density_effective} reduces to Eq.~\eqref{eq:spectral_density_def} in the limit $L\to \infty$. We therefore obtain a controlled approximation of the bath spectral function as the lead size $L$ increases.

In order to obtain a tractable description of the augmented system-lead configuration, we use the fact that both the damping rates $\gamma_k$ and the coupling constants $\kappa_{kp}$ are small in the large-$L$ limit. Tracing out the baths, we derive a master equation describing the joint state of $\tt S$ and $\tt L$, valid  for times $ t\gg \gamma_k^{-1},\tau_{\rm rel}$ and up to second order in both the lead-bath and system-lead coupling (see Appendix~\ref{app:meso_equivalence}). We emphasise that the assumption that \textit{individual} modes of the lead couple weakly to the system does not imply that the overall system-bath coupling $\Gamma$ is weak. The quantum master equation is
\begin{align}
    \label{eq:Lindblad}
    \frac{{\rm d}\hat{\rho}}{{\rm d}t} & = \ii[\hat{\rho},\hat{H}] + \mathcal{L}_{\tt L}\{\hat{\rho}\},
\end{align}
where $\hat{H} = \hat{H}_{\tt S} + \hat{H}_{\tt L} + \hat{H}_{\tt SL}$ denotes the Hamiltonian of the system and lead, while thermalisation of the lead is described by the Lindblad dissipator
\begin{align}
\label{eq:dissipator}
\mathcal{L}_{\tt L}\{\hat{\rho}\} = \sum_{k=1}^{L} \gamma_k(1 - f_k) \left[\hat{a}_k \hat{\rho} \hat{a}^{\dagger}_k - \tfrac{1}{2}\{ \hat{a}^{\dagger}_k \hat{a}_k, \hat{\rho} \} \right] + \sum_{k=1}^{L} \gamma_k f_k \left[\hat{a}^{\dagger}_k \hat{\rho} \hat{a}_k - \tfrac{1}{2}\{ \hat{a}_k \hat{a}^{\dagger}_k, \hat{\rho} \} \right].
\end{align}
with $f_k = f(\varepsilon_k)$ denoting the sampling of the Fermi distribution by the lead modes. This master equation configuration is illustrated in Fig.~\ref{fig:2.3.4}. 

The above representation does not simplify the problem a priori, since it is strictly valid only in the large-$L$ limit. However, a simplification may arise if the expectation values of operators converge with increasing $L$. We show numerically in later sections that this convergence occurs rapidly in several examples of interest for quantum thermodynamics. In such cases, a tractable number of lead sites $L$ can be used to obtain a good approximation of an infinite bath with a continuous spectral density. For this, it is crucial that $\gamma_k$ remains the smallest energy scale in the physical configuration, to both model the spectral function correctly and accurately approximate the baths via the Lindblad equation~\cite{Landi:2016,Reichental2018}. 

So far we have considered a single bath coupled to a particular site of the system. However, the above results are easily generalised to describe the situation of several sites connected to multiple baths at different temperatures and chemical potentials. The steps of the above analysis are carried out independently for each bath, leading to additive contributions to the master equation. 

\section{Landauer-B\"uttiker theory}
\label{sec:landauer_buttiker}

Also known as scattering method, the Landauer-B\"uttiker formalism is a well-established theory to describe transport of electrons across quantum junctions from the wave-functions in quantum mechanics. In its own right, it provides a very simple and powerful framework to study currents of non-interacting particles. Furthermore, the requirement of interacting effects could in principle be relaxed if an effective quasi-particle description suffices, whenever a mean-field approximation is valid such that inelastic scattering effects can be neglected. The second condition associated to the applicability of Landauer-B\"uttiker theory is the absence of environmental effects that destroy coherence. In this sense then, the formalism can be applied whenever {\em coherent} transport is at play~\cite{ryndyk2016nano}.

Crucially, however, the theory is fully consistent from the perspective of thermodynamics whenever a non-interacting system is connected to reservoirs with an infinite number of degrees of freedom. This implies that temperature-dependent effects are fully captured by the theory in such configurations. In our work, Landauer-B\"uttiker theory shall be employed as a benchmark, to understand if the mesoscopic lead construction described in Sec.~\ref{sec:mesoscopic_leads} is consistent even when the reservoir is composed of a finite number of damped modes. 

The theory is based on the scattering effects induced by a central system, when it is coupled locally to reservoirs on each side. Let us assume, without any loss in generality, that we want to study currents in a configuration such as the one depicted in Fig.~\ref{fig:2.3.1}. Consider the fixed temperature configuration $T_{\tt L} = T_{\tt R}$ and a chemical potential difference such that $\mu_{\tt L} > \mu_{\tt R}$. With these parameters, a steady-state flow of particles from left to right should be achieved. The main point of Landauer-B\"uttiker theory is that the central system with Hamiltonian $\hat{H}_{\tt S}$ acts as scatterer for the flow of particles between the two reservoirs.

The main assumption within the theory~\cite{ryndyk2016nano} is that the particles moving from left to right are populated with the equilibrium distribution of the reservoir on the left, while the particles flowing from right to left are populated, accordingly, with the equilibrium distribution of the reservoir on the right. Under this assumption, it is straightforward to derive (in favour of plausible arguments~\cite{ryndyk2016nano}) that the total particle current flowing in the configuration is given by~($\hbar = e = k_{\textrm{B}} = 1$)
\begin{align}
\label{eq:partlb}
J^{\textrm{P}}_{\textrm{LB}} & = \frac{1}{2\pi}\int_{-W}^{W} {\rm d} \omega\, \tau(\omega) [ f_{\tt L}(\omega) - f_{\tt R}(\omega) ],
\end{align}
where $\tau(\omega)$ is an energy-dependent {\em global} transmission function that describes the probability of a particle to flow across the scatterer. It is natural to assume that such function depends on the microscopic details of $\hat{H}_{\tt S}$. On the other hand, there exists a concomitant energy current, which follows trivially from the particle current since the lack of incoherent effects implies that the particles are the only carriers of energy, given by
\begin{align}
\label{eq:enerlb}
J^{\textrm{E}}_{\textrm{LB}} & = \frac{1}{2\pi}\int_{-W}^{W} {\rm d} \omega\,  \omega\tau(\omega) [ f_{\tt L}(\omega) - f_{\tt R}(\omega) ].
\end{align}
In principle, the limits of integration should be taken to infinity, i.e., $W \to \infty$. However, the only channels available for transport are those dictated by the energies of the central system~\cite{ryndyk2016nano}, which implies that the limits of integration can be kept finite as long as $W$ is much larger than the bandwidth of $\hat{H}_{\tt S}$.

\subsection{Transmission functions in Landauer-B\"uttiker theory}
\label{sec:transmission}

As long as we can provide a description of the transmission functions $\tau(\omega)$, we can fully characterise transport from Landauer-B\"uttiker theory. In this section we briefly introduce the methodology to compute the transmission functions $\tau(\omega)$ from Eqs.~\eqref{eq:partlb} and \eqref{eq:enerlb}. As remarked before, these functions are required to compute the currents in Landauer-B\"uttiker theory which correspond to our point of comparison for non-interacting systems.

The transmission function can be obtained in terms of the non-equilibrium Green's function~\cite{ryndyk2016nano,Purkayastha:2019}
\begin{align}
\mathbf{G}(\omega) = \mathbf{M}^{-1}(\omega).
\end{align}
For the specific case of a system composed of $D$ fermionic sites connected to leads on sites $j = 1$ and $j = D$, $\mathbf{M}(\varepsilon)$ can be expressed as
\begin{align}
\mathbf{M}(\omega) = \omega \mathds{1} - \mathbf{H}_{\tt S} - \mathbf{\Sigma}^{(1)}(\omega) - \mathbf{\Sigma}^{(D)}(\omega),
\end{align}
where $\mathbf{H}_S$ is the Hamiltonian matrix of the system and $\mathbf{\Sigma}(\omega)$ corresponds to the self-energy matrices of the leads. We remark that, in order to employ Landauer-B\"uttiker theory, $\hat{H}_S$ must either be non-interacting or fall within the range of applicability of the mean-field approximation. In other words, for fermionic systems, $\hat{H}_S$ must be composed of, at most, quadratic fermionic field operators.
 
For a system with $D$ fermionic sites, all these matrices have $D \times D$ elements. For the specific case of $\mathbf{\Sigma}^{(1)}(\omega)$ and $\mathbf{\Sigma}^{(D)}(\omega)$, the only non-zero matrix elements (due to local coupling) are given by
\begin{align}
[\mathbf{\Sigma}^{(j)}]_{jj} (\omega) = \frac{1}{2\pi} \textrm{P.V.}\int d \omega^{\prime} \frac{\mathcal{J}(\omega^{\prime})}{(\omega^{\prime} - \omega)} - \frac{\textrm{i}}{2}\mathcal{J}(\omega),\; \forall j = 1, D;
\end{align}
where P.V. denotes principal value and $\mathcal{J}(\omega)$ is the spectral function of the leads. In our configuration, both leads are of equivalent form. For the sake of comparison between L-B theory and mesoscopic reservoirs, we employ the wide-band approximation in which 
\begin{align}
\mathcal{J}(\omega) = \begin{cases} \Gamma,\; \forall\, \omega \in [-W, W] \\ 0,\; \textrm{otherwise} \end{cases}
\end{align}
where $\Gamma$ is the coupling strength between the system and the leads. Let us know consider a specific case, in which the central system is a collection of $D$ fermionic modes in a one-dimensional chain, such that
\begin{align}
\label{eq:h_s_m}
\hat{H}_{\tt S} = \sum_{j = 1}^{D} \epsilon_j \hat{c}^{\dagger}_j \hat{c}_j - \sum_{j = 1}^{D - 1} t_{\tt S} \left( \hat{c}^{\dagger}_{j+1} \hat{c}_{j} + \textrm{H.c.} \right),
\end{align}
Under these considerations, the transmission function for a system composed of $D$ fermionic sites with $\hat{H}_{\tt S}$ from Eq.~\eqref{eq:h_s_m} is given by
\begin{align}
\label{eq:transm}
\tau(\varepsilon) = \mathcal{J}^2(\varepsilon) | [\mathbf{G}(\varepsilon)]_{1D} | ^2 = \frac{\mathcal{J}^2(\varepsilon)}{| \textrm{det} [\mathbf{M}] |^2} \prod_{i=1}^{D-1}t_{S,i}^2.
\end{align}
When the central system is a single-level with $\hat{H}_{\tt S} = \epsilon \hat{c}^{\dagger} \hat{c}$, the transmission function can be proven to be of Lorentzian form and equivalent to
\begin{align}
\label{eq:lorentzian_transmission}
\tau_{\rm SL}(\varepsilon) = \frac{\mathcal{J}^2(\varepsilon)}{| \textrm{det} [\mathbf{M}] |^2},
\end{align}
while a central system composed of $D$ fermionic sites with $\hat{H}_S$ from Eq.~\eqref{eq:h_s_m} has a transmission function which corresponds to a convolution of Lorentzian functions whose form depends on the site energies $\epsilon$ and hopping amplitudes $t_{\tt S}$, as observed from Eq.~\eqref{eq:transm}. With the previous expressions for $\tau(\varepsilon)$, Eqs.~\eqref{eq:partlb} and \eqref{eq:enerlb} can then be evaluated numerically to obtain particle and energy currents for a given system.

\begin{figure}[t]
\fontsize{13}{10}\selectfont 
\centering
\includegraphics[width=0.7\columnwidth]{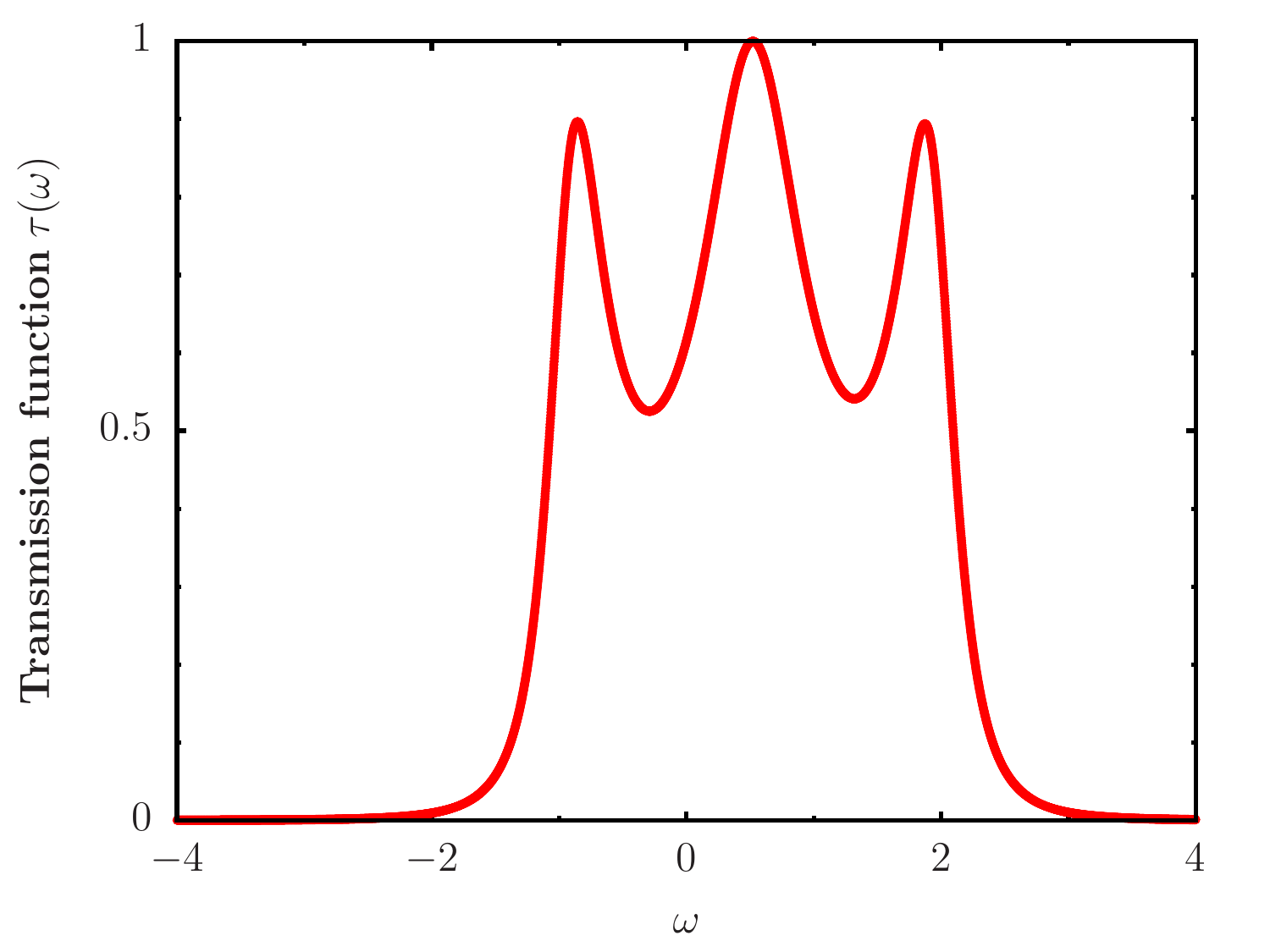}
\caption[Example of a transmission function for a fermionic chain]{Example of a transmission function for a fermionic chain, $\hat{H}_{\tt S}$ from Eq.~\eqref{eq:h_s_m} with $D = 3$ and $t_{\tt S} = 1.0$. For this example we chose $\epsilon_1 = 1/4$, $\epsilon_2 = 1/2$ and $\epsilon_3 = 3/4$.}
\label{fig:2.3.5}
\end{figure}

In Fig.~\ref{fig:2.3.5} we show an example of a transmission function obtained for a fermionic chain with three sites using the procedure denoted above. From Eq.~\eqref{eq:h_s_m}, we chose $t_{\tt S} = 1.0$, $\epsilon_1 = 1/4$, $\epsilon_2 = 1/2$ and $\epsilon_3 = 3/4$. The overall shape of the transmission function will depend on the parameters chosen. It can be observed, as well, that the transmission function decays rather quickly as a function of $\omega$, which in turn depends on the energy bandwidth of the Hamiltonian. It demonstrates that the integrals Eq.~\eqref{eq:partlb} and Eq.~\eqref{eq:enerlb} can be truncated at a finite $W$ since the contributions will decay as $\omega$ increases (decreases). This procedure shall be employed in multiple instances in the following sections, to evaluate currents from Landauer-B\"uttiker theory as a benchmark for our mesoscopic configuration in Sec.~\ref{sec:mesoscopic_leads}.

\section{Superfermion approach for the dynamics of non-equilibrium configurations}
\label{sec:superfermion}

In order to solve the dissipative dynamics under a master equation of the form in Eq.~\eqref{eq:Lindblad}, we use the superfermion formalism introduced in Ref.~\cite{Dzhioev2011}. For a non-interacting (quadratic) open system, this method provides numerically tractable analytical expressions for steady-state quantities. The superfermion representation is also central to our approach to simulating interacting systems. Here, we limit ourselves to a concise review of the formalism; for more details, see Appendix~\ref{ap:superf}.

The superfermion approach is akin to a purification or thermofield scheme for open systems. It doubles the system size by introducing a new fermionic ancilla mode for each of the modes present in the system and leads. To describe the formalism succinctly we stick for now to the single-lead setup of Eq.~\eqref{eq:Lindblad}. In order to distinguish clearly between the ancillary modes and the physical modes of the system and lead, we introduce a unified notation for the latter. In this single-lead setup the total number of system and lead modes is $M = D+L$ and so we define $M$ fermion mode operators
\begin{eqnarray}
    \hat{d}_k \defeq \left\{
\begin{array}{cl}
 \hat{a}_k & \quad k=1,\ldots,L   \\
 \hat{c}_k & \quad k=(L+1),\ldots,M 
\end{array}
\right. \label{eq:unified_notation}
\end{eqnarray}
The ancillary modes are described by $M$ additional canonical creation and annihilation operators $\hat{s}^{\dagger}_k$ and $\hat{s}_k$. We use an interleaved ordering for the physical and ancillary operators, so that the Fock basis of the combined Hilbert space is defined by
\begin{align}
\label{eq:superfermion_Fock_state}
\ket{\underline{n}| \underline{m}} = ( \hat{d}_1^{\dagger} )^{n_1}& ( \hat{s}_1^{\dagger} )^{m_1} \cdots ( \hat{d}_{M}^{\dagger} )^{n_{M}} ( \hat{s}_{M}^{\dagger} )^{m_{M}} \ket{\textrm{vac}}.
\end{align}
Here $\underline{n}$ are $\underline{m}$ are binary strings of length $M$ that describe occupation numbers for the physical and ancillary modes, respectively. While the ordering used for the Fock basis is entirely arbitrary, we shall see shortly that interleaving has useful locality properties, which we will exploit later when we introduce interacting systems. We now define a new (unnormalised) ket vector called the {\em left vacuum} as
\begin{align}
\ket{I} \defeq \sum_{\underline{n}} \ket{\underline{n} |\underline{n}},
\end{align}
where the sum runs over all $2^{M}$ binary strings $\underline{n}$. Using this ket, we can define a quantum state representing the system-lead density operator as
\begin{align}
\hat{\rho}(t)\ket{I} = \ket{\hat{\rho}(t)},
\end{align}
and the expectation values of any system or lead operator $\hat{A}$ as
\begin{align}
\label{eq:expec}
\braket{ I | \hat{A} | \hat{\rho}(t) } = \langle \hat{A}(t) \rangle.
\end{align}

A key aspect of this formalism are the conjugation relations allowing physical creation (annihilation) operators to be swapped for ancillary annihilation (creation) operators. For the interleaved Fock ordering these conjugation relations are given by
\begin{align}
\hat{d}^{\dagger}_j \ket{I} &= -\hat{s}_j \ket{I},\quad \bra{I}\hat{d}_j = -\bra{I}\hat{s}^{\dagger}_j, \nonumber \\
\hat{d}_j \ket{I} &= \phantom{-}\hat{s}^{\dagger}_j \ket{I},\quad \bra{I}\hat{d}^{\dagger}_j = \phantom{-}\bra{I}\hat{s}_j.
\end{align}
Acting the master equation Eq.~\eqref{eq:Lindblad} on $\ket{I}$ and using the conjugation relations yields a Schr\"odinger-type equation for the state,
\begin{align}
\frac{\rm d}{{\rm d}t} \ket{\hat{\rho}(t)} = -\textrm{i}\hat{L}\ket{\hat{\rho}(t)}, \label{eq:superfermion_evolve}
\end{align}
with the (non-Hermitian) generator of time evolution given by
\begin{align}
\label{eq:l_single}
\hat{L} &= \hat{H} - \hat{H}_{d\Leftrightarrow s} \nonumber - \textrm{i}\sum_{k=1}^{L} \gamma_k f_k\\
&\quad-\frac{\textrm{i}}{2}\sum_{k=1}^{L} \gamma_k (1 - 2f_k) \left( \hat{d}^{\dagger}_{k} \hat{d}_{k} + \hat{s}^{\dagger}_{k} \hat{s}_{k} \right)  \nonumber \\
&\quad+\textrm{i}\sum_{k=1}^{L} \gamma_k \left( f_k \hat{d}^{\dagger}_{k} \hat{s}^{\dagger}_{k} -(1-f_k) \hat{d}_{k} \hat{s}_{k}  \right),
\end{align}
where $\hat{H}_{d\Leftrightarrow s}$ is the same as the system-lead Hamiltonian $\hat{H}$ but with all physical operators replaced by their ancillary counterparts, $\hat{d}_k \to \hat{s}_k$. Crucially, dissipative processes are now described by non-Hermitian quadratic operators that, according to the interleaved mode ordering of Eq.~\eqref{eq:superfermion_Fock_state}, couple only nearest neighbours $\hat{d}_{k}$ and $\hat{s}_{k}$. The formalism generalises straightforwardly to multiple leads by introducing an additional ancilla mode needed for each additional lead mode.

So far the superfermion formalism is entirely general. In the special case where the system Hamiltonian $\hat{H}_{\tt S}$ is non-interacting the formalism provides a compact expression for the exact solution of the NESS. In this case the system-lead Hamiltonian is quadratic with the form  
\begin{equation}
    \label{eq:quadratic_Hamiltonian}
    \hat{H} = \sum_{i,j = 1}^M [{\bf H}]_{ij} \hat{d}^\dagger_i \hat{d}_j,
\end{equation}
where ${\bf H}$ is an Hermitian $M\times M$ matrix. Next we define $M\times M$ diagonal matrices $\mathbf{\Gamma}_+$ and $\mathbf{\Gamma}_-$ containing the injection and ejection rates of fermions for each site. Specifically, for the single-lead setup the first $L$ follow the thermal damping rates contained in the dissipator Eq.~\eqref{eq:dissipator}, while the last $D$ entries corresponding to the system modes are zero, giving
\begin{eqnarray}
  \mathbf{\Gamma}_+ &=& {\rm diag}\Big(\gamma_1 f_1,\dots,\gamma_L f_L,0,\dots,0\Big), \nonumber \\
  \mathbf{\Gamma}_- &=& {\rm diag}\Big(\gamma_1 (1-f_1),\dots,\gamma_L (1-f_L),0,\dots,0\Big). \nonumber
\end{eqnarray}
Using these we define two additional diagonal matrices $\mathbf{\Lambda} = (\mathbf{\Gamma}_- + \mathbf{\Gamma}_+)/2$ and $\mathbf{\Omega} = (\mathbf{\Gamma}_- - \mathbf{\Gamma}_+)/2$. Consequently, for the case of a non-interacting system the generator $\hat{L}$ is quadratic with the form
\begin{align}
    \label{eq:L_generator_quadratic}
\hat{L} &= \hat{\mathbf{f}}^{\dagger} \begin{pmatrix} \mathbf{H} - \textrm{i}\mathbf{\Omega} & \textrm{i}\mathbf{\Gamma}_+ \\ \textrm{i}\mathbf{\Gamma}_- & \mathbf{H} + \textrm{i}\mathbf{\Omega} \end{pmatrix} \hat{\mathbf{f}} - \textrm{Tr}\left( \mathbf{H} + \textrm{i}\mathbf{\Lambda} \right) \nonumber \\
&= \hat{\mathbf{f}}^{\dagger}\, \mathbf{L}\, \hat{\mathbf{f}} - \eta,
\end{align}
where $\mathbf{\hat{f}}=(\hat{d}_1,\ldots,\hat{d}_M,\hat{s}^\dagger_1,\ldots,\hat{s}^\dagger_M)^{\rm T}$ is the full $2M$-dimensional column vector of all physical and ancillary operators\footnote{Note that the ordering of operators in this vector is completely unrelated to that used to define the Fock basis.}. 

To determine the NESS we diagonalise $\hat{L}$ by a similarity transformation, $\mathbf{L} = \mathbf{V} \,\bm{\epsilon}\, \mathbf{V}^{-1}$, to find the complex eigenvalues $\bm{\epsilon} = \mathrm{diag}(\epsilon_1,\ldots,\epsilon_{2M})$ and the matrix of right eigenvectors $\mathbf{V}$ of $\mathbf{L}$. As shown in Appendix~\ref{ap:superf}, the many-body NESS is a Fermi-sea-like state in which only modes with $\Im(\epsilon_\mu)>0$ are occupied, furnishing us with a complete solution of the problem. In particular, two-point correlation functions of physical modes in the NESS are found to be
\begin{equation}
    \label{eq:RDM_SF_quadratic}
    \langle \hat{d}_i^\dagger \hat{d}_j\rangle = [\mathbf{V}\,\mathbf{D}\,\mathbf{V}^{-1}]_{ji},
\end{equation}
where $\mathbf{D}_{\mu\nu} = \delta_{\mu\nu} \Theta(\Im\{\epsilon_\mu\})$, with $\Theta(x)$ the Heaviside step function. This gives an efficient prescription to find steady state observables such as currents for non-interacting systems, while higher-order correlation functions follow from Wick's theorem.

\section{Non-equilibrium thermodynamics with mesoscopic leads}
\label{sec:thermo_meso}

\begin{figure}[t]
\fontsize{13}{10}\selectfont 
\centering
\includegraphics[width=0.6\columnwidth]{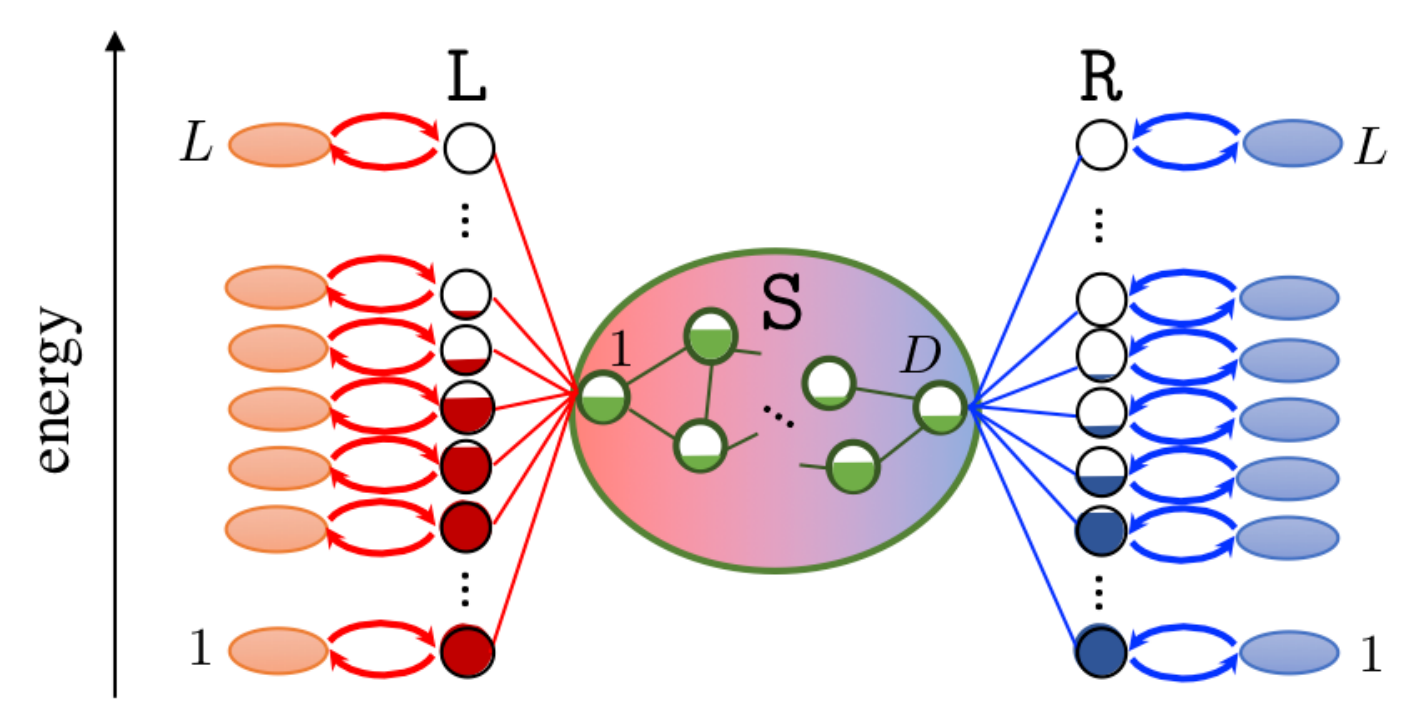}
\caption{The Lindblad mesoscopic lead approximation of the simple thermal machine setup shown in Fig.~\ref{fig:2.3.1} where some generic system $\tt S$ is coupled to two reservoirs with differing chemical potentials and temperatures.}
\label{fig:2.3.6}
\end{figure}

The central focus of our work is autonomous thermal machines in the two-lead configuration illustrated in Fig.~\ref{fig:2.3.6}, with mesoscopic reservoirs labelled by $\alpha = {\tt L},{\tt R}$. These two leads of size $L$ are described by Hamiltonians of the form Eq.~\eqref{eq:H_lead} and Eq.~\eqref{eq:H_lead_sys}, where the left lead couples to the first system site, $p=1$, and the right lead to the last system site, $p=D$. Each lead is also acted on by a dissipator of the form given in Eq.~\eqref{eq:dissipator}. The master equation for this set-up thus reads as
\begin{equation}
    \label{eq:two_terminal_ME}
    \frac{{\rm d}\hat{\rho}}{{\rm d}t} = \ii[\hat{\rho},\hat{H}] + \mathcal{L}_{\tt L}\{\hat{\rho}\}+ \mathcal{L}_{\tt R}\{\hat{\rho}\},
\end{equation}
where $\hat{H} = \hat{H}_{\tt S} + \hat{H}_{\tt L} + \hat{H}_{\tt R} + \hat{H}_{\tt S L}+ \hat{H}_{\tt SR}$. 

To find expressions for the particle and energy currents, we need to consider the continuity equations for the total particle-number operator $\hat{N} =  \hat{N}_{\tt S} + \hat{N}_{\tt L} + \hat{N}_{\tt R}$ and total energy operator $\hat{H}$ for the system and the leads. Since $[\hat{H},\hat{N}] = 0$, we derive
\begin{align}
\label{eq:continuity_equations}
\frac{{\rm d}\langle\hat{N}\rangle}{{\rm d}t}  =  J^{\rm P}_{\tt L} + J^{\rm P}_{\tt R}, \quad
\frac{{\rm d}\langle\hat{H}\rangle}{{\rm d}t}  =  J^{\rm E}_{\tt L} + J^{\rm E}_{\tt R},
\end{align}
where $J^{\rm P}_\alpha$ and $J^{\rm E}_\alpha$ are respectively the particle and energy currents flowing into the entire configuration via lead $\alpha$, given by
\begin{align}
    \label{eq:currents_def}
    J^{\rm P}_\alpha = \Tr \left[ \hat{N} \mathcal{L}_\alpha \{\rho\}\right], \quad {\rm and} \quad  J^{\rm E}_\alpha = \Tr \left[ \hat{H} \mathcal{L}_\alpha \{\rho\}\right].
\end{align}
In the NESS, the time derivatives in Eqs.~\eqref{eq:continuity_equations} vanish. Defining positive currents to flow across the system from left to right, we thus take $J^{\rm P} = J^{\rm P}_{\tt L} = -J^{\rm P}_{\tt R}$ and similarly $J^{\rm E} = J^{\rm E}_{\tt L} = - J^{\rm E}_{\tt R}$. The currents are straightforward to evaluate using the adjoint dissipator $\mathcal{L}_{\alpha}^\dagger$, for $\alpha = L,R$, which satisfies ${\rm Tr}[\hat{A} \mathcal{L}_\alpha\{\hat{B}\}] = {\rm Tr}[\mathcal{L}_\alpha^\dagger\{\hat{B}\} \hat{A}]$ for an arbitrary operator $\hat{A}$. For the Lindblad dissipator in Eq.~\eqref{eq:Lindblad}, we have
\begin{align}
    \mathcal{L}_{\tt L}^\dagger \{ \bullet\} & = \sum_{k=1}^{L} \gamma_k(1 - f(\varepsilon_k)) \left[\hat{a}_k^{\dagger} \bullet \hat{a}_k - \tfrac{1}{2}\{ \hat{a}^{\dagger}_k \hat{a}_k, \bullet \} \right] \nonumber \\
& + \sum_{k=1}^{L} \gamma_k f(\varepsilon_k) \left[\hat{a}_k \bullet \hat{a}^{\dagger}_k - \tfrac{1}{2}\{ \hat{a}_k \hat{a}^{\dagger}_k, \bullet\} \right].
\end{align}
Since this superoperator acts only on the lead degrees of freedom, we find the explicit expressions for the currents flowing from the leads and into the system
\begin{align}
\label{eq:partsf_explicit}
J^{\rm P} & = \sum_{k=1}^L \gamma_k  \left\langle  f_{{\tt L},k} - \hat{a}_k^\dagger \hat{a}_k\right\rangle, \\
\label{eq:enersf_explicit}
J^{\rm E}  & = \sum_{k=1}^L \gamma_k \varepsilon_k  \left \langle f_{{\tt L},k} - \hat{a}_{k}^\dagger \hat{a}_{k} \right \rangle \notag \\
& \qquad -  \frac{1}{2}\sum_{k=1}^L \gamma_k  \left \langle \kappa_{k1}\hat{c}_1^\dagger \hat{a}_{k}  + \kappa_{k1}^* \hat{a}^\dagger_{k} \hat{c}_1 \right\rangle,
\end{align}
where the sum runs over only the modes of the left lead with $f_{\tt L}(\varepsilon) = (\ee^{\beta_{\tt L}(\varepsilon-\mu_{\tt L})} +1 )^{-1}$ being its corresponding equilibrium distribution and $f_{{\tt L},k} = f_{\tt L}(\varepsilon_k)$. Interestingly, the second term of Eq.~\eqref{eq:enersf_explicit}, which describes the effect of thermal dissipation on the system-lead interaction energy, must be taken into account in order to obtain a conserved energy current. 

For sufficiently large systems with short-range interactions\footnote{For nearest-neighbour interactions $D \geq 3$ is sufficient.}, it is possible to define current operators $\hat{J}_{\tt S}^{{\rm P},{\rm E}}$ supported only on ${\tt S}$. In this case, the fermion number and Hamiltonian can be written as
\begin{equation}
    \label{N_H_1D}
    \hat{N}_{\tt S} = \sum_{j=1}^{D} \hat{n}_j, \qquad \hat{H}_{\tt S} = \sum_{j=1}^{D-1} \hat{h}_{j,j+1},
\end{equation}
where $ \hat{n}_j = \hat{c}_j^\dagger \hat{c}_j$ is the local fermion density on site $j$ and $\hat{h}_{j,j+1}$ denotes a local energy density operator. Since $\hat{h}_{j,j+1}$ has support only on sites $j$ and $j+1$, we derive the continuity equation for number density from the Heisenberg equation for $\hat{n}_j$:
\begin{equation}
    \label{cont_n_sys}
    \frac{\rm d}{{\rm d} t} \hat{n}_j = \hat{J}^{\textrm{P}}_{j-1\to j} -  \hat{J}^{\textrm{P}}_{j\to j+1},
\end{equation}
where we defined the particle current operator 
\begin{equation}
\hat{J}^{\textrm{P}}_{j-1\to j} = {\rm i} [\hat{h}_{j-1,j}, \hat{n}_j],    
\end{equation}
 which clearly depends only on system variables. In the steady state, the time derivatives of all expectation values vanish and we find that the current is homogeneous, i.e.\ $\langle \hat{J}^{\textrm{P}}_{j-1\to j}\rangle = \langle \hat{J}^{\textrm{P}}_{j\to j+1}\rangle$. Eq.~\eqref{cont_n_sys} holds only for $j\neq 1,D$. For $j=1$, for example, we have instead that \begin{equation}
    \label{cont_n_site1}
    \frac{\rm d}{{\rm d} t} \hat{n}_1 =  {\rm i}[\hat{H}_{\tt SL}, \hat{n}_{1}] - \hat{J}^{\textrm{P}}_{1\to 2}.
\end{equation}
Meanwhile, the mean number of particles in the left reservoir obeys the equation
\begin{equation}
    \label{cont_N_L}
    \frac{\rm d}{{\rm d} t} \left\langle \hat{N}_{\tt L} \right \rangle = J^{\textrm{P}}_{\tt L} + \left\langle {\rm i}[\hat{H}_{\tt SL}, \hat{n}_{1}]\right\rangle.
\end{equation}
Here we used the fact that $[\hat{H}_{\tt SL}, \hat{N}_{\tt L} + \hat{n}_1] = 0$, which merely reflects the overall conservation of fermion number and the fact that ${\tt L}$ couples only to site $j=1$. Combining Eqs.~\eqref{cont_n_site1} and \eqref{cont_N_L} and assuming steady-state conditions we deduce that
\begin{equation}
    \label{J_P_equiv}
    J^{\textrm{P}}_{\tt L} = \left\langle \hat{J}^{\textrm{P}}_{1\to 2}\right \rangle.
\end{equation}
Therefore, so long as the system comprises $D\geq 2$ sites, the current computed via Eq.~\eqref{eq:partsf_explicit} coincides with the expectation value of a system operator.

For the energy current, one similarly finds in the bulk of the system 
\begin{equation}
    \label{cont_h_sys}
    \frac{\rm d}{{\rm d} t} \hat{h}_{j,j+1} = \hat{J}^{\textrm{E}}_{j-1\to j+1} -  \hat{J}^{\textrm{E}}_{j\to j+2},
\end{equation}
where
\begin{equation}
\hat{J}^{\textrm{E}}_{j-1\to j+1} = {\rm i} [\hat{h}_{j-1,j},\hat{h}_{j,j+1}].
\end{equation}
Considering the leftmost site, on the other hand, 
\begin{equation}
    \label{cont_h1}
    \frac{\rm d}{{\rm d} t} \hat{h}_{1,2} = {\rm i}[\hat{H}_{\tt SL},\hat{h}_{1,2}] - \hat{J}^{\textrm{E}}_{1\to 3}.
\end{equation}
Now, considering the Heisenberg equations for both $\hat{H}_{\tt SL}$ and $\hat{H}_{\tt L}$ and assuming steady-state conditions, we conclude that
\begin{equation}
    \label{J_E_equiv}
    J^{\textrm{E}}_{\tt L} = \left\langle \hat{J}^{\textrm{E}}_{1\to 3}\right \rangle.
\end{equation}
Therefore, the energy current computed from Eq.~\eqref{eq:enersf_explicit} also coincides with the expected value of a system operator, so long as $D\geq 3$. 

The above arguments, although developed for the specific case of two-body interactions in one dimension, are based only on conservation laws and the locality of interactions, which are general principles. Similar arguments can thus be developed for more general $n$-body interacting systems in higher-dimensional geometries, so long as a sufficiently large region of the central system is not directly connected to the baths. However, in some cases, e.g.\ if ${\tt S}$ comprises just a single lattice site, no system operator for the currents can be defined. Nevertheless, whether or not such a system operator exists, the average currents computed from Eqs.~\eqref{eq:partsf_explicit} and~~\eqref{eq:enersf_explicit} converge to the infinite-reservoir prediction when $L\to \infty$.

\subsection{Non-interacting example: the resonant-level heat engine}
\label{sec:non_interacting_example}

In this section, we apply our methods to analyse the performance of an autonomous thermal machine with a non-interacting working medium. Since exact results are available here for the $L\to\infty$ limit, this serves as a benchmark to evaluate the performance of the mesoscopic-reservoir formalism which can also be solved numerically exactly using the superfermion formalism. Using this we estimate the number of lead modes needed to accurately reproduce the continuum limit of infinite baths. 
We take a single resonant level as our working medium, described by the Hamiltonian
\begin{align}
\label{eq:h_s_d}
\hat{H}_{\tt S} = \epsilon\,\hat{c}^{\dagger}\hat{c},
\end{align}
where $\hat{c}^{\dagger}$ and $\hat{c}$ are the fermionic creation and annihilation operators in the system, respectively, and $\epsilon$ is the energy of the level. This models a single quantum dot in the spin-polarised regime running as a heat engine between two baths~\cite{Linke2010}. We note that a quantum-dot heat engine was recently realised experimentally~\cite{Linke2018}.

In principle, our methods can handle structured spectral densities that are different for each bath. For simplicity, however, we take both reservoirs to be characterised by identical, flat spectral densities within a finite energy band, given by
\begin{align}
\label{eq:wideband}
\mathcal{J}(\omega) = \begin{cases} \Gamma,\; \forall\, \omega \in [-W, W] \\ 0,\; \textrm{otherwise} \end{cases}
\end{align}
where $\Gamma$ is the coupling strength between the system and the leads. In the continuum limit of macroscopic baths, the particle and energy currents for a non-interacting system can be computed from the Landauer-B\"uttiker (L-B) formulae
\begin{align}
J^{\textrm{P}}_{\textrm{LB}} & = \frac{1}{2\pi}\int_{-W}^{W} {\rm d} \omega\, \tau(\omega) [ f_{\tt L}(\omega) - f_{\tt R}(\omega) ],\\
J^{\textrm{E}}_{\textrm{LB}} & = \frac{1}{2\pi}\int_{-W}^{W} {\rm d} \omega\,  \omega\tau(\omega) [ f_{\tt L}(\omega) - f_{\tt R}(\omega) ],
\end{align}
where $f_\alpha(\omega)$ denotes the Fermi-Dirac distribution for lead $\alpha = {\tt L,R}$ and $\tau(\omega)$ is the transmission function. The latter is computed using the formalism described in Sec.~\ref{sec:transmission}.

In the mesoscopic-reservoir approach, the spectral density is sampled by a finite number $L$ of lead modes, as in Eq.~\eqref{eq:spectral_density_effective}. Taking the distribution of lead mode energies $\{\varepsilon_k\}$, widths $\{\gamma_k\}$ and couplings $\{\kappa_{kp}\}$ to be identical for each lead, there remains significant freedom to choose these distributions in order to well approximate the continuum limit using moderate values of $L$. In particular, we use the logarithmic-linear discretisation scheme proposed in Refs.~\cite{Schwarz2016,vondelft2009}. Here, $L_{\textrm{lin}}$ modes are placed in the energy window $[-W^*, W^*]$, with equally spaced frequencies, i.e.\ $e_k = \varepsilon_{k+1} - \varepsilon_k = 2W^* / L_{\textrm{lin}}.$ Energies outside of this range are sampled by a smaller set of modes $L_{\textrm{log}}$, with frequencies logarithmically spaced from $W^*$ ($-W^*$) to $W$ ($-W$), with energy intervals $[\varepsilon_{n-1}, \varepsilon_n] = [\pm \Lambda^{-(n-1)}, \pm \Lambda^{-n}]$ for $n = 1, \cdots, L_{\textrm{log}}$ and $\Lambda^{-L_{\rm log}} = W^*$. The dissipation rates are taken equal to these spacings, $\gamma_k = e_k$, while the coupling constants $\kappa_{kp}$ ($p=1,D$) are determined by the equation $\Gamma = 2\pi \kappa_{kp}^2/e_k$~\cite{Dzhioev2011}, in accordance with the considerations of Sec.~\ref{sec:mesoscopic_leads}. For a given number of modes $L = L_{\rm log} + L_{\rm lin}$, this discretisation scheme gives better resolution within a smaller energy window $[-W^*, W^*]$ that includes the most relevant energy scales for the problem at hand. We remark that this discretisation scheme was chosen due to the featureless nature of $\mathcal{J}(\omega)$ in Eq.~\eqref{eq:wideband} to contain more energy modes in a given transport window. If $\mathcal{J}(\omega)$ was structured a different discretisation scheme to resolve its features could provide a better approximation of the spectral function. With respect to smooth spectral functions, however, we expect the chosen discretisation scheme to yield accurate results as the number of modes is increased. In our calculations, we henceforth set $W = 8$ and use this parameter as the overall energy scale, while $W^* = W / 2$. Moreover, we choose $L_{\textrm{log}} / L = 0.2$.

\begin{figure}[t]
\fontsize{13}{10}\selectfont 
\centering
\includegraphics[width=1\columnwidth]{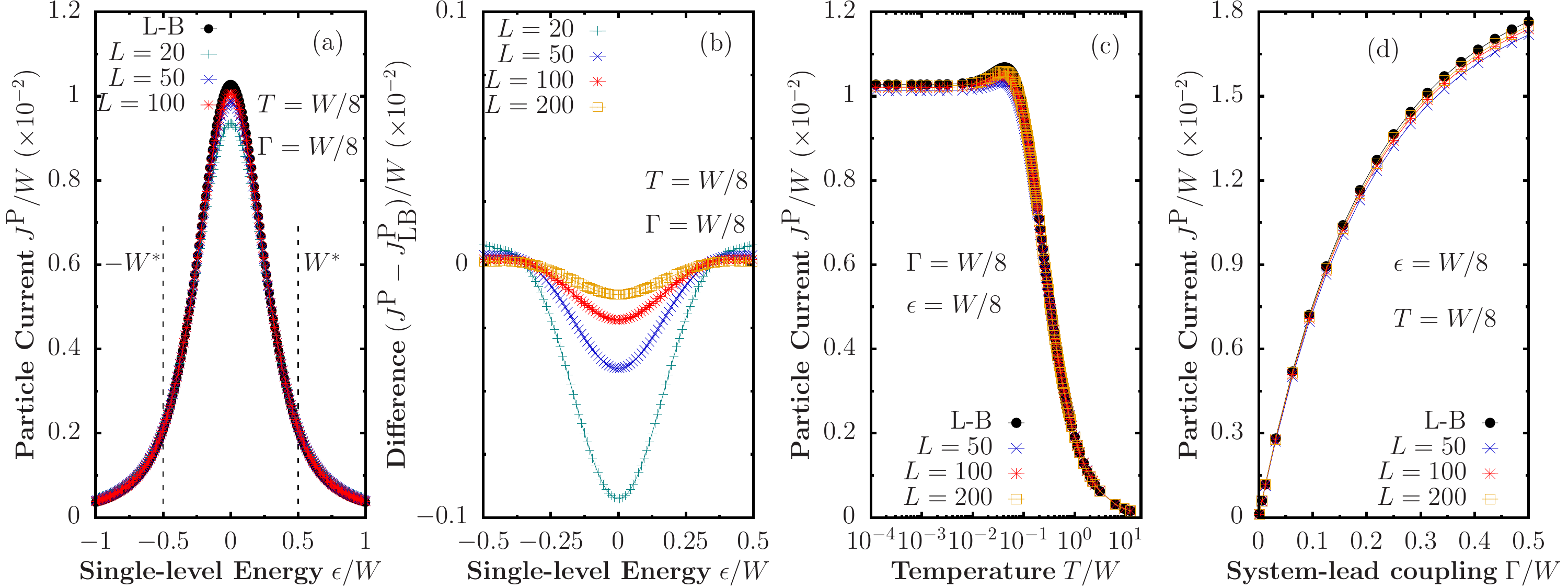}
\caption[Comparison between L-B predictions and the mesoscopic configuration for the total particle current through a single level]{Comparison between L-B predictions and the mesoscopic configuration of the expectation value of the total particle current flowing from the left lead through a single level, (a) as a function of the energy of the level, (b) absolute difference in the predictions from both scenarios with increasing number of modes in the leads $L$, (c) as a function of temperature and (d) as a function of the system-lead coupling strength $\Gamma$. In these calculations we used $\mu_{\texttt L} = -\mu_{\texttt R} = W / 16$, $T_{\texttt L} = T_{\texttt R}$, $L_{\textrm{log}} / L = 0.2$ and $W^* = W / 2$.}
\label{fig:2.3.7}
\end{figure}

Under these conditions, we show in Fig.~\ref{fig:2.3.7} the behaviour of the particle current, where we have set equal temperatures in the leads $T_{\tt L} = T_{\tt R} = W / 8$ but used different chemical potentials $\mu_{\tt L} = -\mu_{\tt R} = W / 16$. In Fig.~\ref{fig:2.3.7}(a) we show the results for the particle current as a function of the system energy $\epsilon$ for different numbers of modes $L$ in the leads and compare it with L-B theory. From both Fig.~\ref{fig:2.3.7}(a) and Fig.~\ref{fig:2.3.7}(b), it can be observed that a good agreement is obtained, the biggest difference observed as $\epsilon \to 0$, when the current reaches its maximum value. As expected, the agreement is improved with increasing $L$, although even moderate values of $L \sim O(10)$ approximately reproduce the continuum. In our calculations, we fixed the bath parameters as we varied the self-energy of the single-level $\epsilon$, however, the approximation could be improved by adapting the mode distribution around the relevant transport window dictated by $\epsilon$. Furthermore, in Fig.~\ref{fig:2.3.7}(c) we fix the energy $\epsilon$ of the level to study the behaviour with increasing $L$ as a function of temperature $T_{\tt L} = T_{\tt R} = T$ with system-lead coupling strength $\Gamma$ fixed, and in Fig.~\ref{fig:2.3.7}(d) the behaviour with $\Gamma$ for fixed $T$. For this particular choice of parameters we find the particle currents are robust to a wide range of $T$ and $\Gamma$. Either low or high temperatures and weak or strong coupling yield similar results in both continuum or mesoscopic scenarios, even for a moderate number of modes in the mesoscopic leads.

\begin{figure}[t]
\fontsize{13}{10}\selectfont 
\centering
\includegraphics[width=1\columnwidth]{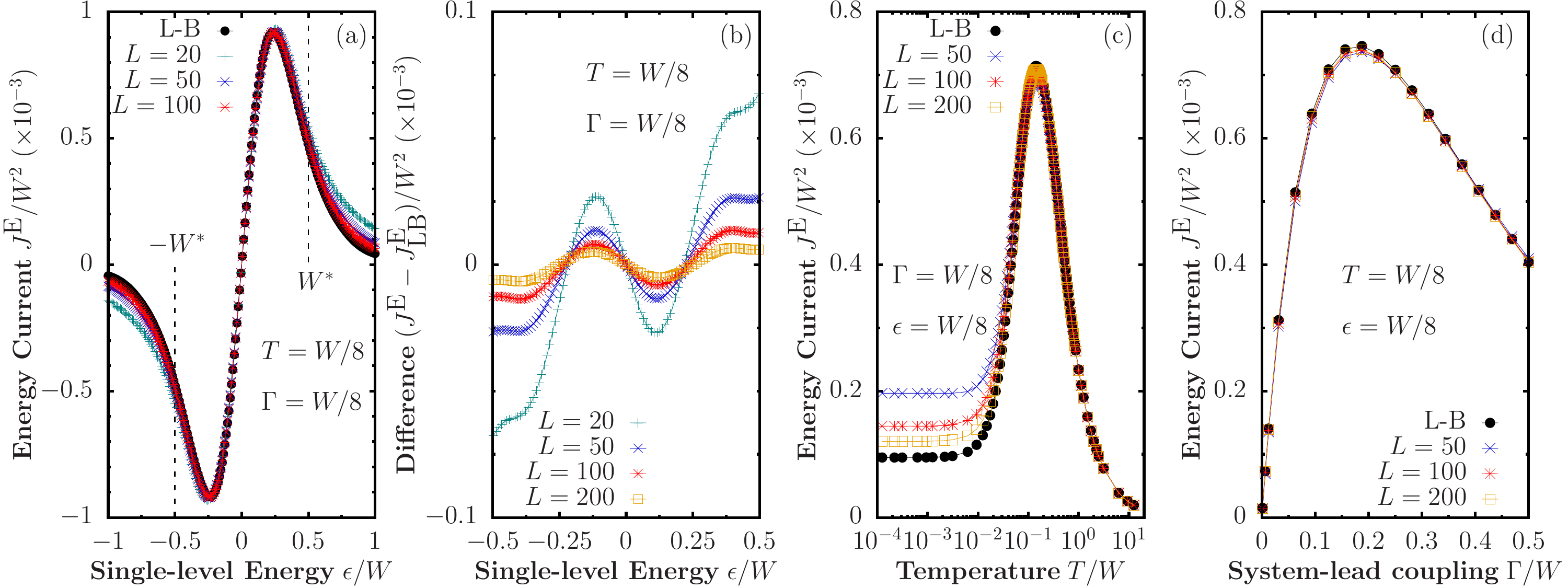}
\caption[Comparison between L-B predictions and the mesoscopic configuration for the total energy current through a single level]{Comparison between L-B predictions and the mesoscopic configuration of the expectation value of the total energy current flowing from the left lead through a single level, (a) as a function of the energy of the level, (b) absolute difference in the predictions from both scenarios with increasing number of modes in the leads $L$, (c) as a function of temperature and (d) as a function of the system-lead coupling strength $\Gamma$. In these calculations we used $\mu_{\texttt L} = -\mu_{\texttt R} = W / 16$, $T_{\texttt L} = T_{\texttt R}$, $L_{\textrm{log}} / L = 0.2$ and $W^* = W / 2$.}
\label{fig:2.3.8}
\end{figure}

In Fig.~\ref{fig:2.3.8} we show the corresponding results for energy current. From Fig.~\ref{fig:2.3.8}(a) it can be observed that a better approximation is obtained when the number of modes in each lead is increased for a fixed set of parameters, with the absolute difference decreasing as a function of $L$, as can be concluded from Fig.~\ref{fig:2.3.8}(b). In Fig.~\ref{fig:2.3.8}(c) a key difference can be observed from the results obtained for particle current. The mesoscopic lead configuration is a good approximation as long as $T$ is kept above a given threshold. This threshold is dictated by the smallest energy spacing in the leads $e_k$ and can be understood as follows. The effective spectral function of the mesoscopic leads is a sum of Lorentzian peaks, as in Eq.~\eqref{eq:spectral_density_effective}. When the temperature is smaller than the minimum energy spacing $e_k$ in the mesoscopic lead, these peaks are too far apart to properly resolve the variation of the Fermi-Dirac distribution. In this regime, the noise statistics given by Eq.~\eqref{eq:noise_correlations} are significantly modified and the approximation is not reliable. It can be observed from Fig.~\ref{fig:2.3.8}(c) that the approximation at lower temperatures is much better for larger leads\footnote{One method that can be used to obtain a better approximation at lower temperatures, that reduces the value of $e_k$ in the leads and without increasing the number of modes, is to change the width and position of the window $[-W^*, W^*]$ depending on the region of the parameter space that needs to be resolved in greater detail.}. 

In Fig.~\ref{fig:2.3.8}(d) we analyse the energy current as a function of the system-lead coupling $\Gamma$. We observe that the approximation for energy current in the mesoscopic lead configuration is quite robust to a wide range of couplings. This provides further evidence that the accuracy of the approximation is primarily determined by the size of $\gamma_k$ and $e_k$ relative to the temperature and voltage bias of the reservoirs~\cite{Gruss2016}.
 
 \begin{figure}[t]
\fontsize{13}{10}\selectfont 
\centering
\includegraphics[width=1\columnwidth]{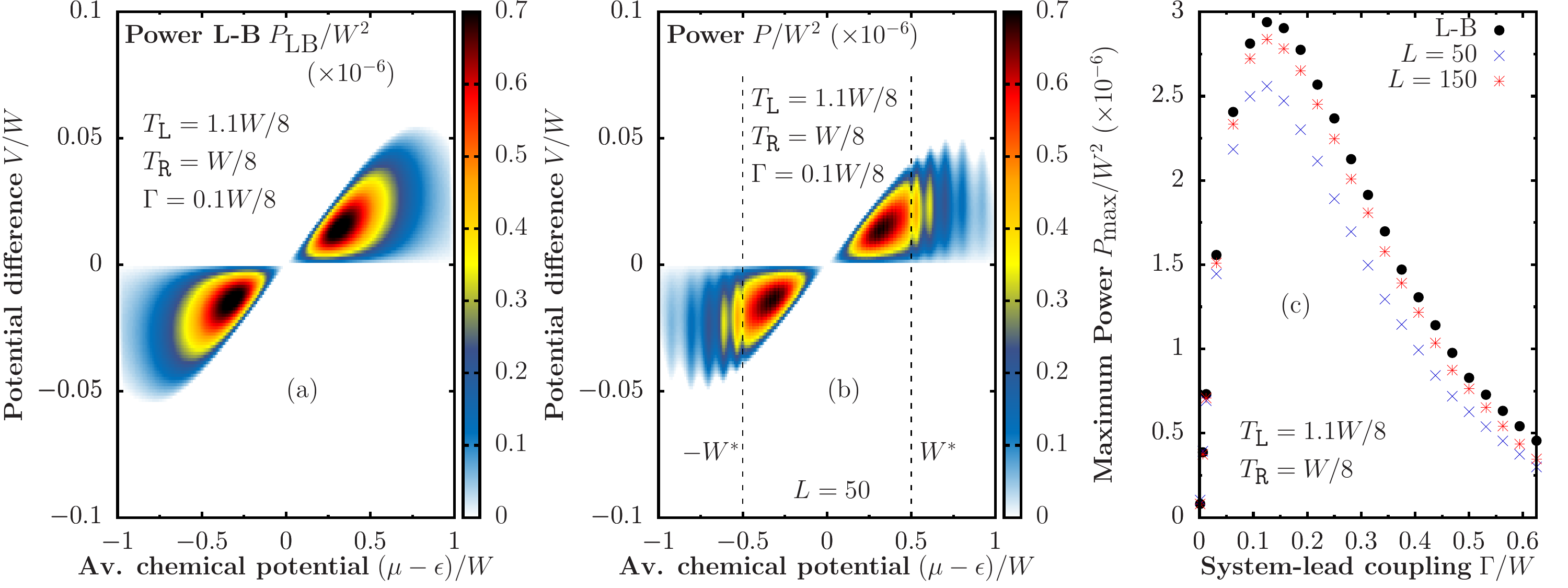}
\caption[Power as a function of potential difference $V$ and average chemical potential $\mu - \epsilon$ for the single-level system]{Power as a function of potential difference $V$ and average chemical potential $\mu - \epsilon$ for the single-level system using (a) continuum leads and (b) mesoscopic reservoirs. In (c) we present the maximum power as a function of the system-lead coupling for both configurations. In (b) and (c) we used $L_{\textrm{log}} / L = 0.2$ and $W^* = W / 2$.}
\label{fig:2.3.9}
\end{figure}
 
Next we evaluate the power and efficiency given by Eqs.~\eqref{eq:power_def} and \eqref{eq:efficiency_def}. In Fig.~\ref{fig:2.3.9}(a) we show the power output as a function of average chemical potential $\mu = (\mu_{\tt L}+\mu_{\tt R})/2$ and the potential difference $V = \mu_{\tt R} - \mu_{\tt L}$ using the L-B prediction for continuum leads. In our calculations we set $T_{\tt L} = 1.1W/8$ and $T_{\tt R} = W/8$ and show the power output results only for the values of $\mu - \epsilon$ and $V$ for which the system acts as a power generator. It can be observed that the power output reaches a maximum value depending on bias and average chemical potential. In Fig.~\ref{fig:2.3.9}(b) we show the results for the same calculation, but instead we substitute the continuum leads with our mesoscopic lead configuration. The results are in good agreement up to the point where $\mu - \epsilon$ reaches the boundary of linearly discretised and logarithmically discretised lead modes. Beyond $W^*$ and $-W^*$, the spectral function is not sampled as finely and the power output results get distorted. We note that the window can be increased to resolve a bigger set of the parameter space, however, this would require more lead modes to resolve the maximum power output with the same accuracy. Alternatively, the range of linearly discretised modes could be adapted for each value of $\epsilon$ to ensure that the relevant energy range for transport is always included within this window. In Fig.~\ref{fig:2.3.9}(c) we show the maximum power output $P_{\textrm{max}}$ as a function of the system-lead coupling for both the L-B and mesoscopic lead predictions, which in turn reveals the value of $\Gamma_\textrm{max}$ for which $P_{\textrm{max}}$ reaches its maximum value. With our choice of parameters, $\Gamma_{\textrm{max}}$ lies very close in both configurations, as well as the overall behaviour as a function of system-lead coupling. The absolute value of the maximum power is better approximated, following the expected behaviour from Fig.~\ref{fig:2.3.7}(a), as the number of lead modes is increased.

\begin{figure}[t]
\fontsize{13}{10}\selectfont 
\centering
\includegraphics[width=1\columnwidth]{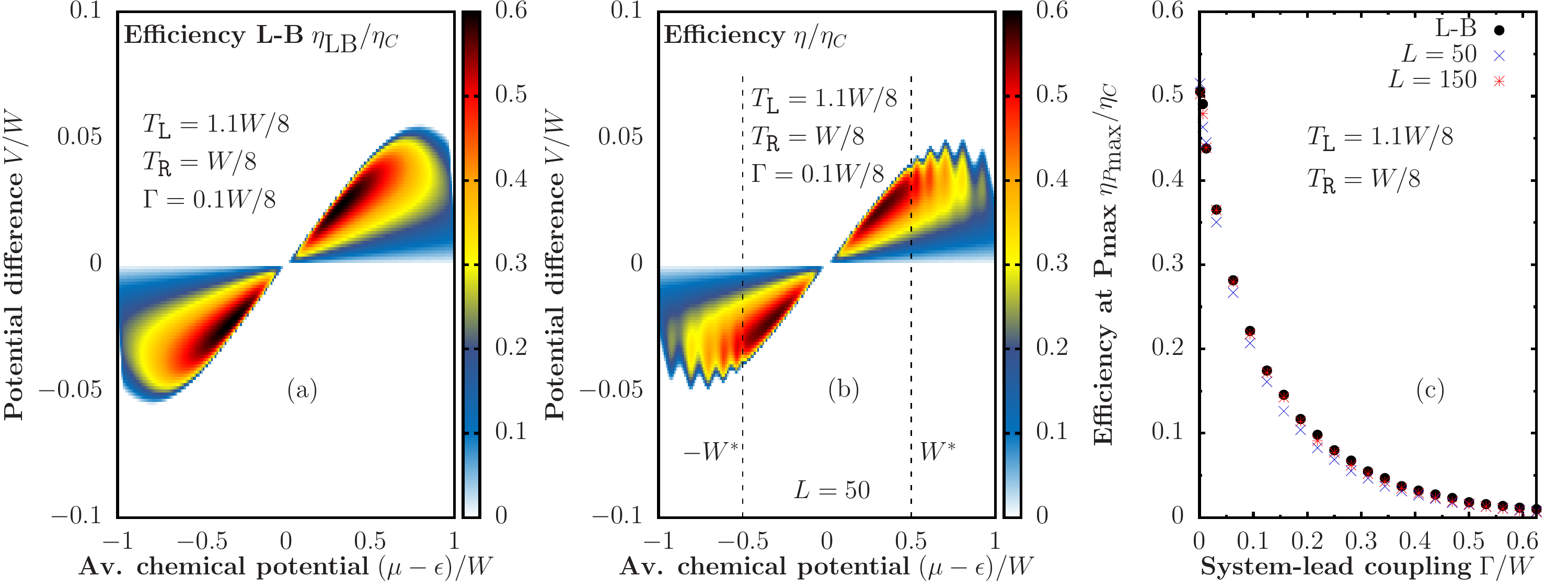}
\caption[Efficiency as a function of potential difference $V$ and average chemical potential $\mu - \epsilon$ for the single-level system]{Efficiency (normalised by the Carnot efficiency $\eta_C$) as a function of potential difference $V$ and average chemical potential $\mu - \epsilon$ for the single-level system using (a) continuum leads and (b) mesoscopic reservoirs. In (c) we present the efficiency at maximum power as a function of the system-lead coupling for both configurations. In (b) and (c) we used $L_{\textrm{log}} / L = 0.2$ and $W^* = W / 2$.}
\label{fig:2.3.10}
\end{figure}

In Fig.~\ref{fig:2.3.10}(a) we show the efficiency obtained using continuum leads, normalised by the Carnot efficiency. It can be observed that the points of maximum efficiency lie close to the boundary where the system stops operating as an engine, i.e., where the potential difference becomes too large for the temperature gradient to drive electrons in the opposite direction of the bias. In Fig.~\ref{fig:2.3.10}(b) we present the results for the mesoscopic lead configuration. As before, we find that both predictions are quantitatively similar up to the point where the boundary of $W^*$ is reached. In Fig.~\ref{fig:2.3.10}(c) we show the efficiency at the point where the maximum power is obtained from the configuration as a function of $\Gamma$, where we observe that both the continuum and mesoscopic lead configurations predict very similar results, even with a moderate number of lead modes. As expected, the approximation becomes more accurate as the lead size is increased. Furthermore, not only is the strong system-lead coupling behaviour well-captured, but so is the Curzon-Ahlborn efficiency limit (approximately given by $\eta_C / 2$) at weak coupling \cite{Curzon1975}.

\section{Tensor network method}
\label{sec:tensor}

Having established that relatively modest sized mesoscopic leads can capture the continuum behaviour of a non-interacting system we now move on to consider the highly non-trivial problem of interacting systems. To do this we introduce in this section a tensor network based numerical method that can efficiently and accurately compute the interacting NESS of the the two-reservoir problem illustrated in Fig.~\ref{fig:2.3.6}. To describe the method we will return briefly to the single-lead configuration shown in Fig.~\ref{fig:2.3.4} in which the first site $p=1$ of the system $\tt S$ is coupled to the mesoscopic lead. Since we will exploit the superfermion formalism we continue to use the unified notation for modes $\hat{d}_k$ given in Eq.~\eqref{eq:unified_notation}.

\subsection{Spin-1/2 representation}
\label{sec:s12_rep}
\begin{figure}[t]
\fontsize{13}{10}\selectfont 
\centering
\includegraphics[width=0.35\columnwidth]{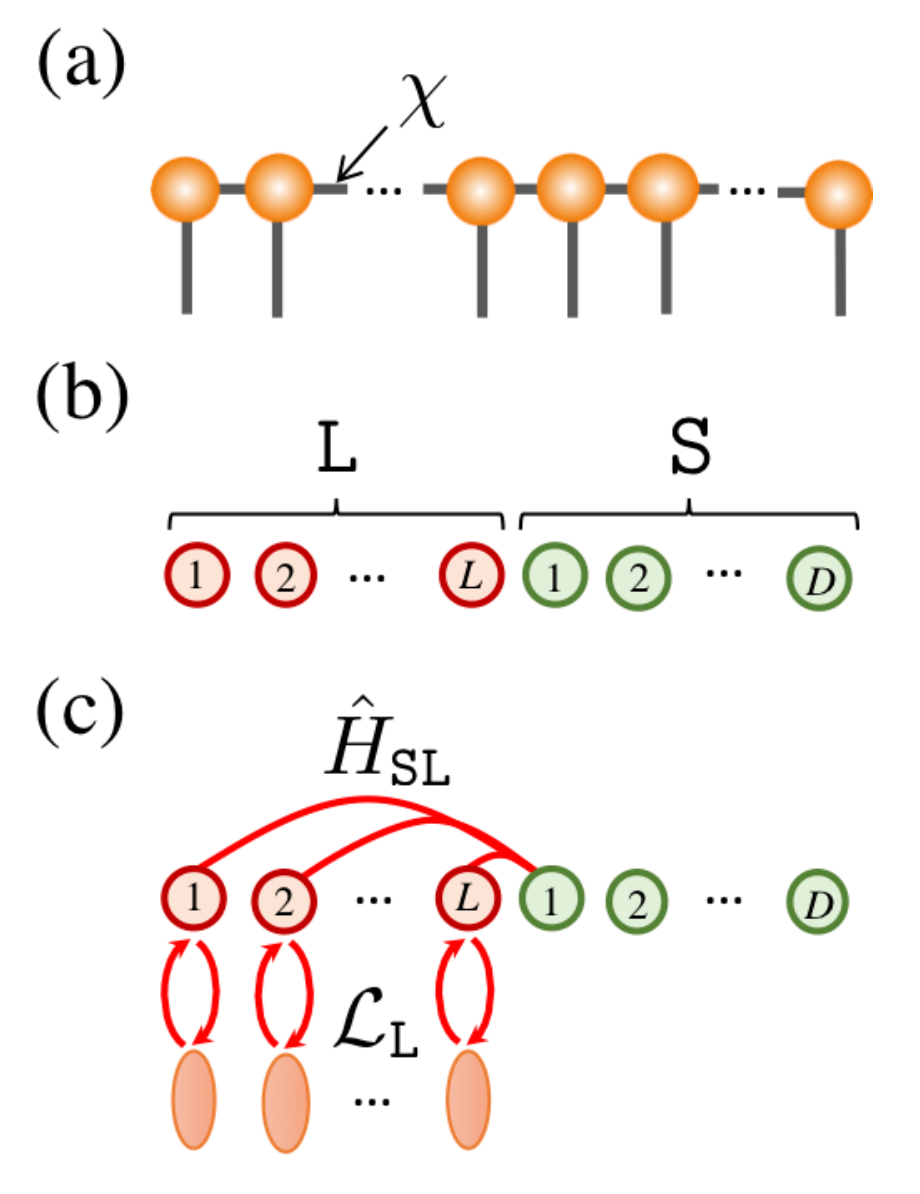}
\caption[MPS network and star geometry in the energy eigenbasis of the leads]{(a) A MPS tensor network in which every site (except the boundaries) have an order-3 tensor associated to it. The vertical dangling legs are the physical indices of the system of dimension 2 in our case, the horizontal contracted legs are the internal bonds of the MPS of dimension $\chi$. (b) The lead and system modes are ordered into a one-dimensional geometry to match the MPS. (c) With this ordering the star geometry system-lead coupling $\hat{H}_{\tt SL}$ is long-ranged and the local fermionic dissipators $\mathcal{L}_{\tt L}$ on the lead also become long-ranged due to JW strings.}
\label{fig:2.3.11}
\end{figure}

Our approach uses the matrix product state (MPS) decomposition that is a tensor network with a one-dimensional chain-like geometry~\cite{Schollwock2011}, as shown in Fig.~\ref{fig:2.3.11}(a). To apply this powerful method to our setup we first map the lead and system modes into a one-dimensional chain. In doing so the coherent coupling between the lead modes and the system become long-ranged within this chain since they corresponding to a so-called {\em star geometry}. Fundamentally this is because we use the energy eigenbasis of the lead.

Additionally, since MPS apply to systems built from a tensor product of local Hilbert spaces, to describe a spinless fermionic system requires that we transform it into a spin-1/2 representation. Our starting point is to introduce Fock states constructed from the unified physical modes with occupation-number vector $\underline{n}$ as
\begin{align} \label{fock_states}
\ket{\underline{n}} = \left( \hat{d}_1^{\dagger} \right)^{n_1} \cdots \left( \hat{d}_{M}^{\dagger} \right)^{n_{M}}  \ket{\textrm{vac}},
\end{align}
which in the single-lead case has $M=L+D$ and is ordered with lead modes first, as shown in Fig.~\ref{fig:2.3.11}(b). A spin-1/2 representation is then obtain via the well-known Jordan-Wigner (JW) transformation involving $M$ spins \cite{Jordan:1928,Coleman:2015}   
\begin{align}
\hat{d}^{\dagger}_j &= \left( \prod_{q=1}^{j-1} \hat{\sigma}^z_q \right) \hat{\sigma}^{-}_j,
\end{align}
where $\hat{\sigma}^z_q$ is the Pauli spin matrix in the $z$ direction and $\hat{\sigma}^{\pm}_q$ are the spin raising/lowering operators for the $q$-th spin. Correspondingly, the Fock states of Eq.~\eqref{fock_states} are equivalent to the spin states
\begin{align}
\ket{\underline{n}} = \left( \hat{\sigma}^-_1 \right)^{n_1} \cdots  \left( \hat{\sigma}^-_{M} \right)^{n_{M}} \ket{\uparrow \cdots \uparrow},
\end{align}
since each JW string vanishes on polarised spins to which it is applied. Transforming the total Hamiltonian $\hat{H} =  \hat{H}_{\tt S} + \hat{H}_{\tt L} + \hat{H}_{\tt SL}$ [from Eqs.~\eqref{eq:H_lead} and \eqref{eq:H_lead_sys}] to this representation gives
\begin{align}
\label{eq:h_jw}
\hat{H} &= \hat{H}_{\tt S} + \sum_{k=1}^L \left\{ \kappa_{k1} \hat{\sigma}^+_{k}\left( \prod_{q=k+1}^{L} \hat{\sigma}^z_q \right)\hat{\sigma}^-_{L+1} \right. \\
&\left. \quad+ \kappa^*_{k1}\hat{\sigma}^-_{k} \left( \prod_{q=k+1}^{L} \hat{\sigma}^z_q \right)\hat{\sigma}^+_{L+1} \right\} + \sum_{k=1}^L \varepsilon_k \hat{\sigma}^-_{k} \hat{\sigma}^+_{k}.\notag
\end{align}
The star geometry, shown in Fig.~\ref{fig:2.3.11}(c), thus introduces JW strings to the lead-system coupling terms making them long-ranged multi-body spin operators. Similarly, the Lindblad dissipator of Eq.~\eqref{eq:dissipator} becomes
\begin{align}
\label{eq:dissipator_jw}
\mathcal{L}_{\tt L}&\{\hat{\rho}\}  = \sum_{k=1}^{L} \gamma_k(1 - f_k) \Biggl[- \frac{1}{2}\{ \hat{\sigma}^-_{k} \hat{\sigma}^+_{k}, \hat{\rho} \} + \hat{\sigma}^+_{k} \left( \prod_{q=1}^{k-1} \hat{\sigma}^z_q \right) \hat{\rho} \left( \prod_{q=1}^{k-1} \hat{\sigma}^z_q \right) \hat{\sigma}^-_{k} \Biggl] \nonumber \\
&\qquad\qquad\, + \sum_{k=1}^{L} \gamma_k f_k \Biggl[- \frac{1}{2}\{ \hat{\sigma}^+_{k} \hat{\sigma}^-_{k}, \hat{\rho} \} + \hat{\sigma}^-_{k} \left( \prod_{q=1}^{k-1} \hat{\sigma}^z_q \right) \hat{\rho} \left( \prod_{q=1}^{k-1} \hat{\sigma}^z_q \right) \hat{\sigma}^+_{k} \Biggl],
\end{align}
showing that the jump operators are now also non-local due to the JW strings.

\subsection{Superfermion representation}
\label{sec:superfermion_spin}

By using the energy eigenbasis of the lead we have arrived at a master equation with a highly non-local multi-body Hamiltonian and dissipator. The JW strings therefore appear to severely frustrate the use of MPS algorithms in this setup. Typically those arising from the star geometry of the Hamiltonian in Eq.~\eqref{eq:h_jw} are dealt with by tridiagonalising the lead Hamiltonian, transforming it into a chain geometry and localising its coupling to the system. However, it is clear that this procedure profoundly complicates the dissipator in Eq.~\eqref{eq:dissipator_jw}. The thermal damping of the lead induced by the dissipator is most naturally described in the lead's energy eigenbasis. 

In the lead energy eigenbasis, the JW strings of the dissipators can be eliminated by exploiting the superfermion representation of the open system introduced in Sec.~\ref{sec:superfermion}. There, an {\em interleaved} physical and ancillary mode ordering was used, resulting in the dissipative processes becoming nearest-neighbour non-Hermitian Hamiltonian terms, as shown in Eq.~\eqref{eq:l_single}. In this form, when moving to a spin-1/2 representation, the JW string of each system or lead site cancels with that of the corresponding ancillary mode, rendering the dissipator terms local. 

To observe this explicitly, first note that the Fock basis of the combined Hilbert space of the physical and ancilla sites, namely Eq.~\eqref{eq:superfermion_Fock_state}, can be written in the spin-1/2 basis as 
\begin{align}
\label{eq:superfermion_Fock_state_spin}
\ket{\underline{n}| \underline{m}} = \left( \hat{\sigma}^-_1 \right)^{n_1} \left( \hat{\sigma}^-_2 \right)^{m_1} \cdots \left( \hat{\sigma}^-_{2M} \right)^{n_{M}} \left( \hat{\sigma}^-_{2M} \right)^{m_{M}} \ket{\uparrow \uparrow \cdots \uparrow \uparrow}.
\end{align}
The non-Hermitian generator of the superfermion time evolution thus becomes
\begin{align}
\label{eq:l_single_spin}
&\hat{L} = \hat{H} - \hat{H}_{d\Leftrightarrow s} \nonumber +\textrm{i}\sum_{k=1}^{L} \gamma_k  (1-f_k) \hat{\sigma}^+_{2k-1} \hat{\sigma}^+_{2k}   \nonumber \\
&\quad+\textrm{i}\sum_{k=1}^{L} \gamma_k  f_k \hat{\sigma}^{-}_{2k-1} \hat{\sigma}^{-}_{2k} - \textrm{i}\sum_{k=1}^{L} \gamma_k f_k \\
&\quad-\frac{\textrm{i}}{2}\sum_{k=1}^{L} \Bigl[\gamma_k (1 - 2f_k)\left( \hat{\sigma}^{-}_{2k-1} \hat{\sigma}^+_{2k-1} + \hat{\sigma}^{-}_{2k}  \hat{\sigma}^+_{2k} \right)\Bigr], \nonumber
\end{align}
showing that the dissipator contribution consists of on-site and nearest-neighbour terms.  

\subsection{Time evolving block decimation with swaps}
\label{sec:tebd}
To efficiently simulate the time evolution of the correlated system Eq.~\eqref{eq:l_single_spin}, we use one of the most well-known algorithms within the tensor network family, namely, the time-evolving block decimation (TEBD)~\cite{VidalTEBD2004,VerstraeteTEBD2004}. Given some system governed by a Hamiltonian $\hat{H}_{\rm loc} = \sum_i \hat{h}_{i,i+1}$, comprising a sum of two-site terms $\hat{h}_{i,i+1}$ along a chain of length $M$, the standard formulation of TEBD computes the MPS approximation of the propagation $\ket{\psi(t)} = \exp(-\ii \hat{H}_{\rm loc}t)\ket{\psi(0)}$. This is done by first breaking up the evolution into many small time-steps $\delta t$ and then performing a second-order Trotter expansion as
\begin{align}
\label{eq:time_step}
e^{-\ii \hat{H}_{\rm loc} \delta t}\approx\left(\prod_{i=1}^{M-1}\hat{U}_{i,i+1}\right)\left(\prod_{i=M-1}^{1}\hat{U}_{i,i+1}\right),
\end{align}
where $\hat{U}_{i,i+1} = \exp(-\frac{\ii}{2} \hat{h}_{i,i+1}\delta t)$. In this way, a time step of propagation is implemented by a staircase circuit of two-site gates sweeping right-to-left and then left-to-right. Each two-site gate can be applied to the MPS and, via a singular value decomposition, the result can be re-factorised and truncated back into MPS form. 

\begin{figure}[t]
\fontsize{13}{10}\selectfont 
\centering
\includegraphics[width=0.3\columnwidth]{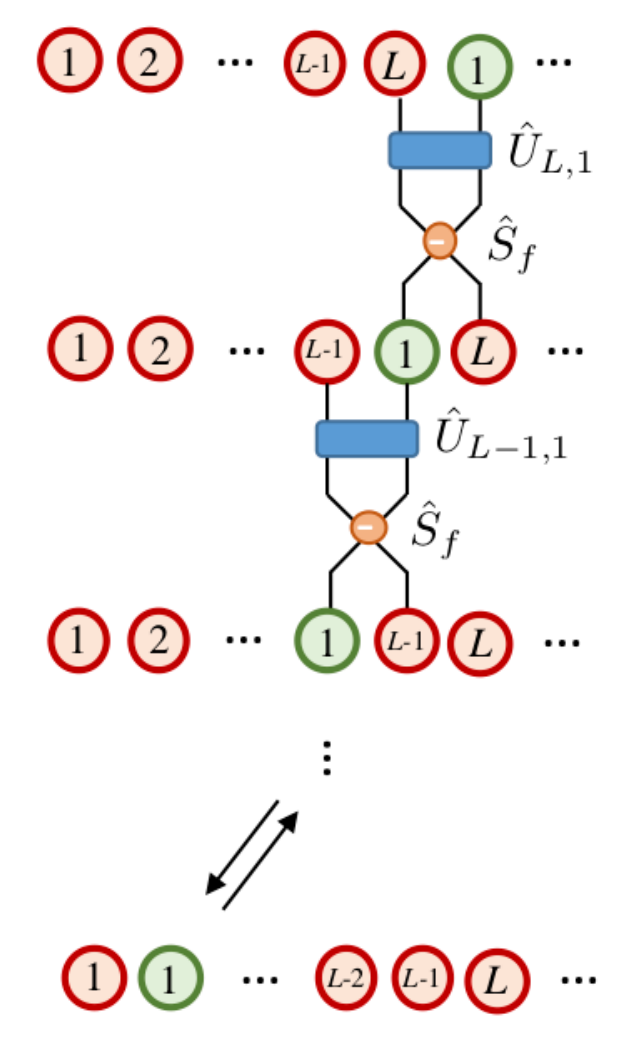}
\caption[The sweeping sequence of two-site gates $\hat{U}_{k,1}$ between the $k$-th lead mode and the first system site along with the fermionic SWAPs $\hat{S}_f$ needed to implement a Trotter step for the star geometry]{The sweeping sequence of two-site gates $\hat{U}_{k,1}$ between the $k$-th lead mode and the first system site along with the fermionic SWAPs $\hat{S}_f$ needed to implement a Trotter step for the star geometry couplings shown in Fig.~\ref{fig:2.3.11}(c).}
\label{fig:2.3.12}
\end{figure}

Here, we use a simple modification of TEBD that allows us to compute the time-evolution under fermionic star-geometry Hamiltonians $\hat{H}_{\rm star} = \sum_i \hat{h}_{i,M}$, where all sites $i<M$ interact with the last site $M$. The key ingredient is the fermionic SWAP gate $\hat{S}_f$, which is a conventional SWAP gate between spins $j$ and $j+1$ that exchanges their spin configurations, but also incorporates the application of the local $\hat{\sigma}_j^z$ operator from the JW string of Eq.~\eqref{eq:h_jw}. For two sites, the gate is given by 
\begin{align} \label{swap2sites}
\hat{S}_f = \begin{pmatrix*}[r] 1 & 0 & 0 & 0 \\ 0 & 0 & \phantom{-}1 & 0 \\ 0 & \phantom{-}1 & 0 & 0 \\ 0 & 0 & 0 & -1 \end{pmatrix*},
\end{align}
where the negative sign accounts for the anticommutation relation between two fermionic creation operators when both sites $j$ and $j+1$ are occupied. By interspersing fermionic SWAP gates within the Trotter expansion, as shown in Fig.~\ref{fig:2.3.12}, distant sites are temporarily made adjacent, allowing the standard nearest-neighbour two-gate gate update to be applied. 

Time-evolution under a long-ranged Hamiltonian is generally considered impractical for tensor network calculations, due to very fast growth of entanglement across the system. This conjecture has been challenged in recent studies of fermionic impurity models, where efficient tensor network calculations have been performed using a star-like geometry~\cite{Wolf:2014,Mendoza:2017}. The proliferation of correlations in these models is curtailed by Pauli exclusion within the majority of the modes of the lead, limiting them to the range of modes around the Fermi energy. This favourable situation persists in the mesoscopic thermal lead setup considered here. Furthermore, it has been recently shown that using a suitable order of the lead modes can significantly enhance the efficiency of tensor network simulations~\cite{Rams2020}. 

\subsection{Non-equilibrium steady state solver}
\label{sec:ness_solve}
The TEBD algorithm works equally well for non-Hermitian Hamiltonians generating non-unitary propagation. Indeed, it has been widely used to study the NESS of incoherently driven chains where the coupling to the reservoirs is localised to one~\cite{Benenti:2009,Znidaric:2010,Znidaric:2010b,Znidaric:2011,Mendoza:2013a,Mendoza:2013b,Mendoza:2014,vznidarivc2016diffusive,vznidarivc2017dephasing} or two sites~\cite{Znidaric:2011c,Mendoza:2015,schulz2018energy,mendoza2018asymmetry} at the boundaries. We have now introduced all the elements required to extend the capabilities of TEBD to simulate the open-system governed by the Hamiltonian Eq.~\eqref{eq:h_jw} and the dissipator Eq.~\eqref{eq:dissipator_jw}.

\begin{figure}[t]
\fontsize{13}{10}\selectfont 
\centering
\includegraphics[width=0.5\columnwidth]{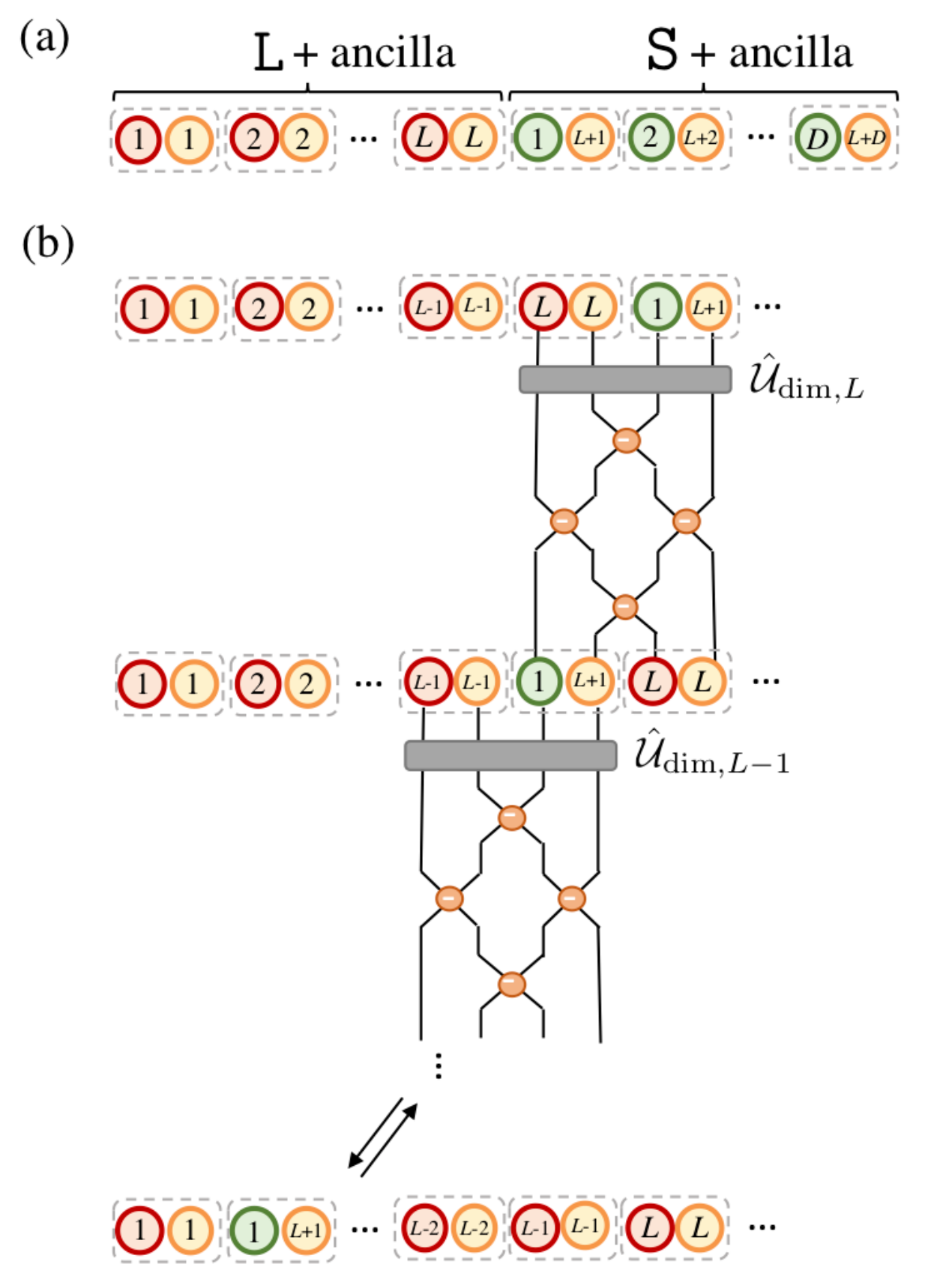}
\caption[Two-site dimer gates for a one-dimensional network for time evolution]{(a) Ancilla modes are interleaved with the system and lead modes they are associated to. Computationally the system or lead site and its ancilla are bundled together as a dimer site. (b) A two-dimer site gate $\hat{\mathcal U}_{{\rm dim},k}$ is applied between the $k$-th lead mode dimer, and the first system site dimer. This is followed by four fermionic SWAPs $\hat{S}_f$ to shuffle the system site and its ancilla through the lead and its ancilla making the next lead mode adjacent. This is repeated all the way along the chain and back to complete one time-step.}
\label{fig:2.3.13}
\end{figure}

First, we move to the superfermion representation where the generator $\hat{L}$ is given by Eq.~\eqref{eq:l_single_spin}. We define dimer sites composed of a physical (system or lead) site and its corresponding ancilla, as shown in Fig.~\ref{fig:2.3.13}(a). This procedure squares the dimension of the local basis. The left vacuum state $\ket{I}$ in this representation is a product state of dimers, with each dimer local to a given site being an equal superposition of $\ket{\uparrow \uparrow}$ and $\ket{\downarrow \downarrow}$.

Next, we identify all the terms in $\hat{L}$ that correlate the dimers located at lead site $k$ and system site $p=1$. Assuming these sites are adjacent to each other through SWAP operations, we express 
\begingroup
\allowdisplaybreaks
\begin{align}
\label{eq:l_dimer}
\hat{L}_{{\rm dim},k} &= \varepsilon_k \left( \hat{\sigma}_1^- \hat{\sigma}_1^+ - \hat{\sigma}_2^- \hat{\sigma}_2^+ \right) + \frac{\epsilon}{L} \left( \hat{\sigma}_3^- \hat{\sigma}_3^+ - \hat{\sigma}_4^- \hat{\sigma}_4^+ \right) \nonumber \\
&+ \kappa_{kL} \hat{\sigma}^-_1 \hat{\sigma}^z_2 \hat{\sigma}^+_3 + \kappa^*_{kL} \hat{\sigma}^+_1 \hat{\sigma}^z_2 \hat{\sigma}^-_3
\nonumber \\
& \qquad -\kappa_{kL} \hat{\sigma}^+_2 \hat{\sigma}^z_3 \hat{\sigma}^-_4 - \kappa^*_{kL} \hat{\sigma}^-_2 \hat{\sigma}^z_3 \hat{\sigma}^+_4  \nonumber \\
& -\frac{\textrm{i}}{2} \gamma_k (1 - 2f_k) \left( \hat{\sigma}^-_1 \hat{\sigma}^+_1 + \hat{\sigma}^-_2 \hat{\sigma}^+_2 \right) -\textrm{i} \gamma_k f_k \nonumber \\
& +\textrm{i}\gamma_k(1 - f_k) \hat{\sigma}^+_1 \hat{\sigma}^+_2 + \textrm{i}\gamma_k f_k \hat{\sigma}^-_1 \hat{\sigma}^-_2.
\end{align}
\endgroup
We identify spin 1 as the $k$-th lead eigenmode with spin 2 being its corresponding ancilla mode. On the other hand, spin 3 is the system site coupled to the lead with spin 4 its corresponding ancilla mode. A JW string appears between interacting spins that are not adjacent, however, they remain local to the dimer pair. The exponential of this operator, $\hat{\mathcal U}_{{\rm dim},k} = \textrm{exp} (-\textrm{i} \hat{L}_{{\rm dim},k} \delta t / 2)$, defines a non-unitary gate for a half time step $\delta t$. This operator accounts for all the coherent interactions and the non-Hermitian terms, describing the dissipation between the lead mode and the system site. We have assumed a Hamiltonian of the form Eq.~\eqref{eq:h_s_d} in Eq.~\eqref{eq:l_dimer}.

\begin{figure}[t]
\fontsize{13}{10}\selectfont 
\centering
\includegraphics[width=0.55\columnwidth]{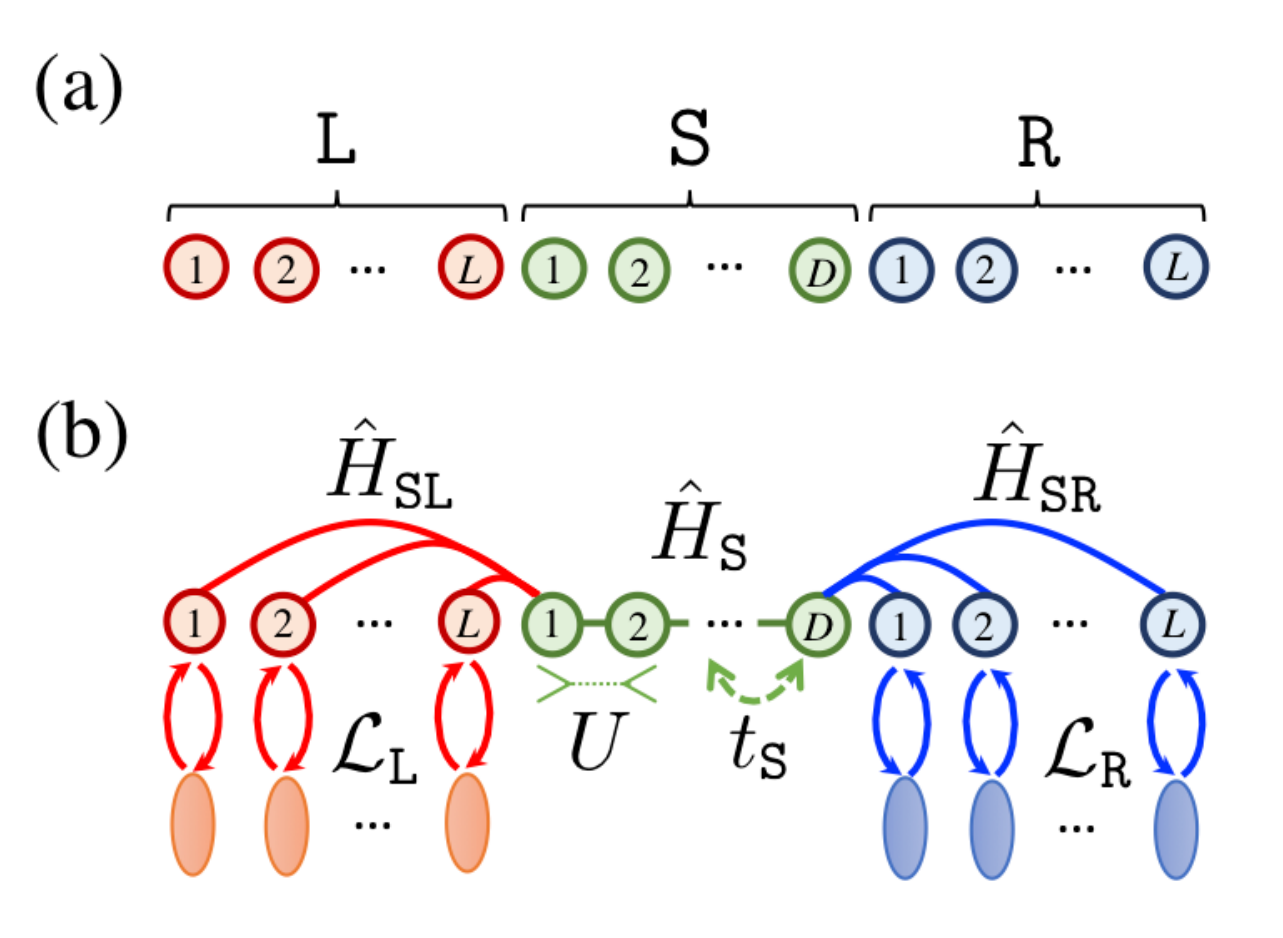}
\caption[The lead and system mode ordering for a two lead setup]{(a) The lead and system mode ordering for a two lead setup. (b) The configuration used for the interacting system examples. Here the system $\tt S$ is a fermionic chain with hopping amplitude $t_{\tt S}$ and nearest-neighbour interaction $U$.}
\label{fig:2.3.14}
\end{figure}

Finally, the non-unitary gates $\hat{\mathcal U}_{{\rm dim},k}$ are then applied along with fermionic SWAP gates that shuffle the system dimer along the chain, as shown in Fig.~\ref{fig:2.3.13}(b). The latter can be defined from the two-site SWAP gates of Eq.~\eqref{swap2sites} in the following way: naming $\hat{A} = I_2 \otimes S_f \otimes I_2$, with $I_2$ the $2\times2$ identity matrix, and $\hat{B} = S_f \otimes S_f$, the two-dimer SWAP gate depicted in Fig.~\ref{fig:2.3.13}(b) is given by $\hat{A}\hat{B}\hat{A}$. Altogether, this sequence of gates computes the action of the propagator $\exp(-\textrm{i}\hat{L}\delta t)$ and formally solves Eq.~\eqref{eq:superfermion_evolve} for a single time-step. We take the initial state to be $\ket{\rho(0)} = \ket{I}$, and find the steady state $\ket{\rho(\infty)}$ by evolving towards the long-time limit. Expectation values and the trace of the density operator follow from the inner product with $\ket{I}$ as given in Eq.~\eqref{eq:expec}.

The same simulation scheme can be readily extended to the two-lead configuration, as shown in Fig.~\ref{fig:2.3.14}(a), with the long-time limit now giving rise to a NESS. The approach to the stationary state is assessed by evaluating the convergence of observables such as the particle and energy currents. For the examples that follow, we used a dynamically-increasing truncation parameter $\chi$ for different time-step parameters $\delta t$. In the standard MPS language~\cite{VidalTEBD2004,VerstraeteTEBD2004}, $\chi$ refers to the maximum MPS bond dimension in between each pair of neighbouring nodes in the network, where each node represents a dimer. To perform the simulation, we chose an initial value of $\chi$ and $\delta t$, and evolved the system up to an intermediate time. The resulting state was then further evolved in time with a larger $\chi$ and an appropriately reduced $\delta t$. This procedure is repeated until the currents obtained converged up to a small tolerance of $1-2 \%$. The largest bond dimension used in our calculations was $\chi_{\textrm{max}} = 220$, showing that a moderate computational effort was required to access the NESS. All MPS calculations in the following examples were performed using the open-source Tensor Network Theory (TNT) library~\cite{tnt,tnt_review1}.

\section{Interacting examples}
\label{sec:interacting}

In this section, we employ the tensor network algorithm from Sec.~\ref{sec:tensor} to study an autonomous thermal machine with an interacting working medium, as depicted in Fig.~\ref{fig:2.3.14}(b). Our methods enable us to consider the challenging problem of simultaneously strong interactions and system-bath coupling, far beyond the linear-response regime. 

\subsection{Interacting three-site engine}

Our first example is an autonomous quantum heat engine with a three-site interacting working medium, which is described by the Hamiltonian
\begin{align}
\label{eq:h_s_i}
\hat{H}_{\tt S} = \sum_{j = 1}^{D} \epsilon_j \hat{n}_j - \sum_{j = 1}^{D - 1} t_{\tt S} \left( \hat{c}^{\dagger}_{j+1} \hat{c}_{j} + \textrm{H.c.} \right) + \sum_{j=1}^{D-1} U \hat{n}_j \hat{n}_{j+1},
\end{align}
where $\hat{n}_j = \hat{c}^{\dagger}_j \hat{c}_j$ is the density operator for site $j$ and $U$ is the interaction strength. The last term in the equation above corresponds to a density-density interaction of neighbouring particles. A small central system composed of $D=3$ interacting fermionic sites can be interpreted as a three-site version of the interacting resonant level model \cite{Kennes2012}.

We set the system hopping $t_{\tt S} = W/8$ and focus on the regime in which the temperature gradient and the difference in chemical potential between the mesoscopic reservoirs is strong. We set $T_{\tt L} = 10t_{\tt S}$, $T_{\tt R} = t_{\tt S}$, $\mu_{\tt L} = -t_{\tt S}/2$, $\mu_{\tt R} = t_{\tt S}/2$ and $\epsilon_j = \epsilon = t_{\tt S}$. With these parameters, the system operates as a heat engine, i.e. particle current flows from the left reservoir to the right reservoir, driven by the temperature gradient against a chemical potential gradient. As in Sec.~\ref{sec:non_interacting_example}, both leads are assumed to have identical, flat spectral densities given by Eq.~\eqref{eq:wideband} and we use the logarithmic-linear discretisation scheme with $W^* = W / 2$ and $L_{\textrm{log}}/L = 0.2$. We remark that the chosen Hamiltonian parameters are far apart from the energy scale dictated by $W$, such that the effect of the finite bandwidth is expected to be negligible. This choice of parameters is thus a useful representative example for exposing the efficacy of the proposed methodology. 

\begin{figure}[t]
\fontsize{13}{10}\selectfont 
\centering
\includegraphics[width=1\columnwidth]{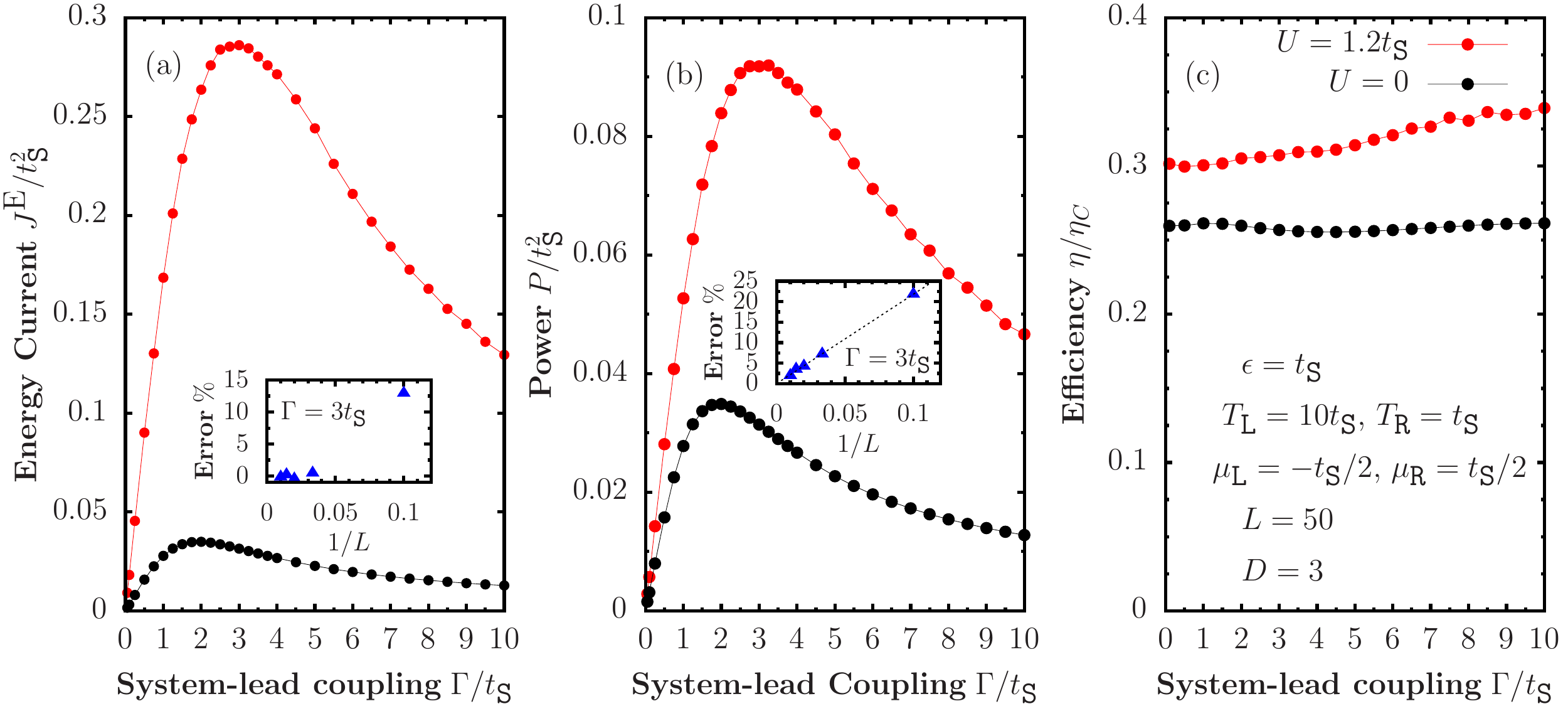}
\caption[Energy current, power and efficiency of the interacting three-site system as a function of the system-lead coupling strength]{(a) Energy current, (b) power and (c) efficiency of the interacting three-site system as a function of the system-lead coupling strength $\Gamma$. The insets in (a) and (b) show the error associated to the finite number of modes in the leads $L$ (up to $L=100$), estimated from extrapolated values of the currents at the point in which the maximum is observed ($\Gamma \approx 3t_{\texttt S}$). In these calculations we used $L_{\textrm{log}}/L = 0.2$, $W^* = W/2$ and $W = 8t_{\texttt S}$.}
\label{fig:2.3.15}
\end{figure}

We first focus on the dependence of the currents on the system-lead coupling $\Gamma$, as shown in Fig.~\ref{fig:2.3.15}. In Fig.~\ref{fig:2.3.15}(a), the energy current for a particular value of the interaction strength $U=1.2t_{\tt S}$ is shown as a function of $\Gamma$. Remarkably, a density-density interaction yields a larger energy current flowing through the system compared to the non-interacting case in the chosen regime. The same observation holds for the particle current in Fig.~\ref{fig:2.3.15}(b), since for our choice of parameters the particle current and the power output are equivalent [see Eq.~\eqref{eq:power_def}]. The efficiency shown in Fig.~\ref{fig:2.3.15}(c), remains approximately constant as a function of system-lead coupling strength just like the non-interacting case. Future work will investigate a larger range of parameters to identify a maximum power output for a given interaction strength.  

The insets in Figs.~\ref{fig:2.3.15}(a) and \ref{fig:2.3.15}(b) show the error associated to employing a finite number of modes in each reservoir for a specific value of $\Gamma = 3t_{\tt S}$, where the currents in the interacting case reach the maximum value. The error is computed from an extrapolated value of the currents to the $L \to \infty$ limit, based on the currents for finite $L$, for each respective case. We define $\textrm{Error \%} \defeq |K(L \to \infty) - K(L)| \cdot 100 / K(L \to \infty)$, where $K = J^{\textrm{E}}, P$ for energy current and power, respectively. The value $K(L \to \infty)$ is taken from an extrapolation following the trend of $K(L)$. A linear extrapolation was made for the power as shown in the inset in Fig.~\ref{fig:2.3.15}(b), while no extrapolation is required for the energy current in Fig.~\ref{fig:2.3.15}(a), as the current has converged for $L$ smaller than the final value of $L = 100$. It can be observed that for the specific choice of parameters, a good approximation can be obtained to a few percent accuracy using $L=50$, compared to larger reservoirs. The energy current converges faster than the particle current (power) in this case. This behaviour is expected, as observing Figs.~\ref{fig:app1} and \ref{fig:app2} for the non-interacting case in Appendix \ref{app:noninteracting}, the largest deviation for the particle current occurs where the maximum value is obtained, while the largest deviation for the energy current is observed near the edges of the band. 

\subsection{High-temperature transport}

The transport properties of spin chains have been studied extensively using standard open-system MPS approaches based on a boundary driving Lindblad master equation. This approach has been successful in accurately describing the high-temperature spin/particle transport behaviour of the integrable anisotropic XXZ Heisenberg model \cite{Znidaric:2010,Znidaric:2010b,Prosen:2011,Znidaric:2011} as well as non-integrable versions of the model when integrability-breaking perturbations are introduced, such as magnetic impurities (see Chapter~\ref{chapter:kubo}) or disorder~\cite{vznidarivc2016diffusive,vznidarivc2017dephasing,schulz2018energy,mendoza2018asymmetry}. However, driving on the boundary spins is formally equivalent to infinite temperature baths. Modelling energy currents therefore requires more elaborate multi-site boundary driving to mimic finite temperature differences. While this approach has proven successful for the very high temperature limit, its reliability as the temperature is lowered is questionable, as we debated in Sec.~\ref{sec:local_v_global}. The mesoscopic leads construction introduced here overcomes this deficiency. 

The system Hamiltonian introduced in Eq.~\eqref{eq:h_s_i} is the spinless fermion equivalent of the anisotropic XXZ Heisenberg model. This model exhibits a range of distinct linear response particle and energy transport behaviour as a function of the anisotropy $U$. Specifically, these include ballistic transport which is characterised by a constant value of the current as a function of system size $D$, as well as diffusive transport, where $J^{\textrm{P}} \propto 1/D^{\nu}$ with $\nu = 1$. Anomalous diffusion is signalled by $0 < \nu < 1$ and $\nu > 1$, corresponding to super-diffusion and sub-diffusion, respectively. A sharp transition in the system's transport properties is known to occur at the isotropic point $U / t_{\tt S} = 2$, with the system displaying ballistic transport for $U / t_{\tt S} < 2$, while for $U / t_{\tt S} > 2$ transport becomes diffusive. Furthermore, precisely at the isotropic point $U / t_{\tt S} = 2$, boundary driving calculations have shown that transport is super-diffusive with $\nu = 1/2$ \cite{Znidaric:2011}. These results are expected to hold only in the linear-response regime at high temperatures, where the structure of the thermal baths becomes irrelevant. We now corroborate these results using our mesoscopic reservoir formalism.

\begin{figure}[t]
\fontsize{13}{10}\selectfont 
\centering
\includegraphics[width=0.55\columnwidth]{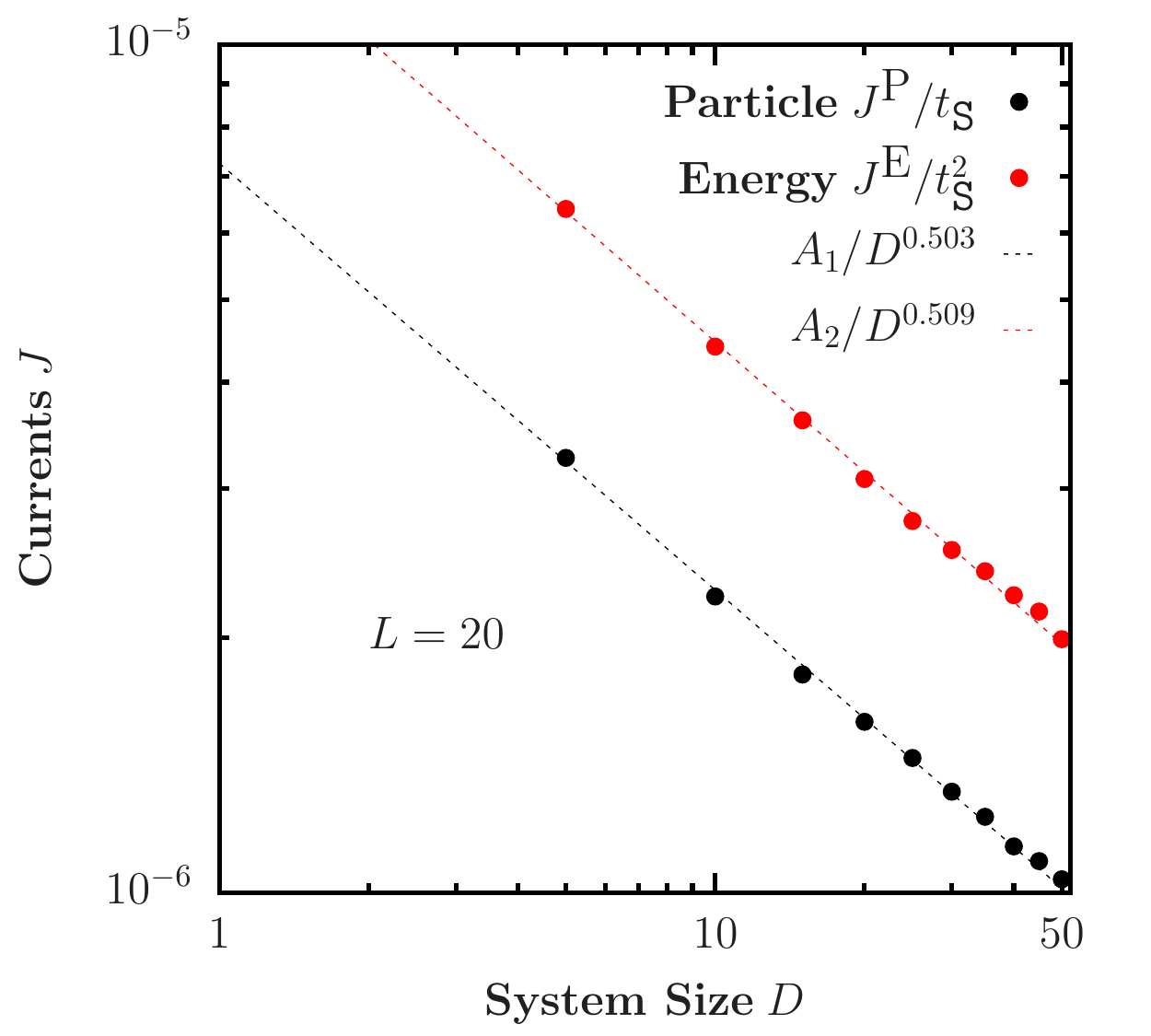}
\caption[Particle and energy currents as a function of system size $D$ for the isotropic Heisenberg model $U / t_{\tt S} = 2$]{Particle and energy currents as a function of system size $D$ for the isotropic Heisenberg model $U / t_{\tt S} = 2$ [see Eq.\eqref{eq:h_s_i}]. The results shown correspond to a very high temperature $T_{\tt L} = T_{\tt R} = 1000t_{\tt S}$ and a small chemical potential bias $\mu_{\tt L} = -\mu_{\tt R} = 0.025t_{\tt S}$, where the system is expected to be in linear response regime.  In these calculations we used $L_{\textrm{log}}/L=0.2$, $W^* = W/2$, $W = 8t_{\texttt S}$ and $\Gamma = \epsilon = t_{\texttt S}$.}
\label{fig:2.3.16}
\end{figure}

As before, we choose the same discretisation scheme and bath structure parameters. We focus on the isotropic point $U / t_{\tt S} = 2$ and set $\epsilon_j / t_{\tt S} = \epsilon / t_{\tt S} = 1$. We set the temperature on each reservoir to a high value of $T_{\tt L} = T_{\tt R} = 1000t_{\tt S}$ and choose a small chemical potential gradient $\mu_{\tt L} = -\mu_{\tt R} = 0.025t_{\tt S}$, where we expect the system to be in linear response regime. In Fig.~\ref{fig:2.3.16} we show both the particle and energy currents as a function of system size $D$. We have used $L = 20$ modes for both left and right reservoirs. As can be observed, the currents fit a power law scaling with an exponent very close to $\nu = 1/2$ in clear indication of super-diffusive behaviour. We remark that at high temperature, fewer reservoir modes can be used to obtain the correct transport exponent, as observed from boundary driving calculations \cite{Znidaric:2011}. 

\subsection{Finite-temperature transport and CP symmetry}

We now test the capabilities of our method to extract transport properties outside of the high-temperature limit. As a benchmark, we focus on the anisotropic Heisenberg Hamiltonian given by Eq.~\eqref{eq:h_s_i} with $U = t_{\tt S}$ and homogeneous on-site energies, $\epsilon_j = \epsilon$. 

\begin{figure}[t]
\fontsize{13}{10}\selectfont 
\centering
\includegraphics[width=0.6\columnwidth]{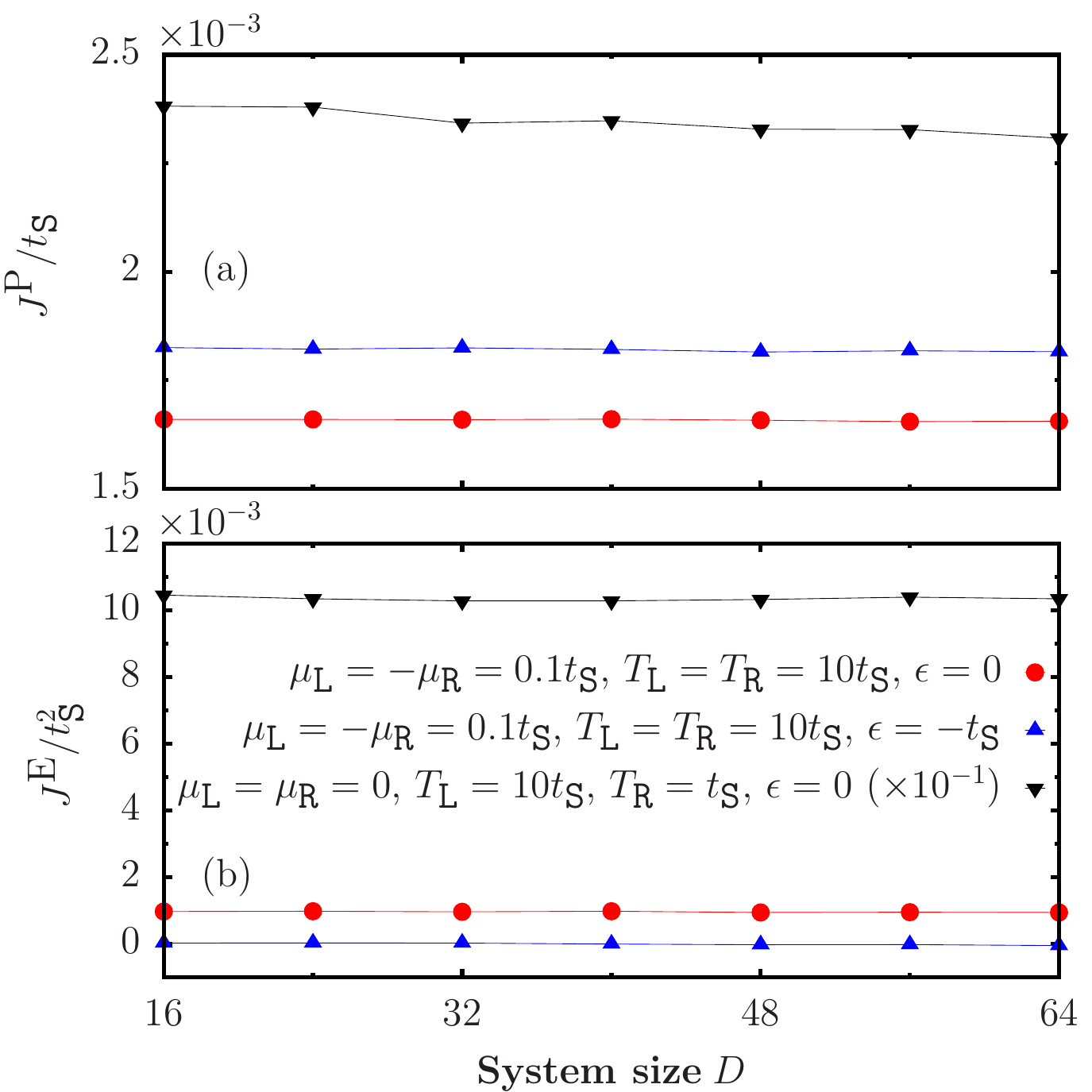}
\caption[Finite-size scaling of particle current and energy current for the anisotropic Heisenberg model with ${U=t_{\tt S}}$]{Finite-size scaling of (a)~particle current and (b)~energy current for the anisotropic Heisenberg model in Eq.~\eqref{eq:h_s_i} with ${U=t_{\tt S}}$. Size-independent currents imply ballistic particle and energy transport under chemical-potential or temperature bias. Data shown by the black triangles are rescaled by a factor of $10^{-1}$ to be visible on the same scale. In these calculations we used $L = 20$, $L_{\textrm{log}}/L=0.2$, $W^* = W/2$, $W = 8t_{\texttt S}$ and $\Gamma = U = t_{\texttt S}$.}
\label{fig:2.3.17}
\end{figure}

In this regime, the Hamiltonian is integrable and the total energy current is conserved, implying ballistic energy transport at all temperatures under linear-response conditions~\cite{zotos1997transport,Bertini:2021}. Ballistic particle conduction is also expected for $U<2t_{\tt S}$, as indicated by extensive numerical calculations~\cite{Bertini:2021} and arguments based on quasi-local conservation laws~\cite{Prosen:2011,Prosen:2013}. We confirm the ballistic nature of transport at finite temperature by a scaling analysis with the system size $D$ of the particle and energy currents, as shown in Fig.~\ref{fig:2.3.17}. We drive the system out of equilibrium either by applying a chemical-potential bias at fixed temperature, or by a temperature gradient applied at fixed chemical potential. In each case we find that the particle and energy currents are essentially independent of system size, as expected. We note that our method can be applied far outside linear response, for example with a large temperature bias $T_{\tt L} - T_{\tt R} \gg T_{\tt R}$, as shown by the black triangles in Fig.~\ref{fig:2.3.17}.

The magnitudes of the currents strongly depend on the bulk Hamiltonian and the thermodynamic potentials of the baths. Configurations that are invariant under a charge conjugation-parity (CP) transformation, i.e., a combined reflection and particle-hole symmetry, are found to exhibit vanishing energy current. More precisely, CP symmetry requires equal bath temperatures, $T_{\tt L}=T_{\tt R}$, opposite chemical potentials, $\mu_{\tt L} = -\mu_{\tt R}$, and bulk Hamiltonian parameters $\epsilon = -U$. As shown by the blue triangles in Fig.~\ref{fig:2.3.17}(b), the energy current is zero in this case, in agreement with exact analytical calculations detailed in Appendix~\ref{app:CP}. A finite energy current emerges whenever the on-site energies of $\hat{H}_{\tt S}$ are moved away from the CP-symmetric point, even when the forcing from the baths remains CP-symmetric (red circles in Fig.~\ref{fig:2.3.17}). This is in stark contrast with the predictions of single-site boundary driving transport calculations on the Heisenberg model, where symmetric driving leads to vanishing energy current independent of the bulk Hamiltonian parameters~\cite{Popkov2013}. This ultimately stems from the fact that boundary driving simulates white noise and thus does not capture the energy dependence of true thermal fluctuations.

\begin{figure}[t]
\fontsize{13}{10}\selectfont 
\centering
\includegraphics[width=0.55\columnwidth]{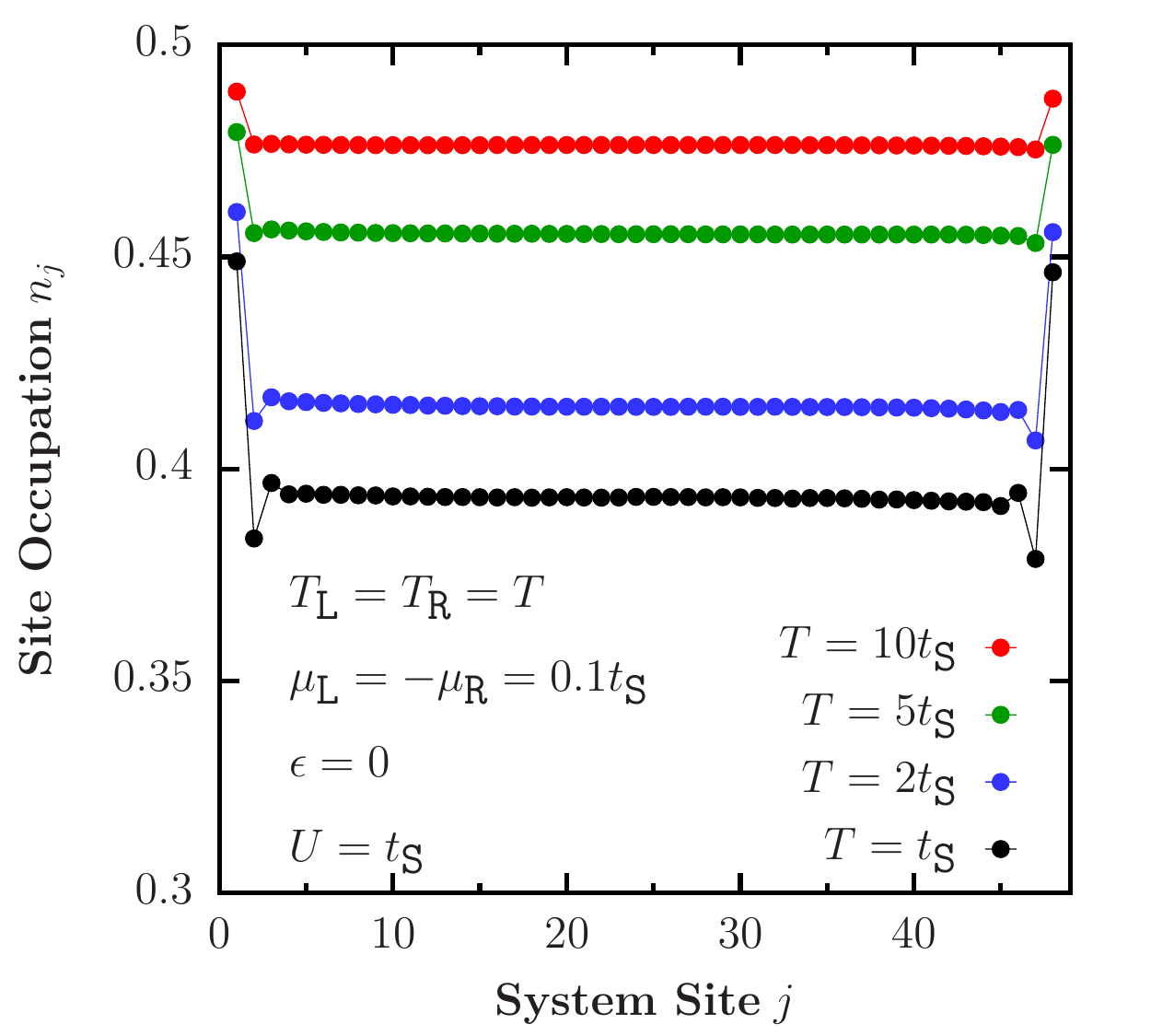}
\caption[Non-equilibrium density profile in the anisotropic Heisenberg model, under symmetric chemical-potential bias and various temperatures]{Non-equilibrium density profile in the anisotropic Heisenberg model, Eq.~\eqref{eq:h_s_i}, under symmetric chemical-potential bias and various temperatures. In these calculations we used $D = 48$, $L = 40, $ $L_{\textrm{log}}/L=0.2$, $W^* = W/2$, $W = 8t_{\texttt S}$ and $\Gamma = t_{\texttt S}$. We checked that the site occupation results are robust towards changes from $L = 20$ to $L = 40$ in all cases shown.}
\label{fig:2.3.18}
\end{figure}

We further explore the effect of temperature by examining the non-equilibrium density profile of the system in Fig.~\ref{fig:2.3.18}. We consider equal reservoir temperatures, $T_{\tt L} = T_{\tt R} = T$, fixed system ($D = 48$) and lead ($L = 40$) sizes, and a symmetric chemical potential bias, $\mu_{\tt L}= -\mu_{\tt R}$. We also take $\epsilon \neq -U$, to break CP-symmetry. Away from the boundaries, we find the flat profile characteristic of ballistic transport, with a density that depends on temperature. Lower temperatures correspond to lower densities and larger currents. As the temperature is increased, the bulk density tends to the CP-symmetric value $\langle \hat{n}_j\rangle \to 0.5$. This shows that the CP symmetry enforced by the single-site boundary driving configuration is indeed recovered in the high-temperature limit.

\section{Summary and outlook}
\label{sec:conclusions}

In this chapter, we have introduced a novel methodology to simulate the heat and particle currents in thermal machines which comprise a complex working medium coupled to fermionic leads at fixed temperatures and chemical potentials. The method is based on the concept of mesoscopic reservoirs whose energy modes are damped in order the replicate the continuum. The method allows for calculations in highly non-equilibrium scenarios such as strong system-lead coupling and large biases. In order to cope with non-quadratic interactions in the working medium, we implemented a novel tensor network algorithm directly in the star geometry using auxiliary modes. 

For the purpose of expounding the method, in this paper we considered only autonomous thermal machines where the working medium is time independent. In order to benchmark our technique we first focused on replicating the steady-state thermodynamics of the resonant-level heat engine. The simplicity of this quadratic model allows for direct comparison with the Landauer-B\"uttiker theory for quantum transport. We observed excellent agreement across a wide parameter regime. We then explored efficiency and power in a strongly interacting three-qubit machine in a parameter regime where other methods are known to struggle. In doing this we observed that, remarkably, the efficiency is enhanced as a function of the system-lead coupling in the presence of non-quadratic interactions. Furthermore, we demonstrated that our technique is capable of highly non-trivial heat and particle transport calculations in strongly correlated many-body systems by performing a scaling analysis at the isotropic point of the paradigmatic Heisenberg model. Finally, we analysed the current scaling and non-equilibrium density profile in the integrable regime of the anisotropic Heisenberg model, confirming the ballistic nature of transport at finite temperature and well beyond linear response.

Due to the flexibility of our technique we expect that the method is extendable further in the direction of steady-state thermodynamics of complex interacting quantum systems. Beyond strong coupling and far-from-equilibrium scenarios, our technique may also find useful applications in the study of time-dependent working media, bulk noise effects and non-trivial spectral densities, thus taking quantum thermodynamics to unexplored horizons.   
\newpage\null\thispagestyle{fancy}\newpage

\appendix
\chapter*{Final remarks and future work}
\addcontentsline{toc}{chapter}{Final remarks and future work}
\markboth{FINAL REMARKS AND FUTURE WORK}{FINAL REMARKS AND FUTURE WORK}

The topics exposed in this thesis were motivated by a very simple question.

A single magnetic impurity is enough to imprint the signatures of chaotic eigenstates onto the integrable Heisenberg model. Yet, intuitively, a $\mathcal{O}(1)$ form of integrability-breaking perturbation does not suffice to render a ballistic system into a diffusive one. We have conciliated this result from the perspective of linear response theory through global conservation laws, where we found translational invariance to be crucial in the determination of spin Drude weights at high temperature. These results were put to the test from the perspective of open systems theory in boundary-driven calculations, which allowed us to overcome finite-size limitations. 

These results prompted the questions about thermodynamics we addressed afterwards, namely, thermalisation and its connection to entanglement and chaos at the fundamental level. The treatment we proposed to address finite-temperature transport in autonomous quantum thermal machines, followed from the necessity to introduce a novel approach to overcome the limitations imposed by the technique of boundary driving. 

Having established these concepts, many open questions remain.

For the sake of the single impurity model and its properties, we open two questions. 

The first one related to thermalisation. We have established that thermalisation in the single impurity model is anomalous, in the sense that the statistical properties of the unperturbed model end up embedded in the perturbed one. Crucially, this holds true for observables away from the site of the impurity or sums of local observables, such as the total spin current. This allowed us to present a complete picture of the nature of coherent transport in the single impurity model. The topic of {\em prethermalisation} has been discussed recently~\cite{Mori:2018,Chengju:2017,Bastianello:2019}, whereby a first stage of relaxation happens to a non-thermal state, followed by true thermalisation as predicted by the ensembles of statistical mechanics at later timescales. Our results suggest that this could be the relaxation mechanism in the single impurity model. This could be explored from the unitary dynamics of typical states in the high-temperature regime, akin to the procedure employed in Sec.~\ref{sec:scaling_otoc}.

The second question we could open in relation to the single impurity is related to transport at finite temperature. For repulsive interactions (such as the ones considered in this thesis), it was shown by Kane and Fisher in Ref.~\cite{Kane:1992} that the probability of electron transmission through a single barrier in a one-dimensional electron gas vanishes at zero temperature. It then follows that the system behaves as an insulator. Yet, from our results, we argued that the system is a perfect conductor at sufficiently high temperature. So, the behaviour between these two temperature limits could shed light to very interesting phenomena. It was argued by Kane and Fisher in Ref.~\cite{Kane:1992} that the conductance vanishes as a power of temperature in the finite-temperature regime, a statement that could be put to the test using the procedure introduced in Chapter~\ref{chapter:finite_temperature}. Moreover, the transport exponents (or lack thereof) could be equally evaluated by employing our method.

This brings us to the domain of finite-temperature transport, around which many open questions remain~\cite{Bertini:2021}. We shall mention spin/particle and particle transport in interacting systems where integrability is broken by the introduction of disorder. This integrability-breaking perturbation introduces a competition between inelastic collisions and coherent effects, giving rise to a phase of matter first described by Basko, Aleiner and Altshuler in Ref.~\cite{Basko:2006}. Its existence has been questioned recently~\cite{Bertini:2018}, though whether there exists a phase transition or not, transport in disordered systems poses some interesting open questions. Particularly in the nature of energy filtering. Whenever a system is exposed to a certain degree of disorder, some energy eigenstates are localised while some are extended, depending on the disorder parameter and the energy density of the relevant eigenstate~\cite{Luitz:2015}. Disordered systems then, could be proposed as energy filters. Suppose we couple a one-dimensional disordered system to mesoscopic reservoirs on both sides kept at different equilibrium states, following the configurations we introduced in Chapter~\ref{chapter:finite_temperature}. Due to the nature of disorder, flow of particles in direction will be favoured depending on the equilibrium states chosen for the reservoirs, as states with lower energy density will not constitute transport channels due to localisation. This energy filtering phenomenon was originally suggested by Basko, Aleiner and Altshuler in Ref.~\cite{Basko:2006}, and it could be probed using our technique.

Finally, another problem that could be addressed with our mesoscopic reservoir approach is related to Kondo physics in quantum impurity models~\cite{Rosch:2003} and the Anderson single-impurity model~\cite{Meisner:2009}, for which most known results have been addressed through the numerical renormalisation group at thermal equilibrium. Away from equilibrium in configurations that lead to non-equilibrium steady states, much less is known, particularly when interactions are treated exactly while resolving very small energy scales~\cite{Schwarz2018}. Recently, a novel scheme was proposed in Ref.~\cite{Schwarz2018} to address part of these issues, with a technique related to the one we introduced in Chapter~\ref{chapter:finite_temperature}. Crucially, the treatment introduced for the reservoirs in Ref.~\cite{Schwarz2018} does not include dissipation, driving each energy mode to thermal equilibrium. For this reason, our intuition is that the methodology introduced in Chapter~\ref{chapter:finite_temperature} will provide better scalability, where we would be able to describe thermal reservoirs via fewer modes. Although the introduction of a spin degree of freedom to the particles to address Kondo physics seems challenging, it could be feasible by introducing auxiliary reservoirs, as suggested in Ref.~\cite{Schwarz2018}, bringing these interesting problems within the reach of our methods.  
\appendix
\chapter*{Acknowledgements}
\addcontentsline{toc}{chapter}{Acknowledgements}
\markboth{ACKNOWLEDGEMENTS}{ACKNOWLEDGEMENTS}
\setcounter{footnote}{0}

I would be willing to bet that most people who embark upon a doctorate training have {\em at least} some idea of the sacrifices the journey entails. Yet, in the words of Morpheus, {\em there is a difference between knowing the path, and walking the path}\footnote{By no means I am comparing myself to {\em The One}. Although, as one of the first members in our research group, when the (ETH) {\em Matrix} (elements) was created, I {\em was} the man born inside it.}. I was blessed to have the best PhD advisor someone could ask for, who made my {\em path} more interesting than I could have ever imagined. Perhaps some people in the past have made a similar claim, but I am convinced that the honour is mine. John Goold allowed me to play the game using my strengths, while making sure I would polish my weaknesses. In retrospective, the numerous nights spent discussing physics and life over countless pints will be sorely missed. Words would fall short to express my gratitude. Nevertheless, my first acknowledgement goes to him, without whom, Dublin would have been a very different place.

Over the course of four years I had the pleasure of working with brilliant scientists. I would like to thank Marcos Rigol, for introducing me to the topic of eigenstate thermalisation and to a level of meticulousness of the greatest proportions. Special thanks Alessandro Silva, whom I greatly respect, for allowing me to work with his extraordinary ideas. I would also like to express my gratitude to Stephen R. Clark for his invaluable support during our project.

A number of my closest collaborators deserve my gratitude, particularly Mark T. Mitchison for his exceptional level of support during the years. I would like to thank Archak Purkayastha for all of our interesting discussions and his support and Juan Jos\'e Mendoza-Arenas, from whom I have learned a lot. I would also like to thank Silvia Pappalardi and Eduardo Mascarenhas for our collaborations. 

I will take the opportunity to thank Marcello Dalmonte, Antonello Scardicchio, Ivan Girotto, Markus Heyl, Vipin K. Varma, scientists who have influenced my work and perspectives in many different ways.

Four years in Dublin has brought me in touch with people I now consider very important in my life. I especially appreciate my friendship with Cecilia Chiaracane, Maria Popovic, Conor Murphy, John Moroney and Giacomo Guarnieri, all of whom suffered through the pain of sharing an office with me. Their friendship has been invaluable to me. 

I thank my mother, Ana Navarro Hidalgo, for teaching me the meaning of true courage and for her invaluable support. Finally, I thank my beloved family, to whom this work is dedicated.  \clearpage

{\setstretch{1.0}
\bibliographystyle{apsrev4-2}
\addcontentsline{toc}{chapter}{Bibliography}
\bibliography{bibliography}
}

\newpage\null\thispagestyle{fancy}\newpage

\appendix
\chapter*{Appendices}
\addcontentsline{toc}{chapter}{Appendices}
\numberwithin{equation}{section}
\renewcommand{\thesection}{\Alph{section}}
\renewcommand{\theequation}{\thesection.\arabic{equation}}
\renewcommand{\thefigure}{\thesection\arabic{equation}}

\section{From macroscopic to mesoscopic reservoirs}
\label{app:meso_equivalence}

In this appendix we give further mathematical details of the connection between mesoscopic and infinite reservoirs described in Sec.~\ref{sec:mesoscopic_leads}. 

\subsection{Infinite-bath configuration}
\label{app:infinite_bath}

We begin by discussing the equations of motion assuming that the system is in contact with an infinite thermal reservoir. The total Hamiltonian is thus $\hat{H} = \hat{H}_{\tt S} + \hat{H}_{\tt B} + \hat{H}_{\tt SB}$, where $\hat{H}_{\tt B}$ and $\hat{H}_{\tt SB}$ are respectively given by
\begin{align}
\label{eq:H_B_infinite_1}
&\hat{H}_{\tt B} = \sum_{m=1}^\infty \omega_m \hat{b}^{\dagger}_m \hat{b}_m,\\
&\label{eq:H_SB_infinite_1}
\hat{H}_{\tt SB} = \sum_{m=1}^{\infty} \left( \lambda_{m} \hat{c}^{\dagger}_p \hat{b}_m + \lambda^*_{m} \hat{b}^{\dagger}_m \hat{c}_p \right),
\end{align}
while $\hat{H}_{\tt S}$ is arbitrary. In the Heisenberg picture, the equations of motion read as
\begin{align}
    \label{b_EOM}
    \frac{\rm d}{{\rm d} t}\hat{b}_m(t) & = -{\rm i}\omega_m \hat{b}_m(t) -{\rm i} \lambda_m^*\hat{c}_p(t),\\
    \label{c_EOM}
\frac{\rm d}{{\rm d} t}\hat{c}_j(t) & = {\rm i}[\hat{H}_{\tt S},\hat{c}_j(t)] - {\rm i}\delta_{jp} \sum_m \lambda_m \hat{b}_m(t),
\end{align}
where $p$ denotes the system site connected to the bath. The formal solution of Eq.~\eqref{b_EOM} reads as
\begin{equation}
    \label{b_formal_solution}
    \hat{b}_m(t) = {\rm e}^{-{\rm i}\omega_m t} \hat{b}_m(0) -{\rm i} \lambda_m^*\int_0^t{\rm d}t'\, {\rm e}^{-{\rm i}\omega_m (t-t')}\hat{c}_p(t'). 
    \end{equation}
Substituting this back into Eq.~\eqref{c_EOM} yields the quantum Langevin equation
\begin{equation}
    \label{QLE}
        \frac{{\rm d}}{{\rm d} t}\hat{c}_j(t) = \ii [\hat{H}_{\tt S}, \hat{c}_j(t)] + \delta_{jp} \left[\hat{\xi}(t) - \int_0^t {\rm d}t'\, \chi(t-t') \hat{c}_p(t')\right].
\end{equation}
Here, the noise operator is $\hat{\xi}(t) = -{\rm i}\sum_m{\rm e}^{-{\rm i}\omega_m t}\lambda_m \hat{b}_m(0)$ and the memory kernel is $\chi(t-t') = \langle \{\hat{\xi}(t),\hat{\xi}^\dagger(t')\}\rangle$.

The solution of Eq.~\eqref{QLE} at time $t$ depends in principle on the entire past history of the noise operator $\hat{\xi}(s)$ for $s<t$. Once found, the solution for $\hat{c}_j(t)$ is sufficient to reconstruct all $n$-point correlation functions of $\tt S$, which together uniquely specify the quantum state (amongst other information). Since the initial bath state is Gaussian, these correlation functions depend on the noise only via its two-time correlations
\begin{align}
    \label{chi_t}
    \langle \{\hat{\xi}(t),\hat{\xi}^\dagger(t')\}\rangle & = \int \frac{{\rm d}\omega}{2\pi} \mathcal{J}(\omega) \ee^{-\ii \omega(t-t')},\\
    \label{phi_t}
    \langle \hat{\xi}^\dagger(t)\hat{\xi}(t')\rangle & = \int \frac{{\rm d}\omega}{2\pi} \mathcal{J}(\omega) f(\omega) \ee^{\ii \omega(t-t')}.
\end{align}

In some cases, like for a single site system, the particle and energy currents from the bath also become important. The particle and energy currents from the bath are given by
\begin{align}
&J^{\textrm{P}} = \textrm{i}\left \langle\sum_{m=1}^{\infty} \left( \lambda_{m} \hat{c}^{\dagger}_p \hat{b}_m  - \lambda^*_{m} \hat{b}^{\dagger}_m \hat{c}_p  \right)\right\rangle, \\
&J^{\textrm{E}} = \textrm{i}\left \langle\sum_{m=1}^{\infty} \omega_m\left( \lambda_{m} \hat{c}^{\dagger}_p \hat{b}_m  - \lambda^*_{m} \hat{b}^{\dagger}_m \hat{c}_p  \right)\right\rangle.
\end{align}
This requires evaluation of the operators $\langle\sum_{m=1}^{\infty} \lambda_{m} \hat{c}^{\dagger}_p \hat{b}_m\rangle$ and $\langle\sum_{m=1}^{\infty}\omega_m\lambda_{m} \hat{c}^{\dagger}_p \hat{b}_m\rangle$. The evolution of these operators can be written down from Eq.~\ref{b_formal_solution} and are given by
\begingroup
\allowdisplaybreaks
\begin{align}
\label{curr_op}
\langle\sum_{m=1}^{\infty} \lambda_{m} \hat{c}^{\dagger}_p(t) \hat{b}_m(t)\rangle = \textrm{i}\langle \hat{c}^{\dagger}_p(t) \hat{\xi}(t) \rangle -\textrm{i}\int_0^t {\rm d}t'\, \chi(t-t') \langle\hat{c}^{\dagger}_p(t)\hat{c}_p(t')\rangle, \\
\label{energy_curr_op}
\langle\sum_{m=1}^{\infty} \omega_m\lambda_{m} \hat{c}^{\dagger}_p(t) \hat{b}_m(t)\rangle = \textrm{i}\langle \hat{c}^{\dagger}_p(t) \hat{\tilde{\xi}}(t) \rangle -\textrm{i}\int_0^t {\rm d}t'\, \tilde{\chi}(t-t') \langle\hat{c}^{\dagger}_p(t)\hat{c}_p(t')\rangle,
\end{align} 
\endgroup
where we have additionally defined  
\begin{align}
\label{xi_tilde_t}
&\hat{\tilde{\xi}}(t)~=-{\rm i}\sum_m{\rm e}^{-{\rm i}\omega_m t}\omega_m\lambda_m \hat{b}_m(0), \\
    \label{chi_tilde_t}
   & \tilde{\chi}(t-t')= \int \frac{{\rm d}\omega}{2\pi} \omega\mathcal{J}(\omega) \ee^{-\ii \omega(t-t')}.
\end{align}
The operator $\hat{\tilde{\xi}}(t)$ satisfies
\begin{align}
    \label{phi_tilde_t1}
    \langle \hat{\tilde{\xi}}^\dagger(t)\hat{\tilde{\xi}}(t')\rangle & = \int \frac{{\rm d}\omega}{2\pi} \omega^2\mathcal{J}(\omega) f(\omega) \ee^{\ii \omega(t-t')}, \\
    \label{phi_tilde_t2}
    \langle \hat{\tilde{\xi}}^\dagger(t)\hat{\xi}(t')\rangle & = \int \frac{{\rm d}\omega}{2\pi} \omega\mathcal{J}(\omega) f(\omega) \ee^{\ii \omega(t-t')}.
\end{align}
Eqs.~(\ref{QLE}), (\ref{chi_tilde_t}), (\ref{phi_t}), (\ref{curr_op}), (\ref{energy_curr_op}), (\ref{phi_tilde_t1}), (\ref{phi_tilde_t2}) completely define time evolution of any operator of the system, as well as that of the energy and particle currents from the baths. In the following, we show that the same equations can be recovered in the mesoscopic-lead configuration, thereby showing their equivalence.

\subsection{Mesoscopic-lead configuration}
\label{app:meso_reservoir}
We now turn to the mesoscopic-reservoir configuration, with total Hamiltonian $\hat{H} = \hat{H}_{\tt S} +  \hat{H}_{\tt SL} + \hat{H}_{\tt L} + \hat{H}_{\tt LB} + \hat{H}_{\tt B}$. Here $\hat{H}_{\tt L}$ and $\hat{H}_{\tt SL}$ describe the lead and its coupling to the system and are given explicitly by
\begin{align}
\label{eq:H_lead_1}
    \hat{H}_{\tt L} & = \sum_{k=1}^L \varepsilon_k \hat{a}^\dagger_k \hat{a}_k,\\
    \label{eq:H_lead_sys_1}
    \hat{H}_{\tt SL} & = \sum_{k=1}^{L} \left( \kappa_{kp} \hat{c}^{\dagger}_p \hat{a}_k + \kappa^*_{kp} \hat{a}^{\dagger}_k\hat{c}_p \right).
\end{align}
Each mode of the lead is further coupled to an infinite reservoir according to 
\begin{align}
    \label{H_bath}
    \hat{H}_{\tt B} &= \sum_{k=1}^L \sum_{q=1}^\infty \Omega_{kq}\hat{b}_{kq}^\dagger \hat{b}_{kq}, \\
    \label{H_lead_bath}
     \hat{H}_{\tt LB} &= \sum_{k=1}^L \sum_{q=1}^\infty \left(\zeta_{kq}\hat{a}_{k}^\dagger \hat{b}_{kq} + \zeta_{kq}^*\hat{b}_{kq}^\dagger \hat{a}_{k}\right),
\end{align}
where $\hat{a}_k$ describes mode $k$ of the lead, while the ladder operators $\hat{b}_{kq}$ describe the bath connected to mode $k$. Each bath is described by the flat spectral density
\begin{equation}\label{flat_spectral_density}
    \mathcal{J}_k(\omega) = 2\pi \sum_q |\zeta_{kq}|^2 \delta(\omega - \Omega_{kq}) = \gamma_k.
\end{equation}
We are interested in the evolution of the joint system-lead state $\rho(t)$ starting from the initial product state Eq.~\eqref{eq:product_state}, where all baths are initialised at the same temperature and chemical potential.

As in Eq~\eqref{b_formal_solution}, we formally solve the Heisenberg equation of motion for the bath variables to find 
\begin{equation}
    \hat{b}_{kq}(t) = {\rm e}^{-{\rm i}\Omega_{kq}t} \hat{b}_{kq}(0) - {\rm i}\zeta_{kq}^*\int_0^t{\rm d}t' \, {\rm e}^{-{\rm i}\Omega_{kq}(t-t')}\hat{a}_k(t').
\end{equation}
Substituting this into the equation of motion for $\hat{a}_k(t)$, we obtain
\begin{align}
\label{a_k_Langevin}
    \frac{\rm d}{{\rm d} t}\hat{a}_k(t) = -{\rm i}\varepsilon_k \hat{a}_k(t) - {\rm i}\kappa_{kp}^* \hat{c}_p(t) + \hat{\xi}_k(t) - \int_0^t{\rm d}t'\, \chi_k(t-t')\hat{a}_k(t').
\end{align}
Here, we defined the noise operators
\begin{equation}
    \label{noise_lead_site_k}
    \hat{\xi}_k(t) = -{\rm i}\sum_q \zeta_{kq} {\rm e}^{-{\rm i}\Omega_{kq}t} \hat{b}_{kq}(0),
\end{equation}
and the memory kernels $\chi_k(t-t') = \langle \{\hat{\xi}_k(t),\hat{\xi}_{k}^\dagger(t')\}\rangle$. For the flat spectral density in Eq.~\eqref{flat_spectral_density}, the noise correlations are given by
\begin{align}
    \label{noise_k_memory}
     \langle \{\hat{\xi}_k(t),\hat{\xi}_{k'}^\dagger(t')\}\rangle & = \delta_{kk'}\gamma_k \delta(t-t'), \\
     \label{noise_k_correlation}
     \langle \hat{\xi}_k^\dagger(t)\hat{\xi}_{k'}(t')\rangle & = \delta_{kk'} \gamma_k \int\frac{{\rm d}\omega}{2\pi}\, f(\omega) {\rm e}^{{\rm i}\omega(t-t')}.
\end{align}

Next we formally solve Eq.~\eqref{a_k_Langevin} to find
\begin{align}
\label{lead_formal_solution}
    \hat{a}_k(t) = {\rm e}^{-{\rm i}\varepsilon_k t - \gamma_k t/2} \hat{a}_k(0) + \int_0^t{\rm d}t'\, {\rm e}^{(-{\rm i}\varepsilon_k - \gamma_k/2)(t-t')}\left[\hat{\xi}_k(t') - {\rm i} \kappa_{kp}^* \hat{c}_p(t') \right].
\end{align}
Considering long times, such that $\gamma_k t \gg 1$, the first term above is negligible and will be ignored in the following. Substituting this solution into the equations of motion for the system variables, we finally obtain an effective quantum Langevin equation
\begin{align}
    \label{Langevin_eff}
        \frac{{\rm d}}{{\rm d} t}\hat{c}_j(t) & = \ii [\hat{H}_{\tt S}, \hat{c}_j(t)] + \delta_{jp} \left[\hat{\xi}_{\rm eff}(t) - \int_0^t {\rm d}t'\, \chi_{\rm eff}(t-t') \hat{c}_p(t')\right].
\end{align}
This is of the same form as Eq.~\eqref{QLE}, but with the noise operator
\begin{equation}\label{noise_eff}
    \hat{\xi}_{\rm eff}(t) = -{\rm i} \sum_{k=1}^L \kappa_{kp}\int_0^t{\rm d}t'\,{\rm e}^{(-{\rm i}\varepsilon_k - \gamma_k/2)(t-t')} \hat{\xi}_k(t'),
\end{equation}
and the memory kernel
\begin{align}
    \chi_{\rm eff}(t-t') & = \sum_{k=1}^L|\kappa_{kp}|^2{\rm e}^{(-{\rm i}\varepsilon_k - \gamma_k/2)(t-t')} \notag \\
    & = \int \frac{{\rm d}\omega}{2\pi} \mathcal{J}^{\rm eff}(\omega) \ee^{-\ii \omega(t-t')},
\end{align}
where the effective spectral density $\mathcal{J}^{\rm eff}(\omega)$ is the sum of Lorentzian functions 
\begin{equation}
    \label{eq:spectral_density_eff}
    \mathcal{J}^{\rm eff}(\omega)  = \sum_{k=1}^L \frac{|\kappa_{kp}|^2 \gamma_k}{(\omega-\varepsilon_k)^2 + (\gamma_k/2)^2}.
\end{equation}
The second equality above follows via an identity which can be proved by contour integration:
\begin{equation}
    \label{exp_identity}
    {\rm e}^{-{\rm i}\varepsilon_k t - \gamma_k t/2} = \int\frac{{\rm d}\omega}{2\pi}\, \frac{\gamma_k{\rm e}^{-{\rm i}\omega t}}{(\omega-\varepsilon_k)^2 + (\gamma_k/2)^2}.
\end{equation}

It remains to check the effective noise correlations. We have, using Eqs.~\eqref{noise_k_memory},~\eqref{noise_k_correlation} and~\eqref{exp_identity}, 
\begin{align}\label{memory_eff}
    \langle \{\hat{\xi}_{\rm eff}(s),\hat{\xi}_{\rm eff}^\dagger(s')\}\rangle &  \approx \int \frac{{\rm d}\omega}{2\pi} \mathcal{J}^{\rm eff}(\omega) \ee^{-\ii \omega(s-s')},\\ 
        \label{noise_eff_corr}
    \langle \hat{\xi}_{\rm eff}^\dagger(s)\hat{\xi}_{\rm eff}(s')\rangle & \approx \int \frac{{\rm d}\omega}{2\pi} \mathcal{J}^{\rm eff}(\omega) f(\omega)\ee^{\ii \omega(s-s')},
\end{align}
where we have neglected all terms proportional to ${\rm e}^{-\gamma_k s}$ or ${\rm e}^{-\gamma_k s'}$. This approximation is valid at long times, so long as the solution of Eq.~\eqref{Langevin_eff} depends only on the past history of $\hat{\xi}_{\rm eff}(s)$ within a time window that is essentially finite. This will generically be the case for any system that relaxes to a steady state when coupled to a bath, since any memory of environmental fluctuations in the far past is eventually lost. In particular, if $\tau_{\rm rel}$ is the (slowest) characteristic timescale of relaxation of ${\tt S}$, then we need consider only arguments of $\hat{\xi}_{\rm eff}(s)$ in the range $t - \tau_{\rm rel}\lesssim s < t$. Hence, the approximations leading to Eqs.~\eqref{memory_eff} and \eqref{noise_eff_corr} are valid for all times such that
\begin{equation}
    \label{noise_eff_approx}
    t \gg \gamma_k^{-1}, \tau_{\rm rel}.
\end{equation}
If this holds, we have shown that the effective noise generated by the mesoscopic lead is equivalent to an infinite bath with a spectral density given by Eq.~\eqref{eq:spectral_density_eff}, giving rise to an identical equation of motion for the system, Eq.~\eqref{Langevin_eff}.

Under this condition, the currents from the mesoscopic leads also become the same as the currents obtained in the infinite bath case. To see this, we write down the expressions for particle and energy currents from the lead,
\begin{align}
\label{curr_op_lead_def}
&J^{\textrm{P}} = \textrm{i}\left \langle \sum_{k=1}^{L} \left( \kappa_{kp} \hat{c}^{\dagger}_p \hat{a}_k - \kappa^*_{kp} \hat{a}^{\dagger}_k\hat{c}_p \right)\right \rangle,  \\
\label{ecurr_op_lead_def}
&J^{\textrm{E}} = \textrm{i}\left \langle\sum_{k=1}^{L} \varepsilon_k\left( \kappa_{kp} \hat{c}^{\dagger}_p \hat{a}_k - \kappa^*_{kp} \hat{a}^{\dagger}_k\hat{c}_p \right)\right \rangle.
\end{align}
This requires evaluation of the operators $\langle \sum_{k=1}^{L} \kappa_{kp} \hat{c}^{\dagger}_p \hat{a}_k \rangle$ and $\langle \sum_{k=1}^{L} \varepsilon_k \kappa_{kp} \hat{c}^{\dagger}_p \hat{a}_k \rangle$. From Eq.~(\ref{lead_formal_solution}), and considering the time regime in Eq.~(\ref{noise_eff_approx}), we have the following equations for evolution of these operators,
\begin{align}
\label{curr_op_lead}
\langle \sum_{k=1}^{L} \kappa_{kp} \hat{c}^{\dagger}_p \hat{a}_k \rangle = \textrm{i}\langle \hat{c}^{\dagger}_p(t) \hat{\xi}_{\rm eff}(t) \rangle -\textrm{i}\int_0^t {\rm d}t'\, \chi_{\rm eff}(t-t') \langle\hat{c}^{\dagger}_p(t)\hat{c}_p(t')\rangle, \\
\label{energy_curr_op_lead}
\langle \sum_{k=1}^{L} \varepsilon_k \kappa_{kp} \hat{c}^{\dagger}_p \hat{a}_k \rangle = \textrm{i}\langle \hat{c}^{\dagger}_p(t) \hat{\tilde{\xi}}_{\rm eff}(t) \rangle -\textrm{i}\int_0^t {\rm d}t'\, \tilde{\chi}_{\rm eff}(t-t') \langle\hat{c}^{\dagger}_p(t)\hat{c}_p(t')\rangle,
\end{align} 
where  
\begin{align}
\label{xi_tilde_t_mesoscopic}
&\hat{\tilde{\xi}}_{\rm eff}(t) = -{\rm i} \sum_{k=1}^L \varepsilon_k \kappa_{kp}\int_0^t{\rm d}t'\,{\rm e}^{(-{\rm i}\varepsilon_k - \gamma_k/2)(t-t')} \hat{\xi}_k(t'), \\
    \label{chi_tilde_t_mesoscopic}
    &\tilde{\chi}_{\rm eff}(t-t')= \int \frac{{\rm d}\omega}{2\pi} \omega\mathcal{J}_{\rm eff}(\omega) \ee^{-\ii \omega(t-t')}.
\end{align}
The operator $\hat{\tilde{\xi}}(t)$ satisfies
\begingroup
\allowdisplaybreaks
\begin{align}
    \label{phi_tilde_t_mesoscopic1}
\langle \hat{\tilde{\xi}}_{\rm eff}^\dagger(t)\hat{\tilde{\xi}}_{\rm eff}(t')\rangle & = \int \frac{{\rm d}\omega}{2\pi} \omega^2\mathcal{J}^{\rm eff}(\omega) f(\omega) \ee^{\ii \omega(t-t')},  \\
\label{phi_tilde_t_mesoscopic2}
\langle \hat{\tilde{\xi}}_{\rm eff}^\dagger(t)\hat{\xi}_{\rm eff}(t')\rangle & = \int \frac{{\rm d}\omega}{2\pi} \omega\mathcal{J}^{\rm eff}(\omega) f(\omega) \ee^{\ii \omega(t-t')}.
\end{align}
\endgroup
Here we have neglected terms proportional to $\ee^{-\gamma_k t}$ and $\ee^{-\gamma_k t'}$, following the same arguments that led to Eqs.~\eqref{memory_eff} and \eqref{noise_eff_corr}. In addition, we have made the approximation $\sum_k \varepsilon_k^n|\kappa_{kp}|^2 \gamma_k/[(\omega - \varepsilon_k)^2+ (\gamma_k/2)^2] \approx \omega^n \mathcal{J}^{\rm eff}(\omega)$, which holds so long as $\gamma_k$ is sufficiently small that the replacement $\varepsilon_k \to \omega$ in the numerator is valid. In this limit, $\mathcal{J}^{\rm eff}(\omega)$ reproduces $\mathcal{J}(\omega)$ faithfully and therefore the above equations become equivalent to Eqs.~\eqref{curr_op}--\eqref{phi_tilde_t2}. 

We note that in Eq.~\eqref{ecurr_op_lead_def} we have considered only the contribution to the current associated with the change in the lead energy, i.e. $J^{\textrm{E}} = -\langle \dd \hat{H}_{\tt L} /\dd t\rangle$. However, due to the Lindblad damping, there is an additional term associated with the change in $\hat{H}_{\tt SL}$, i.e. the second term in Eq.~\eqref{eq:enersf_explicit}. This term is of order $O(\gamma_k\kappa_{kp})$ and therefore becomes negligible in comparison to the first term in the limit $L\to \infty$. Thus, currents from the baths in the infinite bath configuration also become the same as currents from the mesoscopic lead in this regime. 

\subsection{Quantum master equation}

Finally, we briefly discuss the derivation of the quantum master equation. In the limit of large lead size, $L\to \infty$, the energy spacing $e_k=\varepsilon_{k+1}-\varepsilon_k \to 0$. So both the lead-bath couplings $\kappa_{kp} \propto \sqrt{e_k}$ and the system-lead coupling $\gamma_k=e_k$ must tend to zero in order to recover the continuum spectral density $\mathcal{J}(\omega)$ (see the discussion below Eq.~\eqref{eq:spectral_density_effective}). In this limit, we derive a quantum master equation using perturbation theory correct to $O(e_k)$. Following the standard procedure~\cite{BreuerPetruccione}, and working in an interaction picture with respect to the free Hamiltonian $\hat{H}_0 = \hat{H}_{\tt S} + \hat{H}_{\tt SL} + \hat{H}_{\tt L} + \hat{H}_{\tt B}$, we obtain
\begin{equation}
    \label{QME_app}
     \frac{\rm d}{{\rm d} t}\hat{\rho}(t) = -\int_0^\infty {\rm d}t' \,{\rm Tr}_{\tt B}\, [\hat{H}_{\tt LB}(t),[ \hat{H}_{\tt LB}(t-t'),\hat{\rho}(t)\hat{\rho}_{\tt B}]].
\end{equation}
Here, the upper limit of the $t'$ integration is taken to infinity because we consider the long-time limit, i.e. only the Born approximation and not the Markov approximation is invoked in Eq.~\eqref{QME_app}. In the interaction picture, the free evolution of the lead operators is given by
\begin{equation}\label{a_k_interaction}
    \hat{a}_k(t) = {\rm e}^{{\rm i}\hat{H}_0 t}\hat{a}_k{\rm e}^{-{\rm i}\hat{H}_0 t} = {\rm e}^{-{\rm i}\varepsilon_k t}\hat{a}_k + O(\kappa_{kp}).
\end{equation}
Since Eq.~\eqref{QME_app} is already of order $O(\gamma_k)$, we keep only the leading-order term in Eq.~\eqref{a_k_interaction}. Straightforward manipulations then lead to the master equation given by Eq.~\eqref{eq:Lindblad}. Note that the usual Lamb-shift Hamiltonian does not appear here due to the flat spectral densities in Eq.~\eqref{flat_spectral_density}.

The quantum master derived up to $O(e_k)$ is of the form
\begin{align}
\frac{\rm d}{{\rm d} t}\hat{\rho}(t) = \mathcal{L}^{(0)}\hat{\rho} + \mathcal{L}^{(1)}\hat{\rho},
\end{align}
where $\mathcal{L}^{(0)}$ is the $O(1)$ term of the Liouvillian, and $\mathcal{L}^{(1)}$ is the $O(e_k)$ term of the Liouvillian. The solution of this equation is 
\begin{align}
\hat{\rho}(t) = e^{(\mathcal{L}^{(0)} + \mathcal{L}^{(1)})t} \hat{\rho}(0),
\end{align}
which has all orders of $O(e_k)$. Clearly, all orders of $O(e_k)$ are not accurate. Following Ref.~\cite{Fleming2011}, it can be shown that the diagonal elements of $\hat{\rho}(t)$ in the eigenbasis of the system Hamiltonian $\hat{H}_{\tt S}$ are correct to $O(1)$ and error occurs at $O(e_k)$, whereas the off-diagonal elements are correct to $O(e_k)$ and the error occurs at $O(e_k^{3/2})$. Thus, by reducing $e_k$, i.e., by increasing the number of lead modes, it is possible to make results from the quantum master equation arbitrarily close to those obtained from the infinite-bath configuration.  

\newpage

\section[Superfermion formalism]{Superfermion formalism for steady states in configurations kept out of equilibrium}
\label{ap:superf}

In this appendix we give further details the superfermion \cite{Dzhioev2011} steady state solution of the master equation in Eq.~\eqref{eq:Lindblad} for a non-interacting system of size $D$ coupled a single mesoscopic lead of size $L$.

This open system has a quadratic generator $\hat{L} = \hat{\mathbf{f}}^{\dagger}\, \mathbf{L}\, \hat{\mathbf{f}} - \eta$ defined by the $2M \times 2M$ non-Hermitian matrix $\bf L$ where $M=D+L$. To compute its NESS we proceed to diagonalise this matrix as ${\bf L} = {\bf V} \, \bm{\varepsilon} \, {\bf V}^{-1}$ to give a diagonal matrix $\bm{\varepsilon}$ of complex eigenvalues $\varepsilon_\mu$. These eigenvalues come in conjugate pairs and we shall denote the half with ${\rm Im}\{\epsilon_\mu\}>0$ as set $\Xi^+$ and the other half with ${\rm Im}\{\epsilon_\mu\}<0$ as $\Xi^-$. 

We identify the corresponding normal mode operators as $\hat{\bm{\xi}}^{\dagger} = \hat{\mathbf{f}}^{\dagger}\mathbf{V}$ and $\hat{\bm{\chi}} = \mathbf{V}^{-1}\hat{\mathbf{f}}$.
Although $\hat{\chi}_{\mu}$ and $\hat{\xi}_{\mu}$ mix physical $\hat{d}_k$ and ancillary modes $\hat{s}_k$ via a similarity transformation, and so are not Hermitian conjugates of one another, they still obey canonical anticonmmutation relations \cite{Dorda2014}, e.g.
\begin{align}
\{ \hat{\chi}_{\mu}, \hat{\xi}^{\dagger}_{\nu} \} = \delta_{\mu \nu} \mathds{1}.  \label{eq:canonical_normal_modes}
\end{align}
The equations of motion for the normal mode operators follow from the commutator with $\hat{L}$ giving 
\begin{align}
[\hat{L}, \hat{\chi}_{\mu}] &= -\epsilon_{\mu} \hat{\chi}_{\mu}, \quad {\rm and} \quad [\hat{L}, \hat{\xi}^{\dagger}_{\mu}] = \epsilon_{\mu} \hat{\xi}^{\dagger}_{\mu},
\end{align}
so in vector form the time-evolved mode operators are
\begin{align}
\hat{\bm{\xi}}^{\dagger} (t) = \hat{\bm{\xi}}^{\dagger} e^{i \bm{\epsilon} t} \quad {\rm and} \quad
\hat{\bm{\chi}} (t) = e^{-\ii \bm{\epsilon} t} \hat{\bm{\chi}}.
\end{align} 

A defining property of the NESS is $\hat{L}\ket{\rho(\infty)} = 0$. Using this we compute the time-evolution of the NESS when acted upon by a normal mode operator to obtain
\begin{align}
e^{-\ii \hat{L} t}\hat{\xi}^{\dagger}_\mu\ket{\rho(\infty)} = e^{-\ii \epsilon_\mu t}\hat{\xi}^\dagger_\mu\ket{\rho(\infty)},
\end{align} 
and also
\begin{align}
e^{-\ii \hat{L} t}\hat{\chi}_\nu\ket{\rho(\infty)} = e^{\ii \epsilon_\nu t}\hat{\chi}_\nu\ket{\rho(\infty)}.
\end{align} 
For these time-evolved states not to diverge in time we require that $\hat{\xi}^\dagger_\mu\ket{\rho(\infty)} = 0$ when $\mu \in \Xi^+$ and $\hat{\chi}_\nu\ket{\rho(\infty)} = 0$ when $\nu \in \Xi^-$. This pair of constraints is analogous to those of a Fermi sea state $\ket{\rm FS}$ where $\hat{c}^\dagger_j\ket{\rm FS} = 0$ when mode $j$ is occupied, and $\hat{c}_j\ket{\rm FS} = 0$ when it is empty. Similarly for the left vacuum state $\ket{I}$ we get
\begin{align}
\bra{I} \hat{\xi}^{\dagger}_\mu e^{-\ii \hat{L} t} = e^{\ii \varepsilon_\mu t}\bra{I}\hat{\xi}^\dagger_\mu \quad {\rm and} ~\, \bra{I} \hat{\chi}_\nu e^{-\ii \hat{L} t} = e^{-\ii \varepsilon_\mu t}\bra{I}\hat{\chi}_\nu, \nonumber
\end{align} 
implying the complementary constraints $\bra{I}\hat{\xi}^\dagger_\mu = 0$ when $\mu \in \Xi^-$ and $\bra{I}\hat{\chi}_\nu = 0$ when $\nu \in \Xi^+$. Together these relations fully define the $2M \times 2M$ matrix $\mathbf{D}$ of normal mode two-point correlations of the NESS with elements
\begin{align}
D_{\mu\nu} = \bra{I} \hat{\xi}^{\dagger}_\mu\hat{\chi}_\nu\ket{\rho(\infty)}.
\end{align} 
We immediately see that $D_{\mu\nu} = 0$ whenever $\mu \in \Xi^-$ and/or $\nu \in \Xi^-$. The case $\mu,\nu \in \Xi^+$ is then determined using Eq.~\eqref{eq:canonical_normal_modes} to find that $D_{\mu\nu} = \delta_{\mu\nu}$. Hence in general we have 
\begin{eqnarray}
D_{\mu\nu} &=& \delta_{\mu\nu}\Theta({\rm Im}\{\epsilon_\mu\}>0).
\end{eqnarray} 
indicating that the set $\Xi^+$ of normal modes are the unit filled Fermi sea of the NESS.

Using this result we can evaluate physical quantities such as the single-particle Green function $G_{ij}(t,t^{\prime}) = \langle \hat{c}^{\dagger}_i(t) \hat{c}_j(t^{\prime}) \rangle = \braket{I | \hat{c}^{\dagger}_i(t) \hat{c}_j(t^{\prime}) | \rho(\infty)}$ for the system $\tt S$. Transforming back from the normal modes we have 
\begin{align}
\hat{\mathbf{f}}^{\dagger}(t) = \hat{\bm{\xi}}^{\dagger} e^{i \bm{\epsilon} t} \mathbf{V}^{-1}, \quad {\rm and} \quad
\hat{\mathbf{f}}(t) = \mathbf{V} e^{-i \bm{\epsilon} t} \hat{\bm{\chi}},
\end{align}
and thus the Green function follows as
\begingroup
\allowdisplaybreaks
\begin{align}
G_{ij}(t, t^{\prime}) &= \braket{I | \left[\hat{\mathbf{f}}^{\dagger}(t)\right]_i \left[\hat{\mathbf{f}}(t^{\prime})\right]_j | \rho(\infty) } \nonumber \\
&= \sum_{\mu , \nu} \left[ e^{i \bm{\epsilon} t} \mathbf{V}^{-1} \right]_{\mu i} \left[ \mathbf{V} e^{-i \bm{\epsilon} t^{\prime}} \right]_{j\nu} \braket{I | \hat{\xi}^{\dagger}_{\mu} \hat{\chi}_{\nu} | \rho(\infty)},  \nonumber \\
&= \sum_{\mu , \nu} \left[ \mathbf{V} e^{-i \bm{\epsilon} t^{\prime}} \right]_{j\nu} D_{\mu\nu} \left[ e^{i \bm{\epsilon} t} \mathbf{V}^{-1} \right]_{\mu i},  \nonumber \\
&= \left[ \mathbf{V} e^{-i \bm{\epsilon} t^{\prime}} \, \mathbf{D} \, e^{i \bm{\epsilon} t} \mathbf{V}^{-1} \right]_{j i},
\end{align}
\endgroup
where we have used that $\mathbf{D}$ is diagonal and the indices $i,j = (L+1),\dots,M$ give the physical system $\tt S$ modes. This reduces to the NESS expectation value in Eq.~\eqref{eq:RDM_SF_quadratic} once $t=t'=0$. The Fermi sea structure of the NESS allows Wick's theorem to be applied to breakup expectation values for high-order correlations into two-point ones, for example
\begin{align}
\bra{I} \hat{\xi}^{\dagger}_\mu\hat{\chi}_\nu \hat{\xi}^{\dagger}_\tau\hat{\chi}_\sigma \ket{\rho(\infty)} = \bra{I} \hat{\xi}^{\dagger}_\mu\hat{\chi}_\nu \ket{\rho(\infty)}\bra{I} \hat{\xi}^{\dagger}_\tau\hat{\chi}_\sigma \ket{\rho(\infty)}\nonumber \\
+ \bra{I} \hat{\xi}^{\dagger}_\mu\hat{\chi}_\sigma \ket{\rho(\infty)}\bra{I} \hat{\chi}_\nu\hat{\xi}^{\dagger}_\tau \ket{\rho(\infty)}\nonumber \\
- \bra{I} \hat{\xi}^{\dagger}_\mu\hat{\xi}^{\dagger}_\tau \ket{\rho(\infty)}\bra{I} \hat{\chi}_\nu\hat{\chi}_\sigma \nonumber \ket{\rho(\infty)},
\end{align}
leaving products of terms that can be readily evaluated using the NESS normal mode constraints determined above.

\newpage

\section{Many fermionic sites}
\label{app:noninteracting}

Another configuration of interest is a system composed of many fermionic sites, one for which we can express the Hamiltonian as
\begin{align}
\label{eq:h_s_m_app}
\hat{H}_{\tt S} = \sum_{j = 1}^{D} \epsilon_j \hat{c}^{\dagger}_j \hat{c}_j - \sum_{j = 1}^{D - 1} t_{\tt S} \left( \hat{c}^{\dagger}_{j+1} \hat{c}_{j} + \textrm{H.c.} \right),
\end{align}
where $\hat{c}^{\dagger}_j$ and $\hat{c}_j$ are fermionic creation and destruction operators and $D$ is the number of sites in the system. We couple the leftmost and rightmost sites of this system to mesoscopic reservoirs, as shown in Fig.~\ref{fig:2.3.6}. 

Given that our expressions for particle and energy currents in Eqs.~\eqref{eq:partsf_explicit} and \eqref{eq:enersf_explicit} are defined in terms of canonical operators in the leads, the corresponding expressions for the case of a many-fermionic central system are equivalent to those of a single-level system. 

For a sufficiently large amount of sites in the central system, these operators can be defined in terms of just system operators. Here, however, we will use the expressions in Eqs.~\eqref{eq:partsf_explicit} and \eqref{eq:enersf_explicit} which are general for any number of sites $D$.

\begin{figure}[t]
\fontsize{13}{10}\selectfont 
\centering
\includegraphics[width=1\columnwidth]{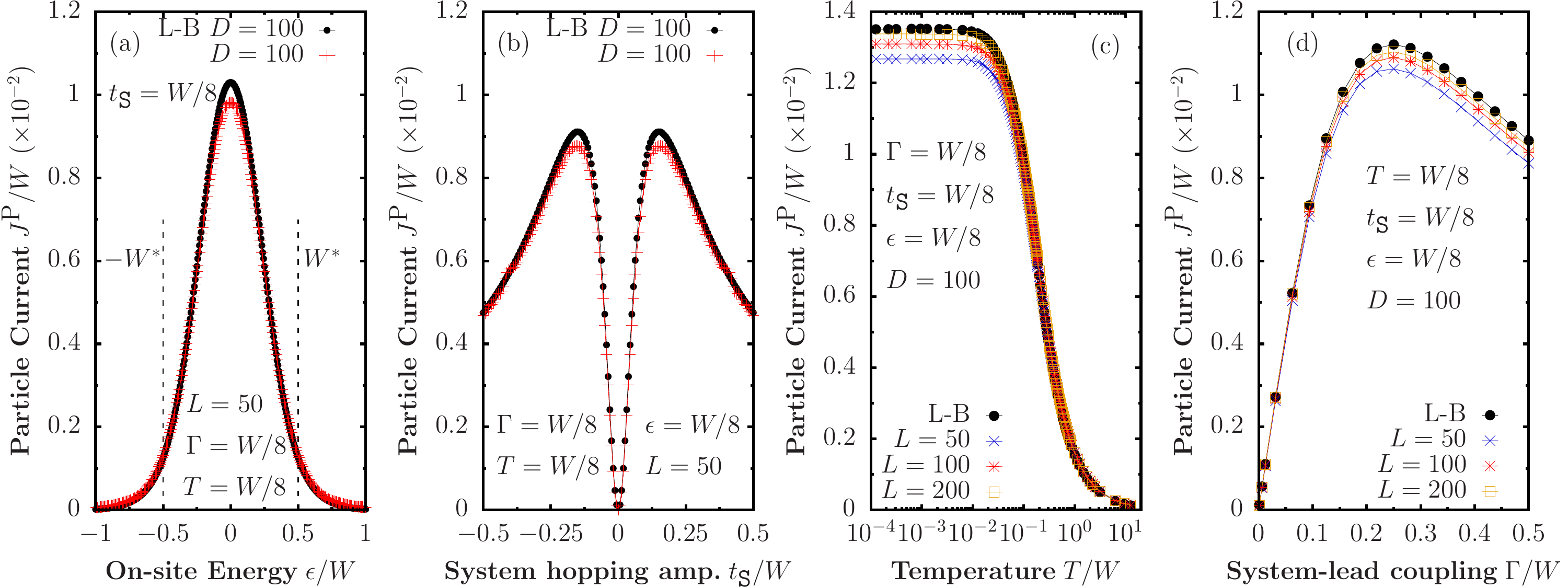}
\caption[Particle current from L-B and mesoscopic reservoir predictions flowing from the left lead and into a system composed of many fermionic sies]{Particle current from L-B and mesoscopic reservoir predictions flowing from the left lead and into the system (a) as a function of the on-site energy (same parameter for every site) for a central system with $D = 100$ sites and a fixed number of modes in the leads $L = 50$, and (b) as a function of the hopping amplitude $t_{\tt S}$ (same parameter for every site). In panels (c) and (d) we fix every parameter and study the particle current as a function of temperature and system-lead coupling, respectively. In these calculations we used $\mu_{\texttt L} = -\mu_{\texttt R} = W / 16$, $T_{\texttt L} = T_{\texttt R}$, $L_{\textrm{log}} / L = 0.2$ and $W^* = W / 2$.}
\label{fig:app1}
\end{figure}

\begin{figure}[t]
\fontsize{13}{10}\selectfont 
\centering
\includegraphics[width=1\columnwidth]{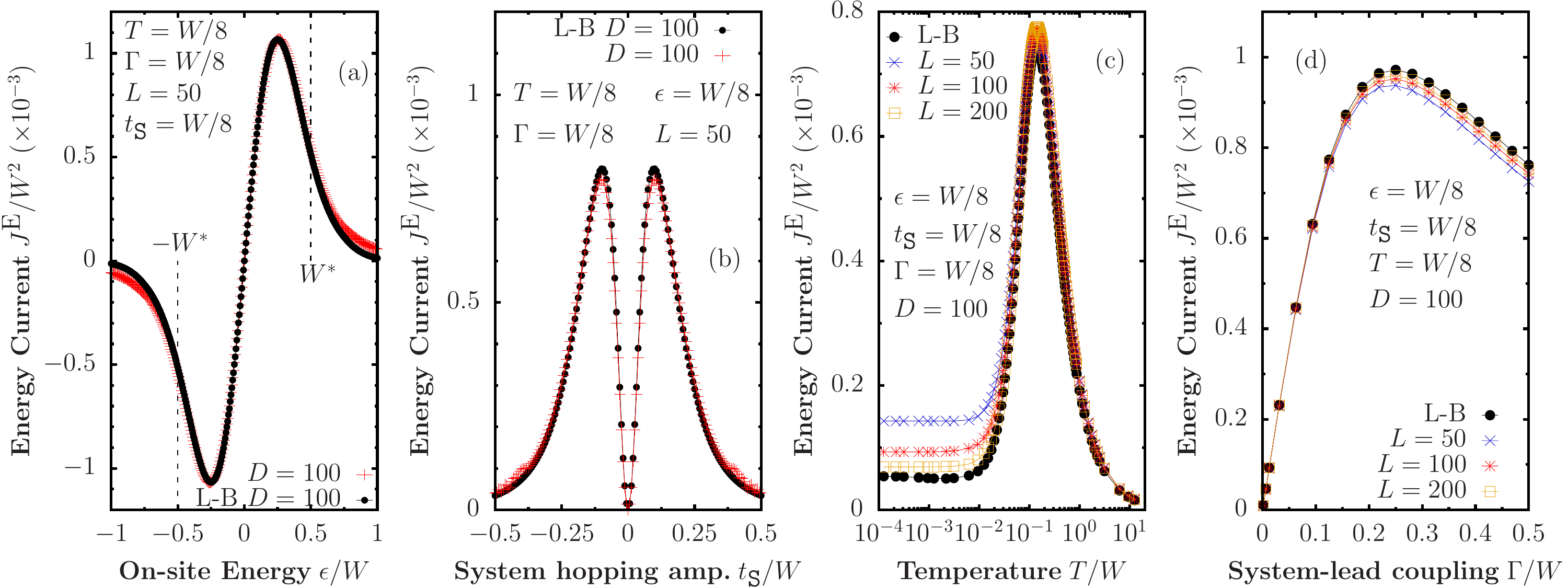}
\caption[Energy current from L-B and mesoscopic reservoir predictions flowing from the left lead and into a system composed of many fermionic sies]{Energy current from L-B and mesoscopic reservoir predictions flowing from the left lead and into the system (a) as a function of the on-site energy (same parameter for every site) for a central system with $D = 100$ sites and a fixed number of modes in the leads $L=50$, and (b) as a function of the hopping amplitude $t_{\tt S}$ (same parameter for every site). In panels (c) and (d) we fix every parameter and study the energy current as a function of temperature and system-lead coupling, respectively. In these calculations we used $\mu_{\texttt L} = -\mu_{\texttt R} = W / 16$, $T_{\texttt L} = T_{\texttt R}$, $L_{\textrm{log}} / L = 0.2$ and $W^* = W / 2$.}
\label{fig:app2}
\end{figure}

We now evaluate whether the mesoscopic lead configuration can provide a good approximation of the continuum even if the central system is composed of many fermionic sites. In a similar fashion as for the single-level system, in Fig.~\ref{fig:app1}(a) we present the particle current flowing from the left lead and into system as a function of the on-site energy $\epsilon = \epsilon_j$ for every site $j$. In our calculations we use the same macroscopic parameters as before, given by $T_{\tt L} = T_{\tt R} = W / 8$ and $\mu_{\tt L} = -\mu_{\tt R} = W/16$. We fix the number of energy modes in each lead to $L = 50$ and the number of sites in the central system to $D = 100$. The Landauer-B\"uttiker calculations are done by evaluating Eq.~\eqref{eq:partlb} using the transmission function obtained as described in Sec~\ref{sec:transmission}. It can be observed that for a fixed number of modes in the leads $L$ and a fixed number of sites in the central system $D$ the approximation to the continuum limit using mesoscopic reservoirs is robust to a wide range of on-site energies. The small oscillations that can be observed near the band edges at $|\epsilon| \gtrapprox |W^*|$ are due to the logarithmic spacing of modes. Furthermore, from Fig.~\ref{fig:app1}(b), the same can be said when $\epsilon$ is fixed and $t_{\tt S}$ is changed to different values. Given that the energies in the central system are bounded by $-2t_{\tt S}$ and $2t_{\tt S}$, the oscillations due to logarithmic discretisation are observed close to $t_{\tt S} \approx W / 2$. The same observations hold for energy current in Figs.~\ref{fig:app2}(a) and \ref{fig:app2}(b)

As a function of temperature, a similar behaviour as for the single-level system can be observed. In particular, for particle current and energy current in Figs.~\ref{fig:app1}(c) and \ref{fig:app2}(c), respectively, the continuum is properly approximated with the exception of the values of temperature that are lower than the minimum energy spacing of the modes in the leads. For these small temperatures, the Fermi-Dirac distributions of the leads resemble a Heaviside step function and the discontinuity can no longer be well-captured by discrete and broadened energy modes. Following from our previous discussion for the single-level system, to obtain a better approximation at lower temperatures one can either increase the number of total energy modes or decrease the width of the window $[-W^*, W^*]$. The former choice comes with the cost of a larger computational complexity, while with the latter one can then only provide a good approximation of the continuum for a smaller range in the parameter space of $\epsilon$, $t_{\tt S}$, $\mu_{\tt L}$ and $\mu_{\tt R}$. If these values are fixed, a good choice of $[-W^*, W^*]$ can be used to obtain better approximations at lower temperatures with its limit, as discussed for the single-level system, related to the minimum value of $e_k$ in the linearly-discretised region. 

As a function of the system-lead coupling, the results are very robust to a wide range of values as observed from Figs.~\ref{fig:app1}(d) and \ref{fig:app2}(d). Because of the ballistic (coherent) nature of transport in the central system, currents become independent of $D$ in the asymptotic regime.

\newpage

\section{CP symmetry}
\label{app:CP}

Here we prove that the energy current vanishes in the Heisenberg model described by Eq.~\eqref{eq:h_s_i} under conditions of combined charge conjugation-parity (CP) symmetry. The symmetry corresponds to a unitary transformation $\hat{\mathcal{C}}\hat{\mathcal{P}}$, with the particle-hole transformation $\hat{\mathcal{C}}$ and the parity transformation $\hat{\mathcal{P}}$.

In the bulk of the system, the parity and particle-hole transformations are respectively defined by
\begin{align}
    \label{P_bulk}
    \hat{\mathcal{P}}\hat{c}_j\hat{\mathcal{P}}^\dagger & = \hat{c}_{D-j+1}.\\
 \label{C_bulk}
    \hat{\mathcal{C}}\hat{c}_j\hat{\mathcal{C}}^\dagger & = (-1)^{j+1}\hat{c}_j^\dagger.
\end{align}
The phase factor in $\hat{\mathcal{C}}$ is defined so that particle excitations are mapped to hole excitations with the same kinetic energy. The bulk Hamiltonian in Eq.~\eqref{eq:h_s_i} is invariant under $\hat{\mathcal{P}}$, i.e.~$\hat{\mathcal{P}}\hat{H}_{\tt S}\hat{\mathcal{P}} = \hat{H}_{\tt S}$, and also invariant under $\hat{\mathcal{C}}$ so long as $\epsilon = -U$. 

The particle-hole transformation for the lead operators that is consistent with the action of $\hat{\mathcal{C}}$ in the bulk is of the form
\begin{align}
    \label{C_bath_L}
    \hat{\mathcal{C}}\hat{a}_{k,\tt L}\hat{\mathcal{C}}^\dagger & = -\hat{a}^\dagger_{L-k,\tt L},\\
    \label{C_bath_R}
    \hat{\mathcal{C}}\hat{a}_{k,\tt R}\hat{\mathcal{C}}^\dagger & = (-1)^D\hat{a}^\dagger_{L-k,\tt R},
\end{align}
while spatial reflection simply consists of the swap ${{\tt L}\leftrightarrow {\tt R}}$. With these conventions, the total Hamiltonian is invariant under $\hat{\mathcal{P}}$ if the left and right leads have identical spectra $\varepsilon_k$ and system-bath couplings $\kappa_{kp}$. The Hamiltonian is also invariant under $\hat{\mathcal{C}}$ if the lead spectra and couplings are symmetric around the centre of the band, i.e. $\varepsilon_k = -\varepsilon_{L-k}$ and $\kappa_{k,p} = \kappa_{L-k,p}$. Finally, the non-equilibrium forcing is CP-symmetric if the bath temperatures are equal, $T_{\tt L} = T_{\tt R}$, and the chemical potentials are opposite, $\mu_{\tt L}=-\mu_{\tt R}$, while the dissipation rates are invariant under spatial reflection and inversion about the centre of the band, i.e.~$\gamma_{k,\tt L} = \gamma_{L-k,\tt L} = \gamma_{k,\tt R}$. 

Under the above assumptions, the generator of the master equation is invariant under a combined CP transformation and therefore so is the steady state, i.e. $\hat{\mathcal{C}}\hat{\mathcal{P}}\hat{\rho}(\infty)(\hat{\mathcal{C}}\hat{\mathcal{P}})^\dagger = \hat{\rho}(\infty)$. At the particle-hole symmetric point of the Hamiltonian, with $\epsilon = -U$, the bulk energy current operator (defined in Sec.~\ref{sec:thermo_meso}) is odd under a CP transformation, in the sense that $\hat{\mathcal{C}}\hat{\mathcal{P}} \hat{J}^{\rm E}_{j-1\to j+1}(\hat{\mathcal{C}}\hat{\mathcal{P}})^\dagger = -\hat{J}^{\rm E}_{D-j\to D-j+2}$. It follows that
\begin{align}
\label{JE_is_zero}
\langle \hat{J}^{\rm E}_{j-1,j+1}\rangle = \langle  \hat{\mathcal{C}}\hat{\mathcal{P}}\hat{J}^{\rm E}_{j-1,j+1}(\hat{\mathcal{C}}\hat{\mathcal{P}})^\dagger\rangle = -\langle \hat{J}^{\rm E}_{D-j,D-j+2}\rangle,
\end{align}
and therefore $\langle \hat{J}^{\rm E}_{j-1,j+1}\rangle = -\langle \hat{J}^{\rm E}_{j-1,j+1}\rangle = 0$ because the mean current is homogeneous in the steady state. Note that the particle current operator is even and therefore is not constrained by CP symmetry. However, the particle density transforms as $\hat{\mathcal{C}}\hat{\mathcal{P}} \hat{n}_j(\hat{\mathcal{C}}\hat{\mathcal{P}})^\dagger = 1-\hat{n}_{D-j+1}$, so that in a CP-symmetric steady state we have $\langle \hat{n}_j\rangle  + \langle \hat{n}_{D-j+1}\rangle = 1$. In a ballistic regime with $\langle \hat{n}_j\rangle  = \rm const.$, we must therefore have $\langle \hat{n}_j\rangle = 0.5$, consistent with the trend in Fig.~\ref{fig:2.3.18} at high temperature. 
\end{document}